%% file: hs-lintimespace.tex
\documentclass[twocolumn,times,5p,sort&compress]{elsarticle}
\pdfoutput=1
\usepackage[utf8]{inputenc}
\usepackage{microtype}
\usepackage[colorlinks]{hyperref}
\usepackage{amsmath,amssymb,amsthm}
\usepackage{hyphenat}
\usepackage{doi}
\usepackage{todonotes}
\usepackage[vlined,linesnumbered,ruled]{algorithm2e}
\DontPrintSemicolon
\usepackage[capitalize,nameinlink]{cleveref}
\usepackage{caption}
\usepackage{subcaption}
\usepackage{booktabs}
\usepackage{paralist}
\usepackage{placeins}
\newcommand{\hs}[1]{$#1$\hyp Hitting Set}

\newcommand{\ecode}[1]{\llcorner #1 \lrcorner}
\newcommand{\eid}[1]{\ensuremath{#1_{\text{id}}}}
\newcommand{\sid}[1]{\ensuremath{#1_{\text{id}}}}
\newcommand{\vid}[1]{\ensuremath{#1_{\text{id}}}}
\newcommand{\arrA}{A}

\newcommand{\arrC}{B}
\newcommand{\arrD}{C}

\newcommand{\cvd}{Cluster Vertex Deletion}
\newcommand{\clusterediting}{Cluster Editing}

\newcommand{\stationcover}{Station Cover}
\newcommand{\bs}{\square}
\newcommand{\fk}{\ref{alg:fk}}
\newcommand{\bev}{\ref{alg:bev}}
\newcommand{\fkp}{\ref{alg:fkp}}
\newcommand{\bevp}{\ref{alg:bevp}}
\newcommand{\wei}{\ref{alg:wei}}
\newcommand{\Ein}{\ensuremath{E_{\text{in}}}}
\newcommand{\Eout}{\ensuremath{E_{\text{out}}}}
\newcommand{\Vin}{\ensuremath{V_{\text{in}}}}
\newcommand{\Vout}{\ensuremath{V_{\text{out}}}}
\newcommand{\Hin}{\ensuremath{H_{\text{in}}}}
\newcommand{\Hout}{\ensuremath{H_{\text{out}}}}
\newcommand{\kin}{\ensuremath{k_{\text{in}}}}
\newcommand{\kout}{\ensuremath{k_{\text{out}}}}

\DeclareMathOperator\bigTh{\Theta}

\DeclareMathOperator\bigO{O}

\DeclareMathOperator\petals{petals}
\DeclareMathOperator\used{used}
\DeclareMathOperator\supersets{supersets}
\DeclareMathOperator\true{true}
\DeclareMathOperator\false{false}
\newcommand{\PH}{the polynomial\hyp time hierarchy}

\newtheorem{theorem}{Theorem}
\newtheorem{problem}[theorem]{Problem}

\newtheorem{definition}[theorem]{Definition}
\newtheoremstyle{iostuff}%
{0pt}%
{0pt}%
{\hangindent=\parindent}%
{}%
{\itshape}%
{:}%
{.5em}%
{}%

\theoremstyle{iostuff}
\newtheorem*{probinstance}{Input}
\newtheorem*{probtask}{Question}

\begin{document}
\begin{frontmatter}
  \title{Optimal-size problem kernels for $d$-Hitting Set in linear time and space\tnoteref{t1}}
  \tnotetext[t1]{This work is based on the Bachelor
    thesis of the second author \cite{Smi18}.}
  \author{René van Bevern\corref{cor1}}
  \ead{rvb@nsu.ru}
  \author{Pavel V.\ Smirnov}
  \ead{p.smirnov@g.nsu.ru}
  \cortext[cor1]{Corresponding author}
  \address{Department of Mechanics and Mathematics,
    Novosibirsk State University, ul.\ Pirogova 1, 630090 Novosibirsk, Russian Federation}
  \begin{abstract}
    The known linear\hyp time kernelizations
    for \hs d
    guarantee linear worst\hyp case running times
    using a quadratic\hyp size data structure
    (that is not fully initialized).
    Getting rid of this data structure,
    we show that
    problem kernels of asymptotically optimal size $\bigO(k^d)$
    for \hs d
    are computable in
    linear time and space.
    Additionally,
    we experimentally compare
    the linear\hyp time kernelizations for \hs d
    to each other and to
    a classical data reduction algorithm
    due to Weihe.
  \end{abstract}

\end{frontmatter}

\section{Introduction}
\thispagestyle{empty}

\noindent
We study data reduction algorithms
for the following combinatorial optimization problem:

\begin{problem}[\hs d, for constant~$d\in\mathbb N$]
  \begin{probinstance}
    A hypergraph~$H=(V,E)$
    with vertex set~$V=\{1,\allowbreak\dots,\allowbreak n\}$,
    edge set~$E\subseteq\{e\subseteq V: |e|\leq d\}$,
    and $k\in\mathbb N$.
  \end{probinstance}
  \begin{probtask}
    Is there a \emph{hitting set}~$S\subseteq V$
    of cardinality at most~$k$,
    that is,
    $\forall e\in E:e\cap S\ne\emptyset$?
  \end{probtask}

  \smallskip
  \noindent
  Throughout this work,
  we denote $n:=|V|$ and $m:=|E|$
  and call $|H|=|V|+|E|$ 
  the \emph{size} of the hypergraph~$H$.
\end{problem}
\noindent
The \hs d problem is an
NP\hyp complete \citep{Kar72}
fundamental
combinatorial optimization problem,
arising in bioinformatics \citep{MK13},
medicine \citep{Vaz09,MPMM10},
clustering \citep{HKMN10,BMN12},
automatic reasoning  \citep{Rei87,FBB18,BTU19},
feature selection \citep{FBNS16,BFL+19},
radio frequency allocation \citep{SMNW14},
software engineering \citep{OC03},
and public transport optimization \citep{Wei98,BFFS19}.

Exact algorithms for NP\hyp complete problems
usually take time exponential in the input size.
Thus,
an important preprocessing step is data reduction,
which has proven to significantly
shrink real\hyp world instances of NP\hyp hard problems
\citep{Wei98,ABN06,MPMM10,BFFS19,BFTxx}.
The main notion
of data reduction
with performance guarantees
is \emph{kernelization} \citep{FLSZ19},
here stated for \hs d:
\begin{definition}
  A \emph{kernelization}
  maps any
  \hs d instance $(\Hin,\kin)$
  to an instance $(\Hout,\kout)$
  in polynomial time such that
\begin{enumerate}
\item $\Hin$ has a hitting set of size~$\kin$
  if and only if $\Hout$ has a hitting set of size~$\kout$,
  
\item $|\Hout|+\kout\leq g(\kin)$
  for a computable function~$g\colon\mathbb N\to\mathbb N$.
\end{enumerate}
One calls $(\Hout,\kout)$ the \emph{problem kernel}
and $g(\kin)$~its \emph{size}.
In the kernelizations studied in our work,
$k:=\kin=\kout$.
\end{definition}

\noindent
\looseness=-1
The existence of problem kernels
of size $\bigO(k^{d-\varepsilon})$
for any $\varepsilon>0$
for \hs d
results in a collapse
of the polynomial\hyp time hierarchy \citep{DM14}.
There are two known $\bigO(n+m)$-time kernelizations
for \hs d
that
yield problem kernels of this optimal size $\bigO(k^d)$ \citep{Bev14b,FK15}.
They can be implemented
\begin{compactenum}[(a)]
\item with expected linear running time
  and linear space using hash tables
  with random hash functions, or
\item with worst\hyp case linear running time
  using a trie data structure,
  which may allocate $\Theta(nm)$~cells of random access memory
  (of which only $O(n+m)$~are initialized) \citep{Bev14c}.
\end{compactenum}
In experiments,
the memory usage of implementation~(b)
proved to be prohibitively large,
yet
implementation~(b)
outperformed implementation~(a)
when enough memory was available
and allocation of zero\hyp initialized
memory was cheap \citep{Bev14c}.

From a theoretical point of view,
a natural question is whether
one can derandomize implementation~(a),
so that it runs in worst\hyp case linear time and space,
or, equivalently,
whether one can lower the memory usage of implementation~(b)
to linear.
From a practical point of view,
it is interesting
how the two kernelizations compare to each other
and to other data reduction algorithms.

\paragraph{Our contributions and organization of this work}
\cref{sec:known}
shows the two known
linear\hyp time \hs d kernelizations.
We then describe our new contributions,
which are two\hyp fold:

In \cref{sec:space},
we resolve the apparent paradox
that the linear worst\hyp case running time
of the known \hs d kernelizations hinges on
quadratic\hyp space data structures.
To this end,
we show how to implement
both of them %
in $\bigO(n+m)$~time and space.
\looseness=-1
In \cref{sec:exp},
for the first time,
we experimentally compare the two kernelizations
to each other
and to
a well\hyp known data reduction algorithm
due to \citet{Wei98},
which runs in superlinear time,
does not yield problem kernels,
but proved to be very effective on
instances of the Station Cover problem.
We will see that
the kernelizations outperform
\citeauthor{Wei98}'s algorithm
for small $d$
and that combinations of kernelization and \citeauthor{Wei98}'s
algorithm may yield significantly stronger data
reduction effects than the individual algorithms.

\paragraph{Related work}
\noindent
There are several
kernelizations for \hs d
\citep{NR03,Dam06,FG06,Abu10,Mos10,Kra12,FK14,BT20,BHRT19}.
\citet{DM14} showed that the existence of a
problem kernel with $\bigO(k^{d-\varepsilon})$~edges for
any~$\varepsilon>0$ for \hs d implies a collapse of \PH{}.
Therefore,
we do not expect polynomial\hyp size
problem kernels for \hs d
if $d$~is \emph{not}
constant.
There are problem kernels
with $\bigO(k^{d-1})$~\emph{vertices},
however \citep{Abu10,Mos10,Bev14c}.

\looseness=-1
Lowering the running time
and space requirements of \hs d kernelizations
both
have been in the focus of research.
The first linear\hyp time kernelization
is due to
\citet{Bev14b,Bev14c}.
The second, due to \citet{FK15},
is simpler and has smaller constant factors:
the problem kernel of
\citet{Bev14b}
has at most
$d!\cdot d^{d+1}\cdot (k+1)^d$~edges,
whereas the problem kernel of \citet{FK15}
has at most $(k+1)^d$~edges.
Both kernelizations work in $\bigO(d\cdot n+2^dd\cdot m)$~time.

Problem kernels of 
size $\bigO(k^d\log k)$ are computable
in $\bigO(k^d\log n)$~space
and $\bigO(k^d m)$~time \citep{FK14},
of size $\bigO(k^d)$ in
logarithmic space and
$\bigO(m^{d+2})$~time \citep{FK15},
and of exponential size
even in constant parallel time \citep{BT20}.

\section{Known linear-time algorithms}
\label{sec:known}
\noindent
\looseness=-1
There are two known
linear\hyp time kernelizations
for \hs d: \ref{alg:fk} and \ref{alg:bev}.
Both 
iterate over each input edge~$e\in\Ein$ once
and
decide whether to add~$e$
to the output edge set~$\Eout$ as follows.

\fk{} does not
add $e$ to~$\Eout$ if $e$~contains
a subset~$s$
that is
contained in $(k+1)^{d-|s|}$ edges of~$\Eout$.
It can be shown that
any hitting set of size~$k$ for $(\Vout,\Eout)$
has to intersect~$s$, and thus~$e$ \citep{FK15}.
Thus,
for each $s\subseteq e\in\Ein$,
\fk{} maintains
the number of supersets of~$s$ in~$\Eout$
in a counter~$\supersets[s]$,
which is updated for each~$s\subseteq e$
whenever adding
an edge~$e$ to~$\Eout$.

\begin{algorithm}[t]\label{FK}
  \small
  \SetAlgoRefName{FK}
  \KwIn{Hypergraph $(\Vin, \Ein)$, $k\in\mathbb N$.}
  \KwOut{Problem kernel $((\Vout,\Eout),k$) with $|\Eout|\leq(k+1)^d$.}
  \tcp{\textrm{Initially, $\forall s\subseteq V:\supersets[s]=0$.}}
  $\Eout \gets \emptyset$\;
  \ForEach{$e \in \Ein$} {
    \If{$\forall s\subseteq e:\supersets[s] < (k+1)^{d-|s|}$} {
      $\Eout \gets \Eout \cup \{e\}$\;
      \lForEach{$s \subseteq e$}{
        $\supersets[s] \gets \supersets[s]+1$
      }
    }
  }
  $\Vout\gets \bigcup_{e\in \Eout}e$\;
  \Return{$((\Vout, \Eout),k)$}\;

  \caption{Algorithm of \citet{FK15}.}
  \label{alg:fk}
\end{algorithm}

\looseness=-1
\bev{} is based on finding \emph{sunflowers}---sets of edges
(called \emph{petals})
with mutually equal intersection (called the \emph{core}).
If~there is a sunflower with $k+1$~petals,
then any hitting set of size~$k$ has to intersect its core.
Thus,
\bev{} does not
add $e$ to~$\Eout$ if it finds
that $e$~contains the core~$s$ of a sunflower
with $k+1$ petals in~$\Eout$.
To this end,
for each $s\subseteq e\in\Ein$,
\bev{}
maintains the information
of \emph{one} sunflower with core~$s$:
$\petals[s]$ is its number of petals
and
$\used[s][v]$ is true if and only if~$v\in V$
is contained in one of them.
Whenever adding an edge~$e$ to~$\Eout$,
\bev{} adds~$e$
to the sunflower with core~$s$
for each~$s\subseteq e$
(if possible).
Depending on the order of the input edges,
\bev{} may not find a sunflower with $k+1$ petals
if it exists,
yet if there is a sunflower with $d(k+1)$~petals,
it finds one with $k+1$~petals for sure \citep{Bev14c}.

\looseness=-1
Access to $\supersets[s]$,
$\used[s]$, and $\petals[s]$
for each $s\subseteq e\in\Ein$
in \fk{} and \bev{}
can be organized in $\bigO(d)$~time
using a trie
that can be initialized
in $\bigO(dn+2^dd\cdot m)$~time \citep[Lemma~5.3]{Bev14c}.
After initialization,
\fk{} and \bev{}
work in $\bigO(2^d\cdot m)$
and $\bigO(2^dd\cdot m)$~time,
respectively.
The culprit is that
the trie can allocate
$\bigTh(nm)$~random access memory,
of which only $\bigO(n+m)$~is initialized
\citep[Lemma~5.3]{Bev14c}.

\begin{algorithm}[t]\label{Bev}
  \small
  \SetAlgoRefName{Bev}
  \KwIn{Hypergraph $(\Vin, \Ein)$, $k\in\mathbb N$.}
  \KwOut{Kernel $((\Vout,\Eout),k)$ with
    $|\Eout|\leq d! d^{d+1} (k+1)^d.$}
  \tcp{\textrm{Initially, $\forall s\subseteq V,v\in V:\petals[s]=0,\used[s][v]=\false$.}}
  $\Eout \gets \emptyset$\;
  \ForEach{$e \in \Ein$} {
    \If{$\forall s\subseteq e:\petals[s] \leq k$}{
      $\Eout \gets \Eout \cup \{e\}$\;
      \ForEach{$s \subseteq e$} {
        \If{$\forall v \in e \setminus s: \used[s][v] = \false$} {
          $\petals[s] \gets \petals[s] + 1$\;
          \lForEach{$v \in e \setminus s$} {
            $\used[s][v] \gets \true$
          }
        }
      }
    }
  }
  $\Vout\gets \bigcup_{e\in \Eout}e$\;
  \Return{$((\Vout, \Eout),k)$}\;
  \caption{Algorithm of \citet{Bev14b}.}
  \label{alg:bev}
\end{algorithm}

\section{Implementing \fk{} and \bev{} in linear space and time}
\label{sec:space}

\subsection{\fk{} in linear space}
\label{sec:fkspace}
\noindent
To implement
\fk{} in linear space,
we apply a series of preprocessing steps.
First,
by iterating over~$\Ein$ once,
simultaneously incrementing a counter,
we assign to each edge~$e\in \Ein$
a unique index $\eid{e}\in\{1,\dots,m\}$,
in $\bigO(m)$~time giving a set
\begin{equation}
\eid{E}:=\{(e,\eid{e}):e\in \Ein\}.\label{eq:eid}
\end{equation}
We will then show how to
compute in linear time and space
a unique index~$\sid{s}\in\{1,\dots,2^dm\}$
for each~$s\subseteq e\in \Ein$
and a size-$m$ array~$\arrA[]$ satisfying
\begin{equation}
\arrA[\eid{e}]=\{(s,\sid{s}):s\subseteq e\}\text{\quad for each~$(e,\eid{e})\in \eid{E}$}.\label{eq:aid}
\end{equation}
\noindent
Then,
instead of iterating over each~$e\in \Ein$,
each $s\subseteq e$,
and looking up $\supersets[s]$ in a trie,
it is enough to
iterate over each~$(e,\eid{e})\in \eid{E}$,
each~$(s,\sid{s})\in\arrA[\eid{e}]$,
and look up $\supersets'[\sid{s}]$
in $\bigO(1)$~time,
where $\supersets'[]$ is an ordinary
array of length~$2^dm$.
The modified algorithm is
shown in \fkp{}.

\begin{theorem}
  \label{thm:fk}
  \fk{}
  can be run in
  $\bigO(nd+2^dd\cdot m)$~time and space.
\end{theorem}

\begin{proof}
  \looseness=-1
  To prove the theorem,
  we show how to
  compute the array~$\arrA[]$
  and a unique index~$\sid{s}$
  for each~$s\subseteq e\in \Ein$
  in linear space and time.
  The tricky bit is
  that $s$~may be a subset of several edges in~$\Ein$,
  yet its index~$\sid{s}$ must be unique.
  Thus,
  we use the following canonical encoding of edges:
  for any subset~$s\subseteq V=\{1,\dots,n\}$
  of size at most~$d$,
  $\ecode{s}\in (V\cup\{\bs\})^d$ is
  a $d$-tuple containing
  the elements of~$s$ in increasing order
  and padded with~$\bs$ at the end.
  For example,
  for $d=4$,
  \(
  \ecode{\{3,1,2\}}=(1,2,3,\bs).
  \)
  Obviously,
  $\ecode{e}$
  for any edge~$e\in\Ein$
  is computable
  in $\bigO(d\log d)$~time.
  The preprocessing for \fk{} now
  consists of three steps.

  \emph{Step 1.} Compute a list
  \(L=[(\ecode{s},\eid{e}):s\subseteq e,(e,\eid{e})\in \eid{E}]\)
  of size at most~$2^dm$
  by first computing
  $\ecode{e}$ for all~$e\in\Ein$
  in $\bigO(m\cdot d\log d)$~time and space
  and then enumerating all substrings of~$\ecode{e}$
  for each~$(e,\eid{e})\in\eid{E}$
  in $\bigO(2^dd\cdot m)$~time and space.

  \emph{Step 2.} Sort $L$ by lexicographically
  non\hyp decreasing~$\ecode{s}$,
  where we assume~$n<\bs{}$.
  Since the $\ecode{s}$
  are $d$-tuples over~$\{1,\dots,n,\bs\}$,
  this works in
  $\bigO(d(n+|L|))=\bigO(nd+2^dd\cdot m)$~time and space
  using radix sort \citep[Section~8.3]{CLRS01}.
  All pairs~$(\ecode{s},\eid{e})$
  belonging to the same subset~$s$
  now occur consecutively in~$L$.
  
  \emph{Step 3.}
  Initialize a size-$m$ array~$\arrA[]$ of empty lists
  and $\sid{s}\gets 1$.
  Iterate over~$L$ as follows.
  For the current pair~$(\ecode{s},\eid{e})$,
  add $(s,\sid{s})$ to $\arrA[\eid{e}]$.
  If
  there is a next pair $(\ecode{s'},\eid{e'})$ on~$L$
  and $\ecode{s}\ne\ecode{s'}$,
  then increment
  $\sid{s}\gets\sid{s}+1$
  and continue.
  Note that $A[\eid{e}]$ does not contain
  duplicates,
  so that the list is actually a set,
  as required by \eqref{eq:aid}.

This concludes
the computation of the $\sid{s}$ and
the array~$\arrA[]$.
The running time and space bottleneck is step 2.
After this preprocessing,
\fk{} can be implemented
to run in
$\bigO(2^d\cdot m)$~time and space
as shown in \fkp{}.
\end{proof}

\begin{algorithm}[t]
  \small
  \SetAlgoRefName{FK'}
  \KwIn{Hypergraph $(\Vin, \Ein)$, $k\in\mathbb N$.}
  \KwOut{Problem kernel $((\Vout,\Eout),k$) with $|\Eout|\leq(k+1)^d$.}
  \tcp{\textrm{$\supersets'[]$ is a zero-initialized size-$2^dm$ array,\\
      $\eid{E}$ and $A[]$ are as in \eqref{eq:eid} and \eqref{eq:aid}.}}
  $\Eout \gets \emptyset$\;
  \ForEach{$(e,\eid{e}) \in \eid{E}$} {
    \If{$\forall (s,\sid{s})\in A[\eid{e}]:\supersets'[\sid{s}] < (k+1)^{d-|s|}$} {
      $\Eout \gets \Eout \cup \{e\}$\;
      \lForEach{$(s,\sid{s}) \in A[\eid{e}]$}{
        $\supersets'[\sid{s}] \gets \supersets[\sid{s}]+1$
      }
    }
  }
  $\Vout\gets \bigcup_{e\in \Eout}e$\;
  \Return{$((\Vout, \Eout),k)$}\;

  \caption{Linear-space version of \fk{}.}
  \label{alg:fkp}
\end{algorithm}

\subsection{\bev{} in linear space}
\label{sec:bevspace}
\noindent
To implement \bev{} in linear time and space,
we replace accesses to tries $\petals[s]$ and $\used[s]$
for each~$s\subseteq e\in\Ein$
by accesses to arrays
$\petals'[\sid{s}]$ and $\used'[\sid{s}]$,
as we did for \fk{}.
However,
while $\petals[s]$~is a counter
that translates
into a counter~$\petals'[\sid{s}]$,
$\used[s][]$ is a size-$n$ array indexed by vertices.
Holding such an array in~$\used'[\sid{s}][]$
would again use $\Omega(nm)$~space.
Instead,
we organize~$\used'[\sid{s}][]$ as follows.
Let
\begin{equation}
  V^s:=\bigcup_{e\in \Ein}(e\setminus s)\quad
\text{for each $s\subseteq e\in\Ein$.}\label{eq:vs}
\end{equation}
We will compute unique indices~$\vid{v}^s\in\{1,\dots,|V^s|\}$
for the vertices $v\in V^s$
for each~$s\subseteq e\in\Ein$,
unique indices~$\sid{s}^e\in\{1,\allowbreak\dots,\allowbreak
2^{|e|}\}$
of the subsets~$s\subseteq e$ of each~$e\in\Ein{}$,
an array~$\arrC[]$ satisfying
\begin{equation}
  \arrC[\eid{e}]=\{(s,\sid{s},\sid{s^e}):s\subseteq e\}\text{\quad for each~$(e,\eid{e})\in \eid{E}$},\label{eq:bid}
\end{equation}
and an array~$\arrD[]$ of arrays satisfying
\begin{align}
  \arrD[\eid{e}][\sid{s^e}]&=\{\vid{v}^s:v\in e\setminus s\}\label{eq:cid}\\
&  \text{\quad for each~$(e,\eid{e})\in \eid{E}$, $(s,\sid{s},\sid{s}^e)\in\arrC[\eid{e}]$.}\notag{}
\end{align}
\bev{} can then be implemented using
arrays~$\petals'[]$ and~$\used'[]$ of size~$2^d m$ each,
where for each $\sid{s}$,
$\used'[\sid{s}][]$ is an array of size~$|V^s|$:
instead of iterating over each~$e\in \Ein$,
each $s\subseteq e$,
each $v\in e\setminus s$,
and looking up $\petals[s]$ and $\used[s][v]$ in tries,
\bev{} can iterate over each $(e,\eid{e})\in \eid{E}$,
each $(s,\sid{s},\sid{s}^e)\in\arrC[\eid{e}]$,
each $\vid{v^s}\in\arrD[\eid{e}][\sid{s^e}]$,
and look up $\petals'[\sid{s}]$
and $\used'[\sid{s}][\vid{v^s}]$.
These are simple array accesses,
each working in constant time.
The modified algorithm is
shown in \bevp{}.

\begin{algorithm}[t]
  \small
  \SetAlgoRefName{Bev'}
  \KwIn{Hypergraph $(\Vin, \Ein)$, $k\in\mathbb N$.}
  \KwOut{Kernel $((\Vout,\Eout),k)$ with
    $|\Eout|\leq d! d^{d+1} (k+1)^d.$}
  \tcp{\textrm{$\eid{E}$, $V^s$, $B[]$, and $C[][]$
      are as in \eqref{eq:eid}, \eqref{eq:vs},
      \eqref{eq:bid}, and \eqref{eq:cid},\\
      $\petals'[]$ is a zero-initialized array of size~$2^dm$,\\
      $\used'[]$ is an array of size~$2^dm$,\\
      $\used'[\sid{s}]$ is a false-initialized array of size $|V^s|$}.}
  $\Eout \gets \emptyset$\;
  \ForEach{$(e,\eid{e}) \in \eid{E}$} {
    \If{$\forall (s,\sid{s},\sid{s^e})\in B[\eid{e}]:\petals'[\sid{s}] \leq k$}{
      $\Eout \gets \Eout \cup \{e\}$\;
      \ForEach{$(s,\sid{s},\sid{s}^e)\in B[\eid{e}]$} {
        \If{$\forall \vid{v^s} \in C[\eid{e}][\sid{s}^e]: \used'[\sid{s}][\vid{v}^s] = \false$} {
          $\petals'[\sid{s}] \gets \petals'[\sid{s}] + 1$\;
          \ForEach{$\vid{v^s}\in C[\eid{e}][\sid{s^e}]$} {
            $\used'[\sid{s}][\vid{v^s}] \gets \true$
          }
        }
      }
    }
  }
  $\Vout\gets \bigcup_{e\in \Eout}e$\;
  \Return{$((\Vout, \Eout),k)$}\;
  \caption{Linear-space version of \bev.}
  \label{alg:bevp}
\end{algorithm}

\begin{theorem}
  \label{thm:bev}
  \bev{}
  can be run in
  $\bigO(nd+2^dd \cdot m)$~time
  and space.
\end{theorem}

\begin{proof}
  We describe how to compute
  the indices~$\vid{v}^s$, $\sid{s}^e$,
  and the arrays~$\arrC[]$ and~$\arrD[]$
  in linear time and space.
  First,
  the indices~$\sid{s}$
  and
  array~$\arrA[]$
  in \eqref{eq:aid}
  are computed as in \cref{thm:fk}
  in $\bigO(nd+2^dd\cdot m)$~time and space.
  For \bev{},
  we use three additional preprocessing steps.

\emph{Step 1.} Initialize a size-$m$ array~$\arrC[]$.
  For each~$(e,\eid{e})\in\eid{E}$,
  compute $\arrC[\eid{e}]$ from $\arrA[\eid{e}]$
  by iterating over each~$(s,\sid{s})\in\arrA[\eid{e}]$,
  simultaneously incrementing a counter~$\sid{s}^e$
  from 0 to~$2^{|e|}$.
  This works in time~$\bigO(2^d\cdot m)$
  and space.

\emph{Step 2.} Iterating over each $(e,\eid{e})\in\eid{E}$
  and each $(s,\sid{s},\sid{s^e})\in\arrC[\eid{e}]$,
  in $\bigO(2^dd\cdot m)$~time,
  generate a list
  \[
    L:=[(\sid{s},v,\sid{s^e},\eid{e})\mid v\in e\setminus s,s\subseteq e,e\in\Ein].
  \]
  Sort the list
  by lexicographically non\hyp decreasing~$(\sid{s},v)$.
  Since these are pairs of numbers in~$\{1,\dots,2^dm\}\cup\{1,\dots,n\}$,
  this works
  in $\bigO(n+2^dm+|L|)=\bigO(n+2^dm)$~time
  using radix sort \citep[Section~8.3]{CLRS01}.
  Thereafter,
  all quadruples
  belonging to the same~$\sid{s}$
  occur consecutively in~$L$.
  Also,
  for each fixed~$\sid{s}$,
  all quadruples belonging to~$\sid{s}$ and the same~$v$
  occur consecutively in~$L$.
  
\emph{Step 3.} Initialize a size-$m$ array~$\arrD[]$,
  and for each~$(e,\eid{e})\in\eid{E}$,
  a size-$2^{|e|}$ array
  $\arrD[\eid{e}][]$ 
  of empty lists.
  Iterate over each~$(\sid{s},v,\sid{s^e},\eid{e})\in L$.
  If there is no predecessor on~$L$
  or the predecessor~$(\sid{s'},v',{\sid{s^e}}',\eid{e'})$
  satisfies $\sid{s}\ne \sid{s'}$,
  then initialize~$\vid{v}^s\gets 1$.
  If $\sid{s}=\sid{s'}$ but $v\ne v'$,
  then increment~$\vid{v}^s\gets\vid{v}^s+1$.
  Add $\vid{v}^s$ to $\arrD[\eid{e}][\sid{s^e}]$
  and continue.

  \looseness=-1
  Note that,
  as a by\hyp product,
  Step 3 also computes~$|V^s|$,
  which is just the largest $\vid{v^s}$.
The running time for the preprocessing
is dominated
by $\bigO(nd+2^dd\cdot m)$
for computing array~$\arrA[]$
as in \cref{thm:fk}.
The space used additionally to \cref{thm:fk}
is the array~$\arrC[]$ of
overall size $\bigO(2^dd\cdot m)$,
and the size-$m$ array~$\arrD[]$.
Each entry of~$\arrD[]$ is
an array of size at most~$2^d$,
whose entries are lists of length at most~$d$.
Thus, $\arrD[]$ takes at most $\bigO(2^dd\cdot m)$ total space.

After preprocessing,
\bev{} can be implemented to
run in $\bigO(2^dd\cdot m)$~time
as shown in \bevp{}:
it uses arrays~$\petals'[]$
and $\used'[]$ with $2^d\cdot m$~entries each.
For each~$s\subseteq e\in\Ein$,
$\used'[\sid{s}][]$ is an array
indexed by~$\{1,\dots,|V^s|\}$,
where
\begin{align*}
  \sum_{s\subseteq e\in \Ein}|V^s|\leq\bigl|\{(s,e,v)\mid s\subseteq e\in\Ein,v\in e\setminus s\}\bigr|\leq 2^dd\cdot m.
\end{align*}
Thus, the total size of~$\used'[][]$
is~$\bigO(2^dd\cdot m)$.
\end{proof}
\section{Experiments}
\label{sec:exp}
\noindent
In this section,
we compare the kernelizations \fk{} and \bev{}
and the well\hyp known data reduction
algorithm \wei{}.
\wei{}
does not yield problem kernels
(as it does not give size bounds),
does not work in linear time,
yet works independently of~$k$.

\begin{algorithm}[t]
  \small
  \SetAlgoRefName{Wei}
  \LinesNotNumbered{}
  \KwIn{Hypergraph $(\Vin, \Ein)$.}
  \KwOut{Hypergraph $(\Vout,\Eout)$
    that has a hitting set of size~$k$
    if and only if $(\Vin, \Ein)$~has.}

  \medskip
  Exhaustively apply the following two
  data reduction rules:
  \begin{enumerate}
  \item If, for some vertex~$v$,
    all edges containing~$v$
    also contain some vertex~$u\ne v$,
    then delete~$v$.
  
  \item If there are
    two edges $e\subseteq e'$,
    then
     delete~$e'$.
\end{enumerate}
 Return the result~$(\Vout,\Eout)$.
  \caption{Algorithm due to \citet{Wei98}}
  \label{alg:wei}
\end{algorithm}

\cref{sec:setup},
describes our experimental setup,
\cref{sec:timespace}
presents time and memory measurements.
We analyze
the effect of data reduction on
instances arising in
data clustering (\cref{sec:cluster})
and
public transportation optimization (\cref{sec:trans}).

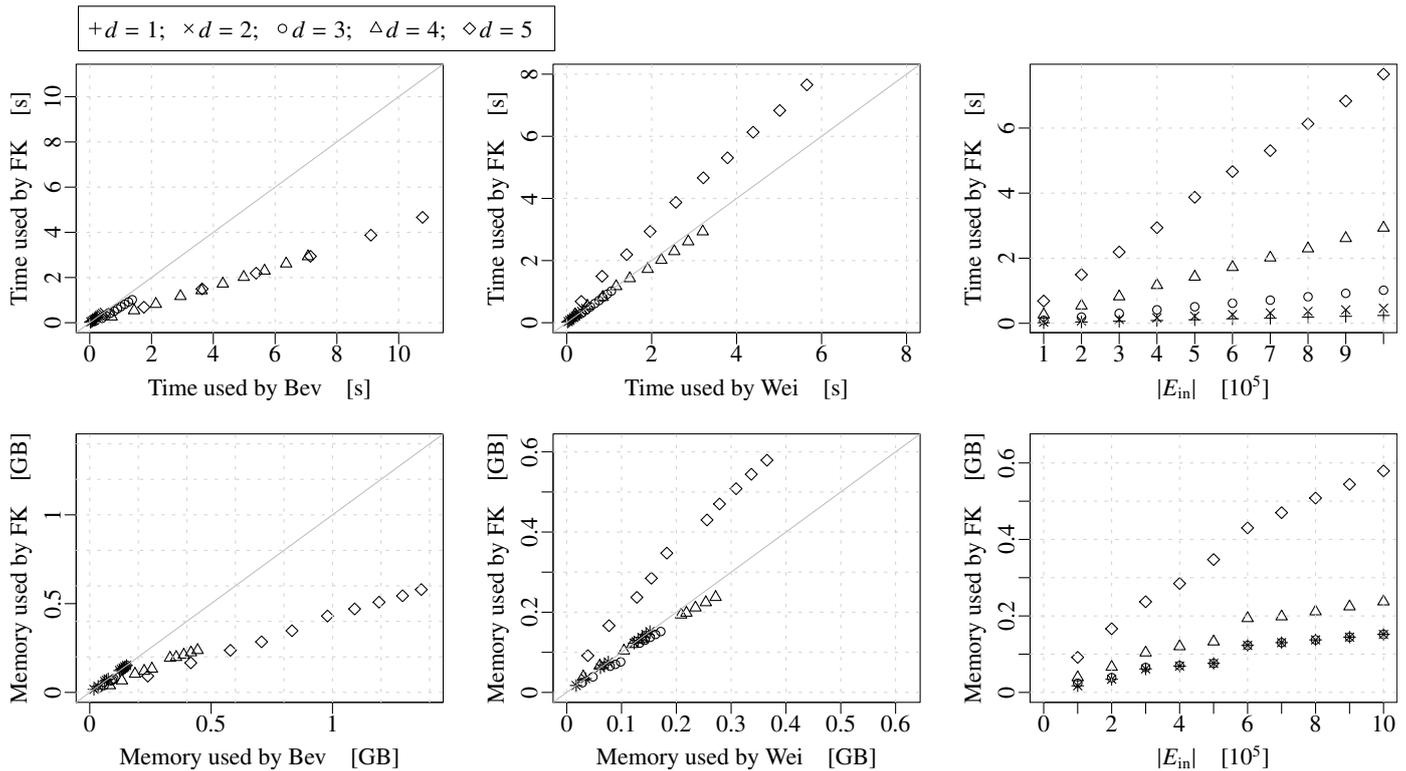
\begin{figure*}[t]
  \raggedright
  \small
  \qquad\input{legend.tex}

  \input{analysis_random-003.tex}
  \hfill\input{analysis_random-001.tex}
  \hfill\input{analysis_random-005.tex}

  \bigskip
  \noindent
  \input{analysis_random-004.tex}
  \hfill\input{analysis_random-002.tex}
  \hfill\input{analysis_random-006.tex}
  \caption{\looseness=-1
    Comparison of running time and memory usage of
    \fk, \bev, and \wei{}
    on the data set described in \cref{sec:timespace}.
    Each point represents an instance.
    The leftmost column shows that
    \fk{} outperforms \bev{}.
    The rightmost column shows time and memory
    usage of \fk{}.
    The middle column compares \fk{} to \wei{}:
    on this particular data set,
    \wei{}
    does run in linear time and outperforms \fk{} for $d=5$.}
  \label{fig:memtime}
\end{figure*}

\subsection{Experimental setup}
\label{sec:setup}
\noindent
\fk{}, \bev{} and \wei{}
were implemented in C++.\footnote{The source code is freely available at \url{https://gitlab.com/PavelSmirnov/hs-lintimespace}.}
The experiments were conducted on a 3.60\,GHz
processor with
16\,GB of RAM.
The running time is measured with the standard C++ library \texttt{ctime}.
The memory consumption is measured using the
\texttt{valgrind} memory measurement tool.
The data reduction effect is measured by
comparing the number~$|\Ein|$ of input edges
to the number~$|\Eout|$ of output edges.

\paragraph{Implementation of \fk{} and \bev{}}
We implemented
three variants of data structures for
$\supersets[]$, $\petals[]$, and $\used[]$:
\begin{itemize}
\item using arrays
  with constant worst\hyp case look\hyp up time
  after linear\hyp time precomputations,
  as described in \cref{sec:space},
\item using hash tables (the \texttt{unordered\_map} type in C++),
  with $\bigO(d)$ expected look\hyp up time
  (we account $\bigO(d)$~time
  for computing the hash value
  of a set of size~$d$),
\item
  using balanced search trees (the \texttt{map} type in C++),
  with $\bigO(d\log 2^d m)\subseteq\bigO(d^2\log n)$
  worst\hyp case look\hyp up time
  (a factor~$d$ is for
  lexicographically comparing sets of size~$d$).
\end{itemize}
All implementations use linear space.
Additionally to
the data reduction described in
\cref{sec:space}, \fk{} and \bev{}
also delete all edges
that are supersets of other edges
(it is easy to add this data reduction
without increasing their
running time
or space usage).

\paragraph{Implementation of \wei{}}
Our implementation of \wei{} uses hash tables
and runs in $\bigO(z\cdot (n^2+nm))$~time,
where $z$~is the number of vertices deleted by \wei{}
and we omit factors depending on hash\hyp table look\hyp up
and~$d$.

\paragraph{Combinations of algorithms}

We analyze the data reduction effect
of the individual algorithms
as well as of their combinations,
applying one data reduction algorithm
to the output of previous data reduction algorithms.
Since the data reduction effect of the algorithms
may depend on the processing order of the edges
\citep[Fig.\ 5.3]{Bev14c},
each algorithm is applied to
a \emph{random permutation} of edges.
This excludes the possibility that
instance generators or previous data reduction algorithms
generate particularly ``friendly'' input orders.
The order of combining the algorithms
is determined by their running times:
since the running time of \wei{} is non\hyp linear,
it is applied last,
so that it is run on a problem kernel
with size independent of~$n+m$.
\bev{} is slower than \fk{}, %
so it is applied after \fk{}.

\paragraph{Computing $k$}
The algorithms \fk{} and \bev{}
require an upper bound~$k$ on the minimum hitting set size
as input.
We compute~$k$ using a greedy approach:
repeatedly pick a vertex with a maximum number of incident edges,
add it to the hitting set,
and remove all incident edges,
until all edges are hit.

\subsection{Time and memory measurements}
\label{sec:timespace}
\noindent
In the following,
we present measurements
of the running time and memory usage.

\paragraph{Data generation}
We randomly
generate one instance for each
combination of
$d\in\{1,\dots,5\}$
and
$m\in\{i\cdot 10^5\mid i\in\{1,\dots,10\}\}$,
consisting of $n=100$~vertices
and $m$~edges of size~$d$,
each chosen with equal probability.
We observed these instances to be reluctant to data reduction:
the result of applying \fk{}, \bev{}, and \wei{}
to these hypergraphs was exactly the same
and was limited to deleting edges
that are supersets of other edges.
The greedily computed upper bound~$k$
was too high for \fk{} and \bev{}
to apply any $k$-dependent data reduction.
Thus,
the random data gives a pessimistic estimate
of the running time of \fk{} and \bev{}:
both algorithms iterate over each subset of
each input edge
and add almost all of them
to the output hypergraph,
updating their data structures.
However,
the random data gives
a too optimistic estimate of the
running time of \wei{},
since the number~$n$ of vertices is constant
and it does not delete any vertex.
Thus,
on this data set,
\wei{} also exhibits linear running\hyp time behavior.

\paragraph{Results}
\looseness=-1
On \emph{each} generated instance,
an implementation of \fk{} and \bev{}
using hash tables
ran about three times faster
than the implementation using arrays
and uses about four times less memory.
This is not surprising since
the intricate precomputation
described in \cref{sec:space},
iterates over the input hypergraph
three to four times,
building four helper arrays and lists
whose size each is
linear in that of the input hypergraph.
Thus,
it seems
that using tries (with $\bigO(nm)$ pre\hyp initialized memory)
remains the only way to outperform hash tables \cite{Bev14c}.
The balanced tree variant
on \emph{each} instance uses
roughly the same amount of memory
as the hash table variant
and is only about 1.5 times faster than the array variant.
Thus,
hash tables are the most reasonable
choice to implement \fk{} and \bev{} in linear space.
In the following,
we only compare the hash table implementations of \fk{},
\bev{}, and \wei{}.
The resource consumption of
these implementations is shown in   \cref{fig:memtime}.

\subsection{Cluster Vertex Deletion}
\label{sec:cluster}
\noindent
In this section,
we analyze the data reduction effect of \fk, \bev{}, and \wei{}
on \hs 3 instances arising from the \cvd{} problem:
the task is to delete at most $k$~vertices from a graph
so that each connected component in the remaining graph is a clique
\citep{HKMN10}.
A \cvd{}
instance $(G,k)$ with $G = (V, E)$ can be reduced %
to a \hs{3} instance $(H,k)$ with
$H = (V, \{e \subseteq V : G[e]\text{ is a path on three vertices}\})$
\citep{HKMN10}.

\paragraph{Data acquisition}
We used \cvd{} instances arising
when clustering real\hyp world protein similarity graphs
initially used by \citet{RWT+07}.%
\footnote{Available at \url{https://bio.informatik.uni-jena.de/data/}
as \texttt{biological\_bielefeld.zip}}
In fact,
they used these graphs as instances for the
weighted \clusterediting{} problem,
where one adds and deletes edges instead of deleting vertices.
As suggested by \citet{RWT+07},
we create an edge between two proteins
if their similarity score is positive.
Since our problem is unweighted,
we ignore edge weights.
The data set contains hypergraphs
up to $10^8$~edges.
However,
due to the high running time of \wei{},
we only used those with up to $10^6$~edges in our comparison
(discarding 12~hypergraphs).

\paragraph{Results}
\fk{} processed each instance
in under 1.1 seconds,
\bev{} in under 2.5 seconds,
\wei{} in under 7.3 seconds.

\begin{figure*}[p]
  \centering
  \small
  \input{analysis_bio-003.tex}
  \hfill  \input{analysis_bio-004.tex}
  \caption{Comparison between \bev{}, \fk{}, and \wei{}
    on the \hs 3 instances described in \cref{sec:cluster}.
    Each point represents an instance.
    The left\hyp hand plot
    shows that \fk{} and \bev{} are about on par.
    Thus,
    the right\hyp hand plot
    shows that \fk{} and \bev{} outperform \wei{}.}
  \label{fig:bio_pairwise}

  \bigskip
  \begin{subfigure}{\textwidth}
    \input{analysis_bio-012.tex}
    \hfill\input{analysis_bio-009.tex}
    \caption{$1\,000\leq|\Ein|\leq 10\,000$.}
  \end{subfigure}

  \bigskip
  \begin{subfigure}{\textwidth}
    \input{analysis_bio-011.tex}
    \hfill\input{analysis_bio-008.tex}
    \caption{$10\,000\leq|\Ein|\leq 100\,000$.}
  \end{subfigure}

  \bigskip
  \begin{subfigure}{\textwidth}
    \input{analysis_bio-010.tex}
    \hfill\input{analysis_bio-007.tex}
    \caption{$100\,000\leq|\Ein|\leq 1\,000\,000$.}
  \end{subfigure}

  \caption{Compression of the \hs 3 instances
    described in \cref{sec:cluster}
    for different orders of magnitude of~$|\Ein|$.
    Each dot represents an instance.
    The boxes show the first quartile, the median, and the third quartile.
    The whiskers extend up to 1.5 times the interquartile range.
    The columns are: N --- no data reduction
    (repeats the distribution of $|\Ein|$); W --- \wei{}; F --- \fk{}; B --- \bev{}; FW --- \fk{} followed by \wei{}; BW --- \bev{} followed by \wei{}; FBW --- \fk{}, \bev{}, and \wei{} applied in this order.}
  \label{fig:bio_boxplots}
\end{figure*}

\cref{fig:bio_pairwise}
shows that the data reduction effect of
\fk{} and \bev{}
is about the same---stronger than that of \wei{}.
Indeed,
it shows that
\fk{} and \bev{} work well
on the very same instances,
which is surprising,
since they are based
on different data reduction criteria
and it seems that \bev{}
should be able to apply data reduction earlier
than \fk{}.

\begin{figure}
  \small
  \input{analysis_bio-005.tex}
  \vspace{-1em}
  \caption{Data reduction effect on the \hs d instances
    described in \cref{sec:cluster}, in dependence of~$k$.}
  \label{fig:bio_k}
\end{figure}
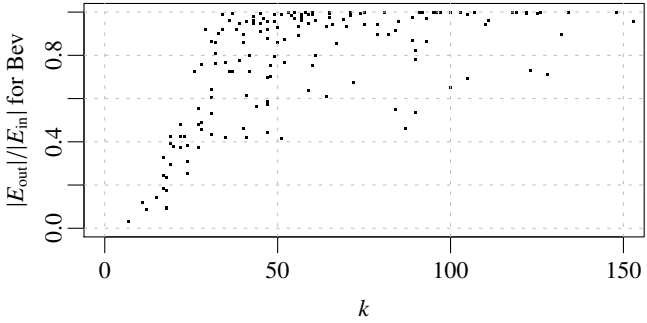

\cref{fig:bio_boxplots}
shows the absolute and relative
data reduction effect of \fk{}, \bev{}, \wei{},
and their combinations.
Combining any of \fk{} and \bev{} with \wei{}
significantly improves the data reduction effect
compared to the single algorithms,
whereas combining all three algorithms
shows no improvement.
Also,
the data reduction effect on larger instances
can be seen to be
lower than that on smaller instances.
As \cref{fig:bio_k} shows,
this effect
is mainly determined by~$k$:
not much data reduction is happening
for~$k\geq 50$,
and we observed that
almost all instances with $|\Ein|\geq 10^5$ have $k\geq 50$
for our greedily computed value of~$k$.

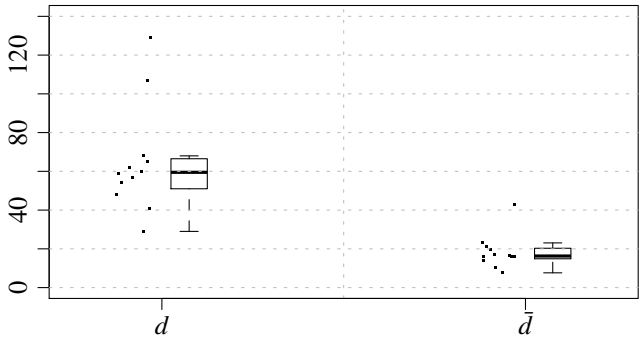
\begin{figure}[t]
  \input{analysis_trans-003.tex}
  \caption{Maximum and average edge cardinalities, $d$ and $\bar d$, respectively, of the instances in \cref{sec:trans}.
    The boxes show the first quartile, the median, and the third quartile.
    The whiskers extend up to 1.5 times the interquartile range.}
  \label{fig:station-d}
\end{figure}

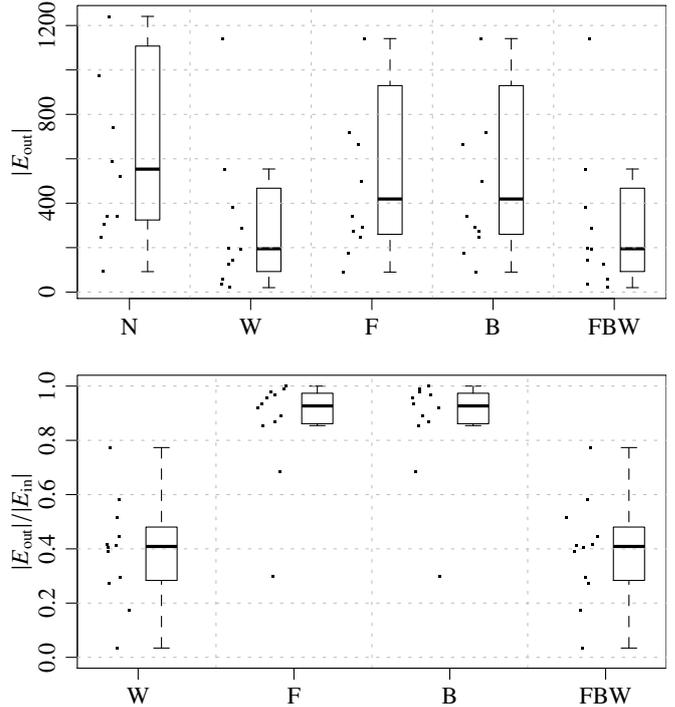
\begin{figure}
  \small
  \input{analysis_trans-002.tex}

  \bigskip
  \noindent
  \input{analysis_trans-001.tex}
  \caption{Compression of the instances described in \cref{sec:trans}.
    Each dot represents an instance.
    The boxes show the first quartile, the median, and the third quartile.
    The whiskers extend up to 1.5 times the interquartile range.
    The columns are as follows: N --- no data reduction (repeats the distribution of $|\Ein|$); W --- \wei{}; F --- \fk{}; B --- \bev{}; FBW --- \fk{}, \bev{}, and \wei{} applied in this order.}
  \label{fig:stationcover}
\end{figure}

\subsection{Station Cover}
\label{sec:trans}
\noindent
\wei{} is well known
for its data reduction effect on
the \stationcover{} problem
in real\hyp world
transportation networks \citep{Wei98}.
In this section,
we compare \fk{}, \bev{}, and \wei{}
on the corresponding \hs d instances.

\paragraph{Data acquisition}
\looseness=-1
We applied \fk{}, \bev{}, and \wei{}
to twelve \hs d instances
modeling the \stationcover{} problem
in European transportation networks
(of cities, rural areas, and countries)
that were kindly made available to us by \citet{BFFS19}

\paragraph{Results}
As shown by \cref{fig:station-d},
the hypergraphs in this data set have
much larger edges
than the ones arising from Cluster Vertex Deletion
in \cref{sec:cluster}.
Thus,
as shown in \cref{fig:stationcover},
\wei{} obviously outperforms \fk{} and \bev{}.
Moreover,
\fk{} and \bev{} work equally bad
and even combining all three algorithms did not
yield any additional data reduction effect
compared to \wei{}.

\looseness=-1
Indeed,
due to high values of~$d$,
\wei{} outperforms \fk{} and \bev{}
even with respect to running time.
\fk{} and \bev{}
are not even applicable to the shown instances right away:
it is infeasible to iterate over all $2^d$~subsets
of an edge~$e$ of size~$d$ when $d$~is large (say, 10).
In such cases,
we do not iterate over all subsets of~$e$,
but only over
intersections of~$e$ with other edges
and its subsets of size bounded by~$d'\leq d$
(we used $d'=1$).
That is,
we do not iterate over $2^d$~subsets,
but only over $m$~subsets of~$e$.
Using this implementation trick,
the running times of \fk{} and \bev{} were
under six seconds on each instance,
the running time of \wei{} was under one second
on each instance.

\section{Conclusion}
\noindent
We presented the first linear\hyp time
and linear\hyp space kernelizations for \hs d,
and thus resolved the apparent paradox that
worst\hyp case linear running times of \hs d kernelizations
hinge on quadratic\hyp size and partly initialized memory
\citep{Bev14c,FK15}.

\looseness=-1
We also conducted the first experimental evaluation of
\fk{},
significantly extended previous experimental results
for \bev{},
and compared them to the well\hyp known \wei{}
data reduction algorithm.
The experiments show that \wei{}
is outperformed by \fk{} and \bev{}
on hypergraphs of small edge cardinality
when one has good upper bounds on the hitting set size.
Otherwise,
\wei{} outperforms \fk{} and \bev{}.
The data reduction effect of \wei{}
can be strengthened by applying \fk{} and \bev{} in advance.
Thus,
the algorithms complement each other.
We have also seen that,
the data reduction of \fk{} and \bev{}
is about equal,
yet \fk{} is significantly faster.

\looseness=-1
Given that \fk{} often kernelizes instances
within a fraction of a second
and yields good data reduction results
when $k$~is small,
it seems to be a good candidate
for effectively applying the technique
of \emph{interleaving}
kernelization and branching \citep{NR00}:
in a branch\hyp and\hyp bound algorithm
for computing minimum hitting sets,
\fk{} can be applied with $k$ set
to the size of the smallest hitting set found so far,
giving a good upper bound.
Herein,
there is no need to
apply \fk{} to the input hypergraph each time,
but to the hypergraphs already kernelized
on higher levels of the search tree.

\paragraph{Acknowledgments}
We thank the anonymous referees
for their comments
and \citet{BFFS19}
for making available their data.

\paragraph{Funding}
\looseness=-1
The work in \cref{sec:space,sec:timespace,sec:cluster}
was supported by the
Russian Foundation for Basic Research,
project 18-501-12031 NNIO\textunderscore a,
work in \cref{sec:trans}
by Mathematical Center in Akademgorodok,
agreement No.\ 075-15-2019-167 with the Ministry of Science and
Higher Education of the Russian Federation.

\vspace{-0.25em}
\bibliographystyle{hs-lintimespace}
\bibliography{hs-lintimespace}

\end{document}

%% file: legend.tex
\begin{tikzpicture}[x=1pt,y=1pt]
\definecolor{fillColor}{RGB}{255,255,255}
\path[use as bounding box,fill=fillColor,fill opacity=0.00] (0,0) rectangle (216.81, 21.68);
\begin{scope}
\path[clip] (  0.00,  0.00) rectangle (216.81, 21.68);
\definecolor{drawColor}{RGB}{0,0,0}

\path[draw=drawColor,line width= 0.4pt,line join=round,line cap=round] (  8.03, 20.88) rectangle (186.41,  4.08);

\path[draw=drawColor,line width= 0.4pt,line join=round,line cap=round] ( 12.10, 12.48) -- ( 16.56, 12.48);

\path[draw=drawColor,line width= 0.4pt,line join=round,line cap=round] ( 14.33, 10.25) -- ( 14.33, 14.71);

\path[draw=drawColor,line width= 0.4pt,line join=round,line cap=round] ( 47.80, 10.90) -- ( 50.95, 14.05);

\path[draw=drawColor,line width= 0.4pt,line join=round,line cap=round] ( 47.80, 14.05) -- ( 50.95, 10.90);

\path[draw=drawColor,line width= 0.4pt,line join=round,line cap=round] ( 84.42, 12.48) circle (  1.58);

\path[draw=drawColor,line width= 0.4pt,line join=round,line cap=round] (119.47, 14.93) --
	(121.59, 11.25) --
	(117.35, 11.25) --
	(119.47, 14.93);

\path[draw=drawColor,line width= 0.4pt,line join=round,line cap=round] (152.28, 12.48) --
	(154.51, 14.71) --
	(156.74, 12.48) --
	(154.51, 10.25) --
	(152.28, 12.48);

\node[text=drawColor,anchor=base west,inner sep=0pt, outer sep=0pt, scale=  1.00] at ( 18.11,  9.10) {$d=1$;};

\node[text=drawColor,anchor=base west,inner sep=0pt, outer sep=0pt, scale=  1.00] at ( 53.16,  9.10) {$d=2$;};

\node[text=drawColor,anchor=base west,inner sep=0pt, outer sep=0pt, scale=  1.00] at ( 88.20,  9.10) {$d=3$;};

\node[text=drawColor,anchor=base west,inner sep=0pt, outer sep=0pt, scale=  1.00] at (123.25,  9.10) {$d=4$;};

\node[text=drawColor,anchor=base west,inner sep=0pt, outer sep=0pt, scale=  1.00] at (158.29,  9.10) {$d=5$};
\end{scope}
\end{tikzpicture}

%% file: analysis_random-003.tex
\begin{tikzpicture}[x=1pt,y=1pt]
\definecolor{fillColor}{RGB}{255,255,255}
\path[use as bounding box,fill=fillColor,fill opacity=0.00] (0,0) rectangle (162.61,126.47);
\begin{scope}
\path[clip] ( 25.20, 25.20) rectangle (162.61,126.47);
\definecolor{drawColor}{RGB}{0,0,0}

\path[draw=drawColor,line width= 0.4pt,line join=round,line cap=round] ( 28.27, 29.11) -- ( 32.72, 29.11);

\path[draw=drawColor,line width= 0.4pt,line join=round,line cap=round] ( 30.49, 26.88) -- ( 30.49, 31.33);

\path[draw=drawColor,line width= 0.4pt,line join=round,line cap=round] ( 28.52, 29.29) -- ( 32.97, 29.29);

\path[draw=drawColor,line width= 0.4pt,line join=round,line cap=round] ( 30.74, 27.06) -- ( 30.74, 31.52);

\path[draw=drawColor,line width= 0.4pt,line join=round,line cap=round] ( 28.79, 29.49) -- ( 33.25, 29.49);

\path[draw=drawColor,line width= 0.4pt,line join=round,line cap=round] ( 31.02, 27.26) -- ( 31.02, 31.72);

\path[draw=drawColor,line width= 0.4pt,line join=round,line cap=round] ( 29.04, 29.68) -- ( 33.50, 29.68);

\path[draw=drawColor,line width= 0.4pt,line join=round,line cap=round] ( 31.27, 27.46) -- ( 31.27, 31.91);

\path[draw=drawColor,line width= 0.4pt,line join=round,line cap=round] ( 29.36, 29.88) -- ( 33.82, 29.88);

\path[draw=drawColor,line width= 0.4pt,line join=round,line cap=round] ( 31.59, 27.65) -- ( 31.59, 32.11);

\path[draw=drawColor,line width= 0.4pt,line join=round,line cap=round] ( 29.64, 30.13) -- ( 34.09, 30.13);

\path[draw=drawColor,line width= 0.4pt,line join=round,line cap=round] ( 31.86, 27.91) -- ( 31.86, 32.36);

\path[draw=drawColor,line width= 0.4pt,line join=round,line cap=round] ( 29.86, 30.28) -- ( 34.31, 30.28);

\path[draw=drawColor,line width= 0.4pt,line join=round,line cap=round] ( 32.08, 28.05) -- ( 32.08, 32.51);

\path[draw=drawColor,line width= 0.4pt,line join=round,line cap=round] ( 30.20, 30.56) -- ( 34.65, 30.56);

\path[draw=drawColor,line width= 0.4pt,line join=round,line cap=round] ( 32.43, 28.33) -- ( 32.43, 32.79);

\path[draw=drawColor,line width= 0.4pt,line join=round,line cap=round] ( 30.48, 30.78) -- ( 34.93, 30.78);

\path[draw=drawColor,line width= 0.4pt,line join=round,line cap=round] ( 32.71, 28.56) -- ( 32.71, 33.01);

\path[draw=drawColor,line width= 0.4pt,line join=round,line cap=round] ( 30.80, 31.00) -- ( 35.25, 31.00);

\path[draw=drawColor,line width= 0.4pt,line join=round,line cap=round] ( 33.02, 28.77) -- ( 33.02, 33.23);

\path[draw=drawColor,line width= 0.4pt,line join=round,line cap=round] ( 29.12, 27.67) -- ( 32.27, 30.82);

\path[draw=drawColor,line width= 0.4pt,line join=round,line cap=round] ( 29.12, 30.82) -- ( 32.27, 27.67);

\path[draw=drawColor,line width= 0.4pt,line join=round,line cap=round] ( 29.59, 28.01) -- ( 32.74, 31.16);

\path[draw=drawColor,line width= 0.4pt,line join=round,line cap=round] ( 29.59, 31.16) -- ( 32.74, 28.01);

\path[draw=drawColor,line width= 0.4pt,line join=round,line cap=round] ( 30.18, 28.40) -- ( 33.33, 31.55);

\path[draw=drawColor,line width= 0.4pt,line join=round,line cap=round] ( 30.18, 31.55) -- ( 33.33, 28.40);

\path[draw=drawColor,line width= 0.4pt,line join=round,line cap=round] ( 30.68, 28.79) -- ( 33.83, 31.94);

\path[draw=drawColor,line width= 0.4pt,line join=round,line cap=round] ( 30.68, 31.94) -- ( 33.83, 28.79);

\path[draw=drawColor,line width= 0.4pt,line join=round,line cap=round] ( 31.25, 29.24) -- ( 34.40, 32.39);

\path[draw=drawColor,line width= 0.4pt,line join=round,line cap=round] ( 31.25, 32.39) -- ( 34.40, 29.24);

\path[draw=drawColor,line width= 0.4pt,line join=round,line cap=round] ( 31.92, 29.66) -- ( 35.07, 32.81);

\path[draw=drawColor,line width= 0.4pt,line join=round,line cap=round] ( 31.92, 32.81) -- ( 35.07, 29.66);

\path[draw=drawColor,line width= 0.4pt,line join=round,line cap=round] ( 32.39, 30.08) -- ( 35.54, 33.23);

\path[draw=drawColor,line width= 0.4pt,line join=round,line cap=round] ( 32.39, 33.23) -- ( 35.54, 30.08);

\path[draw=drawColor,line width= 0.4pt,line join=round,line cap=round] ( 32.98, 30.48) -- ( 36.13, 33.63);

\path[draw=drawColor,line width= 0.4pt,line join=round,line cap=round] ( 32.98, 33.63) -- ( 36.13, 30.48);

\path[draw=drawColor,line width= 0.4pt,line join=round,line cap=round] ( 33.54, 30.90) -- ( 36.69, 34.05);

\path[draw=drawColor,line width= 0.4pt,line join=round,line cap=round] ( 33.54, 34.05) -- ( 36.69, 30.90);

\path[draw=drawColor,line width= 0.4pt,line join=round,line cap=round] ( 34.19, 31.32) -- ( 37.34, 34.47);

\path[draw=drawColor,line width= 0.4pt,line join=round,line cap=round] ( 34.19, 34.47) -- ( 37.34, 31.32);

\path[draw=drawColor,line width= 0.4pt,line join=round,line cap=round] ( 32.55, 29.74) circle (  1.58);

\path[draw=drawColor,line width= 0.4pt,line join=round,line cap=round] ( 34.85, 30.59) circle (  1.58);

\path[draw=drawColor,line width= 0.4pt,line join=round,line cap=round] ( 36.36, 31.54) circle (  1.58);

\path[draw=drawColor,line width= 0.4pt,line join=round,line cap=round] ( 37.92, 32.48) circle (  1.58);

\path[draw=drawColor,line width= 0.4pt,line join=round,line cap=round] ( 39.47, 33.30) circle (  1.58);

\path[draw=drawColor,line width= 0.4pt,line join=round,line cap=round] ( 40.67, 34.21) circle (  1.58);

\path[draw=drawColor,line width= 0.4pt,line join=round,line cap=round] ( 41.99, 35.01) circle (  1.58);

\path[draw=drawColor,line width= 0.4pt,line join=round,line cap=round] ( 43.38, 35.90) circle (  1.58);

\path[draw=drawColor,line width= 0.4pt,line join=round,line cap=round] ( 44.81, 36.79) circle (  1.58);

\path[draw=drawColor,line width= 0.4pt,line join=round,line cap=round] ( 46.21, 37.61) circle (  1.58);

\path[draw=drawColor,line width= 0.4pt,line join=round,line cap=round] ( 38.50, 33.66) --
	( 40.62, 29.98) --
	( 36.38, 29.98) --
	( 38.50, 33.66);

\path[draw=drawColor,line width= 0.4pt,line join=round,line cap=round] ( 46.89, 36.00) --
	( 49.01, 32.33) --
	( 44.77, 32.33) --
	( 46.89, 36.00);

\path[draw=drawColor,line width= 0.4pt,line join=round,line cap=round] ( 55.04, 38.41) --
	( 57.16, 34.73) --
	( 52.91, 34.73) --
	( 55.04, 38.41);

\path[draw=drawColor,line width= 0.4pt,line join=round,line cap=round] ( 64.27, 41.37) --
	( 66.39, 37.70) --
	( 62.15, 37.70) --
	( 64.27, 41.37);

\path[draw=drawColor,line width= 0.4pt,line join=round,line cap=round] ( 72.20, 43.56) --
	( 74.32, 39.88) --
	( 70.08, 39.88) --
	( 72.20, 43.56);

\path[draw=drawColor,line width= 0.4pt,line join=round,line cap=round] ( 80.10, 46.08) --
	( 82.22, 42.41) --
	( 77.98, 42.41) --
	( 80.10, 46.08);

\path[draw=drawColor,line width= 0.4pt,line join=round,line cap=round] ( 87.94, 48.58) --
	( 90.06, 44.91) --
	( 85.82, 44.91) --
	( 87.94, 48.58);

\path[draw=drawColor,line width= 0.4pt,line join=round,line cap=round] ( 95.73, 50.97) --
	( 97.85, 47.30) --
	( 93.61, 47.30) --
	( 95.73, 50.97);

\path[draw=drawColor,line width= 0.4pt,line join=round,line cap=round] (103.83, 53.67) --
	(105.95, 49.99) --
	(101.71, 49.99) --
	(103.83, 53.67);

\path[draw=drawColor,line width= 0.4pt,line join=round,line cap=round] (112.00, 56.40) --
	(114.12, 52.72) --
	(109.88, 52.72) --
	(112.00, 56.40);

\path[draw=drawColor,line width= 0.4pt,line join=round,line cap=round] ( 48.34, 34.79) --
	( 50.57, 37.02) --
	( 52.79, 34.79) --
	( 50.57, 32.56) --
	( 48.34, 34.79);

\path[draw=drawColor,line width= 0.4pt,line join=round,line cap=round] ( 70.14, 41.70) --
	( 72.36, 43.92) --
	( 74.59, 41.70) --
	( 72.36, 39.47) --
	( 70.14, 41.70);

\path[draw=drawColor,line width= 0.4pt,line join=round,line cap=round] ( 90.36, 47.63) --
	( 92.59, 49.85) --
	( 94.81, 47.63) --
	( 92.59, 45.40) --
	( 90.36, 47.63);

\path[draw=drawColor,line width= 0.4pt,line join=round,line cap=round] (110.68, 54.00) --
	(112.91, 56.23) --
	(115.14, 54.00) --
	(112.91, 51.77) --
	(110.68, 54.00);

\path[draw=drawColor,line width= 0.4pt,line join=round,line cap=round] (133.32, 61.96) --
	(135.55, 64.19) --
	(137.78, 61.96) --
	(135.55, 59.74) --
	(133.32, 61.96);

\path[draw=drawColor,line width= 0.4pt,line join=round,line cap=round] (152.67, 68.70) --
	(154.89, 70.93) --
	(157.12, 68.70) --
	(154.89, 66.47) --
	(152.67, 68.70);
\end{scope}
\begin{scope}
\path[clip] (  0.00,  0.00) rectangle (162.61,126.47);
\definecolor{drawColor}{RGB}{0,0,0}

\path[draw=drawColor,line width= 0.4pt,line join=round,line cap=round] ( 30.29, 25.20) -- (145.95, 25.20);

\path[draw=drawColor,line width= 0.4pt,line join=round,line cap=round] ( 30.29, 25.20) -- ( 30.29, 21.00);

\path[draw=drawColor,line width= 0.4pt,line join=round,line cap=round] ( 53.42, 25.20) -- ( 53.42, 21.00);

\path[draw=drawColor,line width= 0.4pt,line join=round,line cap=round] ( 76.55, 25.20) -- ( 76.55, 21.00);

\path[draw=drawColor,line width= 0.4pt,line join=round,line cap=round] ( 99.69, 25.20) -- ( 99.69, 21.00);

\path[draw=drawColor,line width= 0.4pt,line join=round,line cap=round] (122.82, 25.20) -- (122.82, 21.00);

\path[draw=drawColor,line width= 0.4pt,line join=round,line cap=round] (145.95, 25.20) -- (145.95, 21.00);

\node[text=drawColor,anchor=base,inner sep=0pt, outer sep=0pt, scale=  1.00] at ( 30.29, 14.28) {0};

\node[text=drawColor,anchor=base,inner sep=0pt, outer sep=0pt, scale=  1.00] at ( 53.42, 14.28) {2};

\node[text=drawColor,anchor=base,inner sep=0pt, outer sep=0pt, scale=  1.00] at ( 76.55, 14.28) {4};

\node[text=drawColor,anchor=base,inner sep=0pt, outer sep=0pt, scale=  1.00] at ( 99.69, 14.28) {6};

\node[text=drawColor,anchor=base,inner sep=0pt, outer sep=0pt, scale=  1.00] at (122.82, 14.28) {8};

\node[text=drawColor,anchor=base,inner sep=0pt, outer sep=0pt, scale=  1.00] at (145.95, 14.28) {10};

\path[draw=drawColor,line width= 0.4pt,line join=round,line cap=round] ( 25.20, 28.95) -- ( 25.20,114.20);

\path[draw=drawColor,line width= 0.4pt,line join=round,line cap=round] ( 25.20, 28.95) -- ( 21.00, 28.95);

\path[draw=drawColor,line width= 0.4pt,line join=round,line cap=round] ( 25.20, 46.00) -- ( 21.00, 46.00);

\path[draw=drawColor,line width= 0.4pt,line join=round,line cap=round] ( 25.20, 63.05) -- ( 21.00, 63.05);

\path[draw=drawColor,line width= 0.4pt,line join=round,line cap=round] ( 25.20, 80.10) -- ( 21.00, 80.10);

\path[draw=drawColor,line width= 0.4pt,line join=round,line cap=round] ( 25.20, 97.15) -- ( 21.00, 97.15);

\path[draw=drawColor,line width= 0.4pt,line join=round,line cap=round] ( 25.20,114.20) -- ( 21.00,114.20);

\node[text=drawColor,rotate= 90.00,anchor=base,inner sep=0pt, outer sep=0pt, scale=  1.00] at ( 19.32, 28.95) {0};

\node[text=drawColor,rotate= 90.00,anchor=base,inner sep=0pt, outer sep=0pt, scale=  1.00] at ( 19.32, 46.00) {2};

\node[text=drawColor,rotate= 90.00,anchor=base,inner sep=0pt, outer sep=0pt, scale=  1.00] at ( 19.32, 63.05) {4};

\node[text=drawColor,rotate= 90.00,anchor=base,inner sep=0pt, outer sep=0pt, scale=  1.00] at ( 19.32, 80.10) {6};

\node[text=drawColor,rotate= 90.00,anchor=base,inner sep=0pt, outer sep=0pt, scale=  1.00] at ( 19.32, 97.15) {8};

\node[text=drawColor,rotate= 90.00,anchor=base,inner sep=0pt, outer sep=0pt, scale=  1.00] at ( 19.32,114.20) {10};

\path[draw=drawColor,line width= 0.4pt,line join=round,line cap=round] ( 25.20, 25.20) --
	(162.61, 25.20) --
	(162.61,126.47) --
	( 25.20,126.47) --
	( 25.20, 25.20);
\end{scope}
\begin{scope}
\path[clip] (  0.00,  0.00) rectangle (162.61,126.47);
\definecolor{drawColor}{RGB}{0,0,0}

\node[text=drawColor,anchor=base,inner sep=0pt, outer sep=0pt, scale=  1.00] at ( 93.90,  1.68) {Time used by Bev\quad[s]};

\node[text=drawColor,rotate= 90.00,anchor=base,inner sep=0pt, outer sep=0pt, scale=  1.00] at (  6.72, 75.84) {Time used by FK\quad[s]};
\end{scope}
\begin{scope}
\path[clip] ( 25.20, 25.20) rectangle (162.61,126.47);
\definecolor{drawColor}{RGB}{190,190,190}

\path[draw=drawColor,line width= 0.4pt,line join=round,line cap=round] ( 25.20, 25.20) -- (162.61,126.47);
\definecolor{drawColor}{RGB}{211,211,211}

\path[draw=drawColor,line width= 0.4pt,dash pattern=on 1pt off 3pt ,line join=round,line cap=round] ( 30.29, 25.20) -- ( 30.29,126.47);

\path[draw=drawColor,line width= 0.4pt,dash pattern=on 1pt off 3pt ,line join=round,line cap=round] ( 53.42, 25.20) -- ( 53.42,126.47);

\path[draw=drawColor,line width= 0.4pt,dash pattern=on 1pt off 3pt ,line join=round,line cap=round] ( 76.55, 25.20) -- ( 76.55,126.47);

\path[draw=drawColor,line width= 0.4pt,dash pattern=on 1pt off 3pt ,line join=round,line cap=round] ( 99.69, 25.20) -- ( 99.69,126.47);

\path[draw=drawColor,line width= 0.4pt,dash pattern=on 1pt off 3pt ,line join=round,line cap=round] (122.82, 25.20) -- (122.82,126.47);

\path[draw=drawColor,line width= 0.4pt,dash pattern=on 1pt off 3pt ,line join=round,line cap=round] (145.95, 25.20) -- (145.95,126.47);

\path[draw=drawColor,line width= 0.4pt,dash pattern=on 1pt off 3pt ,line join=round,line cap=round] ( 25.20, 28.95) -- (162.61, 28.95);

\path[draw=drawColor,line width= 0.4pt,dash pattern=on 1pt off 3pt ,line join=round,line cap=round] ( 25.20, 46.00) -- (162.61, 46.00);

\path[draw=drawColor,line width= 0.4pt,dash pattern=on 1pt off 3pt ,line join=round,line cap=round] ( 25.20, 63.05) -- (162.61, 63.05);

\path[draw=drawColor,line width= 0.4pt,dash pattern=on 1pt off 3pt ,line join=round,line cap=round] ( 25.20, 80.10) -- (162.61, 80.10);

\path[draw=drawColor,line width= 0.4pt,dash pattern=on 1pt off 3pt ,line join=round,line cap=round] ( 25.20, 97.15) -- (162.61, 97.15);

\path[draw=drawColor,line width= 0.4pt,dash pattern=on 1pt off 3pt ,line join=round,line cap=round] ( 25.20,114.20) -- (162.61,114.20);
\end{scope}
\end{tikzpicture}

%% file: analysis_random-001.tex
\begin{tikzpicture}[x=1pt,y=1pt]
\definecolor{fillColor}{RGB}{255,255,255}
\path[use as bounding box,fill=fillColor,fill opacity=0.00] (0,0) rectangle (162.61,126.47);
\begin{scope}
\path[clip] ( 25.20, 25.20) rectangle (162.61,126.47);
\definecolor{drawColor}{RGB}{0,0,0}

\path[draw=drawColor,line width= 0.4pt,line join=round,line cap=round] ( 28.36, 29.17) -- ( 32.81, 29.17);

\path[draw=drawColor,line width= 0.4pt,line join=round,line cap=round] ( 30.58, 26.94) -- ( 30.58, 31.39);

\path[draw=drawColor,line width= 0.4pt,line join=round,line cap=round] ( 28.67, 29.42) -- ( 33.13, 29.42);

\path[draw=drawColor,line width= 0.4pt,line join=round,line cap=round] ( 30.90, 27.19) -- ( 30.90, 31.64);

\path[draw=drawColor,line width= 0.4pt,line join=round,line cap=round] ( 29.03, 29.69) -- ( 33.49, 29.69);

\path[draw=drawColor,line width= 0.4pt,line join=round,line cap=round] ( 31.26, 27.46) -- ( 31.26, 31.92);

\path[draw=drawColor,line width= 0.4pt,line join=round,line cap=round] ( 29.34, 29.96) -- ( 33.79, 29.96);

\path[draw=drawColor,line width= 0.4pt,line join=round,line cap=round] ( 31.57, 27.73) -- ( 31.57, 32.19);

\path[draw=drawColor,line width= 0.4pt,line join=round,line cap=round] ( 29.76, 30.23) -- ( 34.21, 30.23);

\path[draw=drawColor,line width= 0.4pt,line join=round,line cap=round] ( 31.99, 28.00) -- ( 31.99, 32.46);

\path[draw=drawColor,line width= 0.4pt,line join=round,line cap=round] ( 30.18, 30.58) -- ( 34.64, 30.58);

\path[draw=drawColor,line width= 0.4pt,line join=round,line cap=round] ( 32.41, 28.35) -- ( 32.41, 32.81);

\path[draw=drawColor,line width= 0.4pt,line join=round,line cap=round] ( 30.58, 30.78) -- ( 35.03, 30.78);

\path[draw=drawColor,line width= 0.4pt,line join=round,line cap=round] ( 32.81, 28.55) -- ( 32.81, 33.01);

\path[draw=drawColor,line width= 0.4pt,line join=round,line cap=round] ( 30.97, 31.16) -- ( 35.42, 31.16);

\path[draw=drawColor,line width= 0.4pt,line join=round,line cap=round] ( 33.19, 28.94) -- ( 33.19, 33.39);

\path[draw=drawColor,line width= 0.4pt,line join=round,line cap=round] ( 31.37, 31.47) -- ( 35.82, 31.47);

\path[draw=drawColor,line width= 0.4pt,line join=round,line cap=round] ( 33.60, 29.24) -- ( 33.60, 33.70);

\path[draw=drawColor,line width= 0.4pt,line join=round,line cap=round] ( 31.61, 31.77) -- ( 36.06, 31.77);

\path[draw=drawColor,line width= 0.4pt,line join=round,line cap=round] ( 33.83, 29.54) -- ( 33.83, 34.00);

\path[draw=drawColor,line width= 0.4pt,line join=round,line cap=round] ( 29.27, 27.78) -- ( 32.42, 30.93);

\path[draw=drawColor,line width= 0.4pt,line join=round,line cap=round] ( 29.27, 30.93) -- ( 32.42, 27.78);

\path[draw=drawColor,line width= 0.4pt,line join=round,line cap=round] ( 29.84, 28.25) -- ( 32.99, 31.40);

\path[draw=drawColor,line width= 0.4pt,line join=round,line cap=round] ( 29.84, 31.40) -- ( 32.99, 28.25);

\path[draw=drawColor,line width= 0.4pt,line join=round,line cap=round] ( 30.53, 28.79) -- ( 33.68, 31.94);

\path[draw=drawColor,line width= 0.4pt,line join=round,line cap=round] ( 30.53, 31.94) -- ( 33.68, 28.79);

\path[draw=drawColor,line width= 0.4pt,line join=round,line cap=round] ( 31.30, 29.33) -- ( 34.45, 32.48);

\path[draw=drawColor,line width= 0.4pt,line join=round,line cap=round] ( 31.30, 32.48) -- ( 34.45, 29.33);

\path[draw=drawColor,line width= 0.4pt,line join=round,line cap=round] ( 32.01, 29.94) -- ( 35.16, 33.09);

\path[draw=drawColor,line width= 0.4pt,line join=round,line cap=round] ( 32.01, 33.09) -- ( 35.16, 29.94);

\path[draw=drawColor,line width= 0.4pt,line join=round,line cap=round] ( 32.82, 30.52) -- ( 35.97, 33.67);

\path[draw=drawColor,line width= 0.4pt,line join=round,line cap=round] ( 32.82, 33.67) -- ( 35.97, 30.52);

\path[draw=drawColor,line width= 0.4pt,line join=round,line cap=round] ( 33.55, 31.10) -- ( 36.70, 34.25);

\path[draw=drawColor,line width= 0.4pt,line join=round,line cap=round] ( 33.55, 34.25) -- ( 36.70, 31.10);

\path[draw=drawColor,line width= 0.4pt,line join=round,line cap=round] ( 34.32, 31.65) -- ( 37.47, 34.80);

\path[draw=drawColor,line width= 0.4pt,line join=round,line cap=round] ( 34.32, 34.80) -- ( 37.47, 31.65);

\path[draw=drawColor,line width= 0.4pt,line join=round,line cap=round] ( 35.09, 32.22) -- ( 38.24, 35.37);

\path[draw=drawColor,line width= 0.4pt,line join=round,line cap=round] ( 35.09, 35.37) -- ( 38.24, 32.22);

\path[draw=drawColor,line width= 0.4pt,line join=round,line cap=round] ( 35.81, 32.79) -- ( 38.96, 35.94);

\path[draw=drawColor,line width= 0.4pt,line join=round,line cap=round] ( 35.81, 35.94) -- ( 38.96, 32.79);

\path[draw=drawColor,line width= 0.4pt,line join=round,line cap=round] ( 32.01, 30.04) circle (  1.58);

\path[draw=drawColor,line width= 0.4pt,line join=round,line cap=round] ( 33.82, 31.20) circle (  1.58);

\path[draw=drawColor,line width= 0.4pt,line join=round,line cap=round] ( 35.86, 32.52) circle (  1.58);

\path[draw=drawColor,line width= 0.4pt,line join=round,line cap=round] ( 37.59, 33.80) circle (  1.58);

\path[draw=drawColor,line width= 0.4pt,line join=round,line cap=round] ( 39.16, 34.92) circle (  1.58);

\path[draw=drawColor,line width= 0.4pt,line join=round,line cap=round] ( 40.79, 36.18) circle (  1.58);

\path[draw=drawColor,line width= 0.4pt,line join=round,line cap=round] ( 42.36, 37.29) circle (  1.58);

\path[draw=drawColor,line width= 0.4pt,line join=round,line cap=round] ( 43.77, 38.51) circle (  1.58);

\path[draw=drawColor,line width= 0.4pt,line join=round,line cap=round] ( 45.46, 39.73) circle (  1.58);

\path[draw=drawColor,line width= 0.4pt,line join=round,line cap=round] ( 46.99, 40.86) circle (  1.58);

\path[draw=drawColor,line width= 0.4pt,line join=round,line cap=round] ( 33.38, 34.50) --
	( 35.50, 30.83) --
	( 31.26, 30.83) --
	( 33.38, 34.50);

\path[draw=drawColor,line width= 0.4pt,line join=round,line cap=round] ( 38.03, 37.72) --
	( 40.15, 34.05) --
	( 35.90, 34.05) --
	( 38.03, 37.72);

\path[draw=drawColor,line width= 0.4pt,line join=round,line cap=round] ( 43.76, 41.03) --
	( 45.89, 37.36) --
	( 41.64, 37.36) --
	( 43.76, 41.03);

\path[draw=drawColor,line width= 0.4pt,line join=round,line cap=round] ( 48.79, 45.11) --
	( 50.92, 41.44) --
	( 46.67, 41.44) --
	( 48.79, 45.11);

\path[draw=drawColor,line width= 0.4pt,line join=round,line cap=round] ( 53.90, 48.11) --
	( 56.02, 44.44) --
	( 51.78, 44.44) --
	( 53.90, 48.11);

\path[draw=drawColor,line width= 0.4pt,line join=round,line cap=round] ( 60.60, 51.59) --
	( 62.72, 47.92) --
	( 58.47, 47.92) --
	( 60.60, 51.59);

\path[draw=drawColor,line width= 0.4pt,line join=round,line cap=round] ( 65.71, 55.03) --
	( 67.83, 51.35) --
	( 63.59, 51.35) --
	( 65.71, 55.03);

\path[draw=drawColor,line width= 0.4pt,line join=round,line cap=round] ( 70.55, 58.31) --
	( 72.67, 54.64) --
	( 68.43, 54.64) --
	( 70.55, 58.31);

\path[draw=drawColor,line width= 0.4pt,line join=round,line cap=round] ( 75.78, 62.01) --
	( 77.90, 58.34) --
	( 73.66, 58.34) --
	( 75.78, 62.01);

\path[draw=drawColor,line width= 0.4pt,line join=round,line cap=round] ( 81.20, 65.77) --
	( 83.32, 62.10) --
	( 79.08, 62.10) --
	( 81.20, 65.77);

\path[draw=drawColor,line width= 0.4pt,line join=round,line cap=round] ( 33.52, 36.98) --
	( 35.75, 39.21) --
	( 37.98, 36.98) --
	( 35.75, 34.75) --
	( 33.52, 36.98);

\path[draw=drawColor,line width= 0.4pt,line join=round,line cap=round] ( 41.32, 46.48) --
	( 43.54, 48.70) --
	( 45.77, 46.48) --
	( 43.54, 44.25) --
	( 41.32, 46.48);

\path[draw=drawColor,line width= 0.4pt,line join=round,line cap=round] ( 50.51, 54.63) --
	( 52.74, 56.86) --
	( 54.97, 54.63) --
	( 52.74, 52.40) --
	( 50.51, 54.63);

\path[draw=drawColor,line width= 0.4pt,line join=round,line cap=round] ( 59.29, 63.39) --
	( 61.51, 65.62) --
	( 63.74, 63.39) --
	( 61.51, 61.16) --
	( 59.29, 63.39);

\path[draw=drawColor,line width= 0.4pt,line join=round,line cap=round] ( 68.92, 74.34) --
	( 71.14, 76.57) --
	( 73.37, 74.34) --
	( 71.14, 72.12) --
	( 68.92, 74.34);

\path[draw=drawColor,line width= 0.4pt,line join=round,line cap=round] ( 79.21, 83.61) --
	( 81.44, 85.83) --
	( 83.66, 83.61) --
	( 81.44, 81.38) --
	( 79.21, 83.61);

\path[draw=drawColor,line width= 0.4pt,line join=round,line cap=round] ( 88.26, 91.15) --
	( 90.49, 93.38) --
	( 92.72, 91.15) --
	( 90.49, 88.92) --
	( 88.26, 91.15);

\path[draw=drawColor,line width= 0.4pt,line join=round,line cap=round] ( 97.85,100.82) --
	(100.08,103.05) --
	(102.31,100.82) --
	(100.08, 98.59) --
	( 97.85,100.82);

\path[draw=drawColor,line width= 0.4pt,line join=round,line cap=round] (107.82,109.02) --
	(110.05,111.25) --
	(112.28,109.02) --
	(110.05,106.79) --
	(107.82,109.02);

\path[draw=drawColor,line width= 0.4pt,line join=round,line cap=round] (117.97,118.69) --
	(120.20,120.92) --
	(122.43,118.69) --
	(120.20,116.46) --
	(117.97,118.69);
\end{scope}
\begin{scope}
\path[clip] (  0.00,  0.00) rectangle (162.61,126.47);
\definecolor{drawColor}{RGB}{0,0,0}

\path[draw=drawColor,line width= 0.4pt,line join=round,line cap=round] ( 30.29, 25.20) -- (157.52, 25.20);

\path[draw=drawColor,line width= 0.4pt,line join=round,line cap=round] ( 30.29, 25.20) -- ( 30.29, 21.00);

\path[draw=drawColor,line width= 0.4pt,line join=round,line cap=round] ( 62.10, 25.20) -- ( 62.10, 21.00);

\path[draw=drawColor,line width= 0.4pt,line join=round,line cap=round] ( 93.90, 25.20) -- ( 93.90, 21.00);

\path[draw=drawColor,line width= 0.4pt,line join=round,line cap=round] (125.71, 25.20) -- (125.71, 21.00);

\path[draw=drawColor,line width= 0.4pt,line join=round,line cap=round] (157.52, 25.20) -- (157.52, 21.00);

\node[text=drawColor,anchor=base,inner sep=0pt, outer sep=0pt, scale=  1.00] at ( 30.29, 14.28) {0};

\node[text=drawColor,anchor=base,inner sep=0pt, outer sep=0pt, scale=  1.00] at ( 62.10, 14.28) {2};

\node[text=drawColor,anchor=base,inner sep=0pt, outer sep=0pt, scale=  1.00] at ( 93.90, 14.28) {4};

\node[text=drawColor,anchor=base,inner sep=0pt, outer sep=0pt, scale=  1.00] at (125.71, 14.28) {6};

\node[text=drawColor,anchor=base,inner sep=0pt, outer sep=0pt, scale=  1.00] at (157.52, 14.28) {8};

\path[draw=drawColor,line width= 0.4pt,line join=round,line cap=round] ( 25.20, 28.95) -- ( 25.20,122.72);

\path[draw=drawColor,line width= 0.4pt,line join=round,line cap=round] ( 25.20, 28.95) -- ( 21.00, 28.95);

\path[draw=drawColor,line width= 0.4pt,line join=round,line cap=round] ( 25.20, 52.39) -- ( 21.00, 52.39);

\path[draw=drawColor,line width= 0.4pt,line join=round,line cap=round] ( 25.20, 75.84) -- ( 21.00, 75.84);

\path[draw=drawColor,line width= 0.4pt,line join=round,line cap=round] ( 25.20, 99.28) -- ( 21.00, 99.28);

\path[draw=drawColor,line width= 0.4pt,line join=round,line cap=round] ( 25.20,122.72) -- ( 21.00,122.72);

\node[text=drawColor,rotate= 90.00,anchor=base,inner sep=0pt, outer sep=0pt, scale=  1.00] at ( 19.32, 28.95) {0};

\node[text=drawColor,rotate= 90.00,anchor=base,inner sep=0pt, outer sep=0pt, scale=  1.00] at ( 19.32, 52.39) {2};

\node[text=drawColor,rotate= 90.00,anchor=base,inner sep=0pt, outer sep=0pt, scale=  1.00] at ( 19.32, 75.84) {4};

\node[text=drawColor,rotate= 90.00,anchor=base,inner sep=0pt, outer sep=0pt, scale=  1.00] at ( 19.32, 99.28) {6};

\node[text=drawColor,rotate= 90.00,anchor=base,inner sep=0pt, outer sep=0pt, scale=  1.00] at ( 19.32,122.72) {8};

\path[draw=drawColor,line width= 0.4pt,line join=round,line cap=round] ( 25.20, 25.20) --
	(162.61, 25.20) --
	(162.61,126.47) --
	( 25.20,126.47) --
	( 25.20, 25.20);
\end{scope}
\begin{scope}
\path[clip] (  0.00,  0.00) rectangle (162.61,126.47);
\definecolor{drawColor}{RGB}{0,0,0}

\node[text=drawColor,anchor=base,inner sep=0pt, outer sep=0pt, scale=  1.00] at ( 93.90,  1.68) {Time used by Wei\quad[s]};

\node[text=drawColor,rotate= 90.00,anchor=base,inner sep=0pt, outer sep=0pt, scale=  1.00] at (  6.72, 75.84) {Time used by FK\quad[s]};
\end{scope}
\begin{scope}
\path[clip] ( 25.20, 25.20) rectangle (162.61,126.47);
\definecolor{drawColor}{RGB}{190,190,190}

\path[draw=drawColor,line width= 0.4pt,line join=round,line cap=round] ( 25.20, 25.20) -- (162.61,126.47);
\definecolor{drawColor}{RGB}{211,211,211}

\path[draw=drawColor,line width= 0.4pt,dash pattern=on 1pt off 3pt ,line join=round,line cap=round] ( 30.29, 25.20) -- ( 30.29,126.47);

\path[draw=drawColor,line width= 0.4pt,dash pattern=on 1pt off 3pt ,line join=round,line cap=round] ( 62.10, 25.20) -- ( 62.10,126.47);

\path[draw=drawColor,line width= 0.4pt,dash pattern=on 1pt off 3pt ,line join=round,line cap=round] ( 93.90, 25.20) -- ( 93.90,126.47);

\path[draw=drawColor,line width= 0.4pt,dash pattern=on 1pt off 3pt ,line join=round,line cap=round] (125.71, 25.20) -- (125.71,126.47);

\path[draw=drawColor,line width= 0.4pt,dash pattern=on 1pt off 3pt ,line join=round,line cap=round] (157.52, 25.20) -- (157.52,126.47);

\path[draw=drawColor,line width= 0.4pt,dash pattern=on 1pt off 3pt ,line join=round,line cap=round] ( 25.20, 28.95) -- (162.61, 28.95);

\path[draw=drawColor,line width= 0.4pt,dash pattern=on 1pt off 3pt ,line join=round,line cap=round] ( 25.20, 52.39) -- (162.61, 52.39);

\path[draw=drawColor,line width= 0.4pt,dash pattern=on 1pt off 3pt ,line join=round,line cap=round] ( 25.20, 75.84) -- (162.61, 75.84);

\path[draw=drawColor,line width= 0.4pt,dash pattern=on 1pt off 3pt ,line join=round,line cap=round] ( 25.20, 99.28) -- (162.61, 99.28);

\path[draw=drawColor,line width= 0.4pt,dash pattern=on 1pt off 3pt ,line join=round,line cap=round] ( 25.20,122.72) -- (162.61,122.72);
\end{scope}
\end{tikzpicture}

%% file: analysis_random-005.tex
\begin{tikzpicture}[x=1pt,y=1pt]
\definecolor{fillColor}{RGB}{255,255,255}
\path[use as bounding box,fill=fillColor,fill opacity=0.00] (0,0) rectangle (162.61,126.47);
\begin{scope}
\path[clip] ( 25.20, 25.20) rectangle (162.61,126.47);
\definecolor{drawColor}{RGB}{0,0,0}

\path[draw=drawColor,line width= 0.4pt,line join=round,line cap=round] ( 28.06, 28.95) -- ( 32.52, 28.95);

\path[draw=drawColor,line width= 0.4pt,line join=round,line cap=round] ( 30.29, 26.72) -- ( 30.29, 31.18);

\path[draw=drawColor,line width= 0.4pt,line join=round,line cap=round] ( 42.20, 29.22) -- ( 46.65, 29.22);

\path[draw=drawColor,line width= 0.4pt,line join=round,line cap=round] ( 44.43, 26.99) -- ( 44.43, 31.44);

\path[draw=drawColor,line width= 0.4pt,line join=round,line cap=round] ( 56.33, 29.50) -- ( 60.79, 29.50);

\path[draw=drawColor,line width= 0.4pt,line join=round,line cap=round] ( 58.56, 27.27) -- ( 58.56, 31.73);

\path[draw=drawColor,line width= 0.4pt,line join=round,line cap=round] ( 70.47, 29.78) -- ( 74.93, 29.78);

\path[draw=drawColor,line width= 0.4pt,line join=round,line cap=round] ( 72.70, 27.55) -- ( 72.70, 32.01);

\path[draw=drawColor,line width= 0.4pt,line join=round,line cap=round] ( 84.61, 30.07) -- ( 89.06, 30.07);

\path[draw=drawColor,line width= 0.4pt,line join=round,line cap=round] ( 86.84, 27.84) -- ( 86.84, 32.29);

\path[draw=drawColor,line width= 0.4pt,line join=round,line cap=round] ( 98.74, 30.43) -- (103.20, 30.43);

\path[draw=drawColor,line width= 0.4pt,line join=round,line cap=round] (100.97, 28.20) -- (100.97, 32.66);

\path[draw=drawColor,line width= 0.4pt,line join=round,line cap=round] (112.88, 30.64) -- (117.34, 30.64);

\path[draw=drawColor,line width= 0.4pt,line join=round,line cap=round] (115.11, 28.41) -- (115.11, 32.87);

\path[draw=drawColor,line width= 0.4pt,line join=round,line cap=round] (127.02, 31.04) -- (131.47, 31.04);

\path[draw=drawColor,line width= 0.4pt,line join=round,line cap=round] (129.25, 28.82) -- (129.25, 33.27);

\path[draw=drawColor,line width= 0.4pt,line join=round,line cap=round] (141.15, 31.37) -- (145.61, 31.37);

\path[draw=drawColor,line width= 0.4pt,line join=round,line cap=round] (143.38, 29.14) -- (143.38, 33.59);

\path[draw=drawColor,line width= 0.4pt,line join=round,line cap=round] (155.29, 31.68) -- (159.75, 31.68);

\path[draw=drawColor,line width= 0.4pt,line join=round,line cap=round] (157.52, 29.45) -- (157.52, 33.91);

\path[draw=drawColor,line width= 0.4pt,line join=round,line cap=round] ( 28.71, 27.58) -- ( 31.86, 30.73);

\path[draw=drawColor,line width= 0.4pt,line join=round,line cap=round] ( 28.71, 30.73) -- ( 31.86, 27.58);

\path[draw=drawColor,line width= 0.4pt,line join=round,line cap=round] ( 42.85, 28.06) -- ( 46.00, 31.21);

\path[draw=drawColor,line width= 0.4pt,line join=round,line cap=round] ( 42.85, 31.21) -- ( 46.00, 28.06);

\path[draw=drawColor,line width= 0.4pt,line join=round,line cap=round] ( 56.99, 28.63) -- ( 60.14, 31.78);

\path[draw=drawColor,line width= 0.4pt,line join=round,line cap=round] ( 56.99, 31.78) -- ( 60.14, 28.63);

\path[draw=drawColor,line width= 0.4pt,line join=round,line cap=round] ( 71.12, 29.19) -- ( 74.27, 32.34);

\path[draw=drawColor,line width= 0.4pt,line join=round,line cap=round] ( 71.12, 32.34) -- ( 74.27, 29.19);

\path[draw=drawColor,line width= 0.4pt,line join=round,line cap=round] ( 85.26, 29.84) -- ( 88.41, 32.99);

\path[draw=drawColor,line width= 0.4pt,line join=round,line cap=round] ( 85.26, 32.99) -- ( 88.41, 29.84);

\path[draw=drawColor,line width= 0.4pt,line join=round,line cap=round] ( 99.40, 30.45) -- (102.55, 33.60);

\path[draw=drawColor,line width= 0.4pt,line join=round,line cap=round] ( 99.40, 33.60) -- (102.55, 30.45);

\path[draw=drawColor,line width= 0.4pt,line join=round,line cap=round] (113.53, 31.05) -- (116.68, 34.20);

\path[draw=drawColor,line width= 0.4pt,line join=round,line cap=round] (113.53, 34.20) -- (116.68, 31.05);

\path[draw=drawColor,line width= 0.4pt,line join=round,line cap=round] (127.67, 31.63) -- (130.82, 34.78);

\path[draw=drawColor,line width= 0.4pt,line join=round,line cap=round] (127.67, 34.78) -- (130.82, 31.63);

\path[draw=drawColor,line width= 0.4pt,line join=round,line cap=round] (141.81, 32.23) -- (144.96, 35.38);

\path[draw=drawColor,line width= 0.4pt,line join=round,line cap=round] (141.81, 35.38) -- (144.96, 32.23);

\path[draw=drawColor,line width= 0.4pt,line join=round,line cap=round] (155.94, 32.83) -- (159.09, 35.98);

\path[draw=drawColor,line width= 0.4pt,line join=round,line cap=round] (155.94, 35.98) -- (159.09, 32.83);

\path[draw=drawColor,line width= 0.4pt,line join=round,line cap=round] ( 30.29, 29.86) circle (  1.58);

\path[draw=drawColor,line width= 0.4pt,line join=round,line cap=round] ( 44.43, 31.08) circle (  1.58);

\path[draw=drawColor,line width= 0.4pt,line join=round,line cap=round] ( 58.56, 32.46) circle (  1.58);

\path[draw=drawColor,line width= 0.4pt,line join=round,line cap=round] ( 72.70, 33.81) circle (  1.58);

\path[draw=drawColor,line width= 0.4pt,line join=round,line cap=round] ( 86.84, 34.98) circle (  1.58);

\path[draw=drawColor,line width= 0.4pt,line join=round,line cap=round] (100.97, 36.30) circle (  1.58);

\path[draw=drawColor,line width= 0.4pt,line join=round,line cap=round] (115.11, 37.46) circle (  1.58);

\path[draw=drawColor,line width= 0.4pt,line join=round,line cap=round] (129.25, 38.74) circle (  1.58);

\path[draw=drawColor,line width= 0.4pt,line join=round,line cap=round] (143.38, 40.02) circle (  1.58);

\path[draw=drawColor,line width= 0.4pt,line join=round,line cap=round] (157.52, 41.20) circle (  1.58);

\path[draw=drawColor,line width= 0.4pt,line join=round,line cap=round] ( 30.29, 34.42) --
	( 32.41, 30.75) --
	( 28.17, 30.75) --
	( 30.29, 34.42);

\path[draw=drawColor,line width= 0.4pt,line join=round,line cap=round] ( 44.43, 37.80) --
	( 46.55, 34.13) --
	( 42.30, 34.13) --
	( 44.43, 37.80);

\path[draw=drawColor,line width= 0.4pt,line join=round,line cap=round] ( 58.56, 41.27) --
	( 60.68, 37.59) --
	( 56.44, 37.59) --
	( 58.56, 41.27);

\path[draw=drawColor,line width= 0.4pt,line join=round,line cap=round] ( 72.70, 45.54) --
	( 74.82, 41.86) --
	( 70.58, 41.86) --
	( 72.70, 45.54);

\path[draw=drawColor,line width= 0.4pt,line join=round,line cap=round] ( 86.84, 48.68) --
	( 88.96, 45.01) --
	( 84.71, 45.01) --
	( 86.84, 48.68);

\path[draw=drawColor,line width= 0.4pt,line join=round,line cap=round] (100.97, 52.32) --
	(103.09, 48.65) --
	( 98.85, 48.65) --
	(100.97, 52.32);

\path[draw=drawColor,line width= 0.4pt,line join=round,line cap=round] (115.11, 55.92) --
	(117.23, 52.25) --
	(112.99, 52.25) --
	(115.11, 55.92);

\path[draw=drawColor,line width= 0.4pt,line join=round,line cap=round] (129.25, 59.36) --
	(131.37, 55.69) --
	(127.12, 55.69) --
	(129.25, 59.36);

\path[draw=drawColor,line width= 0.4pt,line join=round,line cap=round] (143.38, 63.24) --
	(145.50, 59.57) --
	(141.26, 59.57) --
	(143.38, 63.24);

\path[draw=drawColor,line width= 0.4pt,line join=round,line cap=round] (157.52, 67.18) --
	(159.64, 63.50) --
	(155.40, 63.50) --
	(157.52, 67.18);

\path[draw=drawColor,line width= 0.4pt,line join=round,line cap=round] ( 28.06, 37.13) --
	( 30.29, 39.36) --
	( 32.52, 37.13) --
	( 30.29, 34.91) --
	( 28.06, 37.13);

\path[draw=drawColor,line width= 0.4pt,line join=round,line cap=round] ( 42.20, 47.08) --
	( 44.43, 49.31) --
	( 46.65, 47.08) --
	( 44.43, 44.85) --
	( 42.20, 47.08);

\path[draw=drawColor,line width= 0.4pt,line join=round,line cap=round] ( 56.33, 55.62) --
	( 58.56, 57.85) --
	( 60.79, 55.62) --
	( 58.56, 53.40) --
	( 56.33, 55.62);

\path[draw=drawColor,line width= 0.4pt,line join=round,line cap=round] ( 70.47, 64.80) --
	( 72.70, 67.03) --
	( 74.93, 64.80) --
	( 72.70, 62.57) --
	( 70.47, 64.80);

\path[draw=drawColor,line width= 0.4pt,line join=round,line cap=round] ( 84.61, 76.27) --
	( 86.84, 78.50) --
	( 89.06, 76.27) --
	( 86.84, 74.04) --
	( 84.61, 76.27);

\path[draw=drawColor,line width= 0.4pt,line join=round,line cap=round] ( 98.74, 85.97) --
	(100.97, 88.20) --
	(103.20, 85.97) --
	(100.97, 83.75) --
	( 98.74, 85.97);

\path[draw=drawColor,line width= 0.4pt,line join=round,line cap=round] (112.88, 93.87) --
	(115.11, 96.10) --
	(117.34, 93.87) --
	(115.11, 91.65) --
	(112.88, 93.87);

\path[draw=drawColor,line width= 0.4pt,line join=round,line cap=round] (127.02,104.00) --
	(129.25,106.23) --
	(131.47,104.00) --
	(129.25,101.78) --
	(127.02,104.00);

\path[draw=drawColor,line width= 0.4pt,line join=round,line cap=round] (141.15,112.60) --
	(143.38,114.82) --
	(145.61,112.60) --
	(143.38,110.37) --
	(141.15,112.60);

\path[draw=drawColor,line width= 0.4pt,line join=round,line cap=round] (155.29,122.72) --
	(157.52,124.95) --
	(159.75,122.72) --
	(157.52,120.49) --
	(155.29,122.72);
\end{scope}
\begin{scope}
\path[clip] (  0.00,  0.00) rectangle (162.61,126.47);
\definecolor{drawColor}{RGB}{0,0,0}

\path[draw=drawColor,line width= 0.4pt,line join=round,line cap=round] ( 25.20, 28.73) -- ( 25.20,102.39);

\path[draw=drawColor,line width= 0.4pt,line join=round,line cap=round] ( 25.20, 28.73) -- ( 21.00, 28.73);

\path[draw=drawColor,line width= 0.4pt,line join=round,line cap=round] ( 25.20, 53.28) -- ( 21.00, 53.28);

\path[draw=drawColor,line width= 0.4pt,line join=round,line cap=round] ( 25.20, 77.84) -- ( 21.00, 77.84);

\path[draw=drawColor,line width= 0.4pt,line join=round,line cap=round] ( 25.20,102.39) -- ( 21.00,102.39);

\node[text=drawColor,rotate= 90.00,anchor=base,inner sep=0pt, outer sep=0pt, scale=  1.00] at ( 19.32, 28.73) {0};

\node[text=drawColor,rotate= 90.00,anchor=base,inner sep=0pt, outer sep=0pt, scale=  1.00] at ( 19.32, 53.28) {2};

\node[text=drawColor,rotate= 90.00,anchor=base,inner sep=0pt, outer sep=0pt, scale=  1.00] at ( 19.32, 77.84) {4};

\node[text=drawColor,rotate= 90.00,anchor=base,inner sep=0pt, outer sep=0pt, scale=  1.00] at ( 19.32,102.39) {6};

\path[draw=drawColor,line width= 0.4pt,line join=round,line cap=round] ( 25.20, 25.20) --
	(162.61, 25.20) --
	(162.61,126.47) --
	( 25.20,126.47) --
	( 25.20, 25.20);
\end{scope}
\begin{scope}
\path[clip] (  0.00,  0.00) rectangle (162.61,126.47);
\definecolor{drawColor}{RGB}{0,0,0}

\node[text=drawColor,anchor=base,inner sep=0pt, outer sep=0pt, scale=  1.00] at ( 93.90,  1.68) {$|E_{\mathrm{in}}|\quad[10^5]$};

\node[text=drawColor,rotate= 90.00,anchor=base,inner sep=0pt, outer sep=0pt, scale=  1.00] at (  6.72, 75.84) {Time used by FK\quad[s]};
\end{scope}
\begin{scope}
\path[clip] (  0.00,  0.00) rectangle (162.61,126.47);
\definecolor{drawColor}{RGB}{0,0,0}

\path[draw=drawColor,line width= 0.4pt,line join=round,line cap=round] ( 25.20, 25.20) -- (157.52, 25.20);

\path[draw=drawColor,line width= 0.4pt,line join=round,line cap=round] ( 30.29, 25.20) -- ( 30.29, 21.00);

\path[draw=drawColor,line width= 0.4pt,line join=round,line cap=round] ( 44.43, 25.20) -- ( 44.43, 21.00);

\path[draw=drawColor,line width= 0.4pt,line join=round,line cap=round] ( 58.56, 25.20) -- ( 58.56, 21.00);

\path[draw=drawColor,line width= 0.4pt,line join=round,line cap=round] ( 72.70, 25.20) -- ( 72.70, 21.00);

\path[draw=drawColor,line width= 0.4pt,line join=round,line cap=round] ( 86.84, 25.20) -- ( 86.84, 21.00);

\path[draw=drawColor,line width= 0.4pt,line join=round,line cap=round] (100.97, 25.20) -- (100.97, 21.00);

\path[draw=drawColor,line width= 0.4pt,line join=round,line cap=round] (115.11, 25.20) -- (115.11, 21.00);

\path[draw=drawColor,line width= 0.4pt,line join=round,line cap=round] (129.25, 25.20) -- (129.25, 21.00);

\path[draw=drawColor,line width= 0.4pt,line join=round,line cap=round] (143.38, 25.20) -- (143.38, 21.00);

\path[draw=drawColor,line width= 0.4pt,line join=round,line cap=round] (157.52, 25.20) -- (157.52, 21.00);

\node[text=drawColor,anchor=base,inner sep=0pt, outer sep=0pt, scale=  1.00] at ( 30.29, 14.28) {1};

\node[text=drawColor,anchor=base,inner sep=0pt, outer sep=0pt, scale=  1.00] at ( 44.43, 14.28) {2};

\node[text=drawColor,anchor=base,inner sep=0pt, outer sep=0pt, scale=  1.00] at ( 58.56, 14.28) {3};

\node[text=drawColor,anchor=base,inner sep=0pt, outer sep=0pt, scale=  1.00] at ( 72.70, 14.28) {4};

\node[text=drawColor,anchor=base,inner sep=0pt, outer sep=0pt, scale=  1.00] at ( 86.84, 14.28) {5};

\node[text=drawColor,anchor=base,inner sep=0pt, outer sep=0pt, scale=  1.00] at (100.97, 14.28) {6};

\node[text=drawColor,anchor=base,inner sep=0pt, outer sep=0pt, scale=  1.00] at (115.11, 14.28) {7};

\node[text=drawColor,anchor=base,inner sep=0pt, outer sep=0pt, scale=  1.00] at (129.25, 14.28) {8};

\node[text=drawColor,anchor=base,inner sep=0pt, outer sep=0pt, scale=  1.00] at (143.38, 14.28) {9};
\end{scope}
\begin{scope}
\path[clip] ( 25.20, 25.20) rectangle (162.61,126.47);
\definecolor{drawColor}{RGB}{211,211,211}

\path[draw=drawColor,line width= 0.4pt,dash pattern=on 1pt off 3pt ,line join=round,line cap=round] ( 44.43, 25.20) -- ( 44.43,126.47);

\path[draw=drawColor,line width= 0.4pt,dash pattern=on 1pt off 3pt ,line join=round,line cap=round] ( 72.70, 25.20) -- ( 72.70,126.47);

\path[draw=drawColor,line width= 0.4pt,dash pattern=on 1pt off 3pt ,line join=round,line cap=round] (100.97, 25.20) -- (100.97,126.47);

\path[draw=drawColor,line width= 0.4pt,dash pattern=on 1pt off 3pt ,line join=round,line cap=round] (129.25, 25.20) -- (129.25,126.47);

\path[draw=drawColor,line width= 0.4pt,dash pattern=on 1pt off 3pt ,line join=round,line cap=round] (157.52, 25.20) -- (157.52,126.47);

\path[draw=drawColor,line width= 0.4pt,dash pattern=on 1pt off 3pt ,line join=round,line cap=round] ( 25.20, 28.73) -- (162.61, 28.73);

\path[draw=drawColor,line width= 0.4pt,dash pattern=on 1pt off 3pt ,line join=round,line cap=round] ( 25.20, 53.28) -- (162.61, 53.28);

\path[draw=drawColor,line width= 0.4pt,dash pattern=on 1pt off 3pt ,line join=round,line cap=round] ( 25.20, 77.84) -- (162.61, 77.84);

\path[draw=drawColor,line width= 0.4pt,dash pattern=on 1pt off 3pt ,line join=round,line cap=round] ( 25.20,102.39) -- (162.61,102.39);
\end{scope}
\end{tikzpicture}

%% file: analysis_random-004.tex
\begin{tikzpicture}[x=1pt,y=1pt]
\definecolor{fillColor}{RGB}{255,255,255}
\path[use as bounding box,fill=fillColor,fill opacity=0.00] (0,0) rectangle (162.61,126.47);
\begin{scope}
\path[clip] ( 25.20, 25.20) rectangle (162.61,126.47);
\definecolor{drawColor}{RGB}{0,0,0}

\path[draw=drawColor,line width= 0.4pt,line join=round,line cap=round] ( 29.63, 30.11) -- ( 34.08, 30.11);

\path[draw=drawColor,line width= 0.4pt,line join=round,line cap=round] ( 31.86, 27.88) -- ( 31.86, 32.33);

\path[draw=drawColor,line width= 0.4pt,line join=round,line cap=round] ( 31.19, 31.26) -- ( 35.64, 31.26);

\path[draw=drawColor,line width= 0.4pt,line join=round,line cap=round] ( 33.42, 29.03) -- ( 33.42, 33.48);

\path[draw=drawColor,line width= 0.4pt,line join=round,line cap=round] ( 33.65, 33.07) -- ( 38.11, 33.07);

\path[draw=drawColor,line width= 0.4pt,line join=round,line cap=round] ( 35.88, 30.84) -- ( 35.88, 35.30);

\path[draw=drawColor,line width= 0.4pt,line join=round,line cap=round] ( 34.31, 33.55) -- ( 38.76, 33.55);

\path[draw=drawColor,line width= 0.4pt,line join=round,line cap=round] ( 36.54, 31.33) -- ( 36.54, 35.78);

\path[draw=drawColor,line width= 0.4pt,line join=round,line cap=round] ( 34.96, 34.04) -- ( 39.42, 34.04);

\path[draw=drawColor,line width= 0.4pt,line join=round,line cap=round] ( 37.19, 31.81) -- ( 37.19, 36.26);

\path[draw=drawColor,line width= 0.4pt,line join=round,line cap=round] ( 39.24, 37.19) -- ( 43.69, 37.19);

\path[draw=drawColor,line width= 0.4pt,line join=round,line cap=round] ( 41.47, 34.96) -- ( 41.47, 39.42);

\path[draw=drawColor,line width= 0.4pt,line join=round,line cap=round] ( 39.89, 37.67) -- ( 44.35, 37.67);

\path[draw=drawColor,line width= 0.4pt,line join=round,line cap=round] ( 42.12, 35.44) -- ( 42.12, 39.90);

\path[draw=drawColor,line width= 0.4pt,line join=round,line cap=round] ( 40.55, 38.15) -- ( 45.00, 38.15);

\path[draw=drawColor,line width= 0.4pt,line join=round,line cap=round] ( 42.77, 35.93) -- ( 42.77, 40.38);

\path[draw=drawColor,line width= 0.4pt,line join=round,line cap=round] ( 41.20, 38.63) -- ( 45.66, 38.63);

\path[draw=drawColor,line width= 0.4pt,line join=round,line cap=round] ( 43.43, 36.41) -- ( 43.43, 40.86);

\path[draw=drawColor,line width= 0.4pt,line join=round,line cap=round] ( 41.86, 39.12) -- ( 46.31, 39.12);

\path[draw=drawColor,line width= 0.4pt,line join=round,line cap=round] ( 44.08, 36.89) -- ( 44.08, 41.34);

\path[draw=drawColor,line width= 0.4pt,line join=round,line cap=round] ( 30.28, 28.53) -- ( 33.43, 31.68);

\path[draw=drawColor,line width= 0.4pt,line join=round,line cap=round] ( 30.28, 31.68) -- ( 33.43, 28.53);

\path[draw=drawColor,line width= 0.4pt,line join=round,line cap=round] ( 31.84, 29.68) -- ( 34.99, 32.83);

\path[draw=drawColor,line width= 0.4pt,line join=round,line cap=round] ( 31.84, 32.83) -- ( 34.99, 29.68);

\path[draw=drawColor,line width= 0.4pt,line join=round,line cap=round] ( 34.31, 31.50) -- ( 37.46, 34.65);

\path[draw=drawColor,line width= 0.4pt,line join=round,line cap=round] ( 34.31, 34.65) -- ( 37.46, 31.50);

\path[draw=drawColor,line width= 0.4pt,line join=round,line cap=round] ( 34.96, 31.98) -- ( 38.11, 35.13);

\path[draw=drawColor,line width= 0.4pt,line join=round,line cap=round] ( 34.96, 35.13) -- ( 38.11, 31.98);

\path[draw=drawColor,line width= 0.4pt,line join=round,line cap=round] ( 35.61, 32.46) -- ( 38.76, 35.61);

\path[draw=drawColor,line width= 0.4pt,line join=round,line cap=round] ( 35.61, 35.61) -- ( 38.76, 32.46);

\path[draw=drawColor,line width= 0.4pt,line join=round,line cap=round] ( 39.89, 35.61) -- ( 43.04, 38.76);

\path[draw=drawColor,line width= 0.4pt,line join=round,line cap=round] ( 39.89, 38.76) -- ( 43.04, 35.61);

\path[draw=drawColor,line width= 0.4pt,line join=round,line cap=round] ( 40.54, 36.10) -- ( 43.69, 39.25);

\path[draw=drawColor,line width= 0.4pt,line join=round,line cap=round] ( 40.54, 39.25) -- ( 43.69, 36.10);

\path[draw=drawColor,line width= 0.4pt,line join=round,line cap=round] ( 41.20, 36.58) -- ( 44.35, 39.73);

\path[draw=drawColor,line width= 0.4pt,line join=round,line cap=round] ( 41.20, 39.73) -- ( 44.35, 36.58);

\path[draw=drawColor,line width= 0.4pt,line join=round,line cap=round] ( 41.85, 37.06) -- ( 45.00, 40.21);

\path[draw=drawColor,line width= 0.4pt,line join=round,line cap=round] ( 41.85, 40.21) -- ( 45.00, 37.06);

\path[draw=drawColor,line width= 0.4pt,line join=round,line cap=round] ( 42.51, 37.54) -- ( 45.66, 40.69);

\path[draw=drawColor,line width= 0.4pt,line join=round,line cap=round] ( 42.51, 40.69) -- ( 45.66, 37.54);

\path[draw=drawColor,line width= 0.4pt,line join=round,line cap=round] ( 33.64, 30.57) circle (  1.58);

\path[draw=drawColor,line width= 0.4pt,line join=round,line cap=round] ( 35.64, 31.54) circle (  1.58);

\path[draw=drawColor,line width= 0.4pt,line join=round,line cap=round] ( 38.26, 33.32) circle (  1.58);

\path[draw=drawColor,line width= 0.4pt,line join=round,line cap=round] ( 38.78, 33.64) circle (  1.58);

\path[draw=drawColor,line width= 0.4pt,line join=round,line cap=round] ( 39.33, 34.04) circle (  1.58);

\path[draw=drawColor,line width= 0.4pt,line join=round,line cap=round] ( 42.16, 37.19) circle (  1.58);

\path[draw=drawColor,line width= 0.4pt,line join=round,line cap=round] ( 42.72, 37.67) circle (  1.58);

\path[draw=drawColor,line width= 0.4pt,line join=round,line cap=round] ( 43.26, 38.15) circle (  1.58);

\path[draw=drawColor,line width= 0.4pt,line join=round,line cap=round] ( 43.77, 38.63) circle (  1.58);

\path[draw=drawColor,line width= 0.4pt,line join=round,line cap=round] ( 44.26, 39.12) circle (  1.58);

\path[draw=drawColor,line width= 0.4pt,line join=round,line cap=round] ( 37.81, 34.02) --
	( 39.93, 30.35) --
	( 35.68, 30.35) --
	( 37.81, 34.02);

\path[draw=drawColor,line width= 0.4pt,line join=round,line cap=round] ( 42.34, 35.85) --
	( 44.46, 32.17) --
	( 40.22, 32.17) --
	( 42.34, 35.85);

\path[draw=drawColor,line width= 0.4pt,line join=round,line cap=round] ( 47.23, 38.36) --
	( 49.36, 34.68) --
	( 45.11, 34.68) --
	( 47.23, 38.36);

\path[draw=drawColor,line width= 0.4pt,line join=round,line cap=round] ( 50.67, 39.45) --
	( 52.79, 35.78) --
	( 48.55, 35.78) --
	( 50.67, 39.45);

\path[draw=drawColor,line width= 0.4pt,line join=round,line cap=round] ( 53.58, 40.28) --
	( 55.70, 36.61) --
	( 51.46, 36.61) --
	( 53.58, 40.28);

\path[draw=drawColor,line width= 0.4pt,line join=round,line cap=round] ( 60.35, 44.37) --
	( 62.47, 40.70) --
	( 58.23, 40.70) --
	( 60.35, 44.37);

\path[draw=drawColor,line width= 0.4pt,line join=round,line cap=round] ( 62.66, 44.68) --
	( 64.78, 41.01) --
	( 60.54, 41.01) --
	( 62.66, 44.68);

\path[draw=drawColor,line width= 0.4pt,line join=round,line cap=round] ( 65.49, 45.55) --
	( 67.61, 41.88) --
	( 63.37, 41.88) --
	( 65.49, 45.55);

\path[draw=drawColor,line width= 0.4pt,line join=round,line cap=round] ( 68.19, 46.43) --
	( 70.31, 42.76) --
	( 66.06, 42.76) --
	( 68.19, 46.43);

\path[draw=drawColor,line width= 0.4pt,line join=round,line cap=round] ( 70.67, 47.31) --
	( 72.79, 43.63) --
	( 68.55, 43.63) --
	( 70.67, 47.31);

\path[draw=drawColor,line width= 0.4pt,line join=round,line cap=round] ( 49.79, 35.09) --
	( 52.02, 37.32) --
	( 54.25, 35.09) --
	( 52.02, 32.86) --
	( 49.79, 35.09);

\path[draw=drawColor,line width= 0.4pt,line join=round,line cap=round] ( 65.91, 40.10) --
	( 68.13, 42.33) --
	( 70.36, 40.10) --
	( 68.13, 37.87) --
	( 65.91, 40.10);

\path[draw=drawColor,line width= 0.4pt,line join=round,line cap=round] ( 80.72, 44.83) --
	( 82.95, 47.06) --
	( 85.17, 44.83) --
	( 82.95, 42.60) --
	( 80.72, 44.83);

\path[draw=drawColor,line width= 0.4pt,line join=round,line cap=round] ( 92.40, 48.03) --
	( 94.62, 50.25) --
	( 96.85, 48.03) --
	( 94.62, 45.80) --
	( 92.40, 48.03);

\path[draw=drawColor,line width= 0.4pt,line join=round,line cap=round] (103.79, 52.22) --
	(106.01, 54.44) --
	(108.24, 52.22) --
	(106.01, 49.99) --
	(103.79, 52.22);

\path[draw=drawColor,line width= 0.4pt,line join=round,line cap=round] (117.07, 57.76) --
	(119.30, 59.99) --
	(121.53, 57.76) --
	(119.30, 55.53) --
	(117.07, 57.76);

\path[draw=drawColor,line width= 0.4pt,line join=round,line cap=round] (127.24, 60.43) --
	(129.47, 62.66) --
	(131.69, 60.43) --
	(129.47, 58.20) --
	(127.24, 60.43);

\path[draw=drawColor,line width= 0.4pt,line join=round,line cap=round] (136.38, 62.99) --
	(138.61, 65.22) --
	(140.84, 62.99) --
	(138.61, 60.76) --
	(136.38, 62.99);

\path[draw=drawColor,line width= 0.4pt,line join=round,line cap=round] (145.18, 65.39) --
	(147.41, 67.62) --
	(149.64, 65.39) --
	(147.41, 63.17) --
	(145.18, 65.39);

\path[draw=drawColor,line width= 0.4pt,line join=round,line cap=round] (152.20, 67.74) --
	(154.43, 69.97) --
	(156.65, 67.74) --
	(154.43, 65.52) --
	(152.20, 67.74);
\end{scope}
\begin{scope}
\path[clip] (  0.00,  0.00) rectangle (162.61,126.47);
\definecolor{drawColor}{RGB}{0,0,0}

\path[draw=drawColor,line width= 0.4pt,line join=round,line cap=round] ( 25.20, 25.20) --
	(162.61, 25.20) --
	(162.61,126.47) --
	( 25.20,126.47) --
	( 25.20, 25.20);
\end{scope}
\begin{scope}
\path[clip] (  0.00,  0.00) rectangle (162.61,126.47);
\definecolor{drawColor}{RGB}{0,0,0}

\node[text=drawColor,anchor=base,inner sep=0pt, outer sep=0pt, scale=  1.00] at ( 93.90,  1.68) {Memory used by Bev\quad[GB]};

\node[text=drawColor,rotate= 90.00,anchor=base,inner sep=0pt, outer sep=0pt, scale=  1.00] at (  6.72, 75.84) {Memory used by FK\quad[GB]};
\end{scope}
\begin{scope}
\path[clip] (  0.00,  0.00) rectangle (162.61,126.47);
\definecolor{drawColor}{RGB}{0,0,0}

\path[draw=drawColor,line width= 0.4pt,line join=round,line cap=round] ( 30.29, 25.20) -- (162.61, 25.20);

\path[draw=drawColor,line width= 0.4pt,line join=round,line cap=round] ( 30.29, 25.20) -- ( 30.29, 21.00);

\path[draw=drawColor,line width= 0.4pt,line join=round,line cap=round] ( 75.73, 25.20) -- ( 75.73, 21.00);

\path[draw=drawColor,line width= 0.4pt,line join=round,line cap=round] (121.17, 25.20) -- (121.17, 21.00);

\node[text=drawColor,anchor=base,inner sep=0pt, outer sep=0pt, scale=  1.00] at ( 30.29, 14.28) {0};

\node[text=drawColor,anchor=base,inner sep=0pt, outer sep=0pt, scale=  1.00] at ( 75.73, 14.28) {0.5};

\node[text=drawColor,anchor=base,inner sep=0pt, outer sep=0pt, scale=  1.00] at (121.17, 14.28) {1};

\path[draw=drawColor,line width= 0.4pt,line join=round,line cap=round] ( 25.20, 28.95) -- ( 25.20,126.47);

\path[draw=drawColor,line width= 0.4pt,line join=round,line cap=round] ( 25.20, 28.95) -- ( 21.00, 28.95);

\path[draw=drawColor,line width= 0.4pt,line join=round,line cap=round] ( 25.20, 62.44) -- ( 21.00, 62.44);

\path[draw=drawColor,line width= 0.4pt,line join=round,line cap=round] ( 25.20, 95.93) -- ( 21.00, 95.93);

\node[text=drawColor,rotate= 90.00,anchor=base,inner sep=0pt, outer sep=0pt, scale=  1.00] at ( 19.32, 28.95) {0};

\node[text=drawColor,rotate= 90.00,anchor=base,inner sep=0pt, outer sep=0pt, scale=  1.00] at ( 19.32, 62.44) {0.5};

\node[text=drawColor,rotate= 90.00,anchor=base,inner sep=0pt, outer sep=0pt, scale=  1.00] at ( 19.32, 95.93) {1};
\end{scope}
\begin{scope}
\path[clip] ( 25.20, 25.20) rectangle (162.61,126.47);
\definecolor{drawColor}{RGB}{190,190,190}

\path[draw=drawColor,line width= 0.4pt,line join=round,line cap=round] ( 25.20, 25.20) -- (162.61,126.47);
\definecolor{drawColor}{RGB}{211,211,211}

\path[draw=drawColor,line width= 0.4pt,dash pattern=on 1pt off 3pt ,line join=round,line cap=round] ( 30.29, 25.20) -- ( 30.29,126.47);

\path[draw=drawColor,line width= 0.4pt,dash pattern=on 1pt off 3pt ,line join=round,line cap=round] ( 48.46, 25.20) -- ( 48.46,126.47);

\path[draw=drawColor,line width= 0.4pt,dash pattern=on 1pt off 3pt ,line join=round,line cap=round] ( 66.64, 25.20) -- ( 66.64,126.47);

\path[draw=drawColor,line width= 0.4pt,dash pattern=on 1pt off 3pt ,line join=round,line cap=round] ( 84.82, 25.20) -- ( 84.82,126.47);

\path[draw=drawColor,line width= 0.4pt,dash pattern=on 1pt off 3pt ,line join=round,line cap=round] (102.99, 25.20) -- (102.99,126.47);

\path[draw=drawColor,line width= 0.4pt,dash pattern=on 1pt off 3pt ,line join=round,line cap=round] (121.17, 25.20) -- (121.17,126.47);

\path[draw=drawColor,line width= 0.4pt,dash pattern=on 1pt off 3pt ,line join=round,line cap=round] (139.34, 25.20) -- (139.34,126.47);

\path[draw=drawColor,line width= 0.4pt,dash pattern=on 1pt off 3pt ,line join=round,line cap=round] (157.52, 25.20) -- (157.52,126.47);

\path[draw=drawColor,line width= 0.4pt,dash pattern=on 1pt off 3pt ,line join=round,line cap=round] ( 25.20, 28.95) -- (162.61, 28.95);

\path[draw=drawColor,line width= 0.4pt,dash pattern=on 1pt off 3pt ,line join=round,line cap=round] ( 25.20, 42.35) -- (162.61, 42.35);

\path[draw=drawColor,line width= 0.4pt,dash pattern=on 1pt off 3pt ,line join=round,line cap=round] ( 25.20, 55.74) -- (162.61, 55.74);

\path[draw=drawColor,line width= 0.4pt,dash pattern=on 1pt off 3pt ,line join=round,line cap=round] ( 25.20, 69.14) -- (162.61, 69.14);

\path[draw=drawColor,line width= 0.4pt,dash pattern=on 1pt off 3pt ,line join=round,line cap=round] ( 25.20, 82.53) -- (162.61, 82.53);

\path[draw=drawColor,line width= 0.4pt,dash pattern=on 1pt off 3pt ,line join=round,line cap=round] ( 25.20, 95.93) -- (162.61, 95.93);

\path[draw=drawColor,line width= 0.4pt,dash pattern=on 1pt off 3pt ,line join=round,line cap=round] ( 25.20,109.33) -- (162.61,109.33);

\path[draw=drawColor,line width= 0.4pt,dash pattern=on 1pt off 3pt ,line join=round,line cap=round] ( 25.20,122.72) -- (162.61,122.72);
\end{scope}
\end{tikzpicture}

%% file: analysis_random-002.tex
\begin{tikzpicture}[x=1pt,y=1pt]
\definecolor{fillColor}{RGB}{255,255,255}
\path[use as bounding box,fill=fillColor,fill opacity=0.00] (0,0) rectangle (162.61,126.47);
\begin{scope}
\path[clip] ( 25.20, 25.20) rectangle (162.61,126.47);
\definecolor{drawColor}{RGB}{0,0,0}

\path[draw=drawColor,line width= 0.4pt,line join=round,line cap=round] ( 31.60, 31.56) -- ( 36.05, 31.56);

\path[draw=drawColor,line width= 0.4pt,line join=round,line cap=round] ( 33.83, 29.33) -- ( 33.83, 33.78);

\path[draw=drawColor,line width= 0.4pt,line join=round,line cap=round] ( 35.12, 34.15) -- ( 39.58, 34.15);

\path[draw=drawColor,line width= 0.4pt,line join=round,line cap=round] ( 37.35, 31.93) -- ( 37.35, 36.38);

\path[draw=drawColor,line width= 0.4pt,line join=round,line cap=round] ( 40.69, 38.26) -- ( 45.14, 38.26);

\path[draw=drawColor,line width= 0.4pt,line join=round,line cap=round] ( 42.92, 36.03) -- ( 42.92, 40.48);

\path[draw=drawColor,line width= 0.4pt,line join=round,line cap=round] ( 42.17, 39.35) -- ( 46.62, 39.35);

\path[draw=drawColor,line width= 0.4pt,line join=round,line cap=round] ( 44.39, 37.12) -- ( 44.39, 41.57);

\path[draw=drawColor,line width= 0.4pt,line join=round,line cap=round] ( 43.64, 40.44) -- ( 48.10, 40.44);

\path[draw=drawColor,line width= 0.4pt,line join=round,line cap=round] ( 45.87, 38.21) -- ( 45.87, 42.66);

\path[draw=drawColor,line width= 0.4pt,line join=round,line cap=round] ( 53.30, 47.55) -- ( 57.75, 47.55);

\path[draw=drawColor,line width= 0.4pt,line join=round,line cap=round] ( 55.53, 45.32) -- ( 55.53, 49.78);

\path[draw=drawColor,line width= 0.4pt,line join=round,line cap=round] ( 54.78, 48.64) -- ( 59.23, 48.64);

\path[draw=drawColor,line width= 0.4pt,line join=round,line cap=round] ( 57.00, 46.41) -- ( 57.00, 50.87);

\path[draw=drawColor,line width= 0.4pt,line join=round,line cap=round] ( 56.25, 49.73) -- ( 60.71, 49.73);

\path[draw=drawColor,line width= 0.4pt,line join=round,line cap=round] ( 58.48, 47.50) -- ( 58.48, 51.96);

\path[draw=drawColor,line width= 0.4pt,line join=round,line cap=round] ( 57.73, 50.82) -- ( 62.19, 50.82);

\path[draw=drawColor,line width= 0.4pt,line join=round,line cap=round] ( 59.96, 48.59) -- ( 59.96, 53.05);

\path[draw=drawColor,line width= 0.4pt,line join=round,line cap=round] ( 59.21, 51.91) -- ( 63.66, 51.91);

\path[draw=drawColor,line width= 0.4pt,line join=round,line cap=round] ( 61.44, 49.68) -- ( 61.44, 54.13);

\path[draw=drawColor,line width= 0.4pt,line join=round,line cap=round] ( 32.25, 29.98) -- ( 35.40, 33.13);

\path[draw=drawColor,line width= 0.4pt,line join=round,line cap=round] ( 32.25, 33.13) -- ( 35.40, 29.98);

\path[draw=drawColor,line width= 0.4pt,line join=round,line cap=round] ( 35.77, 32.58) -- ( 38.92, 35.73);

\path[draw=drawColor,line width= 0.4pt,line join=round,line cap=round] ( 35.77, 35.73) -- ( 38.92, 32.58);

\path[draw=drawColor,line width= 0.4pt,line join=round,line cap=round] ( 41.34, 36.68) -- ( 44.49, 39.83);

\path[draw=drawColor,line width= 0.4pt,line join=round,line cap=round] ( 41.34, 39.83) -- ( 44.49, 36.68);

\path[draw=drawColor,line width= 0.4pt,line join=round,line cap=round] ( 42.82, 37.77) -- ( 45.97, 40.92);

\path[draw=drawColor,line width= 0.4pt,line join=round,line cap=round] ( 42.82, 40.92) -- ( 45.97, 37.77);

\path[draw=drawColor,line width= 0.4pt,line join=round,line cap=round] ( 44.30, 38.86) -- ( 47.45, 42.01);

\path[draw=drawColor,line width= 0.4pt,line join=round,line cap=round] ( 44.30, 42.01) -- ( 47.45, 38.86);

\path[draw=drawColor,line width= 0.4pt,line join=round,line cap=round] ( 53.95, 45.98) -- ( 57.10, 49.13);

\path[draw=drawColor,line width= 0.4pt,line join=round,line cap=round] ( 53.95, 49.13) -- ( 57.10, 45.98);

\path[draw=drawColor,line width= 0.4pt,line join=round,line cap=round] ( 55.43, 47.07) -- ( 58.58, 50.22);

\path[draw=drawColor,line width= 0.4pt,line join=round,line cap=round] ( 55.43, 50.22) -- ( 58.58, 47.07);

\path[draw=drawColor,line width= 0.4pt,line join=round,line cap=round] ( 56.91, 48.15) -- ( 60.06, 51.30);

\path[draw=drawColor,line width= 0.4pt,line join=round,line cap=round] ( 56.91, 51.30) -- ( 60.06, 48.15);

\path[draw=drawColor,line width= 0.4pt,line join=round,line cap=round] ( 58.38, 49.24) -- ( 61.53, 52.39);

\path[draw=drawColor,line width= 0.4pt,line join=round,line cap=round] ( 58.38, 52.39) -- ( 61.53, 49.24);

\path[draw=drawColor,line width= 0.4pt,line join=round,line cap=round] ( 59.86, 50.33) -- ( 63.01, 53.48);

\path[draw=drawColor,line width= 0.4pt,line join=round,line cap=round] ( 59.86, 53.48) -- ( 63.01, 50.33);

\path[draw=drawColor,line width= 0.4pt,line join=round,line cap=round] ( 36.14, 32.60) circle (  1.58);

\path[draw=drawColor,line width= 0.4pt,line join=round,line cap=round] ( 40.10, 34.80) circle (  1.58);

\path[draw=drawColor,line width= 0.4pt,line join=round,line cap=round] ( 46.54, 38.81) circle (  1.58);

\path[draw=drawColor,line width= 0.4pt,line join=round,line cap=round] ( 48.51, 39.53) circle (  1.58);

\path[draw=drawColor,line width= 0.4pt,line join=round,line cap=round] ( 50.48, 40.44) circle (  1.58);

\path[draw=drawColor,line width= 0.4pt,line join=round,line cap=round] ( 57.61, 47.55) circle (  1.58);

\path[draw=drawColor,line width= 0.4pt,line join=round,line cap=round] ( 59.58, 48.64) circle (  1.58);

\path[draw=drawColor,line width= 0.4pt,line join=round,line cap=round] ( 61.55, 49.73) circle (  1.58);

\path[draw=drawColor,line width= 0.4pt,line join=round,line cap=round] ( 63.52, 50.82) circle (  1.58);

\path[draw=drawColor,line width= 0.4pt,line join=round,line cap=round] ( 65.49, 51.91) circle (  1.58);

\path[draw=drawColor,line width= 0.4pt,line join=round,line cap=round] ( 36.50, 37.32) --
	( 38.62, 33.65) --
	( 34.38, 33.65) --
	( 36.50, 37.32);

\path[draw=drawColor,line width= 0.4pt,line join=round,line cap=round] ( 42.67, 41.44) --
	( 44.79, 37.77) --
	( 40.54, 37.77) --
	( 42.67, 41.44);

\path[draw=drawColor,line width= 0.4pt,line join=round,line cap=round] ( 51.70, 47.11) --
	( 53.83, 43.44) --
	( 49.58, 43.44) --
	( 51.70, 47.11);

\path[draw=drawColor,line width= 0.4pt,line join=round,line cap=round] ( 54.87, 49.58) --
	( 56.99, 45.90) --
	( 52.75, 45.90) --
	( 54.87, 49.58);

\path[draw=drawColor,line width= 0.4pt,line join=round,line cap=round] ( 58.58, 51.46) --
	( 60.70, 47.79) --
	( 56.46, 47.79) --
	( 58.58, 51.46);

\path[draw=drawColor,line width= 0.4pt,line join=round,line cap=round] ( 73.13, 60.69) --
	( 75.25, 57.02) --
	( 71.01, 57.02) --
	( 73.13, 60.69);

\path[draw=drawColor,line width= 0.4pt,line join=round,line cap=round] ( 75.10, 61.39) --
	( 77.22, 57.72) --
	( 72.98, 57.72) --
	( 75.10, 61.39);

\path[draw=drawColor,line width= 0.4pt,line join=round,line cap=round] ( 78.48, 63.36) --
	( 80.61, 59.68) --
	( 76.36, 59.68) --
	( 78.48, 63.36);

\path[draw=drawColor,line width= 0.4pt,line join=round,line cap=round] ( 82.36, 65.34) --
	( 84.48, 61.67) --
	( 80.24, 61.67) --
	( 82.36, 65.34);

\path[draw=drawColor,line width= 0.4pt,line join=round,line cap=round] ( 86.04, 67.31) --
	( 88.16, 63.64) --
	( 83.92, 63.64) --
	( 86.04, 67.31);

\path[draw=drawColor,line width= 0.4pt,line join=round,line cap=round] ( 35.99, 42.82) --
	( 38.22, 45.04) --
	( 40.45, 42.82) --
	( 38.22, 40.59) --
	( 35.99, 42.82);

\path[draw=drawColor,line width= 0.4pt,line join=round,line cap=round] ( 43.91, 54.12) --
	( 46.14, 56.35) --
	( 48.37, 54.12) --
	( 46.14, 51.89) --
	( 43.91, 54.12);

\path[draw=drawColor,line width= 0.4pt,line join=round,line cap=round] ( 54.32, 64.81) --
	( 56.55, 67.03) --
	( 58.78, 64.81) --
	( 56.55, 62.58) --
	( 54.32, 64.81);

\path[draw=drawColor,line width= 0.4pt,line join=round,line cap=round] ( 59.76, 72.02) --
	( 61.99, 74.25) --
	( 64.22, 72.02) --
	( 61.99, 69.79) --
	( 59.76, 72.02);

\path[draw=drawColor,line width= 0.4pt,line join=round,line cap=round] ( 65.54, 81.48) --
	( 67.77, 83.71) --
	( 70.00, 81.48) --
	( 67.77, 79.26) --
	( 65.54, 81.48);

\path[draw=drawColor,line width= 0.4pt,line join=round,line cap=round] ( 80.60, 94.01) --
	( 82.82, 96.24) --
	( 85.05, 94.01) --
	( 82.82, 91.78) --
	( 80.60, 94.01);

\path[draw=drawColor,line width= 0.4pt,line join=round,line cap=round] ( 85.29,100.04) --
	( 87.52,102.26) --
	( 89.74,100.04) --
	( 87.52, 97.81) --
	( 85.29,100.04);

\path[draw=drawColor,line width= 0.4pt,line join=round,line cap=round] ( 91.47,105.81) --
	( 93.70,108.04) --
	( 95.92,105.81) --
	( 93.70,103.58) --
	( 91.47,105.81);

\path[draw=drawColor,line width= 0.4pt,line join=round,line cap=round] ( 97.17,111.24) --
	( 99.40,113.47) --
	(101.63,111.24) --
	( 99.40,109.01) --
	( 97.17,111.24);

\path[draw=drawColor,line width= 0.4pt,line join=round,line cap=round] (103.04,116.55) --
	(105.27,118.78) --
	(107.49,116.55) --
	(105.27,114.32) --
	(103.04,116.55);
\end{scope}
\begin{scope}
\path[clip] (  0.00,  0.00) rectangle (162.61,126.47);
\definecolor{drawColor}{RGB}{0,0,0}

\path[draw=drawColor,line width= 0.4pt,line join=round,line cap=round] ( 25.20, 25.20) --
	(162.61, 25.20) --
	(162.61,126.47) --
	( 25.20,126.47) --
	( 25.20, 25.20);
\end{scope}
\begin{scope}
\path[clip] (  0.00,  0.00) rectangle (162.61,126.47);
\definecolor{drawColor}{RGB}{0,0,0}

\node[text=drawColor,anchor=base,inner sep=0pt, outer sep=0pt, scale=  1.00] at ( 93.90,  1.68) {Memory used by Wei\quad[GB]};

\node[text=drawColor,rotate= 90.00,anchor=base,inner sep=0pt, outer sep=0pt, scale=  1.00] at (  6.72, 75.84) {Memory used by FK\quad[GB]};
\end{scope}
\begin{scope}
\path[clip] (  0.00,  0.00) rectangle (162.61,126.47);
\definecolor{drawColor}{RGB}{0,0,0}

\path[draw=drawColor,line width= 0.4pt,line join=round,line cap=round] ( 30.29, 25.20) -- (153.41, 25.20);

\path[draw=drawColor,line width= 0.4pt,line join=round,line cap=round] ( 30.29, 25.20) -- ( 30.29, 21.00);

\path[draw=drawColor,line width= 0.4pt,line join=round,line cap=round] ( 50.81, 25.20) -- ( 50.81, 21.00);

\path[draw=drawColor,line width= 0.4pt,line join=round,line cap=round] ( 71.33, 25.20) -- ( 71.33, 21.00);

\path[draw=drawColor,line width= 0.4pt,line join=round,line cap=round] ( 91.85, 25.20) -- ( 91.85, 21.00);

\path[draw=drawColor,line width= 0.4pt,line join=round,line cap=round] (112.37, 25.20) -- (112.37, 21.00);

\path[draw=drawColor,line width= 0.4pt,line join=round,line cap=round] (132.89, 25.20) -- (132.89, 21.00);

\path[draw=drawColor,line width= 0.4pt,line join=round,line cap=round] (153.41, 25.20) -- (153.41, 21.00);

\node[text=drawColor,anchor=base,inner sep=0pt, outer sep=0pt, scale=  1.00] at ( 30.29, 14.28) {0};

\node[text=drawColor,anchor=base,inner sep=0pt, outer sep=0pt, scale=  1.00] at ( 50.81, 14.28) {0.1};

\node[text=drawColor,anchor=base,inner sep=0pt, outer sep=0pt, scale=  1.00] at ( 71.33, 14.28) {0.2};

\node[text=drawColor,anchor=base,inner sep=0pt, outer sep=0pt, scale=  1.00] at ( 91.85, 14.28) {0.3};

\node[text=drawColor,anchor=base,inner sep=0pt, outer sep=0pt, scale=  1.00] at (112.37, 14.28) {0.4};

\node[text=drawColor,anchor=base,inner sep=0pt, outer sep=0pt, scale=  1.00] at (132.89, 14.28) {0.5};

\node[text=drawColor,anchor=base,inner sep=0pt, outer sep=0pt, scale=  1.00] at (153.41, 14.28) {0.6};

\path[draw=drawColor,line width= 0.4pt,line join=round,line cap=round] ( 25.20, 28.95) -- ( 25.20,119.70);

\path[draw=drawColor,line width= 0.4pt,line join=round,line cap=round] ( 25.20, 28.95) -- ( 21.00, 28.95);

\path[draw=drawColor,line width= 0.4pt,line join=round,line cap=round] ( 25.20, 44.08) -- ( 21.00, 44.08);

\path[draw=drawColor,line width= 0.4pt,line join=round,line cap=round] ( 25.20, 59.20) -- ( 21.00, 59.20);

\path[draw=drawColor,line width= 0.4pt,line join=round,line cap=round] ( 25.20, 74.32) -- ( 21.00, 74.32);

\path[draw=drawColor,line width= 0.4pt,line join=round,line cap=round] ( 25.20, 89.45) -- ( 21.00, 89.45);

\path[draw=drawColor,line width= 0.4pt,line join=round,line cap=round] ( 25.20,104.57) -- ( 21.00,104.57);

\path[draw=drawColor,line width= 0.4pt,line join=round,line cap=round] ( 25.20,119.70) -- ( 21.00,119.70);

\node[text=drawColor,rotate= 90.00,anchor=base,inner sep=0pt, outer sep=0pt, scale=  1.00] at ( 19.32, 28.95) {0};

\node[text=drawColor,rotate= 90.00,anchor=base,inner sep=0pt, outer sep=0pt, scale=  1.00] at ( 19.32, 59.20) {0.2};

\node[text=drawColor,rotate= 90.00,anchor=base,inner sep=0pt, outer sep=0pt, scale=  1.00] at ( 19.32, 89.45) {0.4};

\node[text=drawColor,rotate= 90.00,anchor=base,inner sep=0pt, outer sep=0pt, scale=  1.00] at ( 19.32,119.70) {0.6};
\end{scope}
\begin{scope}
\path[clip] ( 25.20, 25.20) rectangle (162.61,126.47);
\definecolor{drawColor}{RGB}{190,190,190}

\path[draw=drawColor,line width= 0.4pt,line join=round,line cap=round] ( 25.20, 25.20) -- (162.61,126.47);
\definecolor{drawColor}{RGB}{211,211,211}

\path[draw=drawColor,line width= 0.4pt,dash pattern=on 1pt off 3pt ,line join=round,line cap=round] ( 30.29, 25.20) -- ( 30.29,126.47);

\path[draw=drawColor,line width= 0.4pt,dash pattern=on 1pt off 3pt ,line join=round,line cap=round] ( 50.81, 25.20) -- ( 50.81,126.47);

\path[draw=drawColor,line width= 0.4pt,dash pattern=on 1pt off 3pt ,line join=round,line cap=round] ( 71.33, 25.20) -- ( 71.33,126.47);

\path[draw=drawColor,line width= 0.4pt,dash pattern=on 1pt off 3pt ,line join=round,line cap=round] ( 91.85, 25.20) -- ( 91.85,126.47);

\path[draw=drawColor,line width= 0.4pt,dash pattern=on 1pt off 3pt ,line join=round,line cap=round] (112.37, 25.20) -- (112.37,126.47);

\path[draw=drawColor,line width= 0.4pt,dash pattern=on 1pt off 3pt ,line join=round,line cap=round] (132.89, 25.20) -- (132.89,126.47);

\path[draw=drawColor,line width= 0.4pt,dash pattern=on 1pt off 3pt ,line join=round,line cap=round] (153.41, 25.20) -- (153.41,126.47);

\path[draw=drawColor,line width= 0.4pt,dash pattern=on 1pt off 3pt ,line join=round,line cap=round] ( 25.20, 28.95) -- (162.61, 28.95);

\path[draw=drawColor,line width= 0.4pt,dash pattern=on 1pt off 3pt ,line join=round,line cap=round] ( 25.20, 44.08) -- (162.61, 44.08);

\path[draw=drawColor,line width= 0.4pt,dash pattern=on 1pt off 3pt ,line join=round,line cap=round] ( 25.20, 59.20) -- (162.61, 59.20);

\path[draw=drawColor,line width= 0.4pt,dash pattern=on 1pt off 3pt ,line join=round,line cap=round] ( 25.20, 74.32) -- (162.61, 74.32);

\path[draw=drawColor,line width= 0.4pt,dash pattern=on 1pt off 3pt ,line join=round,line cap=round] ( 25.20, 89.45) -- (162.61, 89.45);

\path[draw=drawColor,line width= 0.4pt,dash pattern=on 1pt off 3pt ,line join=round,line cap=round] ( 25.20,104.57) -- (162.61,104.57);

\path[draw=drawColor,line width= 0.4pt,dash pattern=on 1pt off 3pt ,line join=round,line cap=round] ( 25.20,119.70) -- (162.61,119.70);
\end{scope}
\end{tikzpicture}

%% file: analysis_random-006.tex
\begin{tikzpicture}[x=1pt,y=1pt]
\definecolor{fillColor}{RGB}{255,255,255}
\path[use as bounding box,fill=fillColor,fill opacity=0.00] (0,0) rectangle (162.61,126.47);
\begin{scope}
\path[clip] ( 25.20, 25.20) rectangle (162.61,126.47);
\definecolor{drawColor}{RGB}{0,0,0}

\path[draw=drawColor,line width= 0.4pt,line join=round,line cap=round] ( 40.78, 31.44) -- ( 45.24, 31.44);

\path[draw=drawColor,line width= 0.4pt,line join=round,line cap=round] ( 43.01, 29.21) -- ( 43.01, 33.66);

\path[draw=drawColor,line width= 0.4pt,line join=round,line cap=round] ( 53.51, 33.91) -- ( 57.96, 33.91);

\path[draw=drawColor,line width= 0.4pt,line join=round,line cap=round] ( 55.73, 31.69) -- ( 55.73, 36.14);

\path[draw=drawColor,line width= 0.4pt,line join=round,line cap=round] ( 66.23, 37.83) -- ( 70.69, 37.83);

\path[draw=drawColor,line width= 0.4pt,line join=round,line cap=round] ( 68.46, 35.60) -- ( 68.46, 40.06);

\path[draw=drawColor,line width= 0.4pt,line join=round,line cap=round] ( 78.95, 38.87) -- ( 83.41, 38.87);

\path[draw=drawColor,line width= 0.4pt,line join=round,line cap=round] ( 81.18, 36.64) -- ( 81.18, 41.09);

\path[draw=drawColor,line width= 0.4pt,line join=round,line cap=round] ( 91.68, 39.91) -- ( 96.13, 39.91);

\path[draw=drawColor,line width= 0.4pt,line join=round,line cap=round] ( 93.90, 37.68) -- ( 93.90, 42.13);

\path[draw=drawColor,line width= 0.4pt,line join=round,line cap=round] (104.40, 46.69) -- (108.85, 46.69);

\path[draw=drawColor,line width= 0.4pt,line join=round,line cap=round] (106.63, 44.47) -- (106.63, 48.92);

\path[draw=drawColor,line width= 0.4pt,line join=round,line cap=round] (117.12, 47.73) -- (121.58, 47.73);

\path[draw=drawColor,line width= 0.4pt,line join=round,line cap=round] (119.35, 45.50) -- (119.35, 49.96);

\path[draw=drawColor,line width= 0.4pt,line join=round,line cap=round] (129.85, 48.77) -- (134.30, 48.77);

\path[draw=drawColor,line width= 0.4pt,line join=round,line cap=round] (132.07, 46.54) -- (132.07, 51.00);

\path[draw=drawColor,line width= 0.4pt,line join=round,line cap=round] (142.57, 49.81) -- (147.02, 49.81);

\path[draw=drawColor,line width= 0.4pt,line join=round,line cap=round] (144.80, 47.58) -- (144.80, 52.04);

\path[draw=drawColor,line width= 0.4pt,line join=round,line cap=round] (155.29, 50.85) -- (159.75, 50.85);

\path[draw=drawColor,line width= 0.4pt,line join=round,line cap=round] (157.52, 48.62) -- (157.52, 53.08);

\path[draw=drawColor,line width= 0.4pt,line join=round,line cap=round] ( 41.44, 29.86) -- ( 44.59, 33.01);

\path[draw=drawColor,line width= 0.4pt,line join=round,line cap=round] ( 41.44, 33.01) -- ( 44.59, 29.86);

\path[draw=drawColor,line width= 0.4pt,line join=round,line cap=round] ( 54.16, 32.34) -- ( 57.31, 35.49);

\path[draw=drawColor,line width= 0.4pt,line join=round,line cap=round] ( 54.16, 35.49) -- ( 57.31, 32.34);

\path[draw=drawColor,line width= 0.4pt,line join=round,line cap=round] ( 66.88, 36.25) -- ( 70.03, 39.40);

\path[draw=drawColor,line width= 0.4pt,line join=round,line cap=round] ( 66.88, 39.40) -- ( 70.03, 36.25);

\path[draw=drawColor,line width= 0.4pt,line join=round,line cap=round] ( 79.61, 37.29) -- ( 82.76, 40.44);

\path[draw=drawColor,line width= 0.4pt,line join=round,line cap=round] ( 79.61, 40.44) -- ( 82.76, 37.29);

\path[draw=drawColor,line width= 0.4pt,line join=round,line cap=round] ( 92.33, 38.33) -- ( 95.48, 41.48);

\path[draw=drawColor,line width= 0.4pt,line join=round,line cap=round] ( 92.33, 41.48) -- ( 95.48, 38.33);

\path[draw=drawColor,line width= 0.4pt,line join=round,line cap=round] (105.05, 45.12) -- (108.20, 48.27);

\path[draw=drawColor,line width= 0.4pt,line join=round,line cap=round] (105.05, 48.27) -- (108.20, 45.12);

\path[draw=drawColor,line width= 0.4pt,line join=round,line cap=round] (117.77, 46.16) -- (120.92, 49.31);

\path[draw=drawColor,line width= 0.4pt,line join=round,line cap=round] (117.77, 49.31) -- (120.92, 46.16);

\path[draw=drawColor,line width= 0.4pt,line join=round,line cap=round] (130.50, 47.20) -- (133.65, 50.35);

\path[draw=drawColor,line width= 0.4pt,line join=round,line cap=round] (130.50, 50.35) -- (133.65, 47.20);

\path[draw=drawColor,line width= 0.4pt,line join=round,line cap=round] (143.22, 48.23) -- (146.37, 51.38);

\path[draw=drawColor,line width= 0.4pt,line join=round,line cap=round] (143.22, 51.38) -- (146.37, 48.23);

\path[draw=drawColor,line width= 0.4pt,line join=round,line cap=round] (155.94, 49.27) -- (159.09, 52.42);

\path[draw=drawColor,line width= 0.4pt,line join=round,line cap=round] (155.94, 52.42) -- (159.09, 49.27);

\path[draw=drawColor,line width= 0.4pt,line join=round,line cap=round] ( 43.01, 32.43) circle (  1.58);

\path[draw=drawColor,line width= 0.4pt,line join=round,line cap=round] ( 55.73, 34.53) circle (  1.58);

\path[draw=drawColor,line width= 0.4pt,line join=round,line cap=round] ( 68.46, 38.35) circle (  1.58);

\path[draw=drawColor,line width= 0.4pt,line join=round,line cap=round] ( 81.18, 39.05) circle (  1.58);

\path[draw=drawColor,line width= 0.4pt,line join=round,line cap=round] ( 93.90, 39.91) circle (  1.58);

\path[draw=drawColor,line width= 0.4pt,line join=round,line cap=round] (106.63, 46.69) circle (  1.58);

\path[draw=drawColor,line width= 0.4pt,line join=round,line cap=round] (119.35, 47.73) circle (  1.58);

\path[draw=drawColor,line width= 0.4pt,line join=round,line cap=round] (132.07, 48.77) circle (  1.58);

\path[draw=drawColor,line width= 0.4pt,line join=round,line cap=round] (144.80, 49.81) circle (  1.58);

\path[draw=drawColor,line width= 0.4pt,line join=round,line cap=round] (157.52, 50.85) circle (  1.58);

\path[draw=drawColor,line width= 0.4pt,line join=round,line cap=round] ( 43.01, 37.05) --
	( 45.13, 33.37) --
	( 40.89, 33.37) --
	( 43.01, 37.05);

\path[draw=drawColor,line width= 0.4pt,line join=round,line cap=round] ( 55.73, 40.98) --
	( 57.86, 37.31) --
	( 53.61, 37.31) --
	( 55.73, 40.98);

\path[draw=drawColor,line width= 0.4pt,line join=round,line cap=round] ( 68.46, 46.38) --
	( 70.58, 42.71) --
	( 66.34, 42.71) --
	( 68.46, 46.38);

\path[draw=drawColor,line width= 0.4pt,line join=round,line cap=round] ( 81.18, 48.74) --
	( 83.30, 45.06) --
	( 79.06, 45.06) --
	( 81.18, 48.74);

\path[draw=drawColor,line width= 0.4pt,line join=round,line cap=round] ( 93.90, 50.54) --
	( 96.02, 46.86) --
	( 91.78, 46.86) --
	( 93.90, 50.54);

\path[draw=drawColor,line width= 0.4pt,line join=round,line cap=round] (106.63, 59.34) --
	(108.75, 55.67) --
	(104.51, 55.67) --
	(106.63, 59.34);

\path[draw=drawColor,line width= 0.4pt,line join=round,line cap=round] (119.35, 60.01) --
	(121.47, 56.33) --
	(117.23, 56.33) --
	(119.35, 60.01);

\path[draw=drawColor,line width= 0.4pt,line join=round,line cap=round] (132.07, 61.88) --
	(134.19, 58.21) --
	(129.95, 58.21) --
	(132.07, 61.88);

\path[draw=drawColor,line width= 0.4pt,line join=round,line cap=round] (144.80, 63.77) --
	(146.92, 60.10) --
	(142.67, 60.10) --
	(144.80, 63.77);

\path[draw=drawColor,line width= 0.4pt,line join=round,line cap=round] (157.52, 65.66) --
	(159.64, 61.98) --
	(155.40, 61.98) --
	(157.52, 65.66);

\path[draw=drawColor,line width= 0.4pt,line join=round,line cap=round] ( 40.78, 42.18) --
	( 43.01, 44.40) --
	( 45.24, 42.18) --
	( 43.01, 39.95) --
	( 40.78, 42.18);

\path[draw=drawColor,line width= 0.4pt,line join=round,line cap=round] ( 53.51, 52.96) --
	( 55.73, 55.19) --
	( 57.96, 52.96) --
	( 55.73, 50.73) --
	( 53.51, 52.96);

\path[draw=drawColor,line width= 0.4pt,line join=round,line cap=round] ( 66.23, 63.15) --
	( 68.46, 65.38) --
	( 70.69, 63.15) --
	( 68.46, 60.92) --
	( 66.23, 63.15);

\path[draw=drawColor,line width= 0.4pt,line join=round,line cap=round] ( 78.95, 70.03) --
	( 81.18, 72.26) --
	( 83.41, 70.03) --
	( 81.18, 67.81) --
	( 78.95, 70.03);

\path[draw=drawColor,line width= 0.4pt,line join=round,line cap=round] ( 91.68, 79.06) --
	( 93.90, 81.29) --
	( 96.13, 79.06) --
	( 93.90, 76.83) --
	( 91.68, 79.06);

\path[draw=drawColor,line width= 0.4pt,line join=round,line cap=round] (104.40, 91.01) --
	(106.63, 93.23) --
	(108.85, 91.01) --
	(106.63, 88.78) --
	(104.40, 91.01);

\path[draw=drawColor,line width= 0.4pt,line join=round,line cap=round] (117.12, 96.76) --
	(119.35, 98.98) --
	(121.58, 96.76) --
	(119.35, 94.53) --
	(117.12, 96.76);

\path[draw=drawColor,line width= 0.4pt,line join=round,line cap=round] (129.85,102.26) --
	(132.07,104.49) --
	(134.30,102.26) --
	(132.07,100.04) --
	(129.85,102.26);

\path[draw=drawColor,line width= 0.4pt,line join=round,line cap=round] (142.57,107.44) --
	(144.80,109.67) --
	(147.02,107.44) --
	(144.80,105.22) --
	(142.57,107.44);

\path[draw=drawColor,line width= 0.4pt,line join=round,line cap=round] (155.29,112.51) --
	(157.52,114.73) --
	(159.75,112.51) --
	(157.52,110.28) --
	(155.29,112.51);
\end{scope}
\begin{scope}
\path[clip] (  0.00,  0.00) rectangle (162.61,126.47);
\definecolor{drawColor}{RGB}{0,0,0}

\path[draw=drawColor,line width= 0.4pt,line join=round,line cap=round] ( 25.20, 25.20) --
	(162.61, 25.20) --
	(162.61,126.47) --
	( 25.20,126.47) --
	( 25.20, 25.20);
\end{scope}
\begin{scope}
\path[clip] (  0.00,  0.00) rectangle (162.61,126.47);
\definecolor{drawColor}{RGB}{0,0,0}

\node[text=drawColor,anchor=base,inner sep=0pt, outer sep=0pt, scale=  1.00] at ( 93.90,  1.68) {$|E_{\mathrm{in}}|\quad[10^5]$};

\node[text=drawColor,rotate= 90.00,anchor=base,inner sep=0pt, outer sep=0pt, scale=  1.00] at (  6.72, 75.84) {Memory used by FK\quad[GB]};
\end{scope}
\begin{scope}
\path[clip] (  0.00,  0.00) rectangle (162.61,126.47);
\definecolor{drawColor}{RGB}{0,0,0}

\path[draw=drawColor,line width= 0.4pt,line join=round,line cap=round] ( 30.29, 25.20) -- (157.52, 25.20);

\path[draw=drawColor,line width= 0.4pt,line join=round,line cap=round] ( 30.29, 25.20) -- ( 30.29, 21.00);

\path[draw=drawColor,line width= 0.4pt,line join=round,line cap=round] ( 43.01, 25.20) -- ( 43.01, 21.00);

\path[draw=drawColor,line width= 0.4pt,line join=round,line cap=round] ( 55.73, 25.20) -- ( 55.73, 21.00);

\path[draw=drawColor,line width= 0.4pt,line join=round,line cap=round] ( 68.46, 25.20) -- ( 68.46, 21.00);

\path[draw=drawColor,line width= 0.4pt,line join=round,line cap=round] ( 81.18, 25.20) -- ( 81.18, 21.00);

\path[draw=drawColor,line width= 0.4pt,line join=round,line cap=round] ( 93.90, 25.20) -- ( 93.90, 21.00);

\path[draw=drawColor,line width= 0.4pt,line join=round,line cap=round] (106.63, 25.20) -- (106.63, 21.00);

\path[draw=drawColor,line width= 0.4pt,line join=round,line cap=round] (119.35, 25.20) -- (119.35, 21.00);

\path[draw=drawColor,line width= 0.4pt,line join=round,line cap=round] (132.07, 25.20) -- (132.07, 21.00);

\path[draw=drawColor,line width= 0.4pt,line join=round,line cap=round] (144.80, 25.20) -- (144.80, 21.00);

\path[draw=drawColor,line width= 0.4pt,line join=round,line cap=round] (157.52, 25.20) -- (157.52, 21.00);

\node[text=drawColor,anchor=base,inner sep=0pt, outer sep=0pt, scale=  1.00] at ( 30.29, 14.28) {0};

\node[text=drawColor,anchor=base,inner sep=0pt, outer sep=0pt, scale=  1.00] at ( 55.73, 14.28) {2};

\node[text=drawColor,anchor=base,inner sep=0pt, outer sep=0pt, scale=  1.00] at ( 81.18, 14.28) {4};

\node[text=drawColor,anchor=base,inner sep=0pt, outer sep=0pt, scale=  1.00] at (106.63, 14.28) {6};

\node[text=drawColor,anchor=base,inner sep=0pt, outer sep=0pt, scale=  1.00] at (132.07, 14.28) {8};

\node[text=drawColor,anchor=base,inner sep=0pt, outer sep=0pt, scale=  1.00] at (157.52, 14.28) {10};

\path[draw=drawColor,line width= 0.4pt,line join=round,line cap=round] ( 25.20, 28.95) -- ( 25.20,126.47);

\path[draw=drawColor,line width= 0.4pt,line join=round,line cap=round] ( 25.20, 28.95) -- ( 21.00, 28.95);

\path[draw=drawColor,line width= 0.4pt,line join=round,line cap=round] ( 25.20, 43.38) -- ( 21.00, 43.38);

\path[draw=drawColor,line width= 0.4pt,line join=round,line cap=round] ( 25.20, 57.80) -- ( 21.00, 57.80);

\path[draw=drawColor,line width= 0.4pt,line join=round,line cap=round] ( 25.20, 72.23) -- ( 21.00, 72.23);

\path[draw=drawColor,line width= 0.4pt,line join=round,line cap=round] ( 25.20, 86.66) -- ( 21.00, 86.66);

\path[draw=drawColor,line width= 0.4pt,line join=round,line cap=round] ( 25.20,101.08) -- ( 21.00,101.08);

\path[draw=drawColor,line width= 0.4pt,line join=round,line cap=round] ( 25.20,115.51) -- ( 21.00,115.51);

\node[text=drawColor,rotate= 90.00,anchor=base,inner sep=0pt, outer sep=0pt, scale=  1.00] at ( 19.32, 28.95) {0};

\node[text=drawColor,rotate= 90.00,anchor=base,inner sep=0pt, outer sep=0pt, scale=  1.00] at ( 19.32, 57.80) {0.2};

\node[text=drawColor,rotate= 90.00,anchor=base,inner sep=0pt, outer sep=0pt, scale=  1.00] at ( 19.32, 86.66) {0.4};

\node[text=drawColor,rotate= 90.00,anchor=base,inner sep=0pt, outer sep=0pt, scale=  1.00] at ( 19.32,115.51) {0.6};
\end{scope}
\begin{scope}
\path[clip] ( 25.20, 25.20) rectangle (162.61,126.47);
\definecolor{drawColor}{RGB}{211,211,211}

\path[draw=drawColor,line width= 0.4pt,dash pattern=on 1pt off 3pt ,line join=round,line cap=round] ( 30.29, 25.20) -- ( 30.29,126.47);

\path[draw=drawColor,line width= 0.4pt,dash pattern=on 1pt off 3pt ,line join=round,line cap=round] ( 55.73, 25.20) -- ( 55.73,126.47);

\path[draw=drawColor,line width= 0.4pt,dash pattern=on 1pt off 3pt ,line join=round,line cap=round] ( 81.18, 25.20) -- ( 81.18,126.47);

\path[draw=drawColor,line width= 0.4pt,dash pattern=on 1pt off 3pt ,line join=round,line cap=round] (106.63, 25.20) -- (106.63,126.47);

\path[draw=drawColor,line width= 0.4pt,dash pattern=on 1pt off 3pt ,line join=round,line cap=round] (132.07, 25.20) -- (132.07,126.47);

\path[draw=drawColor,line width= 0.4pt,dash pattern=on 1pt off 3pt ,line join=round,line cap=round] (157.52, 25.20) -- (157.52,126.47);

\path[draw=drawColor,line width= 0.4pt,dash pattern=on 1pt off 3pt ,line join=round,line cap=round] ( 25.20, 28.95) -- (162.61, 28.95);

\path[draw=drawColor,line width= 0.4pt,dash pattern=on 1pt off 3pt ,line join=round,line cap=round] ( 25.20, 43.38) -- (162.61, 43.38);

\path[draw=drawColor,line width= 0.4pt,dash pattern=on 1pt off 3pt ,line join=round,line cap=round] ( 25.20, 57.80) -- (162.61, 57.80);

\path[draw=drawColor,line width= 0.4pt,dash pattern=on 1pt off 3pt ,line join=round,line cap=round] ( 25.20, 72.23) -- (162.61, 72.23);

\path[draw=drawColor,line width= 0.4pt,dash pattern=on 1pt off 3pt ,line join=round,line cap=round] ( 25.20, 86.66) -- (162.61, 86.66);

\path[draw=drawColor,line width= 0.4pt,dash pattern=on 1pt off 3pt ,line join=round,line cap=round] ( 25.20,101.08) -- (162.61,101.08);

\path[draw=drawColor,line width= 0.4pt,dash pattern=on 1pt off 3pt ,line join=round,line cap=round] ( 25.20,115.51) -- (162.61,115.51);
\end{scope}
\end{tikzpicture}

%% file: analysis_bio-003.tex
\begin{tikzpicture}[x=1pt,y=1pt]
\definecolor{fillColor}{RGB}{255,255,255}
\path[use as bounding box,fill=fillColor,fill opacity=0.00] (0,0) rectangle (245.72,126.47);
\begin{scope}
\path[clip] ( 36.00, 36.00) rectangle (245.72,124.07);
\definecolor{fillColor}{RGB}{0,0,0}

\path[fill=fillColor] (170.19, 92.06) rectangle (171.19, 93.06);

\path[fill=fillColor] (130.24, 75.29) rectangle (131.24, 76.29);

\path[fill=fillColor] (163.24, 89.15) rectangle (164.24, 90.15);

\path[fill=fillColor] (150.60, 83.84) rectangle (151.60, 84.84);

\path[fill=fillColor] (127.73, 74.49) rectangle (128.73, 75.49);

\path[fill=fillColor] (139.66, 79.24) rectangle (140.66, 80.24);

\path[fill=fillColor] (166.61, 90.56) rectangle (167.61, 91.56);

\path[fill=fillColor] (152.06, 84.45) rectangle (153.06, 85.45);

\path[fill=fillColor] (160.93, 88.18) rectangle (161.93, 89.18);

\path[fill=fillColor] (129.20, 74.85) rectangle (130.20, 75.85);

\path[fill=fillColor] (153.36, 85.00) rectangle (154.36, 86.00);

\path[fill=fillColor] (128.57, 74.59) rectangle (129.57, 75.59);

\path[fill=fillColor] (135.01, 77.29) rectangle (136.01, 78.29);

\path[fill=fillColor] (155.25, 85.79) rectangle (156.25, 86.79);

\path[fill=fillColor] (167.53, 90.95) rectangle (168.53, 91.95);

\path[fill=fillColor] (127.93, 74.32) rectangle (128.93, 75.32);

\path[fill=fillColor] (150.99, 84.00) rectangle (151.99, 85.00);

\path[fill=fillColor] (137.28, 78.24) rectangle (138.28, 79.24);

\path[fill=fillColor] (143.99, 81.88) rectangle (144.99, 82.88);

\path[fill=fillColor] (156.70, 86.41) rectangle (157.70, 87.41);

\path[fill=fillColor] (136.13, 78.01) rectangle (137.13, 79.01);

\path[fill=fillColor] (136.28, 77.83) rectangle (137.28, 78.83);

\path[fill=fillColor] ( 93.95, 60.05) rectangle ( 94.95, 61.05);

\path[fill=fillColor] (  1.95, 21.41) rectangle (  2.95, 22.41);

\path[fill=fillColor] (152.43, 84.61) rectangle (153.43, 85.61);

\path[fill=fillColor] ( 90.77, 58.71) rectangle ( 91.77, 59.71);

\path[fill=fillColor] (138.69, 78.84) rectangle (139.69, 79.84);

\path[fill=fillColor] (116.07, 69.34) rectangle (117.07, 70.34);

\path[fill=fillColor] (149.89, 83.67) rectangle (150.89, 84.67);

\path[fill=fillColor] ( 88.93, 57.94) rectangle ( 89.93, 58.94);

\path[fill=fillColor] (144.01, 81.07) rectangle (145.01, 82.07);

\path[fill=fillColor] (166.16, 90.37) rectangle (167.16, 91.37);

\path[fill=fillColor] (112.62, 67.89) rectangle (113.62, 68.89);

\path[fill=fillColor] ( 80.56, 54.53) rectangle ( 81.56, 55.53);

\path[fill=fillColor] (107.16, 65.60) rectangle (108.16, 66.60);

\path[fill=fillColor] (111.89, 67.58) rectangle (112.89, 68.58);

\path[fill=fillColor] (123.08, 72.28) rectangle (124.08, 73.28);

\path[fill=fillColor] (133.65, 76.78) rectangle (134.65, 77.78);

\path[fill=fillColor] (134.89, 77.24) rectangle (135.89, 78.24);

\path[fill=fillColor] (149.32, 83.30) rectangle (150.32, 84.30);

\path[fill=fillColor] (131.78, 78.18) rectangle (132.78, 79.18);

\path[fill=fillColor] (153.27, 84.96) rectangle (154.27, 85.96);

\path[fill=fillColor] ( 82.85, 56.45) rectangle ( 83.85, 57.45);

\path[fill=fillColor] (154.79, 85.60) rectangle (155.79, 86.60);

\path[fill=fillColor] ( 84.61, 56.87) rectangle ( 85.61, 57.87);

\path[fill=fillColor] (146.33, 83.18) rectangle (147.33, 84.18);

\path[fill=fillColor] (118.19, 70.23) rectangle (119.19, 71.23);

\path[fill=fillColor] (126.24, 73.61) rectangle (127.24, 74.61);

\path[fill=fillColor] (119.86, 70.93) rectangle (120.86, 71.93);

\path[fill=fillColor] (153.13, 84.90) rectangle (154.13, 85.90);

\path[fill=fillColor] (144.17, 81.14) rectangle (145.17, 82.14);

\path[fill=fillColor] (137.84, 78.48) rectangle (138.84, 79.48);

\path[fill=fillColor] (160.86, 88.15) rectangle (161.86, 89.15);

\path[fill=fillColor] (122.88, 72.20) rectangle (123.88, 73.20);

\path[fill=fillColor] (125.41, 73.42) rectangle (126.41, 74.42);

\path[fill=fillColor] ( 89.96, 58.37) rectangle ( 90.96, 59.37);

\path[fill=fillColor] (149.36, 83.65) rectangle (150.36, 84.65);

\path[fill=fillColor] ( 87.54, 57.39) rectangle ( 88.54, 58.39);

\path[fill=fillColor] (118.41, 70.32) rectangle (119.41, 71.32);

\path[fill=fillColor] ( 92.07, 59.26) rectangle ( 93.07, 60.26);

\path[fill=fillColor] (102.98, 64.00) rectangle (103.98, 65.00);

\path[fill=fillColor] (102.66, 63.71) rectangle (103.66, 64.71);

\path[fill=fillColor] ( 86.80, 57.04) rectangle ( 87.80, 58.04);

\path[fill=fillColor] ( 41.83, 38.73) rectangle ( 42.83, 39.73);

\path[fill=fillColor] (145.39, 83.20) rectangle (146.39, 84.20);

\path[fill=fillColor] (127.15, 73.99) rectangle (128.15, 74.99);

\path[fill=fillColor] ( 79.56, 54.48) rectangle ( 80.56, 55.48);

\path[fill=fillColor] (109.48, 66.57) rectangle (110.48, 67.57);

\path[fill=fillColor] (145.11, 81.69) rectangle (146.11, 82.69);

\path[fill=fillColor] (105.98, 65.10) rectangle (106.98, 66.10);

\path[fill=fillColor] (115.97, 69.29) rectangle (116.97, 70.29);

\path[fill=fillColor] (144.45, 81.46) rectangle (145.45, 82.46);

\path[fill=fillColor] ( 95.46, 60.68) rectangle ( 96.46, 61.68);

\path[fill=fillColor] (120.35, 71.13) rectangle (121.35, 72.13);

\path[fill=fillColor] (130.69, 75.48) rectangle (131.69, 76.48);

\path[fill=fillColor] (151.86, 85.87) rectangle (152.86, 86.87);

\path[fill=fillColor] (110.96, 67.19) rectangle (111.96, 68.19);

\path[fill=fillColor] (154.09, 85.44) rectangle (155.09, 86.44);

\path[fill=fillColor] (155.79, 86.02) rectangle (156.79, 87.02);

\path[fill=fillColor] (128.92, 74.73) rectangle (129.92, 75.73);

\path[fill=fillColor] (106.44, 65.29) rectangle (107.44, 66.29);

\path[fill=fillColor] (153.39, 85.31) rectangle (154.39, 86.31);

\path[fill=fillColor] (161.70, 88.50) rectangle (162.70, 89.50);

\path[fill=fillColor] (107.82, 65.87) rectangle (108.82, 66.87);

\path[fill=fillColor] (129.93, 75.16) rectangle (130.93, 76.16);

\path[fill=fillColor] ( 79.69, 54.26) rectangle ( 80.69, 55.26);

\path[fill=fillColor] (124.80, 73.00) rectangle (125.80, 74.00);

\path[fill=fillColor] (130.73, 75.49) rectangle (131.73, 76.49);

\path[fill=fillColor] ( 67.25, 49.06) rectangle ( 68.25, 50.06);

\path[fill=fillColor] ( 67.21, 48.99) rectangle ( 68.21, 49.99);

\path[fill=fillColor] ( 92.84, 59.58) rectangle ( 93.84, 60.58);

\path[fill=fillColor] (114.80, 68.97) rectangle (115.80, 69.97);

\path[fill=fillColor] (119.05, 70.59) rectangle (120.05, 71.59);

\path[fill=fillColor] (139.32, 79.56) rectangle (140.32, 80.56);

\path[fill=fillColor] (108.73, 66.25) rectangle (109.73, 67.25);

\path[fill=fillColor] (143.89, 81.31) rectangle (144.89, 82.31);

\path[fill=fillColor] ( 95.72, 60.79) rectangle ( 96.72, 61.79);

\path[fill=fillColor] ( 54.91, 43.65) rectangle ( 55.91, 44.65);

\path[fill=fillColor] (152.82, 84.77) rectangle (153.82, 85.77);

\path[fill=fillColor] (110.68, 67.07) rectangle (111.68, 68.07);

\path[fill=fillColor] ( 91.02, 58.81) rectangle ( 92.02, 59.81);

\path[fill=fillColor] ( 81.45, 54.80) rectangle ( 82.45, 55.80);

\path[fill=fillColor] (101.73, 63.31) rectangle (102.73, 64.31);

\path[fill=fillColor] (107.23, 65.75) rectangle (108.23, 66.75);

\path[fill=fillColor] ( 37.10, 36.79) rectangle ( 38.10, 37.79);

\path[fill=fillColor] (122.77, 72.19) rectangle (123.77, 73.19);

\path[fill=fillColor] ( 39.09, 37.19) rectangle ( 40.09, 38.19);

\path[fill=fillColor] ( 86.73, 57.14) rectangle ( 87.73, 58.14);

\path[fill=fillColor] ( 53.18, 42.93) rectangle ( 54.18, 43.93);

\path[fill=fillColor] (127.48, 74.13) rectangle (128.48, 75.13);

\path[fill=fillColor] ( 20.70, 29.28) rectangle ( 21.70, 30.28);

\path[fill=fillColor] (133.31, 76.58) rectangle (134.31, 77.58);

\path[fill=fillColor] ( 91.86, 59.17) rectangle ( 92.86, 60.17);

\path[fill=fillColor] (128.06, 74.37) rectangle (129.06, 75.37);

\path[fill=fillColor] (128.36, 74.50) rectangle (129.36, 75.50);

\path[fill=fillColor] ( 89.40, 58.22) rectangle ( 90.40, 59.22);

\path[fill=fillColor] (142.40, 80.39) rectangle (143.40, 81.39);

\path[fill=fillColor] (136.64, 78.29) rectangle (137.64, 79.29);

\path[fill=fillColor] (145.39, 81.65) rectangle (146.39, 82.65);

\path[fill=fillColor] ( 52.14, 42.49) rectangle ( 53.14, 43.49);

\path[fill=fillColor] (104.03, 64.28) rectangle (105.03, 65.28);

\path[fill=fillColor] ( 66.21, 48.46) rectangle ( 67.21, 49.46);

\path[fill=fillColor] (130.11, 75.56) rectangle (131.11, 76.56);

\path[fill=fillColor] ( 83.18, 55.56) rectangle ( 84.18, 56.56);

\path[fill=fillColor] ( 41.29, 37.93) rectangle ( 42.29, 38.93);

\path[fill=fillColor] (106.92, 65.86) rectangle (107.92, 66.86);

\path[fill=fillColor] ( 76.21, 53.10) rectangle ( 77.21, 54.10);

\path[fill=fillColor] (127.95, 74.32) rectangle (128.95, 75.32);

\path[fill=fillColor] ( 68.34, 49.29) rectangle ( 69.34, 50.29);

\path[fill=fillColor] ( 87.33, 57.27) rectangle ( 88.33, 58.27);

\path[fill=fillColor] (114.14, 68.99) rectangle (115.14, 69.99);

\path[fill=fillColor] (100.69, 62.88) rectangle (101.69, 63.88);

\path[fill=fillColor] (121.31, 72.10) rectangle (122.31, 73.10);

\path[fill=fillColor] ( 92.80, 59.72) rectangle ( 93.80, 60.72);

\path[fill=fillColor] (115.57, 69.41) rectangle (116.57, 70.41);

\path[fill=fillColor] ( 82.22, 55.13) rectangle ( 83.22, 56.13);

\path[fill=fillColor] ( 48.44, 40.95) rectangle ( 49.44, 41.95);

\path[fill=fillColor] ( 83.54, 57.51) rectangle ( 84.54, 58.51);

\path[fill=fillColor] (105.21, 65.38) rectangle (106.21, 66.38);

\path[fill=fillColor] (133.67, 76.73) rectangle (134.67, 77.73);

\path[fill=fillColor] (114.81, 68.81) rectangle (115.81, 69.81);

\path[fill=fillColor] ( 88.87, 58.91) rectangle ( 89.87, 59.91);

\path[fill=fillColor] ( 94.12, 60.62) rectangle ( 95.12, 61.62);

\path[fill=fillColor] ( 55.61, 43.91) rectangle ( 56.61, 44.91);

\path[fill=fillColor] (120.64, 71.26) rectangle (121.64, 72.26);

\path[fill=fillColor] ( 95.66, 60.76) rectangle ( 96.66, 61.76);

\path[fill=fillColor] ( 95.17, 60.56) rectangle ( 96.17, 61.56);

\path[fill=fillColor] (105.96, 66.15) rectangle (106.96, 67.15);

\path[fill=fillColor] (100.66, 63.06) rectangle (101.66, 64.06);

\path[fill=fillColor] ( 84.43, 56.30) rectangle ( 85.43, 57.30);

\path[fill=fillColor] ( 71.58, 50.65) rectangle ( 72.58, 51.65);

\path[fill=fillColor] (103.13, 63.90) rectangle (104.13, 64.90);

\path[fill=fillColor] (102.55, 63.66) rectangle (103.55, 64.66);

\path[fill=fillColor] ( 85.47, 56.48) rectangle ( 86.47, 57.48);

\path[fill=fillColor] ( 91.66, 59.42) rectangle ( 92.66, 60.42);

\path[fill=fillColor] (120.43, 72.37) rectangle (121.43, 73.37);

\path[fill=fillColor] (123.62, 72.51) rectangle (124.62, 73.51);

\path[fill=fillColor] ( 93.99, 60.24) rectangle ( 94.99, 61.24);

\path[fill=fillColor] (104.96, 64.67) rectangle (105.96, 65.67);

\path[fill=fillColor] ( 67.41, 49.96) rectangle ( 68.41, 50.96);

\path[fill=fillColor] (112.80, 68.88) rectangle (113.80, 69.88);

\path[fill=fillColor] ( 74.30, 51.79) rectangle ( 75.30, 52.79);

\path[fill=fillColor] ( 69.75, 50.22) rectangle ( 70.75, 51.22);

\path[fill=fillColor] ( 99.82, 62.51) rectangle (100.82, 63.51);

\path[fill=fillColor] ( 97.13, 61.85) rectangle ( 98.13, 62.85);

\path[fill=fillColor] ( 91.52, 59.20) rectangle ( 92.52, 60.20);

\path[fill=fillColor] ( 87.79, 57.46) rectangle ( 88.79, 58.46);

\path[fill=fillColor] ( 51.81, 43.41) rectangle ( 52.81, 44.41);

\path[fill=fillColor] (104.05, 64.29) rectangle (105.05, 65.29);

\path[fill=fillColor] (113.16, 68.42) rectangle (114.16, 69.42);

\path[fill=fillColor] ( 74.68, 52.05) rectangle ( 75.68, 53.05);

\path[fill=fillColor] ( 26.72, 32.39) rectangle ( 27.72, 33.39);

\path[fill=fillColor] ( 75.31, 52.22) rectangle ( 76.31, 53.22);

\path[fill=fillColor] ( 49.56, 41.92) rectangle ( 50.56, 42.92);

\path[fill=fillColor] ( 82.66, 55.30) rectangle ( 83.66, 56.30);

\path[fill=fillColor] ( 81.42, 54.78) rectangle ( 82.42, 55.78);

\path[fill=fillColor] (127.50, 74.14) rectangle (128.50, 75.14);

\path[fill=fillColor] ( 53.63, 43.31) rectangle ( 54.63, 44.31);

\path[fill=fillColor] ( 61.43, 46.39) rectangle ( 62.43, 47.39);

\path[fill=fillColor] (123.09, 72.28) rectangle (124.09, 73.28);

\path[fill=fillColor] (124.09, 72.70) rectangle (125.09, 73.70);

\path[fill=fillColor] ( 32.67, 34.65) rectangle ( 33.67, 35.65);

\path[fill=fillColor] ( 45.90, 39.91) rectangle ( 46.90, 40.91);

\path[fill=fillColor] ( 41.50, 38.84) rectangle ( 42.50, 39.84);

\path[fill=fillColor] ( 71.28, 50.53) rectangle ( 72.28, 51.53);

\path[fill=fillColor] (129.55, 75.00) rectangle (130.55, 76.00);

\path[fill=fillColor] ( 80.65, 54.46) rectangle ( 81.65, 55.46);

\path[fill=fillColor] (  2.56, 21.96) rectangle (  3.56, 22.96);

\path[fill=fillColor] ( 81.48, 54.81) rectangle ( 82.48, 55.81);

\path[fill=fillColor] (115.81, 69.23) rectangle (116.81, 70.23);

\path[fill=fillColor] ( 46.33, 40.58) rectangle ( 47.33, 41.58);

\path[fill=fillColor] ( 90.44, 58.57) rectangle ( 91.44, 59.57);

\path[fill=fillColor] ( 94.56, 60.84) rectangle ( 95.56, 61.84);

\path[fill=fillColor] ( 55.15, 43.75) rectangle ( 56.15, 44.75);

\path[fill=fillColor] ( 73.72, 51.55) rectangle ( 74.72, 52.55);

\path[fill=fillColor] ( 20.88, 29.39) rectangle ( 21.88, 30.39);

\path[fill=fillColor] ( 29.09, 33.55) rectangle ( 30.09, 34.55);

\path[fill=fillColor] ( 62.54, 46.86) rectangle ( 63.54, 47.86);

\path[fill=fillColor] ( 90.46, 59.21) rectangle ( 91.46, 60.21);

\path[fill=fillColor] (113.57, 68.61) rectangle (114.57, 69.61);

\path[fill=fillColor] ( 80.58, 54.48) rectangle ( 81.58, 55.48);

\path[fill=fillColor] (108.73, 66.35) rectangle (109.73, 67.35);

\path[fill=fillColor] ( 18.20, 28.60) rectangle ( 19.20, 29.60);

\path[fill=fillColor] ( 34.03, 34.88) rectangle ( 35.03, 35.88);

\path[fill=fillColor] (104.50, 64.48) rectangle (105.50, 65.48);

\path[fill=fillColor] ( 17.30, 28.60) rectangle ( 18.30, 29.60);

\path[fill=fillColor] (113.91, 68.44) rectangle (114.91, 69.44);

\path[fill=fillColor] ( 82.84, 55.45) rectangle ( 83.84, 56.45);

\path[fill=fillColor] ( 72.03, 50.84) rectangle ( 73.03, 51.84);

\path[fill=fillColor] ( 54.55, 44.96) rectangle ( 55.55, 45.96);

\path[fill=fillColor] ( 81.65, 55.89) rectangle ( 82.65, 56.89);

\path[fill=fillColor] ( 79.07, 54.86) rectangle ( 80.07, 55.86);

\path[fill=fillColor] ( 72.48, 51.17) rectangle ( 73.48, 52.17);

\path[fill=fillColor] (  4.30, 22.99) rectangle (  5.30, 23.99);

\path[fill=fillColor] ( 47.02, 40.46) rectangle ( 48.02, 41.46);

\path[fill=fillColor] ( 83.52, 55.66) rectangle ( 84.52, 56.66);

\path[fill=fillColor] ( 75.43, 53.30) rectangle ( 76.43, 54.30);

\path[fill=fillColor] (  7.98, 25.82) rectangle (  8.98, 26.82);

\path[fill=fillColor] ( 18.61, 28.96) rectangle ( 19.61, 29.96);

\path[fill=fillColor] ( 84.44, 56.94) rectangle ( 85.44, 57.94);

\path[fill=fillColor] ( 17.23, 28.55) rectangle ( 18.23, 29.55);

\path[fill=fillColor] (  4.96, 22.68) rectangle (  5.96, 23.68);

\path[fill=fillColor] ( 21.86, 30.97) rectangle ( 22.86, 31.97);

\path[fill=fillColor] ( 20.95, 29.49) rectangle ( 21.95, 30.49);

\path[fill=fillColor] ( 25.95, 31.49) rectangle ( 26.95, 32.49);

\path[fill=fillColor] ( 32.96, 34.60) rectangle ( 33.96, 35.60);

\path[fill=fillColor] ( 75.36, 54.34) rectangle ( 76.36, 55.34);

\path[fill=fillColor] ( 37.76, 36.66) rectangle ( 38.76, 37.66);

\path[fill=fillColor] (104.33, 64.52) rectangle (105.33, 65.52);

\path[fill=fillColor] ( 29.28, 33.06) rectangle ( 30.28, 34.06);

\path[fill=fillColor] ( 80.57, 55.08) rectangle ( 81.57, 56.08);

\path[fill=fillColor] ( 57.70, 44.84) rectangle ( 58.70, 45.84);

\path[fill=fillColor] ( 53.60, 44.28) rectangle ( 54.60, 45.28);

\path[fill=fillColor] ( 89.84, 60.04) rectangle ( 90.84, 61.04);

\path[fill=fillColor] ( 23.61, 30.81) rectangle ( 24.61, 31.81);

\path[fill=fillColor] ( 84.27, 56.13) rectangle ( 85.27, 57.13);

\path[fill=fillColor] ( 78.93, 54.04) rectangle ( 79.93, 55.04);

\path[fill=fillColor] ( 35.81, 36.49) rectangle ( 36.81, 37.49);

\path[fill=fillColor] ( 14.78, 28.41) rectangle ( 15.78, 29.41);

\path[fill=fillColor] ( 21.98, 30.77) rectangle ( 22.98, 31.77);

\path[fill=fillColor] ( 83.19, 55.77) rectangle ( 84.19, 56.77);

\path[fill=fillColor] (101.77, 63.71) rectangle (102.77, 64.71);

\path[fill=fillColor] ( 58.66, 46.80) rectangle ( 59.66, 47.80);

\path[fill=fillColor] ( 80.45, 54.38) rectangle ( 81.45, 55.38);

\path[fill=fillColor] ( 25.95, 32.13) rectangle ( 26.95, 33.13);

\path[fill=fillColor] ( 92.59, 59.48) rectangle ( 93.59, 60.48);

\path[fill=fillColor] (110.60, 67.04) rectangle (111.60, 68.04);

\path[fill=fillColor] (111.47, 67.52) rectangle (112.47, 68.52);

\path[fill=fillColor] ( 66.51, 49.25) rectangle ( 67.51, 50.25);

\path[fill=fillColor] ( 87.99, 58.23) rectangle ( 88.99, 59.23);

\path[fill=fillColor] ( 49.31, 42.60) rectangle ( 50.31, 43.60);

\path[fill=fillColor] ( 92.04, 60.25) rectangle ( 93.04, 61.25);

\path[fill=fillColor] ( -0.47, 20.39) rectangle (  0.53, 21.39);

\path[fill=fillColor] ( 14.55, 26.99) rectangle ( 15.55, 27.99);

\path[fill=fillColor] (105.99, 65.10) rectangle (106.99, 66.10);

\path[fill=fillColor] ( 27.02, 31.94) rectangle ( 28.02, 32.94);

\path[fill=fillColor] ( 22.93, 30.39) rectangle ( 23.93, 31.39);

\path[fill=fillColor] ( 94.01, 60.68) rectangle ( 95.01, 61.68);

\path[fill=fillColor] ( 74.88, 52.52) rectangle ( 75.88, 53.52);

\path[fill=fillColor] (109.01, 66.37) rectangle (110.01, 67.37);

\path[fill=fillColor] ( 64.63, 47.77) rectangle ( 65.63, 48.77);

\path[fill=fillColor] ( 92.75, 59.68) rectangle ( 93.75, 60.68);

\path[fill=fillColor] ( 72.93, 51.22) rectangle ( 73.93, 52.22);

\path[fill=fillColor] ( 60.91, 46.41) rectangle ( 61.91, 47.41);

\path[fill=fillColor] ( 66.72, 48.61) rectangle ( 67.72, 49.61);

\path[fill=fillColor] ( 60.04, 47.22) rectangle ( 61.04, 48.22);

\path[fill=fillColor] ( 62.30, 47.04) rectangle ( 63.30, 48.04);

\path[fill=fillColor] ( 40.71, 37.75) rectangle ( 41.71, 38.75);

\path[fill=fillColor] ( 71.04, 50.82) rectangle ( 72.04, 51.82);

\path[fill=fillColor] ( 32.30, 35.55) rectangle ( 33.30, 36.55);

\path[fill=fillColor] ( 62.37, 46.78) rectangle ( 63.37, 47.78);

\path[fill=fillColor] (  0.19, 20.78) rectangle (  1.19, 21.78);

\path[fill=fillColor] (110.95, 67.19) rectangle (111.95, 68.19);

\path[fill=fillColor] ( 94.26, 60.18) rectangle ( 95.26, 61.18);

\path[fill=fillColor] ( 99.82, 62.51) rectangle (100.82, 63.51);

\path[fill=fillColor] ( 98.30, 63.54) rectangle ( 99.30, 64.54);

\path[fill=fillColor] ( 68.66, 49.58) rectangle ( 69.66, 50.58);

\path[fill=fillColor] ( 91.24, 59.14) rectangle ( 92.24, 60.14);

\path[fill=fillColor] ( 50.03, 41.79) rectangle ( 51.03, 42.79);

\path[fill=fillColor] ( 32.88, 34.40) rectangle ( 33.88, 35.40);

\path[fill=fillColor] ( 91.72, 59.21) rectangle ( 92.72, 60.21);

\path[fill=fillColor] ( 18.27, 28.96) rectangle ( 19.27, 29.96);

\path[fill=fillColor] ( 71.33, 50.84) rectangle ( 72.33, 51.84);

\path[fill=fillColor] ( 86.86, 57.47) rectangle ( 87.86, 58.47);

\path[fill=fillColor] ( 18.61, 29.41) rectangle ( 19.61, 30.41);

\path[fill=fillColor] (101.52, 63.22) rectangle (102.52, 64.22);

\path[fill=fillColor] ( 54.65, 43.54) rectangle ( 55.65, 44.54);

\path[fill=fillColor] ( 82.72, 55.33) rectangle ( 83.72, 56.33);

\path[fill=fillColor] ( 87.98, 57.99) rectangle ( 88.98, 58.99);

\path[fill=fillColor] ( 97.12, 62.54) rectangle ( 98.12, 63.54);

\path[fill=fillColor] ( 32.55, 34.45) rectangle ( 33.55, 35.45);

\path[fill=fillColor] (106.82, 65.45) rectangle (107.82, 66.45);

\path[fill=fillColor] ( 51.81, 42.99) rectangle ( 52.81, 43.99);

\path[fill=fillColor] ( 16.22, 27.80) rectangle ( 17.22, 28.80);

\path[fill=fillColor] ( 89.91, 58.35) rectangle ( 90.91, 59.35);

\path[fill=fillColor] ( 61.91, 46.59) rectangle ( 62.91, 47.59);

\path[fill=fillColor] ( 17.01, 27.74) rectangle ( 18.01, 28.74);

\path[fill=fillColor] ( 66.27, 48.88) rectangle ( 67.27, 49.88);

\path[fill=fillColor] ( 29.87, 33.14) rectangle ( 30.87, 34.14);

\path[fill=fillColor] ( 71.70, 50.70) rectangle ( 72.70, 51.70);

\path[fill=fillColor] ( 97.17, 61.40) rectangle ( 98.17, 62.40);

\path[fill=fillColor] ( 82.15, 55.09) rectangle ( 83.15, 56.09);

\path[fill=fillColor] (  1.58, 21.77) rectangle (  2.58, 22.77);

\path[fill=fillColor] ( 73.82, 51.59) rectangle ( 74.82, 52.59);

\path[fill=fillColor] ( 40.80, 39.26) rectangle ( 41.80, 40.26);

\path[fill=fillColor] (103.29, 63.97) rectangle (104.29, 64.97);

\path[fill=fillColor] ( 33.28, 35.03) rectangle ( 34.28, 36.03);

\path[fill=fillColor] (  0.19, 20.67) rectangle (  1.19, 21.67);

\path[fill=fillColor] ( 76.91, 53.87) rectangle ( 77.91, 54.87);

\path[fill=fillColor] ( 63.68, 47.33) rectangle ( 64.68, 48.33);

\path[fill=fillColor] ( 43.35, 38.80) rectangle ( 44.35, 39.80);

\path[fill=fillColor] (  4.07, 22.30) rectangle (  5.07, 23.30);

\path[fill=fillColor] ( -0.60, 20.34) rectangle (  0.40, 21.34);

\path[fill=fillColor] (  6.04, 23.13) rectangle (  7.04, 24.13);

\path[fill=fillColor] ( 59.29, 45.49) rectangle ( 60.29, 46.49);

\path[fill=fillColor] ( 23.61, 31.27) rectangle ( 24.61, 32.27);

\path[fill=fillColor] ( 65.10, 47.93) rectangle ( 66.10, 48.93);

\path[fill=fillColor] ( 37.24, 36.86) rectangle ( 38.24, 37.86);

\path[fill=fillColor] ( 75.69, 53.62) rectangle ( 76.69, 54.62);

\path[fill=fillColor] ( 82.08, 55.53) rectangle ( 83.08, 56.53);

\path[fill=fillColor] ( 92.20, 59.31) rectangle ( 93.20, 60.31);

\path[fill=fillColor] ( 80.99, 55.12) rectangle ( 81.99, 56.12);

\path[fill=fillColor] ( 81.97, 55.58) rectangle ( 82.97, 56.58);

\path[fill=fillColor] ( 74.48, 51.87) rectangle ( 75.48, 52.87);

\path[fill=fillColor] (  6.15, 24.35) rectangle (  7.15, 25.35);

\path[fill=fillColor] ( 45.43, 39.71) rectangle ( 46.43, 40.71);

\path[fill=fillColor] ( 96.79, 61.24) rectangle ( 97.79, 62.24);

\path[fill=fillColor] ( 55.84, 44.04) rectangle ( 56.84, 45.04);

\path[fill=fillColor] ( 78.43, 53.55) rectangle ( 79.43, 54.55);

\path[fill=fillColor] ( 95.54, 60.71) rectangle ( 96.54, 61.71);

\path[fill=fillColor] ( 45.38, 40.52) rectangle ( 46.38, 41.52);

\path[fill=fillColor] ( 87.75, 57.60) rectangle ( 88.75, 58.60);

\path[fill=fillColor] ( 62.91, 47.81) rectangle ( 63.91, 48.81);

\path[fill=fillColor] ( 37.38, 36.53) rectangle ( 38.38, 37.53);

\path[fill=fillColor] ( 97.64, 61.67) rectangle ( 98.64, 62.67);

\path[fill=fillColor] ( 55.51, 43.91) rectangle ( 56.51, 44.91);

\path[fill=fillColor] ( 81.08, 54.77) rectangle ( 82.08, 55.77);

\path[fill=fillColor] ( 88.91, 57.93) rectangle ( 89.91, 58.93);

\path[fill=fillColor] ( 63.36, 47.45) rectangle ( 64.36, 48.45);

\path[fill=fillColor] ( 63.77, 47.52) rectangle ( 64.77, 48.52);

\path[fill=fillColor] ( 32.38, 34.58) rectangle ( 33.38, 35.58);

\path[fill=fillColor] ( 57.68, 44.82) rectangle ( 58.68, 45.82);

\path[fill=fillColor] ( 64.51, 48.04) rectangle ( 65.51, 49.04);

\path[fill=fillColor] ( 29.78, 33.46) rectangle ( 30.78, 34.46);

\path[fill=fillColor] (102.10, 63.47) rectangle (103.10, 64.47);

\path[fill=fillColor] ( 48.32, 41.35) rectangle ( 49.32, 42.35);

\path[fill=fillColor] ( 52.26, 42.54) rectangle ( 53.26, 43.54);

\path[fill=fillColor] ( 89.62, 58.23) rectangle ( 90.62, 59.23);

\path[fill=fillColor] ( 52.89, 42.80) rectangle ( 53.89, 43.80);

\path[fill=fillColor] ( 93.00, 59.65) rectangle ( 94.00, 60.65);

\path[fill=fillColor] ( 46.95, 41.07) rectangle ( 47.95, 42.07);

\path[fill=fillColor] ( 85.82, 57.74) rectangle ( 86.82, 58.74);

\path[fill=fillColor] ( 65.58, 48.52) rectangle ( 66.58, 49.52);

\path[fill=fillColor] ( 57.73, 45.01) rectangle ( 58.73, 46.01);

\path[fill=fillColor] ( 81.46, 55.07) rectangle ( 82.46, 56.07);

\path[fill=fillColor] ( 57.02, 44.61) rectangle ( 58.02, 45.61);

\path[fill=fillColor] ( 75.65, 52.77) rectangle ( 76.65, 53.77);

\path[fill=fillColor] ( 41.32, 38.12) rectangle ( 42.32, 39.12);

\path[fill=fillColor] (  6.87, 23.65) rectangle (  7.87, 24.65);

\path[fill=fillColor] ( 32.51, 34.09) rectangle ( 33.51, 35.09);

\path[fill=fillColor] ( 43.93, 39.52) rectangle ( 44.93, 40.52);

\path[fill=fillColor] ( 84.67, 57.70) rectangle ( 85.67, 58.70);

\path[fill=fillColor] ( 49.88, 41.54) rectangle ( 50.88, 42.54);

\path[fill=fillColor] (  1.21, 21.10) rectangle (  2.21, 22.10);

\path[fill=fillColor] ( 70.50, 50.28) rectangle ( 71.50, 51.28);

\path[fill=fillColor] ( 80.32, 54.32) rectangle ( 81.32, 55.32);

\path[fill=fillColor] ( 35.22, 35.83) rectangle ( 36.22, 36.83);

\path[fill=fillColor] ( 38.83, 36.90) rectangle ( 39.83, 37.90);

\path[fill=fillColor] ( 50.90, 42.12) rectangle ( 51.90, 43.12);

\path[fill=fillColor] ( 79.01, 54.30) rectangle ( 80.01, 55.30);

\path[fill=fillColor] ( 59.11, 46.46) rectangle ( 60.11, 47.46);

\path[fill=fillColor] ( 60.44, 45.97) rectangle ( 61.44, 46.97);

\path[fill=fillColor] (  9.89, 25.12) rectangle ( 10.89, 26.12);

\path[fill=fillColor] ( 80.97, 54.59) rectangle ( 81.97, 55.59);

\path[fill=fillColor] ( 43.85, 39.19) rectangle ( 44.85, 40.19);

\path[fill=fillColor] ( 22.16, 30.20) rectangle ( 23.16, 31.20);

\path[fill=fillColor] ( 88.80, 57.88) rectangle ( 89.80, 58.88);

\path[fill=fillColor] ( 34.50, 35.89) rectangle ( 35.50, 36.89);

\path[fill=fillColor] ( 77.60, 53.58) rectangle ( 78.60, 54.58);

\path[fill=fillColor] ( 83.49, 55.74) rectangle ( 84.49, 56.74);

\path[fill=fillColor] ( 80.86, 54.89) rectangle ( 81.86, 55.89);

\path[fill=fillColor] ( 94.42, 60.24) rectangle ( 95.42, 61.24);

\path[fill=fillColor] ( 39.29, 37.58) rectangle ( 40.29, 38.58);

\path[fill=fillColor] ( 64.98, 49.33) rectangle ( 65.98, 50.33);

\path[fill=fillColor] ( 86.70, 57.00) rectangle ( 87.70, 58.00);

\path[fill=fillColor] (  0.32, 20.83) rectangle (  1.32, 21.83);

\path[fill=fillColor] ( 11.85, 25.85) rectangle ( 12.85, 26.85);

\path[fill=fillColor] ( 55.22, 44.03) rectangle ( 56.22, 45.03);

\path[fill=fillColor] ( 55.28, 43.81) rectangle ( 56.28, 44.81);

\path[fill=fillColor] ( 88.09, 57.66) rectangle ( 89.09, 58.66);

\path[fill=fillColor] ( 80.86, 54.78) rectangle ( 81.86, 55.78);

\path[fill=fillColor] ( 27.47, 32.13) rectangle ( 28.47, 33.13);

\path[fill=fillColor] ( 69.69, 49.86) rectangle ( 70.69, 50.86);

\path[fill=fillColor] ( 67.29, 50.85) rectangle ( 68.29, 51.85);

\path[fill=fillColor] ( 61.87, 46.90) rectangle ( 62.87, 47.90);

\path[fill=fillColor] ( 84.24, 55.97) rectangle ( 85.24, 56.97);

\path[fill=fillColor] ( 88.81, 57.89) rectangle ( 89.81, 58.89);

\path[fill=fillColor] ( 53.34, 43.27) rectangle ( 54.34, 44.27);

\path[fill=fillColor] ( 13.68, 26.40) rectangle ( 14.68, 27.40);

\path[fill=fillColor] ( 55.55, 45.19) rectangle ( 56.55, 46.19);

\path[fill=fillColor] ( 37.93, 37.23) rectangle ( 38.93, 38.23);

\path[fill=fillColor] ( 68.80, 50.94) rectangle ( 69.80, 51.94);

\path[fill=fillColor] ( 81.19, 56.60) rectangle ( 82.19, 57.60);

\path[fill=fillColor] ( 65.11, 47.94) rectangle ( 66.11, 48.94);

\path[fill=fillColor] ( 69.29, 49.69) rectangle ( 70.29, 50.69);

\path[fill=fillColor] ( 84.28, 55.98) rectangle ( 85.28, 56.98);

\path[fill=fillColor] ( 82.38, 55.19) rectangle ( 83.38, 56.19);

\path[fill=fillColor] ( 37.72, 37.33) rectangle ( 38.72, 38.33);

\path[fill=fillColor] ( 63.77, 47.69) rectangle ( 64.77, 48.69);

\path[fill=fillColor] ( 78.99, 54.20) rectangle ( 79.99, 55.20);

\path[fill=fillColor] ( 90.44, 58.58) rectangle ( 91.44, 59.58);

\path[fill=fillColor] ( 25.37, 31.24) rectangle ( 26.37, 32.24);

\path[fill=fillColor] ( 70.35, 50.14) rectangle ( 71.35, 51.14);

\path[fill=fillColor] ( 69.43, 50.25) rectangle ( 70.43, 51.25);

\path[fill=fillColor] ( 25.95, 31.49) rectangle ( 26.95, 32.49);

\path[fill=fillColor] ( 81.45, 54.80) rectangle ( 82.45, 55.80);

\path[fill=fillColor] ( 89.04, 57.99) rectangle ( 90.04, 58.99);

\path[fill=fillColor] ( 38.63, 37.25) rectangle ( 39.63, 38.25);

\path[fill=fillColor] ( 56.95, 44.51) rectangle ( 57.95, 45.51);

\path[fill=fillColor] ( 61.55, 46.87) rectangle ( 62.55, 47.87);

\path[fill=fillColor] ( 72.34, 51.33) rectangle ( 73.34, 52.33);

\path[fill=fillColor] ( 79.09, 53.80) rectangle ( 80.09, 54.80);

\path[fill=fillColor] ( 41.83, 38.16) rectangle ( 42.83, 39.16);

\path[fill=fillColor] ( 68.94, 49.54) rectangle ( 69.94, 50.54);

\path[fill=fillColor] ( 79.41, 53.94) rectangle ( 80.41, 54.94);

\path[fill=fillColor] ( 69.54, 49.87) rectangle ( 70.54, 50.87);

\path[fill=fillColor] ( 71.71, 50.71) rectangle ( 72.71, 51.71);

\path[fill=fillColor] ( 59.65, 45.64) rectangle ( 60.65, 46.64);

\path[fill=fillColor] ( 72.01, 50.83) rectangle ( 73.01, 51.83);

\path[fill=fillColor] ( 65.42, 48.70) rectangle ( 66.42, 49.70);

\path[fill=fillColor] ( 49.70, 41.77) rectangle ( 50.70, 42.77);

\path[fill=fillColor] ( 75.76, 52.41) rectangle ( 76.76, 53.41);

\path[fill=fillColor] ( 30.41, 34.21) rectangle ( 31.41, 35.21);

\path[fill=fillColor] ( 82.60, 55.29) rectangle ( 83.60, 56.29);

\path[fill=fillColor] ( 13.76, 26.40) rectangle ( 14.76, 27.40);

\path[fill=fillColor] ( 35.26, 36.63) rectangle ( 36.26, 37.63);

\path[fill=fillColor] ( 68.77, 49.65) rectangle ( 69.77, 50.65);

\path[fill=fillColor] ( 30.77, 34.24) rectangle ( 31.77, 35.24);

\path[fill=fillColor] ( 67.87, 49.09) rectangle ( 68.87, 50.09);

\path[fill=fillColor] ( 31.84, 34.40) rectangle ( 32.84, 35.40);

\path[fill=fillColor] ( 40.62, 38.39) rectangle ( 41.62, 39.39);

\path[fill=fillColor] ( 25.89, 32.02) rectangle ( 26.89, 33.02);

\path[fill=fillColor] ( 44.85, 39.45) rectangle ( 45.85, 40.45);

\path[fill=fillColor] ( 57.72, 45.43) rectangle ( 58.72, 46.43);

\path[fill=fillColor] ( 67.03, 48.74) rectangle ( 68.03, 49.74);

\path[fill=fillColor] ( 66.79, 48.64) rectangle ( 67.79, 49.64);

\path[fill=fillColor] ( 57.58, 45.60) rectangle ( 58.58, 46.60);

\path[fill=fillColor] ( 45.59, 40.17) rectangle ( 46.59, 41.17);

\path[fill=fillColor] ( 74.59, 51.99) rectangle ( 75.59, 52.99);

\path[fill=fillColor] ( 62.01, 46.91) rectangle ( 63.01, 47.91);

\path[fill=fillColor] ( 47.99, 41.90) rectangle ( 48.99, 42.90);

\path[fill=fillColor] ( 28.53, 32.71) rectangle ( 29.53, 33.71);

\path[fill=fillColor] ( 41.77, 38.13) rectangle ( 42.77, 39.13);

\path[fill=fillColor] ( 62.24, 47.02) rectangle ( 63.24, 48.02);

\path[fill=fillColor] ( 44.72, 39.37) rectangle ( 45.72, 40.37);

\path[fill=fillColor] ( 22.87, 30.20) rectangle ( 23.87, 31.20);

\path[fill=fillColor] ( 73.19, 51.33) rectangle ( 74.19, 52.33);

\path[fill=fillColor] ( 76.01, 52.57) rectangle ( 77.01, 53.57);

\path[fill=fillColor] ( 66.62, 48.79) rectangle ( 67.62, 49.79);

\path[fill=fillColor] ( 59.76, 45.69) rectangle ( 60.76, 46.69);

\path[fill=fillColor] ( 41.77, 38.37) rectangle ( 42.77, 39.37);

\path[fill=fillColor] ( 72.86, 51.19) rectangle ( 73.86, 52.19);

\path[fill=fillColor] ( 67.81, 49.11) rectangle ( 68.81, 50.11);

\path[fill=fillColor] ( 23.21, 30.34) rectangle ( 24.21, 31.34);

\path[fill=fillColor] ( 13.02, 26.70) rectangle ( 14.02, 27.70);

\path[fill=fillColor] ( 66.90, 49.81) rectangle ( 67.90, 50.81);

\path[fill=fillColor] ( 16.22, 27.94) rectangle ( 17.22, 28.94);

\path[fill=fillColor] ( 31.71, 34.02) rectangle ( 32.71, 35.02);

\path[fill=fillColor] ( 63.00, 48.99) rectangle ( 64.00, 49.99);

\path[fill=fillColor] ( 27.61, 32.17) rectangle ( 28.61, 33.17);

\path[fill=fillColor] ( 68.81, 49.49) rectangle ( 69.81, 50.49);

\path[fill=fillColor] ( 67.48, 50.07) rectangle ( 68.48, 51.07);

\path[fill=fillColor] ( 44.69, 39.36) rectangle ( 45.69, 40.36);

\path[fill=fillColor] ( 63.21, 47.14) rectangle ( 64.21, 48.14);

\path[fill=fillColor] ( 56.97, 44.52) rectangle ( 57.97, 45.52);

\path[fill=fillColor] ( 40.40, 39.93) rectangle ( 41.40, 40.93);

\path[fill=fillColor] ( 45.69, 39.78) rectangle ( 46.69, 40.78);

\path[fill=fillColor] ( 73.27, 51.36) rectangle ( 74.27, 52.36);

\path[fill=fillColor] ( 62.84, 46.98) rectangle ( 63.84, 47.98);

\path[fill=fillColor] ( 56.34, 44.89) rectangle ( 57.34, 45.89);

\path[fill=fillColor] ( 31.63, 34.17) rectangle ( 32.63, 35.17);

\path[fill=fillColor] ( 30.10, 34.05) rectangle ( 31.10, 35.05);

\path[fill=fillColor] ( 70.61, 50.24) rectangle ( 71.61, 51.24);

\path[fill=fillColor] ( 56.21, 44.75) rectangle ( 57.21, 45.75);

\path[fill=fillColor] ( 55.91, 44.07) rectangle ( 56.91, 45.07);

\path[fill=fillColor] ( 54.14, 43.81) rectangle ( 55.14, 44.81);

\path[fill=fillColor] ( 62.63, 46.89) rectangle ( 63.63, 47.89);

\path[fill=fillColor] ( 70.41, 50.83) rectangle ( 71.41, 51.83);

\path[fill=fillColor] ( 61.04, 46.23) rectangle ( 62.04, 47.23);

\path[fill=fillColor] ( 37.24, 36.63) rectangle ( 38.24, 37.63);

\path[fill=fillColor] ( 49.74, 41.48) rectangle ( 50.74, 42.48);

\path[fill=fillColor] ( 57.51, 45.04) rectangle ( 58.51, 46.04);

\path[fill=fillColor] ( 47.85, 41.10) rectangle ( 48.85, 42.10);

\path[fill=fillColor] ( 38.03, 36.76) rectangle ( 39.03, 37.76);

\path[fill=fillColor] ( 57.04, 44.54) rectangle ( 58.04, 45.54);

\path[fill=fillColor] ( 62.77, 46.95) rectangle ( 63.77, 47.95);

\path[fill=fillColor] ( 69.92, 49.96) rectangle ( 70.92, 50.96);

\path[fill=fillColor] ( 13.43, 26.44) rectangle ( 14.43, 27.44);

\path[fill=fillColor] ( 42.90, 38.61) rectangle ( 43.90, 39.61);

\path[fill=fillColor] ( 53.65, 43.12) rectangle ( 54.65, 44.12);

\path[fill=fillColor] ( 53.28, 42.97) rectangle ( 54.28, 43.97);

\path[fill=fillColor] ( 52.52, 42.65) rectangle ( 53.52, 43.65);

\path[fill=fillColor] ( 58.00, 45.04) rectangle ( 59.00, 46.04);

\path[fill=fillColor] ( 57.38, 44.97) rectangle ( 58.38, 45.97);

\path[fill=fillColor] ( 41.14, 37.94) rectangle ( 42.14, 38.94);

\path[fill=fillColor] ( 42.79, 39.37) rectangle ( 43.79, 40.37);

\path[fill=fillColor] ( 52.56, 42.67) rectangle ( 53.56, 43.67);

\path[fill=fillColor] ( 42.38, 38.39) rectangle ( 43.38, 39.39);

\path[fill=fillColor] ( 21.86, 30.32) rectangle ( 22.86, 31.32);

\path[fill=fillColor] ( 55.75, 44.38) rectangle ( 56.75, 45.38);

\path[fill=fillColor] ( 55.00, 43.78) rectangle ( 56.00, 44.78);

\path[fill=fillColor] ( 48.42, 40.92) rectangle ( 49.42, 41.92);

\path[fill=fillColor] ( 57.40, 44.70) rectangle ( 58.40, 45.70);

\path[fill=fillColor] ( 51.64, 42.28) rectangle ( 52.64, 43.28);

\path[fill=fillColor] ( 65.19, 47.97) rectangle ( 66.19, 48.97);

\path[fill=fillColor] ( 63.28, 47.39) rectangle ( 64.28, 48.39);

\path[fill=fillColor] ( 29.97, 34.21) rectangle ( 30.97, 35.21);

\path[fill=fillColor] ( 62.11, 46.68) rectangle ( 63.11, 47.68);

\path[fill=fillColor] ( 65.90, 48.36) rectangle ( 66.90, 49.36);

\path[fill=fillColor] ( 52.16, 42.50) rectangle ( 53.16, 43.50);

\path[fill=fillColor] ( 60.36, 45.94) rectangle ( 61.36, 46.94);

\path[fill=fillColor] ( 50.51, 41.80) rectangle ( 51.51, 42.80);

\path[fill=fillColor] ( 22.10, 30.07) rectangle ( 23.10, 31.07);

\path[fill=fillColor] ( 35.77, 36.38) rectangle ( 36.77, 37.38);

\path[fill=fillColor] ( 12.61, 26.96) rectangle ( 13.61, 27.96);

\path[fill=fillColor] ( 57.21, 44.62) rectangle ( 58.21, 45.62);

\path[fill=fillColor] ( 19.80, 28.99) rectangle ( 20.80, 29.99);

\path[fill=fillColor] ( 31.97, 35.13) rectangle ( 32.97, 36.13);

\path[fill=fillColor] ( 71.08, 50.44) rectangle ( 72.08, 51.44);

\path[fill=fillColor] ( 45.84, 39.96) rectangle ( 46.84, 40.96);

\path[fill=fillColor] ( 41.80, 39.32) rectangle ( 42.80, 40.32);

\path[fill=fillColor] ( 40.83, 38.03) rectangle ( 41.83, 39.03);

\path[fill=fillColor] ( 20.32, 29.52) rectangle ( 21.32, 30.52);

\path[fill=fillColor] ( 42.50, 38.44) rectangle ( 43.50, 39.44);

\path[fill=fillColor] ( 60.61, 46.04) rectangle ( 61.61, 47.04);

\path[fill=fillColor] ( 11.76, 25.82) rectangle ( 12.76, 26.82);

\path[fill=fillColor] ( 14.47, 27.25) rectangle ( 15.47, 28.25);

\path[fill=fillColor] ( 62.98, 47.04) rectangle ( 63.98, 48.04);

\path[fill=fillColor] ( 18.48, 28.55) rectangle ( 19.48, 29.55);

\path[fill=fillColor] ( 19.21, 29.34) rectangle ( 20.21, 30.34);

\path[fill=fillColor] ( 41.62, 38.07) rectangle ( 42.62, 39.07);

\path[fill=fillColor] ( 52.97, 44.15) rectangle ( 53.97, 45.15);

\path[fill=fillColor] ( 59.44, 45.56) rectangle ( 60.44, 46.56);

\path[fill=fillColor] ( 28.81, 33.08) rectangle ( 29.81, 34.08);

\path[fill=fillColor] ( 60.91, 46.17) rectangle ( 61.91, 47.17);

\path[fill=fillColor] ( 56.34, 44.25) rectangle ( 57.34, 45.25);

\path[fill=fillColor] ( 21.01, 29.44) rectangle ( 22.01, 30.44);

\path[fill=fillColor] ( 17.23, 29.23) rectangle ( 18.23, 30.23);

\path[fill=fillColor] (  9.05, 24.59) rectangle ( 10.05, 25.59);

\path[fill=fillColor] ( 46.00, 39.91) rectangle ( 47.00, 40.91);

\path[fill=fillColor] ( 22.81, 30.84) rectangle ( 23.81, 31.84);

\path[fill=fillColor] ( 37.24, 36.48) rectangle ( 38.24, 37.48);

\path[fill=fillColor] ( 21.07, 29.75) rectangle ( 22.07, 30.75);

\path[fill=fillColor] ( 30.68, 34.65) rectangle ( 31.68, 35.65);

\path[fill=fillColor] ( 42.06, 39.21) rectangle ( 43.06, 40.21);

\path[fill=fillColor] ( 48.79, 42.57) rectangle ( 49.79, 43.57);

\path[fill=fillColor] ( 25.63, 32.02) rectangle ( 26.63, 33.02);

\path[fill=fillColor] ( 61.03, 46.22) rectangle ( 62.03, 47.22);

\path[fill=fillColor] ( 54.59, 43.52) rectangle ( 55.59, 44.52);

\path[fill=fillColor] ( 51.47, 42.31) rectangle ( 52.47, 43.31);

\path[fill=fillColor] ( 47.97, 41.26) rectangle ( 48.97, 42.26);

\path[fill=fillColor] ( 22.46, 30.67) rectangle ( 23.46, 31.67);

\path[fill=fillColor] ( 45.20, 39.57) rectangle ( 46.20, 40.57);

\path[fill=fillColor] ( 21.56, 30.51) rectangle ( 22.56, 31.51);

\path[fill=fillColor] ( 16.07, 28.27) rectangle ( 17.07, 29.27);

\path[fill=fillColor] ( 39.45, 38.75) rectangle ( 40.45, 39.75);

\path[fill=fillColor] ( 57.29, 44.65) rectangle ( 58.29, 45.65);

\path[fill=fillColor] ( 45.79, 40.42) rectangle ( 46.79, 41.42);

\path[fill=fillColor] ( 35.33, 36.80) rectangle ( 36.33, 37.80);

\path[fill=fillColor] ( 46.10, 40.61) rectangle ( 47.10, 41.61);

\path[fill=fillColor] ( 17.86, 28.52) rectangle ( 18.86, 29.52);

\path[fill=fillColor] ( 50.60, 41.84) rectangle ( 51.60, 42.84);

\path[fill=fillColor] ( 43.52, 38.87) rectangle ( 44.52, 39.87);

\path[fill=fillColor] ( 46.85, 40.27) rectangle ( 47.85, 41.27);

\path[fill=fillColor] ( 40.21, 37.48) rectangle ( 41.21, 38.48);

\path[fill=fillColor] ( 41.53, 38.03) rectangle ( 42.53, 39.03);

\path[fill=fillColor] ( 37.69, 37.24) rectangle ( 38.69, 38.24);

\path[fill=fillColor] ( 47.22, 40.42) rectangle ( 48.22, 41.42);

\path[fill=fillColor] ( 19.35, 29.77) rectangle ( 20.35, 30.77);

\path[fill=fillColor] ( 48.37, 40.90) rectangle ( 49.37, 41.90);

\path[fill=fillColor] ( 44.80, 39.41) rectangle ( 45.80, 40.41);

\path[fill=fillColor] (225.87,115.45) rectangle (226.87,116.45);

\path[fill=fillColor] (198.27,103.86) rectangle (199.27,104.86);

\path[fill=fillColor] (215.51,111.09) rectangle (216.51,112.09);

\path[fill=fillColor] (228.62,116.60) rectangle (229.62,117.60);

\path[fill=fillColor] (197.63,103.59) rectangle (198.63,104.59);

\path[fill=fillColor] (180.49, 96.39) rectangle (181.49, 97.39);

\path[fill=fillColor] (212.14,109.68) rectangle (213.14,110.68);

\path[fill=fillColor] (209.09,108.40) rectangle (210.09,109.40);

\path[fill=fillColor] (204.06,106.36) rectangle (205.06,107.36);

\path[fill=fillColor] (222.23,114.13) rectangle (223.23,115.13);

\path[fill=fillColor] (179.43, 95.94) rectangle (180.43, 96.94);

\path[fill=fillColor] (231.07,117.63) rectangle (232.07,118.63);

\path[fill=fillColor] (175.59, 94.33) rectangle (176.59, 95.33);

\path[fill=fillColor] (184.09, 97.90) rectangle (185.09, 98.90);

\path[fill=fillColor] (209.92,109.10) rectangle (210.92,110.10);

\path[fill=fillColor] (219.24,112.69) rectangle (220.24,113.69);

\path[fill=fillColor] (219.64,112.83) rectangle (220.64,113.83);

\path[fill=fillColor] (201.75,105.37) rectangle (202.75,106.37);

\path[fill=fillColor] (197.89,103.79) rectangle (198.89,104.79);

\path[fill=fillColor] (195.35,103.06) rectangle (196.35,104.06);

\path[fill=fillColor] (219.62,112.82) rectangle (220.62,113.82);

\path[fill=fillColor] (223.14,114.30) rectangle (224.14,115.30);

\path[fill=fillColor] (208.88,108.31) rectangle (209.88,109.31);

\path[fill=fillColor] (193.32,101.96) rectangle (194.32,102.96);

\path[fill=fillColor] (180.85, 96.54) rectangle (181.85, 97.54);

\path[fill=fillColor] (173.56, 93.48) rectangle (174.56, 94.48);

\path[fill=fillColor] (199.44,104.35) rectangle (200.44,105.35);

\path[fill=fillColor] (178.20, 95.43) rectangle (179.20, 96.43);

\path[fill=fillColor] (181.05, 96.66) rectangle (182.05, 97.66);

\path[fill=fillColor] (179.39, 95.93) rectangle (180.39, 96.93);

\path[fill=fillColor] (186.19, 98.91) rectangle (187.19, 99.91);

\path[fill=fillColor] (209.27,108.48) rectangle (210.27,109.48);

\path[fill=fillColor] (203.21,105.93) rectangle (204.21,106.93);

\path[fill=fillColor] (184.92, 98.65) rectangle (185.92, 99.65);

\path[fill=fillColor] (178.03, 95.36) rectangle (179.03, 96.36);

\path[fill=fillColor] (165.64, 90.71) rectangle (166.64, 91.71);

\path[fill=fillColor] (177.20, 95.01) rectangle (178.20, 96.01);

\path[fill=fillColor] (173.80, 93.58) rectangle (174.80, 94.58);

\path[fill=fillColor] (173.33, 93.38) rectangle (174.33, 94.38);

\path[fill=fillColor] (165.61, 90.25) rectangle (166.61, 91.25);
\end{scope}
\begin{scope}
\path[clip] (  0.00,  0.00) rectangle (245.72,126.47);
\definecolor{drawColor}{RGB}{0,0,0}

\path[draw=drawColor,line width= 0.4pt,line join=round,line cap=round] ( 36.00, 36.00) --
	(245.72, 36.00) --
	(245.72,124.07) --
	( 36.00,124.07) --
	( 36.00, 36.00);
\end{scope}
\begin{scope}
\path[clip] (  0.00,  0.00) rectangle (245.72,126.47);
\definecolor{drawColor}{RGB}{0,0,0}

\node[text=drawColor,anchor=base,inner sep=0pt, outer sep=0pt, scale=  1.00] at (140.86,  6.00) {$|E_{\mathrm{out}}|$ for  Bev};

\node[text=drawColor,rotate= 90.00,anchor=base,inner sep=0pt, outer sep=0pt, scale=  1.00] at ( 13.20, 80.04) {$|E_{\mathrm{out}}|$ for  FK};
\end{scope}
\begin{scope}
\path[clip] (  0.00,  0.00) rectangle (245.72,126.47);
\definecolor{drawColor}{RGB}{0,0,0}

\path[draw=drawColor,line width= 0.4pt,line join=round,line cap=round] ( 43.77, 36.00) -- (237.95, 36.00);

\path[draw=drawColor,line width= 0.4pt,line join=round,line cap=round] ( 43.77, 36.00) -- ( 43.77, 30.00);

\path[draw=drawColor,line width= 0.4pt,line join=round,line cap=round] (108.50, 36.00) -- (108.50, 30.00);

\path[draw=drawColor,line width= 0.4pt,line join=round,line cap=round] (173.22, 36.00) -- (173.22, 30.00);

\path[draw=drawColor,line width= 0.4pt,line join=round,line cap=round] (237.95, 36.00) -- (237.95, 30.00);

\node[text=drawColor,anchor=base,inner sep=0pt, outer sep=0pt, scale=  1.00] at ( 43.77, 20.40) {$10^3$};

\node[text=drawColor,anchor=base,inner sep=0pt, outer sep=0pt, scale=  1.00] at (108.50, 20.40) {$10^4$};

\node[text=drawColor,anchor=base,inner sep=0pt, outer sep=0pt, scale=  1.00] at (173.22, 20.40) {$10^5$};

\node[text=drawColor,anchor=base,inner sep=0pt, outer sep=0pt, scale=  1.00] at (237.95, 20.40) {$10^6$};

\path[draw=drawColor,line width= 0.4pt,line join=round,line cap=round] ( 36.00, 39.26) -- ( 36.00,120.81);

\path[draw=drawColor,line width= 0.4pt,line join=round,line cap=round] ( 36.00, 39.26) -- ( 30.00, 39.26);

\path[draw=drawColor,line width= 0.4pt,line join=round,line cap=round] ( 36.00, 66.44) -- ( 30.00, 66.44);

\path[draw=drawColor,line width= 0.4pt,line join=round,line cap=round] ( 36.00, 93.63) -- ( 30.00, 93.63);

\path[draw=drawColor,line width= 0.4pt,line join=round,line cap=round] ( 36.00,120.81) -- ( 30.00,120.81);

\node[text=drawColor,rotate= 90.00,anchor=base,inner sep=0pt, outer sep=0pt, scale=  1.00] at ( 27.60, 39.26) {$10^3$};

\node[text=drawColor,rotate= 90.00,anchor=base,inner sep=0pt, outer sep=0pt, scale=  1.00] at ( 27.60, 66.44) {$10^4$};

\node[text=drawColor,rotate= 90.00,anchor=base,inner sep=0pt, outer sep=0pt, scale=  1.00] at ( 27.60, 93.63) {$10^5$};

\node[text=drawColor,rotate= 90.00,anchor=base,inner sep=0pt, outer sep=0pt, scale=  1.00] at ( 27.60,120.81) {$10^6$};
\end{scope}
\begin{scope}
\path[clip] ( 36.00, 36.00) rectangle (245.72,124.07);
\definecolor{drawColor}{RGB}{190,190,190}

\path[draw=drawColor,line width= 0.4pt,dash pattern=on 1pt off 3pt ,line join=round,line cap=round] ( 43.77, 36.00) -- ( 43.77,124.07);

\path[draw=drawColor,line width= 0.4pt,dash pattern=on 1pt off 3pt ,line join=round,line cap=round] (108.50, 36.00) -- (108.50,124.07);

\path[draw=drawColor,line width= 0.4pt,dash pattern=on 1pt off 3pt ,line join=round,line cap=round] (173.22, 36.00) -- (173.22,124.07);

\path[draw=drawColor,line width= 0.4pt,dash pattern=on 1pt off 3pt ,line join=round,line cap=round] (237.95, 36.00) -- (237.95,124.07);

\path[draw=drawColor,line width= 0.4pt,dash pattern=on 1pt off 3pt ,line join=round,line cap=round] ( 36.00, 39.26) -- (245.72, 39.26);

\path[draw=drawColor,line width= 0.4pt,dash pattern=on 1pt off 3pt ,line join=round,line cap=round] ( 36.00, 66.44) -- (245.72, 66.44);

\path[draw=drawColor,line width= 0.4pt,dash pattern=on 1pt off 3pt ,line join=round,line cap=round] ( 36.00, 93.63) -- (245.72, 93.63);

\path[draw=drawColor,line width= 0.4pt,dash pattern=on 1pt off 3pt ,line join=round,line cap=round] ( 36.00,120.81) -- (245.72,120.81);

\path[draw=drawColor,line width= 0.4pt,line join=round,line cap=round] ( 36.00, 36.00) -- (245.72,124.07);
\end{scope}
\end{tikzpicture}

%% file: analysis_bio-004.tex
\begin{tikzpicture}[x=1pt,y=1pt]
\definecolor{fillColor}{RGB}{255,255,255}
\path[use as bounding box,fill=fillColor,fill opacity=0.00] (0,0) rectangle (245.72,126.47);
\begin{scope}
\path[clip] ( 36.00, 36.00) rectangle (245.72,124.07);
\definecolor{fillColor}{RGB}{0,0,0}

\path[fill=fillColor] (170.19, 90.25) rectangle (171.19, 91.25);

\path[fill=fillColor] (130.24, 73.40) rectangle (131.24, 74.40);

\path[fill=fillColor] (163.24, 89.14) rectangle (164.24, 90.14);

\path[fill=fillColor] (150.60, 85.64) rectangle (151.60, 86.64);

\path[fill=fillColor] (127.73, 76.47) rectangle (128.73, 77.47);

\path[fill=fillColor] (139.66, 77.48) rectangle (140.66, 78.48);

\path[fill=fillColor] (166.61, 89.22) rectangle (167.61, 90.22);

\path[fill=fillColor] (152.06, 85.72) rectangle (153.06, 86.72);

\path[fill=fillColor] (160.93, 87.59) rectangle (161.93, 88.59);

\path[fill=fillColor] (129.20, 74.96) rectangle (130.20, 75.96);

\path[fill=fillColor] (153.36, 84.90) rectangle (154.36, 85.90);

\path[fill=fillColor] (128.57, 73.44) rectangle (129.57, 74.44);

\path[fill=fillColor] (135.01, 75.65) rectangle (136.01, 76.65);

\path[fill=fillColor] (155.25, 86.21) rectangle (156.25, 87.21);

\path[fill=fillColor] (167.53, 91.37) rectangle (168.53, 92.37);

\path[fill=fillColor] (127.93, 72.84) rectangle (128.93, 73.84);

\path[fill=fillColor] (150.99, 83.44) rectangle (151.99, 84.44);

\path[fill=fillColor] (137.28, 79.04) rectangle (138.28, 80.04);

\path[fill=fillColor] (143.99, 90.65) rectangle (144.99, 91.65);

\path[fill=fillColor] (156.70, 91.82) rectangle (157.70, 92.82);

\path[fill=fillColor] (136.13, 84.10) rectangle (137.13, 85.10);

\path[fill=fillColor] (136.28, 77.59) rectangle (137.28, 78.59);

\path[fill=fillColor] ( 93.95, 55.89) rectangle ( 94.95, 56.89);

\path[fill=fillColor] (  1.95, 52.78) rectangle (  2.95, 53.78);

\path[fill=fillColor] (152.43, 84.50) rectangle (153.43, 85.50);

\path[fill=fillColor] ( 90.77, 60.72) rectangle ( 91.77, 61.72);

\path[fill=fillColor] (138.69, 78.55) rectangle (139.69, 79.55);

\path[fill=fillColor] (116.07, 79.56) rectangle (117.07, 80.56);

\path[fill=fillColor] (149.89, 88.82) rectangle (150.89, 89.82);

\path[fill=fillColor] ( 88.93, 56.09) rectangle ( 89.93, 57.09);

\path[fill=fillColor] (144.01, 79.69) rectangle (145.01, 80.69);

\path[fill=fillColor] (166.16, 91.20) rectangle (167.16, 92.20);

\path[fill=fillColor] (112.62, 66.79) rectangle (113.62, 67.79);

\path[fill=fillColor] ( 80.56, 82.90) rectangle ( 81.56, 83.90);

\path[fill=fillColor] (107.16, 65.64) rectangle (108.16, 66.64);

\path[fill=fillColor] (111.89, 67.01) rectangle (112.89, 68.01);

\path[fill=fillColor] (123.08, 72.07) rectangle (124.08, 73.07);

\path[fill=fillColor] (133.65, 80.93) rectangle (134.65, 81.93);

\path[fill=fillColor] (134.89, 78.76) rectangle (135.89, 79.76);

\path[fill=fillColor] (149.32, 84.63) rectangle (150.32, 85.63);

\path[fill=fillColor] (131.78, 81.33) rectangle (132.78, 82.33);

\path[fill=fillColor] (153.27, 84.78) rectangle (154.27, 85.78);

\path[fill=fillColor] ( 82.85, 83.24) rectangle ( 83.85, 84.24);

\path[fill=fillColor] (154.79, 85.32) rectangle (155.79, 86.32);

\path[fill=fillColor] ( 84.61, 67.73) rectangle ( 85.61, 68.73);

\path[fill=fillColor] (146.33, 92.26) rectangle (147.33, 93.26);

\path[fill=fillColor] (118.19, 68.86) rectangle (119.19, 69.86);

\path[fill=fillColor] (126.24, 74.10) rectangle (127.24, 75.10);

\path[fill=fillColor] (119.86, 70.20) rectangle (120.86, 71.20);

\path[fill=fillColor] (153.13, 84.69) rectangle (154.13, 85.69);

\path[fill=fillColor] (144.17, 81.10) rectangle (145.17, 82.10);

\path[fill=fillColor] (137.84, 80.00) rectangle (138.84, 81.00);

\path[fill=fillColor] (160.86, 90.77) rectangle (161.86, 91.77);

\path[fill=fillColor] (122.88, 71.51) rectangle (123.88, 72.51);

\path[fill=fillColor] (125.41, 75.32) rectangle (126.41, 76.32);

\path[fill=fillColor] ( 89.96, 56.42) rectangle ( 90.96, 57.42);

\path[fill=fillColor] (149.36, 89.89) rectangle (150.36, 90.89);

\path[fill=fillColor] ( 87.54, 59.60) rectangle ( 88.54, 60.60);

\path[fill=fillColor] (118.41, 69.90) rectangle (119.41, 70.90);

\path[fill=fillColor] ( 92.07, 58.46) rectangle ( 93.07, 59.46);

\path[fill=fillColor] (102.98, 79.97) rectangle (103.98, 80.97);

\path[fill=fillColor] (102.66, 71.23) rectangle (103.66, 72.23);

\path[fill=fillColor] ( 86.80, 61.68) rectangle ( 87.80, 62.68);

\path[fill=fillColor] ( 41.83, 67.03) rectangle ( 42.83, 68.03);

\path[fill=fillColor] (145.39, 90.81) rectangle (146.39, 91.81);

\path[fill=fillColor] (127.15, 73.24) rectangle (128.15, 74.24);

\path[fill=fillColor] ( 79.56, 68.38) rectangle ( 80.56, 69.38);

\path[fill=fillColor] (109.48, 65.57) rectangle (110.48, 66.57);

\path[fill=fillColor] (145.11, 84.67) rectangle (146.11, 85.67);

\path[fill=fillColor] (105.98, 65.00) rectangle (106.98, 66.00);

\path[fill=fillColor] (115.97, 68.79) rectangle (116.97, 69.79);

\path[fill=fillColor] (144.45, 83.24) rectangle (145.45, 84.24);

\path[fill=fillColor] ( 95.46, 70.19) rectangle ( 96.46, 71.19);

\path[fill=fillColor] (120.35, 70.73) rectangle (121.35, 71.73);

\path[fill=fillColor] (130.69, 75.32) rectangle (131.69, 76.32);

\path[fill=fillColor] (151.86, 88.38) rectangle (152.86, 89.38);

\path[fill=fillColor] (110.96, 64.70) rectangle (111.96, 65.70);

\path[fill=fillColor] (154.09, 86.87) rectangle (155.09, 87.87);

\path[fill=fillColor] (155.79, 86.20) rectangle (156.79, 87.20);

\path[fill=fillColor] (128.92, 74.92) rectangle (129.92, 75.92);

\path[fill=fillColor] (106.44, 64.92) rectangle (107.44, 65.92);

\path[fill=fillColor] (153.39, 87.69) rectangle (154.39, 88.69);

\path[fill=fillColor] (161.70, 89.32) rectangle (162.70, 90.32);

\path[fill=fillColor] (107.82, 66.23) rectangle (108.82, 67.23);

\path[fill=fillColor] (129.93, 74.35) rectangle (130.93, 75.35);

\path[fill=fillColor] ( 79.69, 73.97) rectangle ( 80.69, 74.97);

\path[fill=fillColor] (124.80, 73.62) rectangle (125.80, 74.62);

\path[fill=fillColor] (130.73, 75.79) rectangle (131.73, 76.79);

\path[fill=fillColor] ( 67.25, 70.25) rectangle ( 68.25, 71.25);

\path[fill=fillColor] ( 67.21, 62.02) rectangle ( 68.21, 63.02);

\path[fill=fillColor] ( 92.84, 55.71) rectangle ( 93.84, 56.71);

\path[fill=fillColor] (114.80, 74.03) rectangle (115.80, 75.03);

\path[fill=fillColor] (119.05, 70.87) rectangle (120.05, 71.87);

\path[fill=fillColor] (139.32, 84.83) rectangle (140.32, 85.83);

\path[fill=fillColor] (108.73, 64.60) rectangle (109.73, 65.60);

\path[fill=fillColor] (143.89, 84.52) rectangle (144.89, 85.52);

\path[fill=fillColor] ( 95.72, 60.61) rectangle ( 96.72, 61.61);

\path[fill=fillColor] ( 54.91, 39.35) rectangle ( 55.91, 40.35);

\path[fill=fillColor] (152.82, 84.77) rectangle (153.82, 85.77);

\path[fill=fillColor] (110.68, 65.25) rectangle (111.68, 66.25);

\path[fill=fillColor] ( 91.02, 63.64) rectangle ( 92.02, 64.64);

\path[fill=fillColor] ( 81.45, 53.12) rectangle ( 82.45, 54.12);

\path[fill=fillColor] (101.73, 64.16) rectangle (102.73, 65.16);

\path[fill=fillColor] (107.23, 76.52) rectangle (108.23, 77.52);

\path[fill=fillColor] ( 37.10, 50.78) rectangle ( 38.10, 51.78);

\path[fill=fillColor] (122.77, 73.68) rectangle (123.77, 74.68);

\path[fill=fillColor] ( 39.09, 59.07) rectangle ( 40.09, 60.07);

\path[fill=fillColor] ( 86.73, 53.37) rectangle ( 87.73, 54.37);

\path[fill=fillColor] ( 53.18, 39.39) rectangle ( 54.18, 40.39);

\path[fill=fillColor] (127.48, 74.60) rectangle (128.48, 75.60);

\path[fill=fillColor] ( 20.70, 70.63) rectangle ( 21.70, 71.63);

\path[fill=fillColor] (133.31, 75.30) rectangle (134.31, 76.30);

\path[fill=fillColor] ( 91.86, 57.93) rectangle ( 92.86, 58.93);

\path[fill=fillColor] (128.06, 74.90) rectangle (129.06, 75.90);

\path[fill=fillColor] (128.36, 74.31) rectangle (129.36, 75.31);

\path[fill=fillColor] ( 89.40, 71.95) rectangle ( 90.40, 72.95);

\path[fill=fillColor] (142.40, 80.39) rectangle (143.40, 81.39);

\path[fill=fillColor] (136.64, 79.27) rectangle (137.64, 80.27);

\path[fill=fillColor] (145.39, 81.53) rectangle (146.39, 82.53);

\path[fill=fillColor] ( 52.14, 51.43) rectangle ( 53.14, 52.43);

\path[fill=fillColor] (104.03, 65.47) rectangle (105.03, 66.47);

\path[fill=fillColor] ( 66.21, 56.33) rectangle ( 67.21, 57.33);

\path[fill=fillColor] (130.11, 77.84) rectangle (131.11, 78.84);

\path[fill=fillColor] ( 83.18, 65.93) rectangle ( 84.18, 66.93);

\path[fill=fillColor] ( 41.29, 54.05) rectangle ( 42.29, 55.05);

\path[fill=fillColor] (106.92, 73.52) rectangle (107.92, 74.52);

\path[fill=fillColor] ( 76.21, 71.38) rectangle ( 77.21, 72.38);

\path[fill=fillColor] (127.95, 73.73) rectangle (128.95, 74.73);

\path[fill=fillColor] ( 68.34, 46.02) rectangle ( 69.34, 47.02);

\path[fill=fillColor] ( 87.33, 57.67) rectangle ( 88.33, 58.67);

\path[fill=fillColor] (114.14, 77.24) rectangle (115.14, 78.24);

\path[fill=fillColor] (100.69, 65.52) rectangle (101.69, 66.52);

\path[fill=fillColor] (121.31, 74.28) rectangle (122.31, 75.28);

\path[fill=fillColor] ( 92.80, 65.02) rectangle ( 93.80, 66.02);

\path[fill=fillColor] (115.57, 75.03) rectangle (116.57, 76.03);

\path[fill=fillColor] ( 82.22, 62.02) rectangle ( 83.22, 63.02);

\path[fill=fillColor] ( 48.44, 65.97) rectangle ( 49.44, 66.97);

\path[fill=fillColor] ( 83.54, 72.69) rectangle ( 84.54, 73.69);

\path[fill=fillColor] (105.21, 74.08) rectangle (106.21, 75.08);

\path[fill=fillColor] (133.67, 76.72) rectangle (134.67, 77.72);

\path[fill=fillColor] (114.81, 68.60) rectangle (115.81, 69.60);

\path[fill=fillColor] ( 88.87, 64.02) rectangle ( 89.87, 65.02);

\path[fill=fillColor] ( 94.12, 65.08) rectangle ( 95.12, 66.08);

\path[fill=fillColor] ( 55.61, 46.80) rectangle ( 56.61, 47.80);

\path[fill=fillColor] (120.64, 70.05) rectangle (121.64, 71.05);

\path[fill=fillColor] ( 95.66, 65.29) rectangle ( 96.66, 66.29);

\path[fill=fillColor] ( 95.17, 60.79) rectangle ( 96.17, 61.79);

\path[fill=fillColor] (105.96, 76.38) rectangle (106.96, 77.38);

\path[fill=fillColor] (100.66, 62.28) rectangle (101.66, 63.28);

\path[fill=fillColor] ( 84.43, 72.72) rectangle ( 85.43, 73.72);

\path[fill=fillColor] ( 71.58, 49.56) rectangle ( 72.58, 50.56);

\path[fill=fillColor] (103.13, 62.50) rectangle (104.13, 63.50);

\path[fill=fillColor] (102.55, 63.75) rectangle (103.55, 64.75);

\path[fill=fillColor] ( 85.47, 53.65) rectangle ( 86.47, 54.65);

\path[fill=fillColor] ( 91.66, 69.16) rectangle ( 92.66, 70.16);

\path[fill=fillColor] (120.43, 74.08) rectangle (121.43, 75.08);

\path[fill=fillColor] (123.62, 71.21) rectangle (124.62, 72.21);

\path[fill=fillColor] ( 93.99, 69.79) rectangle ( 94.99, 70.79);

\path[fill=fillColor] (104.96, 64.22) rectangle (105.96, 65.22);

\path[fill=fillColor] ( 67.41, 64.45) rectangle ( 68.41, 65.45);

\path[fill=fillColor] (112.80, 74.92) rectangle (113.80, 75.92);

\path[fill=fillColor] ( 74.30, 52.73) rectangle ( 75.30, 53.73);

\path[fill=fillColor] ( 69.75, 58.89) rectangle ( 70.75, 59.89);

\path[fill=fillColor] ( 99.82, 64.77) rectangle (100.82, 65.77);

\path[fill=fillColor] ( 97.13, 70.62) rectangle ( 98.13, 71.62);

\path[fill=fillColor] ( 91.52, 60.35) rectangle ( 92.52, 61.35);

\path[fill=fillColor] ( 87.79, 55.55) rectangle ( 88.79, 56.55);

\path[fill=fillColor] ( 51.81, 61.25) rectangle ( 52.81, 62.25);

\path[fill=fillColor] (104.05, 62.23) rectangle (105.05, 63.23);

\path[fill=fillColor] (113.16, 69.99) rectangle (114.16, 70.99);

\path[fill=fillColor] ( 74.68, 54.28) rectangle ( 75.68, 55.28);

\path[fill=fillColor] ( 26.72, 60.33) rectangle ( 27.72, 61.33);

\path[fill=fillColor] ( 75.31, 55.60) rectangle ( 76.31, 56.60);

\path[fill=fillColor] ( 49.56, 64.40) rectangle ( 50.56, 65.40);

\path[fill=fillColor] ( 82.66, 56.37) rectangle ( 83.66, 57.37);

\path[fill=fillColor] ( 81.42, 55.45) rectangle ( 82.42, 56.45);

\path[fill=fillColor] (127.50, 74.58) rectangle (128.50, 75.58);

\path[fill=fillColor] ( 53.63, 62.58) rectangle ( 54.63, 63.58);

\path[fill=fillColor] ( 61.43, 45.05) rectangle ( 62.43, 46.05);

\path[fill=fillColor] (123.09, 72.82) rectangle (124.09, 73.82);

\path[fill=fillColor] (124.09, 73.34) rectangle (125.09, 74.34);

\path[fill=fillColor] ( 32.67, 49.96) rectangle ( 33.67, 50.96);

\path[fill=fillColor] ( 45.90, 41.35) rectangle ( 46.90, 42.35);

\path[fill=fillColor] ( 41.50, 46.09) rectangle ( 42.50, 47.09);

\path[fill=fillColor] ( 71.28, 45.65) rectangle ( 72.28, 46.65);

\path[fill=fillColor] (129.55, 75.64) rectangle (130.55, 76.64);

\path[fill=fillColor] ( 80.65, 58.07) rectangle ( 81.65, 59.07);

\path[fill=fillColor] (  2.56, 59.12) rectangle (  3.56, 60.12);

\path[fill=fillColor] ( 81.48, 54.62) rectangle ( 82.48, 55.62);

\path[fill=fillColor] (115.81, 69.18) rectangle (116.81, 70.18);

\path[fill=fillColor] ( 46.33, 61.66) rectangle ( 47.33, 62.66);

\path[fill=fillColor] ( 90.44, 58.06) rectangle ( 91.44, 59.06);

\path[fill=fillColor] ( 94.56, 68.62) rectangle ( 95.56, 69.62);

\path[fill=fillColor] ( 55.15, 45.65) rectangle ( 56.15, 46.65);

\path[fill=fillColor] ( 73.72, 52.87) rectangle ( 74.72, 53.87);

\path[fill=fillColor] ( 20.88, 55.49) rectangle ( 21.88, 56.49);

\path[fill=fillColor] ( 29.09, 39.74) rectangle ( 30.09, 40.74);

\path[fill=fillColor] ( 62.54, 56.13) rectangle ( 63.54, 57.13);

\path[fill=fillColor] ( 90.46, 67.23) rectangle ( 91.46, 68.23);

\path[fill=fillColor] (113.57, 70.03) rectangle (114.57, 71.03);

\path[fill=fillColor] ( 80.58, 55.38) rectangle ( 81.58, 56.38);

\path[fill=fillColor] (108.73, 66.93) rectangle (109.73, 67.93);

\path[fill=fillColor] ( 18.20, 58.66) rectangle ( 19.20, 59.66);

\path[fill=fillColor] ( 34.03, 54.62) rectangle ( 35.03, 55.62);

\path[fill=fillColor] (104.50, 65.19) rectangle (105.50, 66.19);

\path[fill=fillColor] ( 17.30, 51.66) rectangle ( 18.30, 52.66);

\path[fill=fillColor] (113.91, 70.23) rectangle (114.91, 71.23);

\path[fill=fillColor] ( 82.84, 52.39) rectangle ( 83.84, 53.39);

\path[fill=fillColor] ( 72.03, 49.85) rectangle ( 73.03, 50.85);

\path[fill=fillColor] ( 54.55, 64.09) rectangle ( 55.55, 65.09);

\path[fill=fillColor] ( 81.65, 68.07) rectangle ( 82.65, 69.07);

\path[fill=fillColor] ( 79.07, 63.60) rectangle ( 80.07, 64.60);

\path[fill=fillColor] ( 72.48, 64.14) rectangle ( 73.48, 65.14);

\path[fill=fillColor] (  4.30, 48.78) rectangle (  5.30, 49.78);

\path[fill=fillColor] ( 47.02, 59.24) rectangle ( 48.02, 60.24);

\path[fill=fillColor] ( 83.52, 50.99) rectangle ( 84.52, 51.99);

\path[fill=fillColor] ( 75.43, 64.99) rectangle ( 76.43, 65.99);

\path[fill=fillColor] (  7.98, 53.74) rectangle (  8.98, 54.74);

\path[fill=fillColor] ( 18.61, 52.51) rectangle ( 19.61, 53.51);

\path[fill=fillColor] ( 84.44, 67.11) rectangle ( 85.44, 68.11);

\path[fill=fillColor] ( 17.23, 46.09) rectangle ( 18.23, 47.09);

\path[fill=fillColor] (  4.96, 51.18) rectangle (  5.96, 52.18);

\path[fill=fillColor] ( 21.86, 50.35) rectangle ( 22.86, 51.35);

\path[fill=fillColor] ( 20.95, 50.24) rectangle ( 21.95, 51.24);

\path[fill=fillColor] ( 25.95, 50.22) rectangle ( 26.95, 51.22);

\path[fill=fillColor] ( 32.96, 56.58) rectangle ( 33.96, 57.58);

\path[fill=fillColor] ( 75.36, 64.38) rectangle ( 76.36, 65.38);

\path[fill=fillColor] ( 37.76, 56.52) rectangle ( 38.76, 57.52);

\path[fill=fillColor] (104.33, 67.70) rectangle (105.33, 68.70);

\path[fill=fillColor] ( 29.28, 56.11) rectangle ( 30.28, 57.11);

\path[fill=fillColor] ( 80.57, 63.34) rectangle ( 81.57, 64.34);

\path[fill=fillColor] ( 57.70, 58.22) rectangle ( 58.70, 59.22);

\path[fill=fillColor] ( 53.60, 60.13) rectangle ( 54.60, 61.13);

\path[fill=fillColor] ( 89.84, 68.40) rectangle ( 90.84, 69.40);

\path[fill=fillColor] ( 23.61, 52.81) rectangle ( 24.61, 53.81);

\path[fill=fillColor] ( 84.27, 58.21) rectangle ( 85.27, 59.21);

\path[fill=fillColor] ( 78.93, 57.90) rectangle ( 79.93, 58.90);

\path[fill=fillColor] ( 35.81, 54.12) rectangle ( 36.81, 55.12);

\path[fill=fillColor] ( 14.78, 51.67) rectangle ( 15.78, 52.67);

\path[fill=fillColor] ( 21.98, 58.10) rectangle ( 22.98, 59.10);

\path[fill=fillColor] ( 83.19, 63.59) rectangle ( 84.19, 64.59);

\path[fill=fillColor] (101.77, 67.15) rectangle (102.77, 68.15);

\path[fill=fillColor] ( 58.66, 64.26) rectangle ( 59.66, 65.26);

\path[fill=fillColor] ( 80.45, 54.42) rectangle ( 81.45, 55.42);

\path[fill=fillColor] ( 25.95, 54.41) rectangle ( 26.95, 55.41);

\path[fill=fillColor] ( 92.59, 58.66) rectangle ( 93.59, 59.66);

\path[fill=fillColor] (110.60, 67.05) rectangle (111.60, 68.05);

\path[fill=fillColor] (111.47, 67.75) rectangle (112.47, 68.75);

\path[fill=fillColor] ( 66.51, 63.45) rectangle ( 67.51, 64.45);

\path[fill=fillColor] ( 87.99, 63.46) rectangle ( 88.99, 64.46);

\path[fill=fillColor] ( 49.31, 59.79) rectangle ( 50.31, 60.79);

\path[fill=fillColor] ( 92.04, 65.82) rectangle ( 93.04, 66.82);

\path[fill=fillColor] ( -0.47, 50.05) rectangle (  0.53, 51.05);

\path[fill=fillColor] ( 14.55, 52.63) rectangle ( 15.55, 53.63);

\path[fill=fillColor] (105.99, 65.05) rectangle (106.99, 66.05);

\path[fill=fillColor] ( 27.02, 40.50) rectangle ( 28.02, 41.50);

\path[fill=fillColor] ( 22.93, 57.06) rectangle ( 23.93, 58.06);

\path[fill=fillColor] ( 94.01, 65.09) rectangle ( 95.01, 66.09);

\path[fill=fillColor] ( 74.88, 62.34) rectangle ( 75.88, 63.34);

\path[fill=fillColor] (109.01, 67.37) rectangle (110.01, 68.37);

\path[fill=fillColor] ( 64.63, 60.43) rectangle ( 65.63, 61.43);

\path[fill=fillColor] ( 92.75, 60.83) rectangle ( 93.75, 61.83);

\path[fill=fillColor] ( 72.93, 51.46) rectangle ( 73.93, 52.46);

\path[fill=fillColor] ( 60.91, 59.13) rectangle ( 61.91, 60.13);

\path[fill=fillColor] ( 66.72, 48.06) rectangle ( 67.72, 49.06);

\path[fill=fillColor] ( 60.04, 63.28) rectangle ( 61.04, 64.28);

\path[fill=fillColor] ( 62.30, 58.53) rectangle ( 63.30, 59.53);

\path[fill=fillColor] ( 40.71, 55.10) rectangle ( 41.71, 56.10);

\path[fill=fillColor] ( 71.04, 57.52) rectangle ( 72.04, 58.52);

\path[fill=fillColor] ( 32.30, 54.87) rectangle ( 33.30, 55.87);

\path[fill=fillColor] ( 62.37, 46.33) rectangle ( 63.37, 47.33);

\path[fill=fillColor] (  0.19, 48.76) rectangle (  1.19, 49.76);

\path[fill=fillColor] (110.95, 67.29) rectangle (111.95, 68.29);

\path[fill=fillColor] ( 94.26, 60.96) rectangle ( 95.26, 61.96);

\path[fill=fillColor] ( 99.82, 62.69) rectangle (100.82, 63.69);

\path[fill=fillColor] ( 98.30, 65.79) rectangle ( 99.30, 66.79);

\path[fill=fillColor] ( 68.66, 48.18) rectangle ( 69.66, 49.18);

\path[fill=fillColor] ( 91.24, 62.69) rectangle ( 92.24, 63.69);

\path[fill=fillColor] ( 50.03, 47.03) rectangle ( 51.03, 48.03);

\path[fill=fillColor] ( 32.88, 36.98) rectangle ( 33.88, 37.98);

\path[fill=fillColor] ( 91.72, 61.77) rectangle ( 92.72, 62.77);

\path[fill=fillColor] ( 18.27, 52.80) rectangle ( 19.27, 53.80);

\path[fill=fillColor] ( 71.33, 60.83) rectangle ( 72.33, 61.83);

\path[fill=fillColor] ( 86.86, 58.63) rectangle ( 87.86, 59.63);

\path[fill=fillColor] ( 18.61, 55.69) rectangle ( 19.61, 56.69);

\path[fill=fillColor] (101.52, 63.52) rectangle (102.52, 64.52);

\path[fill=fillColor] ( 54.65, 56.67) rectangle ( 55.65, 57.67);

\path[fill=fillColor] ( 82.72, 54.85) rectangle ( 83.72, 55.85);

\path[fill=fillColor] ( 87.98, 61.59) rectangle ( 88.98, 62.59);

\path[fill=fillColor] ( 97.12, 65.12) rectangle ( 98.12, 66.12);

\path[fill=fillColor] ( 32.55, 50.41) rectangle ( 33.55, 51.41);

\path[fill=fillColor] (106.82, 65.15) rectangle (107.82, 66.15);

\path[fill=fillColor] ( 51.81, 53.69) rectangle ( 52.81, 54.69);

\path[fill=fillColor] ( 16.22, 52.46) rectangle ( 17.22, 53.46);

\path[fill=fillColor] ( 89.91, 59.01) rectangle ( 90.91, 60.01);

\path[fill=fillColor] ( 61.91, 49.87) rectangle ( 62.91, 50.87);

\path[fill=fillColor] ( 17.01, 39.69) rectangle ( 18.01, 40.69);

\path[fill=fillColor] ( 66.27, 55.45) rectangle ( 67.27, 56.45);

\path[fill=fillColor] ( 29.87, 41.55) rectangle ( 30.87, 42.55);

\path[fill=fillColor] ( 71.70, 49.08) rectangle ( 72.70, 50.08);

\path[fill=fillColor] ( 97.17, 60.81) rectangle ( 98.17, 61.81);

\path[fill=fillColor] ( 82.15, 53.54) rectangle ( 83.15, 54.54);

\path[fill=fillColor] (  1.58, 51.95) rectangle (  2.58, 52.95);

\path[fill=fillColor] ( 73.82, 53.86) rectangle ( 74.82, 54.86);

\path[fill=fillColor] ( 40.80, 56.17) rectangle ( 41.80, 57.17);

\path[fill=fillColor] (103.29, 64.48) rectangle (104.29, 65.48);

\path[fill=fillColor] ( 33.28, 45.20) rectangle ( 34.28, 46.20);

\path[fill=fillColor] (  0.19, 42.58) rectangle (  1.19, 43.58);

\path[fill=fillColor] ( 76.91, 60.60) rectangle ( 77.91, 61.60);

\path[fill=fillColor] ( 63.68, 45.22) rectangle ( 64.68, 46.22);

\path[fill=fillColor] ( 43.35, 40.15) rectangle ( 44.35, 41.15);

\path[fill=fillColor] (  4.07, 49.66) rectangle (  5.07, 50.66);

\path[fill=fillColor] ( -0.60, 48.70) rectangle (  0.40, 49.70);

\path[fill=fillColor] (  6.04, 51.82) rectangle (  7.04, 52.82);

\path[fill=fillColor] ( 59.29, 46.67) rectangle ( 60.29, 47.67);

\path[fill=fillColor] ( 23.61, 51.79) rectangle ( 24.61, 52.79);

\path[fill=fillColor] ( 65.10, 48.72) rectangle ( 66.10, 49.72);

\path[fill=fillColor] ( 37.24, 23.56) rectangle ( 38.24, 24.56);

\path[fill=fillColor] ( 75.69, 60.13) rectangle ( 76.69, 61.13);

\path[fill=fillColor] ( 82.08, 57.60) rectangle ( 83.08, 58.60);

\path[fill=fillColor] ( 92.20, 57.22) rectangle ( 93.20, 58.22);

\path[fill=fillColor] ( 80.99, 59.20) rectangle ( 81.99, 60.20);

\path[fill=fillColor] ( 81.97, 59.27) rectangle ( 82.97, 60.27);

\path[fill=fillColor] ( 74.48, 52.10) rectangle ( 75.48, 53.10);

\path[fill=fillColor] (  6.15, 45.92) rectangle (  7.15, 46.92);

\path[fill=fillColor] ( 45.43, 56.05) rectangle ( 46.43, 57.05);

\path[fill=fillColor] ( 96.79, 58.80) rectangle ( 97.79, 59.80);

\path[fill=fillColor] ( 55.84, 44.31) rectangle ( 56.84, 45.31);

\path[fill=fillColor] ( 78.43, 56.51) rectangle ( 79.43, 57.51);

\path[fill=fillColor] ( 95.54, 61.08) rectangle ( 96.54, 62.08);

\path[fill=fillColor] ( 45.38, 49.75) rectangle ( 46.38, 50.75);

\path[fill=fillColor] ( 87.75, 58.64) rectangle ( 88.75, 59.64);

\path[fill=fillColor] ( 62.91, 59.44) rectangle ( 63.91, 60.44);

\path[fill=fillColor] ( 37.38, 49.04) rectangle ( 38.38, 50.04);

\path[fill=fillColor] ( 97.64, 62.85) rectangle ( 98.64, 63.85);

\path[fill=fillColor] ( 55.51, 43.71) rectangle ( 56.51, 44.71);

\path[fill=fillColor] ( 81.08, 59.74) rectangle ( 82.08, 60.74);

\path[fill=fillColor] ( 88.91, 55.22) rectangle ( 89.91, 56.22);

\path[fill=fillColor] ( 63.36, 58.23) rectangle ( 64.36, 59.23);

\path[fill=fillColor] ( 63.77, 57.42) rectangle ( 64.77, 58.42);

\path[fill=fillColor] ( 32.38, 53.57) rectangle ( 33.38, 54.57);

\path[fill=fillColor] ( 57.68, 53.86) rectangle ( 58.68, 54.86);

\path[fill=fillColor] ( 64.51, 50.06) rectangle ( 65.51, 51.06);

\path[fill=fillColor] ( 29.78, 48.22) rectangle ( 30.78, 49.22);

\path[fill=fillColor] (102.10, 63.70) rectangle (103.10, 64.70);

\path[fill=fillColor] ( 48.32, 40.53) rectangle ( 49.32, 41.53);

\path[fill=fillColor] ( 52.26, 43.40) rectangle ( 53.26, 44.40);

\path[fill=fillColor] ( 89.62, 57.13) rectangle ( 90.62, 58.13);

\path[fill=fillColor] ( 52.89, 42.16) rectangle ( 53.89, 43.16);

\path[fill=fillColor] ( 93.00, 60.70) rectangle ( 94.00, 61.70);

\path[fill=fillColor] ( 46.95, 43.52) rectangle ( 47.95, 44.52);

\path[fill=fillColor] ( 85.82, 59.33) rectangle ( 86.82, 60.33);

\path[fill=fillColor] ( 65.58, 56.57) rectangle ( 66.58, 57.57);

\path[fill=fillColor] ( 57.73, 47.98) rectangle ( 58.73, 48.98);

\path[fill=fillColor] ( 81.46, 57.10) rectangle ( 82.46, 58.10);

\path[fill=fillColor] ( 57.02, 43.65) rectangle ( 58.02, 44.65);

\path[fill=fillColor] ( 75.65, 59.34) rectangle ( 76.65, 60.34);

\path[fill=fillColor] ( 41.32, 55.22) rectangle ( 42.32, 56.22);

\path[fill=fillColor] (  6.87, 47.88) rectangle (  7.87, 48.88);

\path[fill=fillColor] ( 32.51, 47.90) rectangle ( 33.51, 48.90);

\path[fill=fillColor] ( 43.93, 42.88) rectangle ( 44.93, 43.88);

\path[fill=fillColor] ( 84.67, 60.54) rectangle ( 85.67, 61.54);

\path[fill=fillColor] ( 49.88, 43.74) rectangle ( 50.88, 44.74);

\path[fill=fillColor] (  1.21, 47.14) rectangle (  2.21, 48.14);

\path[fill=fillColor] ( 70.50, 56.85) rectangle ( 71.50, 57.85);

\path[fill=fillColor] ( 80.32, 53.59) rectangle ( 81.32, 54.59);

\path[fill=fillColor] ( 35.22, 50.87) rectangle ( 36.22, 51.87);

\path[fill=fillColor] ( 38.83, 36.95) rectangle ( 39.83, 37.95);

\path[fill=fillColor] ( 50.90, 50.72) rectangle ( 51.90, 51.72);

\path[fill=fillColor] ( 79.01, 59.63) rectangle ( 80.01, 60.63);

\path[fill=fillColor] ( 59.11, 54.66) rectangle ( 60.11, 55.66);

\path[fill=fillColor] ( 60.44, 45.51) rectangle ( 61.44, 46.51);

\path[fill=fillColor] (  9.89, 43.40) rectangle ( 10.89, 44.40);

\path[fill=fillColor] ( 80.97, 56.72) rectangle ( 81.97, 57.72);

\path[fill=fillColor] ( 43.85, 51.82) rectangle ( 44.85, 52.82);

\path[fill=fillColor] ( 22.16, 47.75) rectangle ( 23.16, 48.75);

\path[fill=fillColor] ( 88.80, 58.56) rectangle ( 89.80, 59.56);

\path[fill=fillColor] ( 34.50, 49.11) rectangle ( 35.50, 50.11);

\path[fill=fillColor] ( 77.60, 56.76) rectangle ( 78.60, 57.76);

\path[fill=fillColor] ( 83.49, 58.17) rectangle ( 84.49, 59.17);

\path[fill=fillColor] ( 80.86, 57.65) rectangle ( 81.86, 58.65);

\path[fill=fillColor] ( 94.42, 60.48) rectangle ( 95.42, 61.48);

\path[fill=fillColor] ( 39.29, 53.38) rectangle ( 40.29, 54.38);

\path[fill=fillColor] ( 64.98, 56.44) rectangle ( 65.98, 57.44);

\path[fill=fillColor] ( 86.70, 57.85) rectangle ( 87.70, 58.85);

\path[fill=fillColor] (  0.32, 46.38) rectangle (  1.32, 47.38);

\path[fill=fillColor] ( 11.85, 48.08) rectangle ( 12.85, 49.08);

\path[fill=fillColor] ( 55.22, 32.89) rectangle ( 56.22, 33.89);

\path[fill=fillColor] ( 55.28, 41.94) rectangle ( 56.28, 42.94);

\path[fill=fillColor] ( 88.09, 59.19) rectangle ( 89.09, 60.19);

\path[fill=fillColor] ( 80.86, 56.98) rectangle ( 81.86, 57.98);

\path[fill=fillColor] ( 27.47, 49.60) rectangle ( 28.47, 50.60);

\path[fill=fillColor] ( 69.69, 49.40) rectangle ( 70.69, 50.40);

\path[fill=fillColor] ( 67.29, 54.72) rectangle ( 68.29, 55.72);

\path[fill=fillColor] ( 61.87, 54.53) rectangle ( 62.87, 55.53);

\path[fill=fillColor] ( 84.24, 55.78) rectangle ( 85.24, 56.78);

\path[fill=fillColor] ( 88.81, 58.27) rectangle ( 89.81, 59.27);

\path[fill=fillColor] ( 53.34, 52.77) rectangle ( 54.34, 53.77);

\path[fill=fillColor] ( 13.68, 43.87) rectangle ( 14.68, 44.87);

\path[fill=fillColor] ( 55.55, 56.67) rectangle ( 56.55, 57.67);

\path[fill=fillColor] ( 37.93, 51.21) rectangle ( 38.93, 52.21);

\path[fill=fillColor] ( 68.80, 55.94) rectangle ( 69.80, 56.94);

\path[fill=fillColor] ( 81.19, 57.98) rectangle ( 82.19, 58.98);

\path[fill=fillColor] ( 65.11, 45.93) rectangle ( 66.11, 46.93);

\path[fill=fillColor] ( 69.29, 48.82) rectangle ( 70.29, 49.82);

\path[fill=fillColor] ( 84.28, 55.77) rectangle ( 85.28, 56.77);

\path[fill=fillColor] ( 82.38, 55.10) rectangle ( 83.38, 56.10);

\path[fill=fillColor] ( 37.72, 47.71) rectangle ( 38.72, 48.71);

\path[fill=fillColor] ( 63.77, 53.37) rectangle ( 64.77, 54.37);

\path[fill=fillColor] ( 78.99, 57.25) rectangle ( 79.99, 58.25);

\path[fill=fillColor] ( 90.44, 59.23) rectangle ( 91.44, 60.23);

\path[fill=fillColor] ( 25.37, 47.03) rectangle ( 26.37, 48.03);

\path[fill=fillColor] ( 70.35, 49.97) rectangle ( 71.35, 50.97);

\path[fill=fillColor] ( 69.43, 56.00) rectangle ( 70.43, 57.00);

\path[fill=fillColor] ( 25.95, 49.39) rectangle ( 26.95, 50.39);

\path[fill=fillColor] ( 81.45, 54.65) rectangle ( 82.45, 55.65);

\path[fill=fillColor] ( 89.04, 58.11) rectangle ( 90.04, 59.11);

\path[fill=fillColor] ( 38.63, 45.92) rectangle ( 39.63, 46.92);

\path[fill=fillColor] ( 56.95, 44.23) rectangle ( 57.95, 45.23);

\path[fill=fillColor] ( 61.55, 51.38) rectangle ( 62.55, 52.38);

\path[fill=fillColor] ( 72.34, 55.42) rectangle ( 73.34, 56.42);

\path[fill=fillColor] ( 79.09, 55.61) rectangle ( 80.09, 56.61);

\path[fill=fillColor] ( 41.83, 34.64) rectangle ( 42.83, 35.64);

\path[fill=fillColor] ( 68.94, 50.38) rectangle ( 69.94, 51.38);

\path[fill=fillColor] ( 79.41, 54.83) rectangle ( 80.41, 55.83);

\path[fill=fillColor] ( 69.54, 50.92) rectangle ( 70.54, 51.92);

\path[fill=fillColor] ( 71.71, 50.75) rectangle ( 72.71, 51.75);

\path[fill=fillColor] ( 59.65, 47.62) rectangle ( 60.65, 48.62);

\path[fill=fillColor] ( 72.01, 50.44) rectangle ( 73.01, 51.44);

\path[fill=fillColor] ( 65.42, 49.45) rectangle ( 66.42, 50.45);

\path[fill=fillColor] ( 49.70, 51.57) rectangle ( 50.70, 52.57);

\path[fill=fillColor] ( 75.76, 51.97) rectangle ( 76.76, 52.97);

\path[fill=fillColor] ( 30.41, 47.36) rectangle ( 31.41, 48.36);

\path[fill=fillColor] ( 82.60, 56.59) rectangle ( 83.60, 57.59);

\path[fill=fillColor] ( 13.76, 44.75) rectangle ( 14.76, 45.75);

\path[fill=fillColor] ( 35.26, 50.25) rectangle ( 36.26, 51.25);

\path[fill=fillColor] ( 68.77, 51.32) rectangle ( 69.77, 52.32);

\path[fill=fillColor] ( 30.77, 47.00) rectangle ( 31.77, 48.00);

\path[fill=fillColor] ( 67.87, 48.12) rectangle ( 68.87, 49.12);

\path[fill=fillColor] ( 31.84, 39.12) rectangle ( 32.84, 40.12);

\path[fill=fillColor] ( 40.62, 50.05) rectangle ( 41.62, 51.05);

\path[fill=fillColor] ( 25.89, 47.46) rectangle ( 26.89, 48.46);

\path[fill=fillColor] ( 44.85, 41.42) rectangle ( 45.85, 42.42);

\path[fill=fillColor] ( 57.72, 51.85) rectangle ( 58.72, 52.85);

\path[fill=fillColor] ( 67.03, 47.33) rectangle ( 68.03, 48.33);

\path[fill=fillColor] ( 66.79, 49.33) rectangle ( 67.79, 50.33);

\path[fill=fillColor] ( 57.58, 52.18) rectangle ( 58.58, 53.18);

\path[fill=fillColor] ( 45.59, 49.14) rectangle ( 46.59, 50.14);

\path[fill=fillColor] ( 74.59, 53.94) rectangle ( 75.59, 54.94);

\path[fill=fillColor] ( 62.01, 51.81) rectangle ( 63.01, 52.81);

\path[fill=fillColor] ( 47.99, 41.81) rectangle ( 48.99, 42.81);

\path[fill=fillColor] ( 28.53, 48.42) rectangle ( 29.53, 49.42);

\path[fill=fillColor] ( 41.77, 39.54) rectangle ( 42.77, 40.54);

\path[fill=fillColor] ( 62.24, 52.39) rectangle ( 63.24, 53.39);

\path[fill=fillColor] ( 44.72, 38.60) rectangle ( 45.72, 39.60);

\path[fill=fillColor] ( 22.87, 43.18) rectangle ( 23.87, 44.18);

\path[fill=fillColor] ( 73.19, 50.28) rectangle ( 74.19, 51.28);

\path[fill=fillColor] ( 76.01, 54.18) rectangle ( 77.01, 55.18);

\path[fill=fillColor] ( 66.62, 50.42) rectangle ( 67.62, 51.42);

\path[fill=fillColor] ( 59.76, 46.08) rectangle ( 60.76, 47.08);

\path[fill=fillColor] ( 41.77, 50.20) rectangle ( 42.77, 51.20);

\path[fill=fillColor] ( 72.86, 50.63) rectangle ( 73.86, 51.63);

\path[fill=fillColor] ( 67.81, 48.90) rectangle ( 68.81, 49.90);

\path[fill=fillColor] ( 23.21, 40.85) rectangle ( 24.21, 41.85);

\path[fill=fillColor] ( 13.02, 43.35) rectangle ( 14.02, 44.35);

\path[fill=fillColor] ( 66.90, 52.27) rectangle ( 67.90, 53.27);

\path[fill=fillColor] ( 16.22, 41.93) rectangle ( 17.22, 42.93);

\path[fill=fillColor] ( 31.71, 40.70) rectangle ( 32.71, 41.70);

\path[fill=fillColor] ( 63.00, 53.64) rectangle ( 64.00, 54.64);

\path[fill=fillColor] ( 27.61, 46.45) rectangle ( 28.61, 47.45);

\path[fill=fillColor] ( 68.81, 51.71) rectangle ( 69.81, 52.71);

\path[fill=fillColor] ( 67.48, 53.46) rectangle ( 68.48, 54.46);

\path[fill=fillColor] ( 44.69, 41.51) rectangle ( 45.69, 42.51);

\path[fill=fillColor] ( 63.21, 47.85) rectangle ( 64.21, 48.85);

\path[fill=fillColor] ( 56.97, 45.31) rectangle ( 57.97, 46.31);

\path[fill=fillColor] ( 40.40, 51.03) rectangle ( 41.40, 52.03);

\path[fill=fillColor] ( 45.69, 35.13) rectangle ( 46.69, 36.13);

\path[fill=fillColor] ( 73.27, 51.93) rectangle ( 74.27, 52.93);

\path[fill=fillColor] ( 62.84, 43.54) rectangle ( 63.84, 44.54);

\path[fill=fillColor] ( 56.34, 44.65) rectangle ( 57.34, 45.65);

\path[fill=fillColor] ( 31.63, 42.05) rectangle ( 32.63, 43.05);

\path[fill=fillColor] ( 30.10, 45.39) rectangle ( 31.10, 46.39);

\path[fill=fillColor] ( 70.61, 51.09) rectangle ( 71.61, 52.09);

\path[fill=fillColor] ( 56.21, 50.10) rectangle ( 57.21, 51.10);

\path[fill=fillColor] ( 55.91, 40.62) rectangle ( 56.91, 41.62);

\path[fill=fillColor] ( 54.14, 44.94) rectangle ( 55.14, 45.94);

\path[fill=fillColor] ( 62.63, 46.71) rectangle ( 63.63, 47.71);

\path[fill=fillColor] ( 70.41, 51.94) rectangle ( 71.41, 52.94);

\path[fill=fillColor] ( 61.04, 44.54) rectangle ( 62.04, 45.54);

\path[fill=fillColor] ( 37.24, 44.06) rectangle ( 38.24, 45.06);

\path[fill=fillColor] ( 49.74, 38.30) rectangle ( 50.74, 39.30);

\path[fill=fillColor] ( 57.51, 45.68) rectangle ( 58.51, 46.68);

\path[fill=fillColor] ( 47.85, 41.23) rectangle ( 48.85, 42.23);

\path[fill=fillColor] ( 38.03, 47.50) rectangle ( 39.03, 48.50);

\path[fill=fillColor] ( 57.04, 45.60) rectangle ( 58.04, 46.60);

\path[fill=fillColor] ( 62.77, 47.58) rectangle ( 63.77, 48.58);

\path[fill=fillColor] ( 69.92, 50.26) rectangle ( 70.92, 51.26);

\path[fill=fillColor] ( 13.43, 44.57) rectangle ( 14.43, 45.57);

\path[fill=fillColor] ( 42.90, 37.48) rectangle ( 43.90, 38.48);

\path[fill=fillColor] ( 53.65, 41.46) rectangle ( 54.65, 42.46);

\path[fill=fillColor] ( 53.28, 42.87) rectangle ( 54.28, 43.87);

\path[fill=fillColor] ( 52.52, 39.20) rectangle ( 53.52, 40.20);

\path[fill=fillColor] ( 58.00, 46.78) rectangle ( 59.00, 47.78);

\path[fill=fillColor] ( 57.38, 45.33) rectangle ( 58.38, 46.33);

\path[fill=fillColor] ( 41.14, 38.39) rectangle ( 42.14, 39.39);

\path[fill=fillColor] ( 42.79, 47.95) rectangle ( 43.79, 48.95);

\path[fill=fillColor] ( 52.56, 41.40) rectangle ( 53.56, 42.40);

\path[fill=fillColor] ( 42.38, 34.21) rectangle ( 43.38, 35.21);

\path[fill=fillColor] ( 21.86, 42.71) rectangle ( 22.86, 43.71);

\path[fill=fillColor] ( 55.75, 43.81) rectangle ( 56.75, 44.81);

\path[fill=fillColor] ( 55.00, 48.21) rectangle ( 56.00, 49.21);

\path[fill=fillColor] ( 48.42, 41.53) rectangle ( 49.42, 42.53);

\path[fill=fillColor] ( 57.40, 46.11) rectangle ( 58.40, 47.11);

\path[fill=fillColor] ( 51.64, 40.95) rectangle ( 52.64, 41.95);

\path[fill=fillColor] ( 65.19, 47.49) rectangle ( 66.19, 48.49);

\path[fill=fillColor] ( 63.28, 49.54) rectangle ( 64.28, 50.54);

\path[fill=fillColor] ( 29.97, 39.27) rectangle ( 30.97, 40.27);

\path[fill=fillColor] ( 62.11, 46.86) rectangle ( 63.11, 47.86);

\path[fill=fillColor] ( 65.90, 50.00) rectangle ( 66.90, 51.00);

\path[fill=fillColor] ( 52.16, 42.31) rectangle ( 53.16, 43.31);

\path[fill=fillColor] ( 60.36, 45.19) rectangle ( 61.36, 46.19);

\path[fill=fillColor] ( 50.51, 39.74) rectangle ( 51.51, 40.74);

\path[fill=fillColor] ( 22.10, 43.31) rectangle ( 23.10, 44.31);

\path[fill=fillColor] ( 35.77, 36.58) rectangle ( 36.77, 37.58);

\path[fill=fillColor] ( 12.61, 40.86) rectangle ( 13.61, 41.86);

\path[fill=fillColor] ( 57.21, 45.05) rectangle ( 58.21, 46.05);

\path[fill=fillColor] ( 19.80, 39.51) rectangle ( 20.80, 40.51);

\path[fill=fillColor] ( 31.97, 38.83) rectangle ( 32.97, 39.83);

\path[fill=fillColor] ( 71.08, 50.57) rectangle ( 72.08, 51.57);

\path[fill=fillColor] ( 45.84, 41.22) rectangle ( 46.84, 42.22);

\path[fill=fillColor] ( 41.80, 42.00) rectangle ( 42.80, 43.00);

\path[fill=fillColor] ( 40.83, 40.24) rectangle ( 41.83, 41.24);

\path[fill=fillColor] ( 20.32, 40.75) rectangle ( 21.32, 41.75);

\path[fill=fillColor] ( 42.50, 35.89) rectangle ( 43.50, 36.89);

\path[fill=fillColor] ( 60.61, 46.44) rectangle ( 61.61, 47.44);

\path[fill=fillColor] ( 11.76, 40.94) rectangle ( 12.76, 41.94);

\path[fill=fillColor] ( 14.47, 45.14) rectangle ( 15.47, 46.14);

\path[fill=fillColor] ( 62.98, 46.92) rectangle ( 63.98, 47.92);

\path[fill=fillColor] ( 18.48, 40.99) rectangle ( 19.48, 41.99);

\path[fill=fillColor] ( 19.21, 40.11) rectangle ( 20.21, 41.11);

\path[fill=fillColor] ( 41.62, 38.76) rectangle ( 42.62, 39.76);

\path[fill=fillColor] ( 52.97, 44.94) rectangle ( 53.97, 45.94);

\path[fill=fillColor] ( 59.44, 46.40) rectangle ( 60.44, 47.40);

\path[fill=fillColor] ( 28.81, 43.80) rectangle ( 29.81, 44.80);

\path[fill=fillColor] ( 60.91, 46.66) rectangle ( 61.91, 47.66);

\path[fill=fillColor] ( 56.34, 43.87) rectangle ( 57.34, 44.87);

\path[fill=fillColor] ( 21.01, 40.98) rectangle ( 22.01, 41.98);

\path[fill=fillColor] ( 17.23, 41.55) rectangle ( 18.23, 42.55);

\path[fill=fillColor] (  9.05, 38.82) rectangle ( 10.05, 39.82);

\path[fill=fillColor] ( 46.00, 39.36) rectangle ( 47.00, 40.36);

\path[fill=fillColor] ( 22.81, 42.14) rectangle ( 23.81, 43.14);

\path[fill=fillColor] ( 37.24, 42.56) rectangle ( 38.24, 43.56);

\path[fill=fillColor] ( 21.07, 38.61) rectangle ( 22.07, 39.61);

\path[fill=fillColor] ( 30.68, 42.87) rectangle ( 31.68, 43.87);

\path[fill=fillColor] ( 42.06, 44.52) rectangle ( 43.06, 45.52);

\path[fill=fillColor] ( 48.79, 45.05) rectangle ( 49.79, 46.05);

\path[fill=fillColor] ( 25.63, 43.24) rectangle ( 26.63, 44.24);

\path[fill=fillColor] ( 61.03, 46.41) rectangle ( 62.03, 47.41);

\path[fill=fillColor] ( 54.59, 43.87) rectangle ( 55.59, 44.87);

\path[fill=fillColor] ( 51.47, 44.77) rectangle ( 52.47, 45.77);

\path[fill=fillColor] ( 47.97, 38.66) rectangle ( 48.97, 39.66);

\path[fill=fillColor] ( 22.46, 36.84) rectangle ( 23.46, 37.84);

\path[fill=fillColor] ( 45.20, 37.31) rectangle ( 46.20, 38.31);

\path[fill=fillColor] ( 21.56, 39.91) rectangle ( 22.56, 40.91);

\path[fill=fillColor] ( 16.07, 39.03) rectangle ( 17.07, 40.03);

\path[fill=fillColor] ( 39.45, 42.92) rectangle ( 40.45, 43.92);

\path[fill=fillColor] ( 57.29, 45.17) rectangle ( 58.29, 46.17);

\path[fill=fillColor] ( 45.79, 41.73) rectangle ( 46.79, 42.73);

\path[fill=fillColor] ( 35.33, 40.49) rectangle ( 36.33, 41.49);

\path[fill=fillColor] ( 46.10, 41.72) rectangle ( 47.10, 42.72);

\path[fill=fillColor] ( 17.86, 37.89) rectangle ( 18.86, 38.89);

\path[fill=fillColor] ( 50.60, 42.45) rectangle ( 51.60, 43.45);

\path[fill=fillColor] ( 43.52, 37.99) rectangle ( 44.52, 38.99);

\path[fill=fillColor] ( 46.85, 39.46) rectangle ( 47.85, 40.46);

\path[fill=fillColor] ( 40.21, 38.93) rectangle ( 41.21, 39.93);

\path[fill=fillColor] ( 41.53, 39.42) rectangle ( 42.53, 40.42);

\path[fill=fillColor] ( 37.69, 40.27) rectangle ( 38.69, 41.27);

\path[fill=fillColor] ( 47.22, 40.96) rectangle ( 48.22, 41.96);

\path[fill=fillColor] ( 19.35, 38.76) rectangle ( 20.35, 39.76);

\path[fill=fillColor] ( 48.37, 40.90) rectangle ( 49.37, 41.90);

\path[fill=fillColor] ( 44.80, 39.47) rectangle ( 45.80, 40.47);

\path[fill=fillColor] (225.87,115.35) rectangle (226.87,116.35);

\path[fill=fillColor] (198.27,103.47) rectangle (199.27,104.47);

\path[fill=fillColor] (215.51,110.99) rectangle (216.51,111.99);

\path[fill=fillColor] (228.62,115.73) rectangle (229.62,116.73);

\path[fill=fillColor] (197.63,103.65) rectangle (198.63,104.65);

\path[fill=fillColor] (180.49, 93.85) rectangle (181.49, 94.85);

\path[fill=fillColor] (212.14,109.03) rectangle (213.14,110.03);

\path[fill=fillColor] (209.09,108.43) rectangle (210.09,109.43);

\path[fill=fillColor] (204.06,110.59) rectangle (205.06,111.59);

\path[fill=fillColor] (222.23,117.90) rectangle (223.23,118.90);

\path[fill=fillColor] (179.43, 95.34) rectangle (180.43, 96.34);

\path[fill=fillColor] (231.07,118.04) rectangle (232.07,119.04);

\path[fill=fillColor] (175.59, 92.97) rectangle (176.59, 93.97);

\path[fill=fillColor] (184.09, 98.05) rectangle (185.09, 99.05);

\path[fill=fillColor] (209.92,113.31) rectangle (210.92,114.31);

\path[fill=fillColor] (219.24,116.38) rectangle (220.24,117.38);

\path[fill=fillColor] (219.64,112.44) rectangle (220.64,113.44);

\path[fill=fillColor] (201.75,112.64) rectangle (202.75,113.64);

\path[fill=fillColor] (197.89,106.28) rectangle (198.89,107.28);

\path[fill=fillColor] (195.35,111.72) rectangle (196.35,112.72);

\path[fill=fillColor] (219.62,112.65) rectangle (220.62,113.65);

\path[fill=fillColor] (223.14,115.57) rectangle (224.14,116.57);

\path[fill=fillColor] (208.88,107.76) rectangle (209.88,108.76);

\path[fill=fillColor] (193.32,108.80) rectangle (194.32,109.80);

\path[fill=fillColor] (180.85, 97.04) rectangle (181.85, 98.04);

\path[fill=fillColor] (173.56, 95.23) rectangle (174.56, 96.23);

\path[fill=fillColor] (199.44,104.60) rectangle (200.44,105.60);

\path[fill=fillColor] (178.20, 95.15) rectangle (179.20, 96.15);

\path[fill=fillColor] (181.05, 98.26) rectangle (182.05, 99.26);

\path[fill=fillColor] (179.39, 95.69) rectangle (180.39, 96.69);

\path[fill=fillColor] (186.19,100.63) rectangle (187.19,101.63);

\path[fill=fillColor] (209.27,108.98) rectangle (210.27,109.98);

\path[fill=fillColor] (203.21,105.97) rectangle (204.21,106.97);

\path[fill=fillColor] (184.92,102.79) rectangle (185.92,103.79);

\path[fill=fillColor] (178.03, 94.38) rectangle (179.03, 95.38);

\path[fill=fillColor] (165.64,100.53) rectangle (166.64,101.53);

\path[fill=fillColor] (177.20, 95.79) rectangle (178.20, 96.79);

\path[fill=fillColor] (173.80, 94.05) rectangle (174.80, 95.05);

\path[fill=fillColor] (173.33, 93.42) rectangle (174.33, 94.42);

\path[fill=fillColor] (165.61, 93.47) rectangle (166.61, 94.47);
\end{scope}
\begin{scope}
\path[clip] (  0.00,  0.00) rectangle (245.72,126.47);
\definecolor{drawColor}{RGB}{0,0,0}

\path[draw=drawColor,line width= 0.4pt,line join=round,line cap=round] ( 36.00, 36.00) --
	(245.72, 36.00) --
	(245.72,124.07) --
	( 36.00,124.07) --
	( 36.00, 36.00);
\end{scope}
\begin{scope}
\path[clip] (  0.00,  0.00) rectangle (245.72,126.47);
\definecolor{drawColor}{RGB}{0,0,0}

\node[text=drawColor,anchor=base,inner sep=0pt, outer sep=0pt, scale=  1.00] at (140.86,  6.00) {$|E_{\mathrm{out}}|$ for  Bev};

\node[text=drawColor,rotate= 90.00,anchor=base,inner sep=0pt, outer sep=0pt, scale=  1.00] at ( 13.20, 80.04) {$|E_{\mathrm{out}}|$ for  Wei};
\end{scope}
\begin{scope}
\path[clip] (  0.00,  0.00) rectangle (245.72,126.47);
\definecolor{drawColor}{RGB}{0,0,0}

\path[draw=drawColor,line width= 0.4pt,line join=round,line cap=round] ( 43.77, 36.00) -- (237.95, 36.00);

\path[draw=drawColor,line width= 0.4pt,line join=round,line cap=round] ( 43.77, 36.00) -- ( 43.77, 30.00);

\path[draw=drawColor,line width= 0.4pt,line join=round,line cap=round] (108.50, 36.00) -- (108.50, 30.00);

\path[draw=drawColor,line width= 0.4pt,line join=round,line cap=round] (173.22, 36.00) -- (173.22, 30.00);

\path[draw=drawColor,line width= 0.4pt,line join=round,line cap=round] (237.95, 36.00) -- (237.95, 30.00);

\node[text=drawColor,anchor=base,inner sep=0pt, outer sep=0pt, scale=  1.00] at ( 43.77, 20.40) {$10^3$};

\node[text=drawColor,anchor=base,inner sep=0pt, outer sep=0pt, scale=  1.00] at (108.50, 20.40) {$10^4$};

\node[text=drawColor,anchor=base,inner sep=0pt, outer sep=0pt, scale=  1.00] at (173.22, 20.40) {$10^5$};

\node[text=drawColor,anchor=base,inner sep=0pt, outer sep=0pt, scale=  1.00] at (237.95, 20.40) {$10^6$};

\path[draw=drawColor,line width= 0.4pt,line join=round,line cap=round] ( 36.00, 39.26) -- ( 36.00,120.81);

\path[draw=drawColor,line width= 0.4pt,line join=round,line cap=round] ( 36.00, 39.26) -- ( 30.00, 39.26);

\path[draw=drawColor,line width= 0.4pt,line join=round,line cap=round] ( 36.00, 66.44) -- ( 30.00, 66.44);

\path[draw=drawColor,line width= 0.4pt,line join=round,line cap=round] ( 36.00, 93.63) -- ( 30.00, 93.63);

\path[draw=drawColor,line width= 0.4pt,line join=round,line cap=round] ( 36.00,120.81) -- ( 30.00,120.81);

\node[text=drawColor,rotate= 90.00,anchor=base,inner sep=0pt, outer sep=0pt, scale=  1.00] at ( 27.60, 39.26) {$10^3$};

\node[text=drawColor,rotate= 90.00,anchor=base,inner sep=0pt, outer sep=0pt, scale=  1.00] at ( 27.60, 66.44) {$10^4$};

\node[text=drawColor,rotate= 90.00,anchor=base,inner sep=0pt, outer sep=0pt, scale=  1.00] at ( 27.60, 93.63) {$10^5$};

\node[text=drawColor,rotate= 90.00,anchor=base,inner sep=0pt, outer sep=0pt, scale=  1.00] at ( 27.60,120.81) {$10^6$};
\end{scope}
\begin{scope}
\path[clip] ( 36.00, 36.00) rectangle (245.72,124.07);
\definecolor{drawColor}{RGB}{190,190,190}

\path[draw=drawColor,line width= 0.4pt,dash pattern=on 1pt off 3pt ,line join=round,line cap=round] ( 43.77, 36.00) -- ( 43.77,124.07);

\path[draw=drawColor,line width= 0.4pt,dash pattern=on 1pt off 3pt ,line join=round,line cap=round] (108.50, 36.00) -- (108.50,124.07);

\path[draw=drawColor,line width= 0.4pt,dash pattern=on 1pt off 3pt ,line join=round,line cap=round] (173.22, 36.00) -- (173.22,124.07);

\path[draw=drawColor,line width= 0.4pt,dash pattern=on 1pt off 3pt ,line join=round,line cap=round] (237.95, 36.00) -- (237.95,124.07);

\path[draw=drawColor,line width= 0.4pt,dash pattern=on 1pt off 3pt ,line join=round,line cap=round] ( 36.00, 39.26) -- (245.72, 39.26);

\path[draw=drawColor,line width= 0.4pt,dash pattern=on 1pt off 3pt ,line join=round,line cap=round] ( 36.00, 66.44) -- (245.72, 66.44);

\path[draw=drawColor,line width= 0.4pt,dash pattern=on 1pt off 3pt ,line join=round,line cap=round] ( 36.00, 93.63) -- (245.72, 93.63);

\path[draw=drawColor,line width= 0.4pt,dash pattern=on 1pt off 3pt ,line join=round,line cap=round] ( 36.00,120.81) -- (245.72,120.81);

\path[draw=drawColor,line width= 0.4pt,line join=round,line cap=round] ( 36.00, 36.00) -- (245.72,124.07);
\end{scope}
\end{tikzpicture}

%% file: analysis_bio-010.tex
\begin{tikzpicture}[x=1pt,y=1pt]
\definecolor{fillColor}{RGB}{255,255,255}
\path[use as bounding box,fill=fillColor,fill opacity=0.00] (0,0) rectangle (245.72,126.47);
\begin{scope}
\path[clip] ( 36.00, 20.40) rectangle (245.72,124.07);
\definecolor{drawColor}{RGB}{0,0,0}

\path[draw=drawColor,line width= 1.2pt,line join=round] ( 51.98, 50.39) -- ( 59.45, 50.39);

\path[draw=drawColor,line width= 0.4pt,dash pattern=on 4pt off 4pt ,line join=round,line cap=round] ( 55.72, 27.55) -- ( 55.72, 31.14);

\path[draw=drawColor,line width= 0.4pt,dash pattern=on 4pt off 4pt ,line join=round,line cap=round] ( 55.72,120.23) -- ( 55.72, 78.27);

\path[draw=drawColor,line width= 0.4pt,line join=round,line cap=round] ( 53.85, 27.55) -- ( 57.58, 27.55);

\path[draw=drawColor,line width= 0.4pt,line join=round,line cap=round] ( 53.85,120.23) -- ( 57.58,120.23);

\path[draw=drawColor,line width= 0.4pt,line join=round,line cap=round] ( 51.98, 31.14) --
	( 59.45, 31.14) --
	( 59.45, 78.27) --
	( 51.98, 78.27) --
	( 51.98, 31.14);

\path[draw=drawColor,line width= 1.2pt,line join=round] ( 81.86, 50.11) -- ( 89.33, 50.11);

\path[draw=drawColor,line width= 0.4pt,dash pattern=on 4pt off 4pt ,line join=round,line cap=round] ( 85.59, 27.05) -- ( 85.59, 30.36);

\path[draw=drawColor,line width= 0.4pt,dash pattern=on 4pt off 4pt ,line join=round,line cap=round] ( 85.59,119.20) -- ( 85.59, 77.71);

\path[draw=drawColor,line width= 0.4pt,line join=round,line cap=round] ( 83.72, 27.05) -- ( 87.46, 27.05);

\path[draw=drawColor,line width= 0.4pt,line join=round,line cap=round] ( 83.72,119.20) -- ( 87.46,119.20);

\path[draw=drawColor,line width= 0.4pt,line join=round,line cap=round] ( 81.86, 30.36) --
	( 89.33, 30.36) --
	( 89.33, 77.71) --
	( 81.86, 77.71) --
	( 81.86, 30.36);

\path[draw=drawColor,line width= 1.2pt,line join=round] (111.73, 45.56) -- (119.20, 45.56);

\path[draw=drawColor,line width= 0.4pt,dash pattern=on 4pt off 4pt ,line join=round,line cap=round] (115.47, 24.47) -- (115.47, 30.63);

\path[draw=drawColor,line width= 0.4pt,dash pattern=on 4pt off 4pt ,line join=round,line cap=round] (115.47,115.59) -- (115.47, 64.84);

\path[draw=drawColor,line width= 0.4pt,line join=round,line cap=round] (113.60, 24.47) -- (117.33, 24.47);

\path[draw=drawColor,line width= 0.4pt,line join=round,line cap=round] (113.60,115.59) -- (117.33,115.59);

\path[draw=drawColor,line width= 0.4pt,line join=round,line cap=round] (111.73, 30.63) --
	(119.20, 30.63) --
	(119.20, 64.84) --
	(111.73, 64.84) --
	(111.73, 30.63);

\path[draw=drawColor,line width= 1.2pt,line join=round] (141.61, 45.44) -- (149.07, 45.44);

\path[draw=drawColor,line width= 0.4pt,dash pattern=on 4pt off 4pt ,line join=round,line cap=round] (145.34, 24.39) -- (145.34, 30.63);

\path[draw=drawColor,line width= 0.4pt,dash pattern=on 4pt off 4pt ,line join=round,line cap=round] (145.34,107.17) -- (145.34, 64.12);

\path[draw=drawColor,line width= 0.4pt,line join=round,line cap=round] (143.47, 24.39) -- (147.21, 24.39);

\path[draw=drawColor,line width= 0.4pt,line join=round,line cap=round] (143.47,107.17) -- (147.21,107.17);

\path[draw=drawColor,line width= 0.4pt,line join=round,line cap=round] (141.61, 30.63) --
	(149.07, 30.63) --
	(149.07, 64.12) --
	(141.61, 64.12) --
	(141.61, 30.63);

\path[draw=drawColor,line width= 1.2pt,line join=round] (171.48, 45.08) -- (178.95, 45.08);

\path[draw=drawColor,line width= 0.4pt,dash pattern=on 4pt off 4pt ,line join=round,line cap=round] (175.21, 24.47) -- (175.21, 29.66);

\path[draw=drawColor,line width= 0.4pt,dash pattern=on 4pt off 4pt ,line join=round,line cap=round] (175.21,100.53) -- (175.21, 62.26);

\path[draw=drawColor,line width= 0.4pt,line join=round,line cap=round] (173.35, 24.47) -- (177.08, 24.47);

\path[draw=drawColor,line width= 0.4pt,line join=round,line cap=round] (173.35,100.53) -- (177.08,100.53);

\path[draw=drawColor,line width= 0.4pt,line join=round,line cap=round] (171.48, 29.66) --
	(178.95, 29.66) --
	(178.95, 62.26) --
	(171.48, 62.26) --
	(171.48, 29.66);

\path[draw=drawColor,line width= 1.2pt,line join=round] (201.35, 45.08) -- (208.82, 45.08);

\path[draw=drawColor,line width= 0.4pt,dash pattern=on 4pt off 4pt ,line join=round,line cap=round] (205.09, 24.39) -- (205.09, 29.66);

\path[draw=drawColor,line width= 0.4pt,dash pattern=on 4pt off 4pt ,line join=round,line cap=round] (205.09,100.53) -- (205.09, 61.56);

\path[draw=drawColor,line width= 0.4pt,line join=round,line cap=round] (203.22, 24.39) -- (206.96, 24.39);

\path[draw=drawColor,line width= 0.4pt,line join=round,line cap=round] (203.22,100.53) -- (206.96,100.53);

\path[draw=drawColor,line width= 0.4pt,line join=round,line cap=round] (201.35, 29.66) --
	(208.82, 29.66) --
	(208.82, 61.56) --
	(201.35, 61.56) --
	(201.35, 29.66);

\path[draw=drawColor,line width= 1.2pt,line join=round] (231.23, 45.06) -- (238.70, 45.06);

\path[draw=drawColor,line width= 0.4pt,dash pattern=on 4pt off 4pt ,line join=round,line cap=round] (234.96, 24.24) -- (234.96, 29.66);

\path[draw=drawColor,line width= 0.4pt,dash pattern=on 4pt off 4pt ,line join=round,line cap=round] (234.96,100.53) -- (234.96, 61.55);

\path[draw=drawColor,line width= 0.4pt,line join=round,line cap=round] (233.10, 24.24) -- (236.83, 24.24);

\path[draw=drawColor,line width= 0.4pt,line join=round,line cap=round] (233.10,100.53) -- (236.83,100.53);

\path[draw=drawColor,line width= 0.4pt,line join=round,line cap=round] (231.23, 29.66) --
	(238.70, 29.66) --
	(238.70, 61.55) --
	(231.23, 61.55) --
	(231.23, 29.66);
\end{scope}
\begin{scope}
\path[clip] (  0.00,  0.00) rectangle (245.72,126.47);
\definecolor{drawColor}{RGB}{0,0,0}

\node[text=drawColor,rotate= 90.00,anchor=base,inner sep=0pt, outer sep=0pt, scale=  1.00] at (  9.60, 72.24) {$|E_{\mathrm{out}}|\quad [10^5]$};
\end{scope}
\begin{scope}
\path[clip] (  0.00,  0.00) rectangle (245.72,126.47);
\definecolor{drawColor}{RGB}{0,0,0}

\path[draw=drawColor,line width= 0.4pt,line join=round,line cap=round] ( 36.00, 20.40) --
	(245.72, 20.40) --
	(245.72,124.07) --
	( 36.00,124.07) --
	( 36.00, 20.40);
\end{scope}
\begin{scope}
\path[clip] ( 36.00, 20.40) rectangle (245.72,124.07);
\definecolor{drawColor}{RGB}{190,190,190}

\path[draw=drawColor,line width= 0.4pt,dash pattern=on 1pt off 3pt ,line join=round,line cap=round] ( 66.17, 20.40) -- ( 66.17,124.07);

\path[draw=drawColor,line width= 0.4pt,dash pattern=on 1pt off 3pt ,line join=round,line cap=round] ( 96.05, 20.40) -- ( 96.05,124.07);

\path[draw=drawColor,line width= 0.4pt,dash pattern=on 1pt off 3pt ,line join=round,line cap=round] (125.92, 20.40) -- (125.92,124.07);

\path[draw=drawColor,line width= 0.4pt,dash pattern=on 1pt off 3pt ,line join=round,line cap=round] (155.80, 20.40) -- (155.80,124.07);

\path[draw=drawColor,line width= 0.4pt,dash pattern=on 1pt off 3pt ,line join=round,line cap=round] (185.67, 20.40) -- (185.67,124.07);

\path[draw=drawColor,line width= 0.4pt,dash pattern=on 1pt off 3pt ,line join=round,line cap=round] (215.54, 20.40) -- (215.54,124.07);

\path[draw=drawColor,line width= 0.4pt,dash pattern=on 1pt off 3pt ,line join=round,line cap=round] ( 36.00, 27.22) -- (245.72, 27.22);

\path[draw=drawColor,line width= 0.4pt,dash pattern=on 1pt off 3pt ,line join=round,line cap=round] ( 36.00, 39.90) -- (245.72, 39.90);

\path[draw=drawColor,line width= 0.4pt,dash pattern=on 1pt off 3pt ,line join=round,line cap=round] ( 36.00, 52.58) -- (245.72, 52.58);

\path[draw=drawColor,line width= 0.4pt,dash pattern=on 1pt off 3pt ,line join=round,line cap=round] ( 36.00, 65.26) -- (245.72, 65.26);

\path[draw=drawColor,line width= 0.4pt,dash pattern=on 1pt off 3pt ,line join=round,line cap=round] ( 36.00, 77.94) -- (245.72, 77.94);

\path[draw=drawColor,line width= 0.4pt,dash pattern=on 1pt off 3pt ,line join=round,line cap=round] ( 36.00, 90.62) -- (245.72, 90.62);

\path[draw=drawColor,line width= 0.4pt,dash pattern=on 1pt off 3pt ,line join=round,line cap=round] ( 36.00,103.30) -- (245.72,103.30);

\path[draw=drawColor,line width= 0.4pt,dash pattern=on 1pt off 3pt ,line join=round,line cap=round] ( 36.00,115.98) -- (245.72,115.98);
\definecolor{fillColor}{RGB}{0,0,0}

\path[fill=fillColor] ( 44.11, 98.03) rectangle ( 45.11, 99.03);

\path[fill=fillColor] ( 44.58, 45.50) rectangle ( 45.58, 46.50);

\path[fill=fillColor] ( 47.21, 72.15) rectangle ( 48.21, 73.15);

\path[fill=fillColor] ( 45.38,106.75) rectangle ( 46.38,107.75);

\path[fill=fillColor] ( 48.93, 45.10) rectangle ( 49.93, 46.10);

\path[fill=fillColor] ( 43.55, 30.76) rectangle ( 44.55, 31.76);

\path[fill=fillColor] ( 48.52, 66.89) rectangle ( 49.52, 67.89);

\path[fill=fillColor] ( 45.95, 60.39) rectangle ( 46.95, 61.39);

\path[fill=fillColor] ( 48.58, 69.69) rectangle ( 49.58, 70.69);

\path[fill=fillColor] ( 46.17,117.62) rectangle ( 47.17,118.62);

\path[fill=fillColor] ( 45.22, 30.23) rectangle ( 46.22, 31.23);

\path[fill=fillColor] ( 48.36,119.73) rectangle ( 49.36,120.73);

\path[fill=fillColor] ( 47.65, 28.09) rectangle ( 48.65, 29.09);

\path[fill=fillColor] ( 44.80, 33.29) rectangle ( 45.80, 34.29);

\path[fill=fillColor] ( 46.70, 87.42) rectangle ( 47.70, 88.42);

\path[fill=fillColor] ( 47.95,104.97) rectangle ( 48.95,105.97);

\path[fill=fillColor] ( 46.40, 82.01) rectangle ( 47.40, 83.01);

\path[fill=fillColor] ( 43.40, 80.27) rectangle ( 44.40, 81.27);

\path[fill=fillColor] ( 45.49, 53.71) rectangle ( 46.49, 54.71);

\path[fill=fillColor] ( 44.71, 75.26) rectangle ( 45.71, 76.26);

\path[fill=fillColor] ( 44.73, 82.46) rectangle ( 45.73, 83.46);

\path[fill=fillColor] ( 47.83, 98.89) rectangle ( 48.83, 99.89);

\path[fill=fillColor] ( 48.50, 59.96) rectangle ( 49.50, 60.96);

\path[fill=fillColor] ( 45.02, 61.87) rectangle ( 46.02, 62.87);

\path[fill=fillColor] ( 46.96, 31.71) rectangle ( 47.96, 32.71);

\path[fill=fillColor] ( 48.39, 29.20) rectangle ( 49.39, 30.20);

\path[fill=fillColor] ( 43.95, 48.11) rectangle ( 44.95, 49.11);

\path[fill=fillColor] ( 46.79, 29.45) rectangle ( 47.79, 30.45);

\path[fill=fillColor] ( 48.42, 33.67) rectangle ( 49.42, 34.67);

\path[fill=fillColor] ( 45.20, 30.52) rectangle ( 46.20, 31.52);

\path[fill=fillColor] ( 44.47, 39.00) rectangle ( 45.47, 40.00);

\path[fill=fillColor] ( 48.97, 63.40) rectangle ( 49.97, 64.40);

\path[fill=fillColor] ( 46.58, 51.68) rectangle ( 47.58, 52.68);

\path[fill=fillColor] ( 46.14, 43.15) rectangle ( 47.14, 44.15);

\path[fill=fillColor] ( 45.09, 29.83) rectangle ( 46.09, 30.83);

\path[fill=fillColor] ( 45.47, 37.77) rectangle ( 46.47, 38.77);

\path[fill=fillColor] ( 45.01, 30.27) rectangle ( 46.01, 31.27);

\path[fill=fillColor] ( 47.06, 28.04) rectangle ( 48.06, 29.04);

\path[fill=fillColor] ( 46.86, 27.05) rectangle ( 47.86, 28.05);

\path[fill=fillColor] ( 46.72, 27.09) rectangle ( 47.72, 28.09);

\path[fill=fillColor] ( 73.50, 97.34) rectangle ( 74.50, 98.34);

\path[fill=fillColor] ( 77.71, 44.50) rectangle ( 78.71, 45.50);

\path[fill=fillColor] ( 74.47, 71.64) rectangle ( 75.47, 72.64);

\path[fill=fillColor] ( 76.28,100.09) rectangle ( 77.28,101.09);

\path[fill=fillColor] ( 74.30, 44.96) rectangle ( 75.30, 45.96);

\path[fill=fillColor] ( 73.92, 27.53) rectangle ( 74.92, 28.53);

\path[fill=fillColor] ( 76.55, 62.82) rectangle ( 77.55, 63.82);

\path[fill=fillColor] ( 76.20, 60.39) rectangle ( 77.20, 61.39);

\path[fill=fillColor] ( 73.94, 69.69) rectangle ( 74.94, 70.69);

\path[fill=fillColor] ( 74.66,117.44) rectangle ( 75.66,118.44);

\path[fill=fillColor] ( 76.07, 29.33) rectangle ( 77.07, 30.33);

\path[fill=fillColor] ( 75.94,118.70) rectangle ( 76.94,119.70);

\path[fill=fillColor] ( 76.23, 26.55) rectangle ( 77.23, 27.55);

\path[fill=fillColor] ( 76.75, 33.28) rectangle ( 77.75, 34.28);

\path[fill=fillColor] ( 77.11, 84.12) rectangle ( 78.11, 85.12);

\path[fill=fillColor] ( 75.21,104.97) rectangle ( 76.21,105.97);

\path[fill=fillColor] ( 76.23, 79.15) rectangle ( 77.23, 80.15);

\path[fill=fillColor] ( 78.56, 80.27) rectangle ( 79.56, 81.27);

\path[fill=fillColor] ( 75.03, 52.66) rectangle ( 76.03, 53.66);

\path[fill=fillColor] ( 75.90, 75.26) rectangle ( 76.90, 76.26);

\path[fill=fillColor] ( 77.26, 80.31) rectangle ( 78.26, 81.31);

\path[fill=fillColor] ( 78.47, 98.89) rectangle ( 79.47, 99.89);

\path[fill=fillColor] ( 76.44, 57.82) rectangle ( 77.44, 58.82);

\path[fill=fillColor] ( 73.39, 61.87) rectangle ( 74.39, 62.87);

\path[fill=fillColor] ( 75.22, 31.71) rectangle ( 76.22, 32.71);

\path[fill=fillColor] ( 77.21, 29.20) rectangle ( 78.21, 30.20);

\path[fill=fillColor] ( 78.73, 47.55) rectangle ( 79.73, 48.55);

\path[fill=fillColor] ( 79.04, 29.09) rectangle ( 80.04, 30.09);

\path[fill=fillColor] ( 75.78, 33.63) rectangle ( 76.78, 34.63);

\path[fill=fillColor] ( 77.71, 29.79) rectangle ( 78.71, 30.79);

\path[fill=fillColor] ( 74.21, 37.99) rectangle ( 75.21, 38.99);

\path[fill=fillColor] ( 73.73, 62.60) rectangle ( 74.73, 63.60);

\path[fill=fillColor] ( 75.43, 51.67) rectangle ( 76.43, 52.67);

\path[fill=fillColor] ( 76.97, 42.79) rectangle ( 77.97, 43.79);

\path[fill=fillColor] ( 73.52, 28.14) rectangle ( 74.52, 29.14);

\path[fill=fillColor] ( 75.78, 37.77) rectangle ( 76.78, 38.77);

\path[fill=fillColor] ( 76.02, 29.93) rectangle ( 77.02, 30.93);

\path[fill=fillColor] ( 74.25, 27.75) rectangle ( 75.25, 28.75);

\path[fill=fillColor] ( 73.94, 27.04) rectangle ( 74.94, 28.04);

\path[fill=fillColor] ( 74.29, 27.09) rectangle ( 75.29, 28.09);

\path[fill=fillColor] (108.70, 98.03) rectangle (109.70, 99.03);

\path[fill=fillColor] (107.78, 45.50) rectangle (108.78, 46.50);

\path[fill=fillColor] (105.44, 72.13) rectangle (106.44, 73.13);

\path[fill=fillColor] (106.16,106.67) rectangle (107.16,107.67);

\path[fill=fillColor] (108.16, 44.80) rectangle (109.16, 45.80);

\path[fill=fillColor] (105.79, 30.76) rectangle (106.79, 31.76);

\path[fill=fillColor] (104.72, 65.58) rectangle (105.72, 66.58);

\path[fill=fillColor] (107.03, 60.28) rectangle (108.03, 61.28);

\path[fill=fillColor] (108.29, 52.93) rectangle (109.29, 53.93);

\path[fill=fillColor] (103.95, 89.14) rectangle (104.95, 90.14);

\path[fill=fillColor] (108.34, 30.14) rectangle (109.34, 31.14);

\path[fill=fillColor] (105.78,115.09) rectangle (106.78,116.09);

\path[fill=fillColor] (105.86, 28.08) rectangle (106.86, 29.08);

\path[fill=fillColor] (106.09, 33.04) rectangle (107.09, 34.04);

\path[fill=fillColor] (103.58, 63.10) rectangle (104.58, 64.10);

\path[fill=fillColor] (104.44, 80.54) rectangle (105.44, 81.54);

\path[fill=fillColor] (104.96, 81.34) rectangle (105.96, 82.34);

\path[fill=fillColor] (108.81, 49.82) rectangle (109.81, 50.82);

\path[fill=fillColor] (107.16, 45.33) rectangle (108.16, 46.33);

\path[fill=fillColor] (104.85, 43.46) rectangle (105.85, 44.46);

\path[fill=fillColor] (104.26, 81.29) rectangle (105.26, 82.29);

\path[fill=fillColor] (103.76, 90.26) rectangle (104.76, 91.26);

\path[fill=fillColor] (108.32, 59.93) rectangle (109.32, 60.93);

\path[fill=fillColor] (103.87, 40.85) rectangle (104.87, 41.85);

\path[fill=fillColor] (108.47, 30.97) rectangle (109.47, 31.97);

\path[fill=fillColor] (106.32, 27.10) rectangle (107.32, 28.10);

\path[fill=fillColor] (108.86, 46.84) rectangle (109.86, 47.84);

\path[fill=fillColor] (108.12, 29.45) rectangle (109.12, 30.45);

\path[fill=fillColor] (106.07, 31.15) rectangle (107.07, 32.15);

\path[fill=fillColor] (105.17, 30.11) rectangle (106.17, 31.11);

\path[fill=fillColor] (103.90, 34.74) rectangle (104.90, 35.74);

\path[fill=fillColor] (107.72, 60.58) rectangle (108.72, 61.58);

\path[fill=fillColor] (108.96, 51.54) rectangle (109.96, 52.54);

\path[fill=fillColor] (107.18, 34.28) rectangle (108.18, 35.28);

\path[fill=fillColor] (106.68, 29.36) rectangle (107.68, 30.36);

\path[fill=fillColor] (103.15, 24.38) rectangle (104.15, 25.38);

\path[fill=fillColor] (104.86, 28.91) rectangle (105.86, 29.91);

\path[fill=fillColor] (108.95, 27.21) rectangle (109.95, 28.21);

\path[fill=fillColor] (106.27, 27.00) rectangle (107.27, 28.00);

\path[fill=fillColor] (108.36, 23.97) rectangle (109.36, 24.97);

\path[fill=fillColor] (134.28, 98.03) rectangle (135.28, 99.03);

\path[fill=fillColor] (133.48, 45.50) rectangle (134.48, 46.50);

\path[fill=fillColor] (137.77, 72.13) rectangle (138.77, 73.13);

\path[fill=fillColor] (133.68,106.67) rectangle (134.68,107.67);

\path[fill=fillColor] (134.57, 44.80) rectangle (135.57, 45.80);

\path[fill=fillColor] (137.18, 30.76) rectangle (138.18, 31.76);

\path[fill=fillColor] (136.50, 65.58) rectangle (137.50, 66.58);

\path[fill=fillColor] (136.46, 60.28) rectangle (137.46, 61.28);

\path[fill=fillColor] (136.22, 52.70) rectangle (137.22, 53.70);

\path[fill=fillColor] (136.66, 87.81) rectangle (137.66, 88.81);

\path[fill=fillColor] (137.37, 30.14) rectangle (138.37, 31.14);

\path[fill=fillColor] (134.18,115.09) rectangle (135.18,116.09);

\path[fill=fillColor] (138.39, 28.08) rectangle (139.39, 29.08);

\path[fill=fillColor] (136.48, 33.04) rectangle (137.48, 34.04);

\path[fill=fillColor] (133.91, 61.66) rectangle (134.91, 62.66);

\path[fill=fillColor] (133.07, 80.37) rectangle (134.07, 81.37);

\path[fill=fillColor] (135.72, 81.34) rectangle (136.72, 82.34);

\path[fill=fillColor] (133.72, 49.65) rectangle (134.72, 50.65);

\path[fill=fillColor] (133.73, 45.08) rectangle (134.73, 46.08);

\path[fill=fillColor] (134.48, 42.40) rectangle (135.48, 43.40);

\path[fill=fillColor] (134.38, 81.29) rectangle (135.38, 82.29);

\path[fill=fillColor] (135.21, 90.26) rectangle (136.21, 91.26);

\path[fill=fillColor] (133.58, 59.93) rectangle (134.58, 60.93);

\path[fill=fillColor] (134.80, 40.42) rectangle (135.80, 41.42);

\path[fill=fillColor] (138.11, 30.97) rectangle (139.11, 31.97);

\path[fill=fillColor] (134.31, 27.10) rectangle (135.31, 28.10);

\path[fill=fillColor] (134.17, 46.84) rectangle (135.17, 47.84);

\path[fill=fillColor] (137.17, 29.45) rectangle (138.17, 30.45);

\path[fill=fillColor] (138.76, 31.09) rectangle (139.76, 32.09);

\path[fill=fillColor] (133.18, 30.11) rectangle (134.18, 31.11);

\path[fill=fillColor] (137.31, 34.51) rectangle (138.31, 35.51);

\path[fill=fillColor] (137.44, 60.58) rectangle (138.44, 61.58);

\path[fill=fillColor] (136.41, 51.54) rectangle (137.41, 52.54);

\path[fill=fillColor] (136.27, 33.61) rectangle (137.27, 34.61);

\path[fill=fillColor] (137.40, 29.36) rectangle (138.40, 30.36);

\path[fill=fillColor] (134.99, 23.90) rectangle (135.99, 24.90);

\path[fill=fillColor] (134.91, 28.91) rectangle (135.91, 29.91);

\path[fill=fillColor] (135.85, 27.21) rectangle (136.85, 28.21);

\path[fill=fillColor] (138.13, 27.00) rectangle (139.13, 28.00);

\path[fill=fillColor] (135.35, 23.89) rectangle (136.35, 24.89);

\path[fill=fillColor] (166.86, 97.34) rectangle (167.86, 98.34);

\path[fill=fillColor] (165.36, 44.50) rectangle (166.36, 45.50);

\path[fill=fillColor] (167.39, 71.62) rectangle (168.39, 72.62);

\path[fill=fillColor] (166.50,100.03) rectangle (167.50,101.03);

\path[fill=fillColor] (167.51, 44.66) rectangle (168.51, 45.66);

\path[fill=fillColor] (168.63, 27.53) rectangle (169.63, 28.53);

\path[fill=fillColor] (164.86, 61.83) rectangle (165.86, 62.83);

\path[fill=fillColor] (164.40, 60.28) rectangle (165.40, 61.28);

\path[fill=fillColor] (164.17, 52.87) rectangle (165.17, 53.87);

\path[fill=fillColor] (164.90, 89.06) rectangle (165.90, 90.06);

\path[fill=fillColor] (163.21, 29.24) rectangle (164.21, 30.24);

\path[fill=fillColor] (165.77,114.94) rectangle (166.77,115.94);

\path[fill=fillColor] (163.09, 26.54) rectangle (164.09, 27.54);

\path[fill=fillColor] (165.24, 33.03) rectangle (166.24, 34.03);

\path[fill=fillColor] (164.27, 61.69) rectangle (165.27, 62.69);

\path[fill=fillColor] (164.25, 80.54) rectangle (165.25, 81.54);

\path[fill=fillColor] (165.69, 78.72) rectangle (166.69, 79.72);

\path[fill=fillColor] (164.59, 49.82) rectangle (165.59, 50.82);

\path[fill=fillColor] (163.24, 44.90) rectangle (164.24, 45.90);

\path[fill=fillColor] (163.92, 43.46) rectangle (164.92, 44.46);

\path[fill=fillColor] (168.29, 79.26) rectangle (169.29, 80.26);

\path[fill=fillColor] (162.79, 90.26) rectangle (163.79, 91.26);

\path[fill=fillColor] (165.07, 57.85) rectangle (166.07, 58.85);

\path[fill=fillColor] (165.30, 40.85) rectangle (166.30, 41.85);

\path[fill=fillColor] (164.97, 30.97) rectangle (165.97, 31.97);

\path[fill=fillColor] (165.86, 27.10) rectangle (166.86, 28.10);

\path[fill=fillColor] (164.48, 46.30) rectangle (165.48, 47.30);

\path[fill=fillColor] (168.40, 29.09) rectangle (169.40, 30.09);

\path[fill=fillColor] (165.54, 31.11) rectangle (166.54, 32.11);

\path[fill=fillColor] (166.06, 29.41) rectangle (167.06, 30.41);

\path[fill=fillColor] (167.00, 33.97) rectangle (168.00, 34.97);

\path[fill=fillColor] (164.46, 59.99) rectangle (165.46, 60.99);

\path[fill=fillColor] (164.01, 51.54) rectangle (165.01, 52.54);

\path[fill=fillColor] (168.11, 33.89) rectangle (169.11, 34.89);

\path[fill=fillColor] (164.84, 27.86) rectangle (165.84, 28.86);

\path[fill=fillColor] (164.19, 24.38) rectangle (165.19, 25.38);

\path[fill=fillColor] (164.63, 28.64) rectangle (165.63, 29.64);

\path[fill=fillColor] (165.26, 26.70) rectangle (166.26, 27.70);

\path[fill=fillColor] (166.16, 26.98) rectangle (167.16, 27.98);

\path[fill=fillColor] (168.09, 23.97) rectangle (169.09, 24.97);

\path[fill=fillColor] (192.74, 97.34) rectangle (193.74, 98.34);

\path[fill=fillColor] (197.89, 44.50) rectangle (198.89, 45.50);

\path[fill=fillColor] (194.97, 71.62) rectangle (195.97, 72.62);

\path[fill=fillColor] (194.37,100.03) rectangle (195.37,101.03);

\path[fill=fillColor] (198.39, 44.66) rectangle (199.39, 45.66);

\path[fill=fillColor] (196.53, 27.53) rectangle (197.53, 28.53);

\path[fill=fillColor] (194.21, 61.83) rectangle (195.21, 62.83);

\path[fill=fillColor] (193.57, 60.28) rectangle (194.57, 61.28);

\path[fill=fillColor] (195.73, 52.63) rectangle (196.73, 53.63);

\path[fill=fillColor] (194.31, 87.73) rectangle (195.31, 88.73);

\path[fill=fillColor] (192.92, 29.24) rectangle (193.92, 30.24);

\path[fill=fillColor] (196.41,114.94) rectangle (197.41,115.94);

\path[fill=fillColor] (195.21, 26.54) rectangle (196.21, 27.54);

\path[fill=fillColor] (195.58, 33.03) rectangle (196.58, 34.03);

\path[fill=fillColor] (197.67, 60.30) rectangle (198.67, 61.30);

\path[fill=fillColor] (197.97, 80.37) rectangle (198.97, 81.37);

\path[fill=fillColor] (197.29, 78.72) rectangle (198.29, 79.72);

\path[fill=fillColor] (192.90, 49.65) rectangle (193.90, 50.65);

\path[fill=fillColor] (196.84, 44.65) rectangle (197.84, 45.65);

\path[fill=fillColor] (198.06, 42.40) rectangle (199.06, 43.40);

\path[fill=fillColor] (192.95, 79.26) rectangle (193.95, 80.26);

\path[fill=fillColor] (193.22, 90.26) rectangle (194.22, 91.26);

\path[fill=fillColor] (196.91, 57.85) rectangle (197.91, 58.85);

\path[fill=fillColor] (195.27, 40.42) rectangle (196.27, 41.42);

\path[fill=fillColor] (192.92, 30.97) rectangle (193.92, 31.97);

\path[fill=fillColor] (196.13, 27.10) rectangle (197.13, 28.10);

\path[fill=fillColor] (196.75, 46.30) rectangle (197.75, 47.30);

\path[fill=fillColor] (194.93, 29.09) rectangle (195.93, 30.09);

\path[fill=fillColor] (194.38, 31.06) rectangle (195.38, 32.06);

\path[fill=fillColor] (193.78, 29.41) rectangle (194.78, 30.41);

\path[fill=fillColor] (195.54, 33.75) rectangle (196.54, 34.75);

\path[fill=fillColor] (193.16, 59.99) rectangle (194.16, 60.99);

\path[fill=fillColor] (197.83, 51.54) rectangle (198.83, 52.54);

\path[fill=fillColor] (195.11, 33.22) rectangle (196.11, 34.22);

\path[fill=fillColor] (195.41, 27.86) rectangle (196.41, 28.86);

\path[fill=fillColor] (196.87, 23.90) rectangle (197.87, 24.90);

\path[fill=fillColor] (194.03, 28.64) rectangle (195.03, 29.64);

\path[fill=fillColor] (197.06, 26.70) rectangle (198.06, 27.70);

\path[fill=fillColor] (195.45, 26.98) rectangle (196.45, 27.98);

\path[fill=fillColor] (194.57, 23.89) rectangle (195.57, 24.89);

\path[fill=fillColor] (226.67, 97.34) rectangle (227.67, 98.34);

\path[fill=fillColor] (227.04, 44.50) rectangle (228.04, 45.50);

\path[fill=fillColor] (226.06, 71.62) rectangle (227.06, 72.62);

\path[fill=fillColor] (222.98,100.03) rectangle (223.98,101.03);

\path[fill=fillColor] (227.87, 44.66) rectangle (228.87, 45.66);

\path[fill=fillColor] (227.51, 27.53) rectangle (228.51, 28.53);

\path[fill=fillColor] (222.81, 61.83) rectangle (223.81, 62.83);

\path[fill=fillColor] (227.27, 60.28) rectangle (228.27, 61.28);

\path[fill=fillColor] (228.35, 52.87) rectangle (229.35, 53.87);

\path[fill=fillColor] (227.11, 87.39) rectangle (228.11, 88.39);

\path[fill=fillColor] (227.52, 29.24) rectangle (228.52, 30.24);

\path[fill=fillColor] (224.45,114.94) rectangle (225.45,115.94);

\path[fill=fillColor] (225.83, 26.54) rectangle (226.83, 27.54);

\path[fill=fillColor] (226.12, 33.03) rectangle (227.12, 34.03);

\path[fill=fillColor] (228.47, 60.09) rectangle (229.47, 61.09);

\path[fill=fillColor] (223.24, 79.64) rectangle (224.24, 80.64);

\path[fill=fillColor] (225.22, 78.72) rectangle (226.22, 79.72);

\path[fill=fillColor] (225.57, 49.43) rectangle (226.57, 50.43);

\path[fill=fillColor] (223.27, 44.62) rectangle (224.27, 45.62);

\path[fill=fillColor] (224.72, 43.01) rectangle (225.72, 44.01);

\path[fill=fillColor] (223.40, 79.26) rectangle (224.40, 80.26);

\path[fill=fillColor] (224.27, 90.26) rectangle (225.27, 91.26);

\path[fill=fillColor] (223.88, 57.85) rectangle (224.88, 58.85);

\path[fill=fillColor] (223.59, 39.99) rectangle (224.59, 40.99);

\path[fill=fillColor] (222.75, 30.97) rectangle (223.75, 31.97);

\path[fill=fillColor] (226.39, 27.10) rectangle (227.39, 28.10);

\path[fill=fillColor] (228.05, 46.30) rectangle (229.05, 47.30);

\path[fill=fillColor] (222.76, 29.09) rectangle (223.76, 30.09);

\path[fill=fillColor] (225.68, 31.11) rectangle (226.68, 32.11);

\path[fill=fillColor] (223.08, 29.41) rectangle (224.08, 30.41);

\path[fill=fillColor] (224.71, 33.72) rectangle (225.71, 34.72);

\path[fill=fillColor] (224.25, 59.83) rectangle (225.25, 60.83);

\path[fill=fillColor] (226.54, 51.52) rectangle (227.54, 52.52);

\path[fill=fillColor] (227.24, 32.88) rectangle (228.24, 33.88);

\path[fill=fillColor] (227.51, 27.81) rectangle (228.51, 28.81);

\path[fill=fillColor] (226.20, 23.74) rectangle (227.20, 24.74);

\path[fill=fillColor] (227.94, 28.64) rectangle (228.94, 29.64);

\path[fill=fillColor] (222.95, 26.70) rectangle (223.95, 27.70);

\path[fill=fillColor] (224.77, 26.98) rectangle (225.77, 27.98);

\path[fill=fillColor] (224.50, 23.88) rectangle (225.50, 24.88);
\end{scope}
\begin{scope}
\path[clip] (  0.00,  0.00) rectangle (245.72,126.47);
\definecolor{drawColor}{RGB}{0,0,0}

\path[draw=drawColor,line width= 0.4pt,line join=round,line cap=round] ( 51.24, 20.40) -- (230.48, 20.40);

\path[draw=drawColor,line width= 0.4pt,line join=round,line cap=round] ( 51.24, 20.40) -- ( 51.24, 14.40);

\path[draw=drawColor,line width= 0.4pt,line join=round,line cap=round] ( 81.11, 20.40) -- ( 81.11, 14.40);

\path[draw=drawColor,line width= 0.4pt,line join=round,line cap=round] (110.98, 20.40) -- (110.98, 14.40);

\path[draw=drawColor,line width= 0.4pt,line join=round,line cap=round] (140.86, 20.40) -- (140.86, 14.40);

\path[draw=drawColor,line width= 0.4pt,line join=round,line cap=round] (170.73, 20.40) -- (170.73, 14.40);

\path[draw=drawColor,line width= 0.4pt,line join=round,line cap=round] (200.61, 20.40) -- (200.61, 14.40);

\path[draw=drawColor,line width= 0.4pt,line join=round,line cap=round] (230.48, 20.40) -- (230.48, 14.40);

\node[text=drawColor,anchor=base,inner sep=0pt, outer sep=0pt, scale=  1.00] at ( 51.24,  2.40) {N};

\node[text=drawColor,anchor=base,inner sep=0pt, outer sep=0pt, scale=  1.00] at ( 81.11,  2.40) {W};

\node[text=drawColor,anchor=base,inner sep=0pt, outer sep=0pt, scale=  1.00] at (110.98,  2.40) {F};

\node[text=drawColor,anchor=base,inner sep=0pt, outer sep=0pt, scale=  1.00] at (140.86,  2.40) {B};

\node[text=drawColor,anchor=base,inner sep=0pt, outer sep=0pt, scale=  1.00] at (170.73,  2.40) {FW};

\node[text=drawColor,anchor=base,inner sep=0pt, outer sep=0pt, scale=  1.00] at (200.61,  2.40) {BW};

\node[text=drawColor,anchor=base,inner sep=0pt, outer sep=0pt, scale=  1.00] at (230.48,  2.40) {FBW};

\path[draw=drawColor,line width= 0.4pt,line join=round,line cap=round] ( 36.00, 20.40) -- ( 36.00,124.07);

\path[draw=drawColor,line width= 0.4pt,line join=round,line cap=round] ( 36.00, 27.22) -- ( 30.00, 27.22);

\path[draw=drawColor,line width= 0.4pt,line join=round,line cap=round] ( 36.00, 39.90) -- ( 30.00, 39.90);

\path[draw=drawColor,line width= 0.4pt,line join=round,line cap=round] ( 36.00, 52.58) -- ( 30.00, 52.58);

\path[draw=drawColor,line width= 0.4pt,line join=round,line cap=round] ( 36.00, 65.26) -- ( 30.00, 65.26);

\path[draw=drawColor,line width= 0.4pt,line join=round,line cap=round] ( 36.00, 77.94) -- ( 30.00, 77.94);

\path[draw=drawColor,line width= 0.4pt,line join=round,line cap=round] ( 36.00, 90.62) -- ( 30.00, 90.62);

\path[draw=drawColor,line width= 0.4pt,line join=round,line cap=round] ( 36.00,103.30) -- ( 30.00,103.30);

\path[draw=drawColor,line width= 0.4pt,line join=round,line cap=round] ( 36.00,115.98) -- ( 30.00,115.98);

\node[text=drawColor,rotate= 90.00,anchor=base,inner sep=0pt, outer sep=0pt, scale=  1.00] at ( 25.20, 27.22) {1};

\node[text=drawColor,rotate= 90.00,anchor=base,inner sep=0pt, outer sep=0pt, scale=  1.00] at ( 25.20, 52.58) {3};

\node[text=drawColor,rotate= 90.00,anchor=base,inner sep=0pt, outer sep=0pt, scale=  1.00] at ( 25.20, 77.94) {5};

\node[text=drawColor,rotate= 90.00,anchor=base,inner sep=0pt, outer sep=0pt, scale=  1.00] at ( 25.20,103.30) {7};
\end{scope}
\end{tikzpicture}

%% file: analysis_bio-007.tex
\begin{tikzpicture}[x=1pt,y=1pt]
\definecolor{fillColor}{RGB}{255,255,255}
\path[use as bounding box,fill=fillColor,fill opacity=0.00] (0,0) rectangle (245.72,126.47);
\begin{scope}
\path[clip] ( 36.00, 20.40) rectangle (245.72,124.07);
\definecolor{drawColor}{RGB}{0,0,0}

\path[draw=drawColor,line width= 1.2pt,line join=round] ( 54.36,119.18) -- ( 61.42,119.18);

\path[draw=drawColor,line width= 0.4pt,dash pattern=on 4pt off 4pt ,line join=round,line cap=round] ( 57.89,112.83) -- ( 57.89,116.25);

\path[draw=drawColor,line width= 0.4pt,dash pattern=on 4pt off 4pt ,line join=round,line cap=round] ( 57.89,120.23) -- ( 57.89,120.23);

\path[draw=drawColor,line width= 0.4pt,line join=round,line cap=round] ( 56.12,112.83) -- ( 59.66,112.83);

\path[draw=drawColor,line width= 0.4pt,line join=round,line cap=round] ( 56.12,120.23) -- ( 59.66,120.23);

\path[draw=drawColor,line width= 0.4pt,line join=round,line cap=round] ( 54.36,116.25) --
	( 61.42,116.25) --
	( 61.42,120.23) --
	( 54.36,120.23) --
	( 54.36,116.25);

\path[draw=drawColor,line width= 1.2pt,line join=round] ( 89.67,116.44) -- ( 96.73,116.44);

\path[draw=drawColor,line width= 0.4pt,dash pattern=on 4pt off 4pt ,line join=round,line cap=round] ( 93.20, 70.37) -- ( 93.20, 98.63);

\path[draw=drawColor,line width= 0.4pt,dash pattern=on 4pt off 4pt ,line join=round,line cap=round] ( 93.20,120.23) -- ( 93.20,119.86);

\path[draw=drawColor,line width= 0.4pt,line join=round,line cap=round] ( 91.43, 70.37) -- ( 94.96, 70.37);

\path[draw=drawColor,line width= 0.4pt,line join=round,line cap=round] ( 91.43,120.23) -- ( 94.96,120.23);

\path[draw=drawColor,line width= 0.4pt,line join=round,line cap=round] ( 89.67, 98.63) --
	( 96.73, 98.63) --
	( 96.73,119.86) --
	( 89.67,119.86) --
	( 89.67, 98.63);

\path[draw=drawColor,line width= 1.2pt,line join=round] (124.97,116.44) -- (132.03,116.44);

\path[draw=drawColor,line width= 0.4pt,dash pattern=on 4pt off 4pt ,line join=round,line cap=round] (128.50, 68.71) -- (128.50, 98.00);

\path[draw=drawColor,line width= 0.4pt,dash pattern=on 4pt off 4pt ,line join=round,line cap=round] (128.50,120.23) -- (128.50,119.86);

\path[draw=drawColor,line width= 0.4pt,line join=round,line cap=round] (126.74, 68.71) -- (130.27, 68.71);

\path[draw=drawColor,line width= 0.4pt,line join=round,line cap=round] (126.74,120.23) -- (130.27,120.23);

\path[draw=drawColor,line width= 0.4pt,line join=round,line cap=round] (124.97, 98.00) --
	(132.03, 98.00) --
	(132.03,119.86) --
	(124.97,119.86) --
	(124.97, 98.00);

\path[draw=drawColor,line width= 1.2pt,line join=round] (160.28,111.04) -- (167.34,111.04);

\path[draw=drawColor,line width= 0.4pt,dash pattern=on 4pt off 4pt ,line join=round,line cap=round] (163.81, 76.11) -- (163.81, 98.11);

\path[draw=drawColor,line width= 0.4pt,dash pattern=on 4pt off 4pt ,line join=round,line cap=round] (163.81,120.01) -- (163.81,116.05);

\path[draw=drawColor,line width= 0.4pt,line join=round,line cap=round] (162.04, 76.11) -- (165.57, 76.11);

\path[draw=drawColor,line width= 0.4pt,line join=round,line cap=round] (162.04,120.01) -- (165.57,120.01);

\path[draw=drawColor,line width= 0.4pt,line join=round,line cap=round] (160.28, 98.11) --
	(167.34, 98.11) --
	(167.34,116.05) --
	(160.28,116.05) --
	(160.28, 98.11);

\path[draw=drawColor,line width= 1.2pt,line join=round] (195.58,111.04) -- (202.64,111.04);

\path[draw=drawColor,line width= 0.4pt,dash pattern=on 4pt off 4pt ,line join=round,line cap=round] (199.11, 75.85) -- (199.11, 97.48);

\path[draw=drawColor,line width= 0.4pt,dash pattern=on 4pt off 4pt ,line join=round,line cap=round] (199.11,120.01) -- (199.11,116.05);

\path[draw=drawColor,line width= 0.4pt,line join=round,line cap=round] (197.35, 75.85) -- (200.88, 75.85);

\path[draw=drawColor,line width= 0.4pt,line join=round,line cap=round] (197.35,120.01) -- (200.88,120.01);

\path[draw=drawColor,line width= 0.4pt,line join=round,line cap=round] (195.58, 97.48) --
	(202.64, 97.48) --
	(202.64,116.05) --
	(195.58,116.05) --
	(195.58, 97.48);

\path[draw=drawColor,line width= 1.2pt,line join=round] (230.89,111.04) -- (237.95,111.04);

\path[draw=drawColor,line width= 0.4pt,dash pattern=on 4pt off 4pt ,line join=round,line cap=round] (234.42, 69.66) -- (234.42, 97.40);

\path[draw=drawColor,line width= 0.4pt,dash pattern=on 4pt off 4pt ,line join=round,line cap=round] (234.42,120.01) -- (234.42,116.05);

\path[draw=drawColor,line width= 0.4pt,line join=round,line cap=round] (232.65, 69.66) -- (236.19, 69.66);

\path[draw=drawColor,line width= 0.4pt,line join=round,line cap=round] (232.65,120.01) -- (236.19,120.01);

\path[draw=drawColor,line width= 0.4pt,line join=round,line cap=round] (230.89, 97.40) --
	(237.95, 97.40) --
	(237.95,116.05) --
	(230.89,116.05) --
	(230.89, 97.40);
\end{scope}
\begin{scope}
\path[clip] (  0.00,  0.00) rectangle (245.72,126.47);
\definecolor{drawColor}{RGB}{0,0,0}

\path[draw=drawColor,line width= 0.4pt,line join=round,line cap=round] ( 36.00, 24.24) -- ( 36.00,120.23);

\path[draw=drawColor,line width= 0.4pt,line join=round,line cap=round] ( 36.00, 24.24) -- ( 30.00, 24.24);

\path[draw=drawColor,line width= 0.4pt,line join=round,line cap=round] ( 36.00, 43.44) -- ( 30.00, 43.44);

\path[draw=drawColor,line width= 0.4pt,line join=round,line cap=round] ( 36.00, 62.64) -- ( 30.00, 62.64);

\path[draw=drawColor,line width= 0.4pt,line join=round,line cap=round] ( 36.00, 81.84) -- ( 30.00, 81.84);

\path[draw=drawColor,line width= 0.4pt,line join=round,line cap=round] ( 36.00,101.03) -- ( 30.00,101.03);

\path[draw=drawColor,line width= 0.4pt,line join=round,line cap=round] ( 36.00,120.23) -- ( 30.00,120.23);

\node[text=drawColor,rotate= 90.00,anchor=base,inner sep=0pt, outer sep=0pt, scale=  1.00] at ( 25.20, 24.24) {0.0};

\node[text=drawColor,rotate= 90.00,anchor=base,inner sep=0pt, outer sep=0pt, scale=  1.00] at ( 25.20, 62.64) {0.4};

\node[text=drawColor,rotate= 90.00,anchor=base,inner sep=0pt, outer sep=0pt, scale=  1.00] at ( 25.20,101.03) {0.8};
\end{scope}
\begin{scope}
\path[clip] (  0.00,  0.00) rectangle (245.72,126.47);
\definecolor{drawColor}{RGB}{0,0,0}

\node[text=drawColor,rotate= 90.00,anchor=base,inner sep=0pt, outer sep=0pt, scale=  1.00] at (  9.60, 72.24) {$|E_{\mathrm{out}}|/|E_{\mathrm{in}}|$};
\end{scope}
\begin{scope}
\path[clip] (  0.00,  0.00) rectangle (245.72,126.47);
\definecolor{drawColor}{RGB}{0,0,0}

\path[draw=drawColor,line width= 0.4pt,line join=round,line cap=round] ( 36.00, 20.40) --
	(245.72, 20.40) --
	(245.72,124.07) --
	( 36.00,124.07) --
	( 36.00, 20.40);
\end{scope}
\begin{scope}
\path[clip] ( 36.00, 20.40) rectangle (245.72,124.07);
\definecolor{drawColor}{RGB}{190,190,190}

\path[draw=drawColor,line width= 0.4pt,dash pattern=on 1pt off 3pt ,line join=round,line cap=round] ( 70.25, 20.40) -- ( 70.25,124.07);

\path[draw=drawColor,line width= 0.4pt,dash pattern=on 1pt off 3pt ,line join=round,line cap=round] (105.55, 20.40) -- (105.55,124.07);

\path[draw=drawColor,line width= 0.4pt,dash pattern=on 1pt off 3pt ,line join=round,line cap=round] (140.86, 20.40) -- (140.86,124.07);

\path[draw=drawColor,line width= 0.4pt,dash pattern=on 1pt off 3pt ,line join=round,line cap=round] (176.17, 20.40) -- (176.17,124.07);

\path[draw=drawColor,line width= 0.4pt,dash pattern=on 1pt off 3pt ,line join=round,line cap=round] (211.47, 20.40) -- (211.47,124.07);

\path[draw=drawColor,line width= 0.4pt,dash pattern=on 1pt off 3pt ,line join=round,line cap=round] ( 36.00, 24.24) -- (245.72, 24.24);

\path[draw=drawColor,line width= 0.4pt,dash pattern=on 1pt off 3pt ,line join=round,line cap=round] ( 36.00, 43.44) -- (245.72, 43.44);

\path[draw=drawColor,line width= 0.4pt,dash pattern=on 1pt off 3pt ,line join=round,line cap=round] ( 36.00, 62.64) -- (245.72, 62.64);

\path[draw=drawColor,line width= 0.4pt,dash pattern=on 1pt off 3pt ,line join=round,line cap=round] ( 36.00, 81.84) -- (245.72, 81.84);

\path[draw=drawColor,line width= 0.4pt,dash pattern=on 1pt off 3pt ,line join=round,line cap=round] ( 36.00,101.03) -- (245.72,101.03);

\path[draw=drawColor,line width= 0.4pt,dash pattern=on 1pt off 3pt ,line join=round,line cap=round] ( 36.00,120.23) -- (245.72,120.23);
\definecolor{fillColor}{RGB}{0,0,0}

\path[fill=fillColor] ( 45.78,118.94) rectangle ( 46.78,119.94);

\path[fill=fillColor] ( 44.55,116.68) rectangle ( 45.55,117.68);

\path[fill=fillColor] ( 47.89,118.89) rectangle ( 48.89,119.89);

\path[fill=fillColor] ( 46.11,112.83) rectangle ( 47.11,113.83);

\path[fill=fillColor] ( 46.95,119.31) rectangle ( 47.95,120.31);

\path[fill=fillColor] ( 49.38,101.18) rectangle ( 50.38,102.18);

\path[fill=fillColor] ( 45.84,112.33) rectangle ( 46.84,113.33);

\path[fill=fillColor] ( 49.72,119.73) rectangle ( 50.72,120.73);

\path[fill=fillColor] ( 48.58,119.73) rectangle ( 49.58,120.73);

\path[fill=fillColor] ( 50.02,119.57) rectangle ( 51.02,120.57);

\path[fill=fillColor] ( 43.49,114.39) rectangle ( 44.49,115.39);

\path[fill=fillColor] ( 48.16,118.79) rectangle ( 49.16,119.79);

\path[fill=fillColor] ( 46.05,109.18) rectangle ( 47.05,110.18);

\path[fill=fillColor] ( 47.47,119.68) rectangle ( 48.47,120.68);

\path[fill=fillColor] ( 44.94,115.41) rectangle ( 45.94,116.41);

\path[fill=fillColor] ( 45.27,119.73) rectangle ( 46.27,120.73);

\path[fill=fillColor] ( 49.59,115.68) rectangle ( 50.59,116.68);

\path[fill=fillColor] ( 48.65,119.73) rectangle ( 49.65,120.73);

\path[fill=fillColor] ( 48.00,117.21) rectangle ( 49.00,118.21);

\path[fill=fillColor] ( 49.21,119.73) rectangle ( 50.21,120.73);

\path[fill=fillColor] ( 48.46,116.71) rectangle ( 49.46,117.71);

\path[fill=fillColor] ( 47.73,119.73) rectangle ( 48.73,120.73);

\path[fill=fillColor] ( 46.46,115.26) rectangle ( 47.46,116.26);

\path[fill=fillColor] ( 43.51,119.73) rectangle ( 44.51,120.73);

\path[fill=fillColor] ( 49.37,119.73) rectangle ( 50.37,120.73);

\path[fill=fillColor] ( 44.72,119.73) rectangle ( 45.72,120.73);

\path[fill=fillColor] ( 48.31,118.16) rectangle ( 49.31,119.16);

\path[fill=fillColor] ( 44.60,117.49) rectangle ( 45.60,118.49);

\path[fill=fillColor] ( 48.16,119.54) rectangle ( 49.16,120.54);

\path[fill=fillColor] ( 46.57,115.51) rectangle ( 47.57,116.51);

\path[fill=fillColor] ( 45.55,115.82) rectangle ( 46.55,116.82);

\path[fill=fillColor] ( 48.48,118.18) rectangle ( 49.48,119.18);

\path[fill=fillColor] ( 50.28,119.72) rectangle ( 51.28,120.72);

\path[fill=fillColor] ( 49.60,118.58) rectangle ( 50.60,119.58);

\path[fill=fillColor] ( 44.35,109.45) rectangle ( 45.35,110.45);

\path[fill=fillColor] ( 50.17,119.73) rectangle ( 51.17,120.73);

\path[fill=fillColor] ( 47.77,117.72) rectangle ( 48.77,118.72);

\path[fill=fillColor] ( 46.68,117.73) rectangle ( 47.68,118.73);

\path[fill=fillColor] ( 48.19,119.62) rectangle ( 49.19,120.62);

\path[fill=fillColor] ( 46.45,119.73) rectangle ( 47.45,120.73);

\path[fill=fillColor] ( 80.95,119.73) rectangle ( 81.95,120.73);

\path[fill=fillColor] ( 80.49,119.73) rectangle ( 81.49,120.73);

\path[fill=fillColor] ( 81.09,119.70) rectangle ( 82.09,120.70);

\path[fill=fillColor] ( 78.72,119.65) rectangle ( 79.72,120.65);

\path[fill=fillColor] ( 79.93,118.81) rectangle ( 80.93,119.81);

\path[fill=fillColor] ( 82.18,119.73) rectangle ( 83.18,120.73);

\path[fill=fillColor] ( 82.06,117.35) rectangle ( 83.06,118.35);

\path[fill=fillColor] ( 81.83,119.51) rectangle ( 82.83,120.51);

\path[fill=fillColor] ( 83.99, 90.83) rectangle ( 84.99, 91.83);

\path[fill=fillColor] ( 83.60, 93.34) rectangle ( 84.60, 94.34);

\path[fill=fillColor] ( 85.52,119.16) rectangle ( 86.52,120.16);

\path[fill=fillColor] ( 79.89,115.52) rectangle ( 80.89,116.52);

\path[fill=fillColor] ( 79.26,119.67) rectangle ( 80.26,120.67);

\path[fill=fillColor] ( 81.69,118.49) rectangle ( 82.69,119.49);

\path[fill=fillColor] ( 82.22, 87.91) rectangle ( 83.22, 88.91);

\path[fill=fillColor] ( 81.62, 93.95) rectangle ( 82.62, 94.95);

\path[fill=fillColor] ( 79.46,118.78) rectangle ( 80.46,119.78);

\path[fill=fillColor] ( 82.89, 75.61) rectangle ( 83.89, 76.61);

\path[fill=fillColor] ( 85.29, 99.46) rectangle ( 86.29,100.46);

\path[fill=fillColor] ( 81.66, 69.87) rectangle ( 82.66, 70.87);

\path[fill=fillColor] ( 84.46,118.09) rectangle ( 85.46,119.09);

\path[fill=fillColor] ( 80.06,109.97) rectangle ( 81.06,110.97);

\path[fill=fillColor] ( 79.43,119.68) rectangle ( 80.43,120.68);

\path[fill=fillColor] ( 82.70, 77.53) rectangle ( 83.70, 78.53);

\path[fill=fillColor] ( 79.15,115.73) rectangle ( 80.15,116.73);

\path[fill=fillColor] ( 80.99,106.47) rectangle ( 81.99,107.47);

\path[fill=fillColor] ( 83.30,116.16) rectangle ( 84.30,117.16);

\path[fill=fillColor] ( 83.86,119.71) rectangle ( 84.86,120.71);

\path[fill=fillColor] ( 80.56,107.39) rectangle ( 81.56,108.39);

\path[fill=fillColor] ( 80.01,117.36) rectangle ( 81.01,118.36);

\path[fill=fillColor] ( 82.02,103.35) rectangle ( 83.02,104.35);

\path[fill=fillColor] ( 82.11,114.26) rectangle ( 83.11,115.26);

\path[fill=fillColor] ( 79.62,119.39) rectangle ( 80.62,120.39);

\path[fill=fillColor] ( 84.00, 90.50) rectangle ( 85.00, 91.50);

\path[fill=fillColor] ( 84.98,116.83) rectangle ( 85.98,117.83);

\path[fill=fillColor] ( 79.15, 65.54) rectangle ( 80.15, 66.54);

\path[fill=fillColor] ( 80.47,111.69) rectangle ( 81.47,112.69);

\path[fill=fillColor] ( 82.88,114.08) rectangle ( 83.88,115.08);

\path[fill=fillColor] ( 85.27,119.34) rectangle ( 86.27,120.34);

\path[fill=fillColor] ( 83.31, 96.79) rectangle ( 84.31, 97.79);

\path[fill=fillColor] (116.04,119.73) rectangle (117.04,120.73);

\path[fill=fillColor] (119.46,119.73) rectangle (120.46,120.73);

\path[fill=fillColor] (119.53,119.70) rectangle (120.53,120.70);

\path[fill=fillColor] (118.35,119.65) rectangle (119.35,120.65);

\path[fill=fillColor] (114.64,118.81) rectangle (115.64,119.81);

\path[fill=fillColor] (119.97,119.73) rectangle (120.97,120.73);

\path[fill=fillColor] (116.31,117.35) rectangle (117.31,118.35);

\path[fill=fillColor] (120.75,119.51) rectangle (121.75,120.51);

\path[fill=fillColor] (115.09, 90.42) rectangle (116.09, 91.42);

\path[fill=fillColor] (116.76, 92.11) rectangle (117.76, 93.11);

\path[fill=fillColor] (115.97,119.16) rectangle (116.97,120.16);

\path[fill=fillColor] (114.59,115.52) rectangle (115.59,116.52);

\path[fill=fillColor] (116.03,119.67) rectangle (117.03,120.67);

\path[fill=fillColor] (115.82,118.49) rectangle (116.82,119.49);

\path[fill=fillColor] (117.27, 86.03) rectangle (118.27, 87.03);

\path[fill=fillColor] (116.65, 93.77) rectangle (117.65, 94.77);

\path[fill=fillColor] (114.09,118.78) rectangle (115.09,119.78);

\path[fill=fillColor] (117.40, 75.35) rectangle (118.40, 76.35);

\path[fill=fillColor] (120.81, 98.85) rectangle (121.81, 99.85);

\path[fill=fillColor] (120.39, 68.21) rectangle (121.39, 69.21);

\path[fill=fillColor] (120.29,118.09) rectangle (121.29,119.09);

\path[fill=fillColor] (120.53,109.97) rectangle (121.53,110.97);

\path[fill=fillColor] (115.34,119.68) rectangle (116.34,120.68);

\path[fill=fillColor] (116.26, 76.69) rectangle (117.26, 77.69);

\path[fill=fillColor] (115.10,115.73) rectangle (116.10,116.73);

\path[fill=fillColor] (117.99,106.47) rectangle (118.99,107.47);

\path[fill=fillColor] (115.62,116.16) rectangle (116.62,117.16);

\path[fill=fillColor] (119.68,119.71) rectangle (120.68,120.71);

\path[fill=fillColor] (114.88,107.11) rectangle (115.88,108.11);

\path[fill=fillColor] (116.50,117.36) rectangle (117.50,118.36);

\path[fill=fillColor] (115.16,102.47) rectangle (116.16,103.47);

\path[fill=fillColor] (120.79,114.26) rectangle (121.79,115.26);

\path[fill=fillColor] (118.11,119.39) rectangle (119.11,120.39);

\path[fill=fillColor] (115.31, 88.28) rectangle (116.31, 89.28);

\path[fill=fillColor] (119.62,116.83) rectangle (120.62,117.83);

\path[fill=fillColor] (115.30, 63.61) rectangle (116.30, 64.61);

\path[fill=fillColor] (120.26,111.69) rectangle (121.26,112.69);

\path[fill=fillColor] (117.17,114.08) rectangle (118.17,115.08);

\path[fill=fillColor] (115.28,119.34) rectangle (116.28,120.34);

\path[fill=fillColor] (117.10, 96.15) rectangle (118.10, 97.15);

\path[fill=fillColor] (154.30,118.94) rectangle (155.30,119.94);

\path[fill=fillColor] (152.00,116.68) rectangle (153.00,117.68);

\path[fill=fillColor] (153.63,118.86) rectangle (154.63,119.86);

\path[fill=fillColor] (151.60,112.78) rectangle (152.60,113.78);

\path[fill=fillColor] (150.65,118.39) rectangle (151.65,119.39);

\path[fill=fillColor] (149.60,101.18) rectangle (150.60,102.18);

\path[fill=fillColor] (154.83,110.54) rectangle (155.83,111.54);

\path[fill=fillColor] (153.92,119.51) rectangle (154.92,120.51);

\path[fill=fillColor] (151.00, 90.71) rectangle (152.00, 91.71);

\path[fill=fillColor] (149.62, 93.26) rectangle (150.62, 94.26);

\path[fill=fillColor] (156.02,113.82) rectangle (157.02,114.82);

\path[fill=fillColor] (151.31,115.38) rectangle (152.31,116.38);

\path[fill=fillColor] (151.07,109.14) rectangle (152.07,110.14);

\path[fill=fillColor] (150.23,118.44) rectangle (151.23,119.44);

\path[fill=fillColor] (154.25, 86.06) rectangle (155.25, 87.06);

\path[fill=fillColor] (151.05, 93.95) rectangle (152.05, 94.95);

\path[fill=fillColor] (152.77,115.08) rectangle (153.77,116.08);

\path[fill=fillColor] (149.89, 75.61) rectangle (150.89, 76.61);

\path[fill=fillColor] (151.33, 98.43) rectangle (152.33, 99.43);

\path[fill=fillColor] (154.94, 69.87) rectangle (155.94, 70.87);

\path[fill=fillColor] (149.48,115.24) rectangle (150.48,116.24);

\path[fill=fillColor] (155.48,109.97) rectangle (156.48,110.97);

\path[fill=fillColor] (150.56,115.32) rectangle (151.56,116.32);

\path[fill=fillColor] (153.12, 77.53) rectangle (154.12, 78.53);

\path[fill=fillColor] (155.40,115.73) rectangle (156.40,116.73);

\path[fill=fillColor] (151.93,106.47) rectangle (152.93,107.47);

\path[fill=fillColor] (151.56,114.64) rectangle (152.56,115.64);

\path[fill=fillColor] (154.54,117.46) rectangle (155.54,118.46);

\path[fill=fillColor] (153.26,107.22) rectangle (154.26,108.22);

\path[fill=fillColor] (154.99,113.28) rectangle (155.99,114.28);

\path[fill=fillColor] (153.86,100.39) rectangle (154.86,101.39);

\path[fill=fillColor] (149.21,113.10) rectangle (150.21,114.10);

\path[fill=fillColor] (155.51,119.37) rectangle (156.51,120.37);

\path[fill=fillColor] (155.05, 89.21) rectangle (156.05, 90.21);

\path[fill=fillColor] (153.98,107.73) rectangle (154.98,108.73);

\path[fill=fillColor] (150.34, 65.54) rectangle (151.34, 66.54);

\path[fill=fillColor] (152.34,110.07) rectangle (153.34,111.07);

\path[fill=fillColor] (150.18,110.55) rectangle (151.18,111.55);

\path[fill=fillColor] (153.80,119.23) rectangle (154.80,120.23);

\path[fill=fillColor] (149.41, 96.79) rectangle (150.41, 97.79);

\path[fill=fillColor] (188.73,118.94) rectangle (189.73,119.94);

\path[fill=fillColor] (184.91,116.68) rectangle (185.91,117.68);

\path[fill=fillColor] (187.39,118.86) rectangle (188.39,119.86);

\path[fill=fillColor] (190.38,112.78) rectangle (191.38,113.78);

\path[fill=fillColor] (186.12,118.39) rectangle (187.12,119.39);

\path[fill=fillColor] (189.78,101.18) rectangle (190.78,102.18);

\path[fill=fillColor] (190.95,110.54) rectangle (191.95,111.54);

\path[fill=fillColor] (185.37,119.51) rectangle (186.37,120.51);

\path[fill=fillColor] (189.35, 90.31) rectangle (190.35, 91.31);

\path[fill=fillColor] (186.31, 92.04) rectangle (187.31, 93.04);

\path[fill=fillColor] (186.58,113.82) rectangle (187.58,114.82);

\path[fill=fillColor] (187.41,115.38) rectangle (188.41,116.38);

\path[fill=fillColor] (187.88,109.14) rectangle (188.88,110.14);

\path[fill=fillColor] (185.13,118.44) rectangle (186.13,119.44);

\path[fill=fillColor] (185.88, 84.24) rectangle (186.88, 85.24);

\path[fill=fillColor] (191.36, 93.77) rectangle (192.36, 94.77);

\path[fill=fillColor] (186.77,115.08) rectangle (187.77,116.08);

\path[fill=fillColor] (190.59, 75.35) rectangle (191.59, 76.35);

\path[fill=fillColor] (187.25, 97.82) rectangle (188.25, 98.82);

\path[fill=fillColor] (190.70, 68.21) rectangle (191.70, 69.21);

\path[fill=fillColor] (190.26,115.24) rectangle (191.26,116.24);

\path[fill=fillColor] (185.62,109.97) rectangle (186.62,110.97);

\path[fill=fillColor] (184.72,115.32) rectangle (185.72,116.32);

\path[fill=fillColor] (186.09, 76.69) rectangle (187.09, 77.69);

\path[fill=fillColor] (185.87,115.73) rectangle (186.87,116.73);

\path[fill=fillColor] (189.23,106.47) rectangle (190.23,107.47);

\path[fill=fillColor] (189.26,114.64) rectangle (190.26,115.64);

\path[fill=fillColor] (190.75,117.46) rectangle (191.75,118.46);

\path[fill=fillColor] (190.55,106.95) rectangle (191.55,107.95);

\path[fill=fillColor] (190.81,113.28) rectangle (191.81,114.28);

\path[fill=fillColor] (187.20, 99.52) rectangle (188.20,100.52);

\path[fill=fillColor] (186.97,113.10) rectangle (187.97,114.10);

\path[fill=fillColor] (190.38,119.37) rectangle (191.38,120.37);

\path[fill=fillColor] (184.59, 87.00) rectangle (185.59, 88.00);

\path[fill=fillColor] (189.47,107.73) rectangle (190.47,108.73);

\path[fill=fillColor] (185.68, 63.61) rectangle (186.68, 64.61);

\path[fill=fillColor] (189.27,110.07) rectangle (190.27,111.07);

\path[fill=fillColor] (190.61,110.55) rectangle (191.61,111.55);

\path[fill=fillColor] (187.88,119.23) rectangle (188.88,120.23);

\path[fill=fillColor] (188.61, 96.15) rectangle (189.61, 97.15);

\path[fill=fillColor] (222.10,118.94) rectangle (223.10,119.94);

\path[fill=fillColor] (220.55,116.68) rectangle (221.55,117.68);

\path[fill=fillColor] (224.52,118.86) rectangle (225.52,119.86);

\path[fill=fillColor] (226.76,112.78) rectangle (227.76,113.78);

\path[fill=fillColor] (223.76,118.39) rectangle (224.76,119.39);

\path[fill=fillColor] (223.22,101.18) rectangle (224.22,102.18);

\path[fill=fillColor] (226.19,110.54) rectangle (227.19,111.54);

\path[fill=fillColor] (224.34,119.51) rectangle (225.34,120.51);

\path[fill=fillColor] (220.82, 90.71) rectangle (221.82, 91.71);

\path[fill=fillColor] (223.14, 91.71) rectangle (224.14, 92.71);

\path[fill=fillColor] (222.15,113.82) rectangle (223.15,114.82);

\path[fill=fillColor] (225.56,115.38) rectangle (226.56,116.38);

\path[fill=fillColor] (221.87,109.14) rectangle (222.87,110.14);

\path[fill=fillColor] (223.62,118.44) rectangle (224.62,119.44);

\path[fill=fillColor] (220.29, 83.98) rectangle (221.29, 84.98);

\path[fill=fillColor] (223.12, 92.99) rectangle (224.12, 93.99);

\path[fill=fillColor] (224.18,115.08) rectangle (225.18,116.08);

\path[fill=fillColor] (220.34, 75.04) rectangle (221.34, 76.04);

\path[fill=fillColor] (223.54, 97.73) rectangle (224.54, 98.73);

\path[fill=fillColor] (219.88, 69.16) rectangle (220.88, 70.16);

\path[fill=fillColor] (222.18,115.24) rectangle (223.18,116.24);

\path[fill=fillColor] (226.73,109.97) rectangle (227.73,110.97);

\path[fill=fillColor] (226.21,115.32) rectangle (227.21,116.32);

\path[fill=fillColor] (224.58, 75.82) rectangle (225.58, 76.82);

\path[fill=fillColor] (224.26,115.73) rectangle (225.26,116.73);

\path[fill=fillColor] (221.41,106.47) rectangle (222.41,107.47);

\path[fill=fillColor] (223.39,114.64) rectangle (224.39,115.64);

\path[fill=fillColor] (225.75,117.46) rectangle (226.75,118.46);

\path[fill=fillColor] (223.43,107.22) rectangle (224.43,108.22);

\path[fill=fillColor] (224.05,113.28) rectangle (225.05,114.28);

\path[fill=fillColor] (223.54, 99.44) rectangle (224.54,100.44);

\path[fill=fillColor] (222.08,112.79) rectangle (223.08,113.79);

\path[fill=fillColor] (221.66,119.33) rectangle (222.66,120.33);

\path[fill=fillColor] (221.10, 85.88) rectangle (222.10, 86.88);

\path[fill=fillColor] (224.77,107.44) rectangle (225.77,108.44);

\path[fill=fillColor] (226.45, 62.97) rectangle (227.45, 63.97);

\path[fill=fillColor] (222.72,110.07) rectangle (223.72,111.07);

\path[fill=fillColor] (222.33,110.55) rectangle (223.33,111.55);

\path[fill=fillColor] (223.45,119.23) rectangle (224.45,120.23);

\path[fill=fillColor] (226.42, 96.06) rectangle (227.42, 97.06);
\end{scope}
\begin{scope}
\path[clip] (  0.00,  0.00) rectangle (245.72,126.47);
\definecolor{drawColor}{RGB}{0,0,0}

\path[draw=drawColor,line width= 0.4pt,line join=round,line cap=round] ( 52.59, 20.40) -- (229.12, 20.40);

\path[draw=drawColor,line width= 0.4pt,line join=round,line cap=round] ( 52.59, 20.40) -- ( 52.59, 14.40);

\path[draw=drawColor,line width= 0.4pt,line join=round,line cap=round] ( 87.90, 20.40) -- ( 87.90, 14.40);

\path[draw=drawColor,line width= 0.4pt,line join=round,line cap=round] (123.21, 20.40) -- (123.21, 14.40);

\path[draw=drawColor,line width= 0.4pt,line join=round,line cap=round] (158.51, 20.40) -- (158.51, 14.40);

\path[draw=drawColor,line width= 0.4pt,line join=round,line cap=round] (193.82, 20.40) -- (193.82, 14.40);

\path[draw=drawColor,line width= 0.4pt,line join=round,line cap=round] (229.12, 20.40) -- (229.12, 14.40);

\node[text=drawColor,anchor=base,inner sep=0pt, outer sep=0pt, scale=  1.00] at ( 52.59,  2.40) {W};

\node[text=drawColor,anchor=base,inner sep=0pt, outer sep=0pt, scale=  1.00] at ( 87.90,  2.40) {F};

\node[text=drawColor,anchor=base,inner sep=0pt, outer sep=0pt, scale=  1.00] at (123.21,  2.40) {B};

\node[text=drawColor,anchor=base,inner sep=0pt, outer sep=0pt, scale=  1.00] at (158.51,  2.40) {FW};

\node[text=drawColor,anchor=base,inner sep=0pt, outer sep=0pt, scale=  1.00] at (193.82,  2.40) {BW};

\node[text=drawColor,anchor=base,inner sep=0pt, outer sep=0pt, scale=  1.00] at (229.12,  2.40) {FBW};
\end{scope}
\end{tikzpicture}

%% file: analysis_bio-005.tex
\begin{tikzpicture}[x=1pt,y=1pt]
\definecolor{fillColor}{RGB}{255,255,255}
\path[use as bounding box,fill=fillColor,fill opacity=0.00] (0,0) rectangle (245.72,126.47);
\begin{scope}
\path[clip] ( 36.00, 36.00) rectangle (245.72,124.07);
\definecolor{fillColor}{RGB}{0,0,0}

\path[fill=fillColor] (234.86,120.31) rectangle (235.86,121.31);

\path[fill=fillColor] (197.32,120.31) rectangle (198.32,121.31);

\path[fill=fillColor] (201.20,119.76) rectangle (202.20,120.76);

\path[fill=fillColor] (130.00,108.46) rectangle (131.00,109.46);

\path[fill=fillColor] (105.41, 96.12) rectangle (106.41, 97.12);

\path[fill=fillColor] (166.25,120.26) rectangle (167.25,121.26);

\path[fill=fillColor] (201.20,120.31) rectangle (202.20,121.31);

\path[fill=fillColor] (149.42,111.74) rectangle (150.42,112.74);

\path[fill=fillColor] (196.02,120.31) rectangle (197.02,121.31);

\path[fill=fillColor] (119.65,116.05) rectangle (120.65,117.05);

\path[fill=fillColor] (176.61,120.14) rectangle (177.61,121.14);

\path[fill=fillColor] (168.84,120.31) rectangle (169.84,121.31);

\path[fill=fillColor] (181.78,120.31) rectangle (182.78,121.31);

\path[fill=fillColor] (145.54,112.01) rectangle (146.54,113.01);

\path[fill=fillColor] (145.54,115.82) rectangle (146.54,116.82);

\path[fill=fillColor] (161.07,120.31) rectangle (162.07,121.31);

\path[fill=fillColor] (162.37,120.19) rectangle (163.37,121.19);

\path[fill=fillColor] (126.12,114.96) rectangle (127.12,115.96);

\path[fill=fillColor] (104.11, 74.94) rectangle (105.11, 75.94);

\path[fill=fillColor] (126.12, 88.50) rectangle (127.12, 89.50);

\path[fill=fillColor] (104.11, 86.44) rectangle (105.11, 87.44);

\path[fill=fillColor] (172.72,120.28) rectangle (173.72,121.28);

\path[fill=fillColor] (167.54,120.30) rectangle (168.54,121.30);

\path[fill=fillColor] (128.71,115.49) rectangle (129.71,116.49);

\path[fill=fillColor] ( 88.58, 73.06) rectangle ( 89.58, 74.06);

\path[fill=fillColor] (119.65, 90.90) rectangle (120.65, 91.90);

\path[fill=fillColor] (148.13,120.12) rectangle (149.13,121.12);

\path[fill=fillColor] (133.89,114.79) rectangle (134.89,115.79);

\path[fill=fillColor] (127.41,119.88) rectangle (128.41,120.88);

\path[fill=fillColor] ( 66.57, 46.07) rectangle ( 67.57, 47.07);

\path[fill=fillColor] (118.35,120.13) rectangle (119.35,121.13);

\path[fill=fillColor] (127.41,120.28) rectangle (128.41,121.28);

\path[fill=fillColor] (105.41, 95.84) rectangle (106.41, 96.84);

\path[fill=fillColor] (110.58,109.98) rectangle (111.58,110.98);

\path[fill=fillColor] (115.76,111.46) rectangle (116.76,112.46);

\path[fill=fillColor] (100.23, 84.81) rectangle (101.23, 85.81);

\path[fill=fillColor] (139.06,119.84) rectangle (140.06,120.84);

\path[fill=fillColor] ( 66.57, 46.47) rectangle ( 67.57, 47.47);

\path[fill=fillColor] (132.59,117.45) rectangle (133.59,118.45);

\path[fill=fillColor] ( 71.75, 69.28) rectangle ( 72.75, 70.28);

\path[fill=fillColor] ( 96.34, 73.07) rectangle ( 97.34, 74.07);

\path[fill=fillColor] (101.52,113.03) rectangle (102.52,114.03);

\path[fill=fillColor] (117.06,116.97) rectangle (118.06,117.97);

\path[fill=fillColor] (120.94,120.03) rectangle (121.94,121.03);

\path[fill=fillColor] (139.06,119.72) rectangle (140.06,120.72);

\path[fill=fillColor] (127.41,118.41) rectangle (128.41,119.41);

\path[fill=fillColor] (104.11,110.47) rectangle (105.11,111.47);

\path[fill=fillColor] (122.24,104.03) rectangle (123.24,105.03);

\path[fill=fillColor] (117.06,119.49) rectangle (118.06,120.49);

\path[fill=fillColor] ( 97.64,104.02) rectangle ( 98.64,105.02);

\path[fill=fillColor] (104.11, 85.48) rectangle (105.11, 86.48);

\path[fill=fillColor] (117.06,119.55) rectangle (118.06,120.55);

\path[fill=fillColor] ( 74.34, 59.55) rectangle ( 75.34, 60.55);

\path[fill=fillColor] ( 83.40, 81.88) rectangle ( 84.40, 82.88);

\path[fill=fillColor] ( 58.80, 45.83) rectangle ( 59.80, 46.83);

\path[fill=fillColor] ( 95.05, 76.30) rectangle ( 96.05, 77.30);

\path[fill=fillColor] (114.47,120.00) rectangle (115.47,121.00);

\path[fill=fillColor] ( 67.86, 62.89) rectangle ( 68.86, 63.89);

\path[fill=fillColor] (111.88,120.16) rectangle (112.88,121.16);

\path[fill=fillColor] (110.58,101.29) rectangle (111.58,102.29);

\path[fill=fillColor] (108.00,114.56) rectangle (109.00,115.56);

\path[fill=fillColor] (105.41,100.25) rectangle (106.41,101.25);

\path[fill=fillColor] ( 83.40, 74.08) rectangle ( 84.40, 75.08);

\path[fill=fillColor] (120.94,119.56) rectangle (121.94,120.56);

\path[fill=fillColor] (135.18,120.30) rectangle (136.18,121.30);

\path[fill=fillColor] (104.11, 95.56) rectangle (105.11, 96.56);

\path[fill=fillColor] (111.88,120.31) rectangle (112.88,121.31);

\path[fill=fillColor] (108.00,108.66) rectangle (109.00,109.66);

\path[fill=fillColor] (135.18,119.07) rectangle (136.18,120.07);

\path[fill=fillColor] (104.11,113.56) rectangle (105.11,114.56);

\path[fill=fillColor] (106.70,103.73) rectangle (107.70,104.73);

\path[fill=fillColor] (115.76,114.84) rectangle (116.76,115.84);

\path[fill=fillColor] ( 92.46,113.58) rectangle ( 93.46,114.58);

\path[fill=fillColor] (117.06,119.41) rectangle (118.06,120.41);

\path[fill=fillColor] ( 66.57, 52.98) rectangle ( 67.57, 53.98);

\path[fill=fillColor] ( 98.93,116.14) rectangle ( 99.93,117.14);

\path[fill=fillColor] (114.47,118.05) rectangle (115.47,119.05);

\path[fill=fillColor] ( 62.69, 50.32) rectangle ( 63.69, 51.32);

\path[fill=fillColor] ( 83.40, 91.12) rectangle ( 84.40, 92.12);

\path[fill=fillColor] ( 95.05,111.95) rectangle ( 96.05,112.95);

\path[fill=fillColor] ( 96.34, 88.93) rectangle ( 97.34, 89.93);

\path[fill=fillColor] (109.29,119.59) rectangle (110.29,120.59);

\path[fill=fillColor] ( 97.64, 97.81) rectangle ( 98.64, 98.81);

\path[fill=fillColor] (140.36,120.31) rectangle (141.36,121.31);

\path[fill=fillColor] (106.70,119.54) rectangle (107.70,120.54);

\path[fill=fillColor] ( 78.22, 69.15) rectangle ( 79.22, 70.15);

\path[fill=fillColor] ( 91.17, 97.92) rectangle ( 92.17, 98.92);

\path[fill=fillColor] (122.24,117.14) rectangle (123.24,118.14);

\path[fill=fillColor] ( 52.33, 41.22) rectangle ( 53.33, 42.22);

\path[fill=fillColor] (119.65,119.26) rectangle (120.65,120.26);

\path[fill=fillColor] ( 98.93,116.73) rectangle ( 99.93,117.73);

\path[fill=fillColor] (104.11,116.65) rectangle (105.11,117.65);

\path[fill=fillColor] ( 74.34, 64.07) rectangle ( 75.34, 65.07);

\path[fill=fillColor] (106.70,116.68) rectangle (107.70,117.68);

\path[fill=fillColor] ( 95.05,108.76) rectangle ( 96.05,109.76);

\path[fill=fillColor] (117.06,118.34) rectangle (118.06,119.34);

\path[fill=fillColor] ( 93.75,102.23) rectangle ( 94.75,103.23);

\path[fill=fillColor] ( 69.16, 69.53) rectangle ( 70.16, 70.53);

\path[fill=fillColor] ( 79.51, 78.70) rectangle ( 80.51, 79.70);

\path[fill=fillColor] ( 65.27, 53.95) rectangle ( 66.27, 54.95);

\path[fill=fillColor] (113.17,119.21) rectangle (114.17,120.21);

\path[fill=fillColor] ( 78.22, 77.75) rectangle ( 79.22, 78.75);

\path[fill=fillColor] ( 89.87, 97.98) rectangle ( 90.87, 98.98);

\path[fill=fillColor] ( 88.58,101.43) rectangle ( 89.58,102.43);

\path[fill=fillColor] ( 83.40, 88.23) rectangle ( 84.40, 89.23);

\path[fill=fillColor] ( 57.51, 48.54) rectangle ( 58.51, 49.54);

\path[fill=fillColor] ( 66.57, 58.06) rectangle ( 67.57, 59.06);

\path[fill=fillColor] ( 79.51, 75.86) rectangle ( 80.51, 76.86);

\path[fill=fillColor] (102.82,117.90) rectangle (103.82,118.90);

\path[fill=fillColor] ( 98.93,117.17) rectangle ( 99.93,118.17);

\path[fill=fillColor] (101.52,116.00) rectangle (102.52,117.00);

\path[fill=fillColor] ( 74.34, 70.10) rectangle ( 75.34, 71.10);

\path[fill=fillColor] ( 65.27, 58.63) rectangle ( 66.27, 59.63);

\path[fill=fillColor] ( 73.04, 73.51) rectangle ( 74.04, 74.51);

\path[fill=fillColor] ( 84.69,100.77) rectangle ( 85.69,101.77);

\path[fill=fillColor] ( 98.93,118.83) rectangle ( 99.93,119.83);

\path[fill=fillColor] ( 71.75, 73.24) rectangle ( 72.75, 74.24);

\path[fill=fillColor] ( 78.22, 84.00) rectangle ( 79.22, 85.00);

\path[fill=fillColor] ( 71.75, 73.45) rectangle ( 72.75, 74.45);

\path[fill=fillColor] ( 84.69,104.74) rectangle ( 85.69,105.74);

\path[fill=fillColor] ( 89.87,117.27) rectangle ( 90.87,118.27);

\path[fill=fillColor] ( 96.34,116.66) rectangle ( 97.34,117.66);

\path[fill=fillColor] ( 98.93,116.03) rectangle ( 99.93,117.03);

\path[fill=fillColor] ( 93.75,116.00) rectangle ( 94.75,117.00);

\path[fill=fillColor] (101.52,119.77) rectangle (102.52,120.77);

\path[fill=fillColor] ( 71.75, 77.83) rectangle ( 72.75, 78.83);

\path[fill=fillColor] ( 71.75, 77.97) rectangle ( 72.75, 78.97);

\path[fill=fillColor] ( 83.40,109.13) rectangle ( 84.40,110.13);

\path[fill=fillColor] ( 85.99,112.29) rectangle ( 86.99,113.29);

\path[fill=fillColor] ( 84.69,108.79) rectangle ( 85.69,109.79);

\path[fill=fillColor] ( 65.27, 65.45) rectangle ( 66.27, 66.45);

\path[fill=fillColor] ( 67.86, 70.72) rectangle ( 68.86, 71.72);

\path[fill=fillColor] ( 79.51,100.47) rectangle ( 80.51,101.47);

\path[fill=fillColor] ( 67.86, 73.48) rectangle ( 68.86, 74.48);

\path[fill=fillColor] ( 76.93, 97.79) rectangle ( 77.93, 98.79);

\path[fill=fillColor] ( 87.28,119.53) rectangle ( 88.28,120.53);

\path[fill=fillColor] ( 80.81,113.74) rectangle ( 81.81,114.74);

\path[fill=fillColor] ( 87.28,113.70) rectangle ( 88.28,114.70);

\path[fill=fillColor] ( 91.17,119.60) rectangle ( 92.17,120.60);

\path[fill=fillColor] (179.20, 95.41) rectangle (180.20, 96.41);

\path[fill=fillColor] (208.97, 96.84) rectangle (209.97, 97.84);

\path[fill=fillColor] (205.09,119.83) rectangle (206.09,120.83);

\path[fill=fillColor] (241.33,116.73) rectangle (242.33,117.73);

\path[fill=fillColor] (216.74,120.26) rectangle (217.74,121.26);

\path[fill=fillColor] (179.20,119.26) rectangle (180.20,120.26);

\path[fill=fillColor] (172.72, 91.68) rectangle (173.72, 92.68);

\path[fill=fillColor] (202.50, 98.25) rectangle (203.50, 99.25);

\path[fill=fillColor] (159.78, 82.61) rectangle (160.78, 83.61);

\path[fill=fillColor] (159.78,102.57) rectangle (160.78,103.57);

\path[fill=fillColor] (155.89, 76.54) rectangle (156.89, 77.54);

\path[fill=fillColor] (214.15,112.02) rectangle (215.15,113.02);

\path[fill=fillColor] (152.01, 83.74) rectangle (153.01, 84.74);

\path[fill=fillColor] (153.30,116.91) rectangle (154.30,117.91);

\path[fill=fillColor] (163.66,109.05) rectangle (164.66,110.05);

\path[fill=fillColor] (186.96,117.27) rectangle (187.96,118.27);

\path[fill=fillColor] (206.38,120.29) rectangle (207.38,121.29);

\path[fill=fillColor] (158.48,109.59) rectangle (159.48,110.59);

\path[fill=fillColor] (158.48,118.30) rectangle (159.48,119.30);

\path[fill=fillColor] (159.78,105.65) rectangle (160.78,106.65);

\path[fill=fillColor] (185.67,115.66) rectangle (186.67,116.66);

\path[fill=fillColor] (201.20,120.02) rectangle (202.20,121.02);

\path[fill=fillColor] (136.48, 93.59) rectangle (137.48, 94.59);

\path[fill=fillColor] (163.66,117.85) rectangle (164.66,118.85);

\path[fill=fillColor] (109.29, 72.63) rectangle (110.29, 73.63);

\path[fill=fillColor] (152.01,113.48) rectangle (153.01,114.48);

\path[fill=fillColor] (140.36,115.51) rectangle (141.36,116.51);

\path[fill=fillColor] (167.54,119.97) rectangle (168.54,120.97);

\path[fill=fillColor] (120.94,100.27) rectangle (121.94,101.27);
\end{scope}
\begin{scope}
\path[clip] (  0.00,  0.00) rectangle (245.72,126.47);
\definecolor{drawColor}{RGB}{0,0,0}

\path[draw=drawColor,line width= 0.4pt,line join=round,line cap=round] ( 43.77, 36.00) -- (237.95, 36.00);

\path[draw=drawColor,line width= 0.4pt,line join=round,line cap=round] ( 43.77, 36.00) -- ( 43.77, 30.00);

\path[draw=drawColor,line width= 0.4pt,line join=round,line cap=round] (108.50, 36.00) -- (108.50, 30.00);

\path[draw=drawColor,line width= 0.4pt,line join=round,line cap=round] (173.22, 36.00) -- (173.22, 30.00);

\path[draw=drawColor,line width= 0.4pt,line join=round,line cap=round] (237.95, 36.00) -- (237.95, 30.00);

\node[text=drawColor,anchor=base,inner sep=0pt, outer sep=0pt, scale=  1.00] at ( 43.77, 20.40) {0};

\node[text=drawColor,anchor=base,inner sep=0pt, outer sep=0pt, scale=  1.00] at (108.50, 20.40) {50};

\node[text=drawColor,anchor=base,inner sep=0pt, outer sep=0pt, scale=  1.00] at (173.22, 20.40) {100};

\node[text=drawColor,anchor=base,inner sep=0pt, outer sep=0pt, scale=  1.00] at (237.95, 20.40) {150};

\path[draw=drawColor,line width= 0.4pt,line join=round,line cap=round] ( 36.00, 39.26) -- ( 36.00,120.81);

\path[draw=drawColor,line width= 0.4pt,line join=round,line cap=round] ( 36.00, 39.26) -- ( 30.00, 39.26);

\path[draw=drawColor,line width= 0.4pt,line join=round,line cap=round] ( 36.00, 55.57) -- ( 30.00, 55.57);

\path[draw=drawColor,line width= 0.4pt,line join=round,line cap=round] ( 36.00, 71.88) -- ( 30.00, 71.88);

\path[draw=drawColor,line width= 0.4pt,line join=round,line cap=round] ( 36.00, 88.19) -- ( 30.00, 88.19);

\path[draw=drawColor,line width= 0.4pt,line join=round,line cap=round] ( 36.00,104.50) -- ( 30.00,104.50);

\path[draw=drawColor,line width= 0.4pt,line join=round,line cap=round] ( 36.00,120.81) -- ( 30.00,120.81);

\node[text=drawColor,rotate= 90.00,anchor=base,inner sep=0pt, outer sep=0pt, scale=  1.00] at ( 27.60, 39.26) {0.0};

\node[text=drawColor,rotate= 90.00,anchor=base,inner sep=0pt, outer sep=0pt, scale=  1.00] at ( 27.60, 71.88) {0.4};

\node[text=drawColor,rotate= 90.00,anchor=base,inner sep=0pt, outer sep=0pt, scale=  1.00] at ( 27.60,104.50) {0.8};

\path[draw=drawColor,line width= 0.4pt,line join=round,line cap=round] ( 36.00, 36.00) --
	(245.72, 36.00) --
	(245.72,124.07) --
	( 36.00,124.07) --
	( 36.00, 36.00);
\end{scope}
\begin{scope}
\path[clip] (  0.00,  0.00) rectangle (245.72,126.47);
\definecolor{drawColor}{RGB}{0,0,0}

\node[text=drawColor,anchor=base,inner sep=0pt, outer sep=0pt, scale=  1.00] at (140.86,  6.00) {$k$};

\node[text=drawColor,rotate= 90.00,anchor=base,inner sep=0pt, outer sep=0pt, scale=  1.00] at ( 13.20, 80.04) {$|E_{\mathrm{out}}|/|E_{\mathrm{in}}|$ for Bev};
\end{scope}
\begin{scope}
\path[clip] ( 36.00, 36.00) rectangle (245.72,124.07);
\definecolor{drawColor}{RGB}{190,190,190}

\path[draw=drawColor,line width= 0.4pt,dash pattern=on 1pt off 3pt ,line join=round,line cap=round] ( 43.77, 36.00) -- ( 43.77,124.07);

\path[draw=drawColor,line width= 0.4pt,dash pattern=on 1pt off 3pt ,line join=round,line cap=round] (108.50, 36.00) -- (108.50,124.07);

\path[draw=drawColor,line width= 0.4pt,dash pattern=on 1pt off 3pt ,line join=round,line cap=round] (173.22, 36.00) -- (173.22,124.07);

\path[draw=drawColor,line width= 0.4pt,dash pattern=on 1pt off 3pt ,line join=round,line cap=round] (237.95, 36.00) -- (237.95,124.07);

\path[draw=drawColor,line width= 0.4pt,dash pattern=on 1pt off 3pt ,line join=round,line cap=round] ( 36.00, 39.26) -- (245.72, 39.26);

\path[draw=drawColor,line width= 0.4pt,dash pattern=on 1pt off 3pt ,line join=round,line cap=round] ( 36.00, 55.57) -- (245.72, 55.57);

\path[draw=drawColor,line width= 0.4pt,dash pattern=on 1pt off 3pt ,line join=round,line cap=round] ( 36.00, 71.88) -- (245.72, 71.88);

\path[draw=drawColor,line width= 0.4pt,dash pattern=on 1pt off 3pt ,line join=round,line cap=round] ( 36.00, 88.19) -- (245.72, 88.19);

\path[draw=drawColor,line width= 0.4pt,dash pattern=on 1pt off 3pt ,line join=round,line cap=round] ( 36.00,104.50) -- (245.72,104.50);

\path[draw=drawColor,line width= 0.4pt,dash pattern=on 1pt off 3pt ,line join=round,line cap=round] ( 36.00,120.81) -- (245.72,120.81);
\end{scope}
\end{tikzpicture}

%% file: analysis_trans-003.tex
\begin{tikzpicture}[x=1pt,y=1pt]
\definecolor{fillColor}{RGB}{255,255,255}
\path[use as bounding box,fill=fillColor,fill opacity=0.00] (0,0) rectangle (245.72,126.47);
\begin{scope}
\path[clip] ( 25.20, 14.28) rectangle (245.72,124.79);
\definecolor{drawColor}{RGB}{0,0,0}

\path[draw=drawColor,line width= 1.2pt,line join=round] ( 70.80, 61.86) -- ( 84.41, 61.86);

\path[draw=drawColor,line width= 0.4pt,dash pattern=on 4pt off 4pt ,line join=round,line cap=round] ( 77.61, 39.57) -- ( 77.61, 55.65);

\path[draw=drawColor,line width= 0.4pt,dash pattern=on 4pt off 4pt ,line join=round,line cap=round] ( 77.61, 68.07) -- ( 77.61, 66.98);

\path[draw=drawColor,line width= 0.4pt,line join=round,line cap=round] ( 74.20, 39.57) -- ( 81.01, 39.57);

\path[draw=drawColor,line width= 0.4pt,line join=round,line cap=round] ( 74.20, 68.07) -- ( 81.01, 68.07);

\path[draw=drawColor,line width= 0.4pt,line join=round,line cap=round] ( 70.80, 55.65) --
	( 84.41, 55.65) --
	( 84.41, 66.98) --
	( 70.80, 66.98) --
	( 70.80, 55.65);

\path[draw=drawColor,line width= 1.2pt,line join=round] (206.92, 30.28) -- (220.54, 30.28);

\path[draw=drawColor,line width= 0.4pt,dash pattern=on 4pt off 4pt ,line join=round,line cap=round] (213.73, 23.94) -- (213.73, 29.24);

\path[draw=drawColor,line width= 0.4pt,dash pattern=on 4pt off 4pt ,line join=round,line cap=round] (213.73, 35.23) -- (213.73, 33.21);

\path[draw=drawColor,line width= 0.4pt,line join=round,line cap=round] (210.33, 23.94) -- (217.13, 23.94);

\path[draw=drawColor,line width= 0.4pt,line join=round,line cap=round] (210.33, 35.23) -- (217.13, 35.23);

\path[draw=drawColor,line width= 0.4pt,line join=round,line cap=round] (206.92, 29.24) --
	(220.54, 29.24) --
	(220.54, 33.21) --
	(206.92, 33.21) --
	(206.92, 29.24);
\end{scope}
\begin{scope}
\path[clip] (  0.00,  0.00) rectangle (245.72,126.47);
\definecolor{drawColor}{RGB}{0,0,0}

\path[draw=drawColor,line width= 0.4pt,line join=round,line cap=round] ( 25.20, 18.37) -- ( 25.20,120.70);

\path[draw=drawColor,line width= 0.4pt,line join=round,line cap=round] ( 25.20, 18.37) -- ( 21.00, 18.37);

\path[draw=drawColor,line width= 0.4pt,line join=round,line cap=round] ( 25.20, 32.99) -- ( 21.00, 32.99);

\path[draw=drawColor,line width= 0.4pt,line join=round,line cap=round] ( 25.20, 47.61) -- ( 21.00, 47.61);

\path[draw=drawColor,line width= 0.4pt,line join=round,line cap=round] ( 25.20, 62.23) -- ( 21.00, 62.23);

\path[draw=drawColor,line width= 0.4pt,line join=round,line cap=round] ( 25.20, 76.85) -- ( 21.00, 76.85);

\path[draw=drawColor,line width= 0.4pt,line join=round,line cap=round] ( 25.20, 91.46) -- ( 21.00, 91.46);

\path[draw=drawColor,line width= 0.4pt,line join=round,line cap=round] ( 25.20,106.08) -- ( 21.00,106.08);

\path[draw=drawColor,line width= 0.4pt,line join=round,line cap=round] ( 25.20,120.70) -- ( 21.00,120.70);

\node[text=drawColor,rotate= 90.00,anchor=base,inner sep=0pt, outer sep=0pt, scale=  1.00] at ( 16.80, 18.37) {0};

\node[text=drawColor,rotate= 90.00,anchor=base,inner sep=0pt, outer sep=0pt, scale=  1.00] at ( 16.80, 47.61) {40};

\node[text=drawColor,rotate= 90.00,anchor=base,inner sep=0pt, outer sep=0pt, scale=  1.00] at ( 16.80, 76.85) {80};

\node[text=drawColor,rotate= 90.00,anchor=base,inner sep=0pt, outer sep=0pt, scale=  1.00] at ( 16.80,106.08) {120};

\path[draw=drawColor,line width= 0.4pt,line join=round,line cap=round] ( 25.20, 14.28) --
	(245.72, 14.28) --
	(245.72,124.79) --
	( 25.20,124.79) --
	( 25.20, 14.28);
\end{scope}
\begin{scope}
\path[clip] ( 25.20, 14.28) rectangle (245.72,124.79);
\definecolor{drawColor}{RGB}{190,190,190}

\path[draw=drawColor,line width= 0.4pt,dash pattern=on 1pt off 3pt ,line join=round,line cap=round] (135.46, 14.28) -- (135.46,124.79);
\definecolor{fillColor}{RGB}{0,0,0}

\path[fill=fillColor] ( 61.43, 65.38) rectangle ( 62.43, 66.38);

\path[fill=fillColor] ( 60.12, 39.07) rectangle ( 61.12, 40.07);

\path[fill=fillColor] ( 50.57, 61.00) rectangle ( 51.57, 62.00);

\path[fill=fillColor] ( 60.14, 67.57) rectangle ( 61.14, 68.57);

\path[fill=fillColor] ( 54.69, 63.19) rectangle ( 55.69, 64.19);

\path[fill=fillColor] ( 62.68,112.16) rectangle ( 63.68,113.16);

\path[fill=fillColor] ( 55.89, 59.53) rectangle ( 56.89, 60.53);

\path[fill=fillColor] ( 50.02, 52.96) rectangle ( 51.02, 53.96);

\path[fill=fillColor] ( 51.63, 57.34) rectangle ( 52.63, 58.34);

\path[fill=fillColor] ( 59.33, 61.73) rectangle ( 60.33, 62.73);

\path[fill=fillColor] ( 62.28, 47.84) rectangle ( 63.28, 48.84);

\path[fill=fillColor] ( 61.45, 96.08) rectangle ( 62.45, 97.08);

\path[fill=fillColor] (188.48, 33.19) rectangle (189.48, 34.19);

\path[fill=fillColor] (194.39, 23.44) rectangle (195.39, 24.44);

\path[fill=fillColor] (199.04, 29.66) rectangle (200.04, 30.66);

\path[fill=fillColor] (191.79, 25.62) rectangle (192.79, 26.62);

\path[fill=fillColor] (191.53, 30.43) rectangle (192.53, 31.43);

\path[fill=fillColor] (198.83, 49.03) rectangle (199.83, 50.03);

\path[fill=fillColor] (186.96, 34.73) rectangle (187.96, 35.73);

\path[fill=fillColor] (196.93, 29.89) rectangle (197.93, 30.89);

\path[fill=fillColor] (190.09, 32.22) rectangle (191.09, 33.22);

\path[fill=fillColor] (187.22, 29.55) rectangle (188.22, 30.55);

\path[fill=fillColor] (187.39, 27.93) rectangle (188.39, 28.93);

\path[fill=fillColor] (197.76, 29.66) rectangle (198.76, 30.66);
\end{scope}
\begin{scope}
\path[clip] (  0.00,  0.00) rectangle (245.72,126.47);
\definecolor{drawColor}{RGB}{0,0,0}

\path[draw=drawColor,line width= 0.4pt,line join=round,line cap=round] ( 67.40, 14.28) -- (203.52, 14.28);

\path[draw=drawColor,line width= 0.4pt,line join=round,line cap=round] ( 67.40, 14.28) -- ( 67.40, 10.08);

\path[draw=drawColor,line width= 0.4pt,line join=round,line cap=round] (203.52, 14.28) -- (203.52, 10.08);

\node[text=drawColor,anchor=base,inner sep=0pt, outer sep=0pt, scale=  1.00] at ( 67.40,  0.84) {$d$};

\node[text=drawColor,anchor=base,inner sep=0pt, outer sep=0pt, scale=  1.00] at (203.52,  0.84) {$\bar d$};
\end{scope}
\begin{scope}
\path[clip] ( 25.20, 14.28) rectangle (245.72,124.79);
\definecolor{drawColor}{RGB}{190,190,190}

\path[draw=drawColor,line width= 0.4pt,dash pattern=on 1pt off 3pt ,line join=round,line cap=round] ( 25.20, 18.37) -- (245.72, 18.37);

\path[draw=drawColor,line width= 0.4pt,dash pattern=on 1pt off 3pt ,line join=round,line cap=round] ( 25.20, 32.99) -- (245.72, 32.99);

\path[draw=drawColor,line width= 0.4pt,dash pattern=on 1pt off 3pt ,line join=round,line cap=round] ( 25.20, 47.61) -- (245.72, 47.61);

\path[draw=drawColor,line width= 0.4pt,dash pattern=on 1pt off 3pt ,line join=round,line cap=round] ( 25.20, 62.23) -- (245.72, 62.23);

\path[draw=drawColor,line width= 0.4pt,dash pattern=on 1pt off 3pt ,line join=round,line cap=round] ( 25.20, 76.85) -- (245.72, 76.85);

\path[draw=drawColor,line width= 0.4pt,dash pattern=on 1pt off 3pt ,line join=round,line cap=round] ( 25.20, 91.46) -- (245.72, 91.46);

\path[draw=drawColor,line width= 0.4pt,dash pattern=on 1pt off 3pt ,line join=round,line cap=round] ( 25.20,106.08) -- (245.72,106.08);

\path[draw=drawColor,line width= 0.4pt,dash pattern=on 1pt off 3pt ,line join=round,line cap=round] ( 25.20,120.70) -- (245.72,120.70);
\end{scope}
\end{tikzpicture}

%% file: analysis_trans-002.tex
\begin{tikzpicture}[x=1pt,y=1pt]
\definecolor{fillColor}{RGB}{255,255,255}
\path[use as bounding box,fill=fillColor,fill opacity=0.00] (0,0) rectangle (245.72,126.47);
\begin{scope}
\path[clip] ( 25.20, 14.28) rectangle (245.72,124.79);
\definecolor{drawColor}{RGB}{0,0,0}

\path[draw=drawColor,line width= 1.2pt,line join=round] ( 46.98, 63.08) -- ( 56.05, 63.08);

\path[draw=drawColor,line width= 0.4pt,dash pattern=on 4pt off 4pt ,line join=round,line cap=round] ( 51.52, 24.41) -- ( 51.52, 43.89);

\path[draw=drawColor,line width= 0.4pt,dash pattern=on 4pt off 4pt ,line join=round,line cap=round] ( 51.52,120.70) -- ( 51.52,109.51);

\path[draw=drawColor,line width= 0.4pt,line join=round,line cap=round] ( 49.25, 24.41) -- ( 53.79, 24.41);

\path[draw=drawColor,line width= 0.4pt,line join=round,line cap=round] ( 49.25,120.70) -- ( 53.79,120.70);

\path[draw=drawColor,line width= 0.4pt,line join=round,line cap=round] ( 46.98, 43.89) --
	( 56.05, 43.89) --
	( 56.05,109.51) --
	( 46.98,109.51) --
	( 46.98, 43.89);

\path[draw=drawColor,line width= 1.2pt,line join=round] ( 92.35, 33.04) -- (101.43, 33.04);

\path[draw=drawColor,line width= 0.4pt,dash pattern=on 4pt off 4pt ,line join=round,line cap=round] ( 96.89, 18.37) -- ( 96.89, 24.45);

\path[draw=drawColor,line width= 0.4pt,dash pattern=on 4pt off 4pt ,line join=round,line cap=round] ( 96.89, 63.13) -- ( 96.89, 55.88);

\path[draw=drawColor,line width= 0.4pt,line join=round,line cap=round] ( 94.62, 18.37) -- ( 99.16, 18.37);

\path[draw=drawColor,line width= 0.4pt,line join=round,line cap=round] ( 94.62, 63.13) -- ( 99.16, 63.13);

\path[draw=drawColor,line width= 0.4pt,line join=round,line cap=round] ( 92.35, 24.45) --
	(101.43, 24.45) --
	(101.43, 55.88) --
	( 92.35, 55.88) --
	( 92.35, 24.45);

\path[draw=drawColor,line width= 1.2pt,line join=round] (137.73, 51.81) -- (146.80, 51.81);

\path[draw=drawColor,line width= 0.4pt,dash pattern=on 4pt off 4pt ,line join=round,line cap=round] (142.27, 24.24) -- (142.27, 38.49);

\path[draw=drawColor,line width= 0.4pt,dash pattern=on 4pt off 4pt ,line join=round,line cap=round] (142.27,112.32) -- (142.27, 94.59);

\path[draw=drawColor,line width= 0.4pt,line join=round,line cap=round] (140.00, 24.24) -- (144.53, 24.24);

\path[draw=drawColor,line width= 0.4pt,line join=round,line cap=round] (140.00,112.32) -- (144.53,112.32);

\path[draw=drawColor,line width= 0.4pt,line join=round,line cap=round] (137.73, 38.49) --
	(146.80, 38.49) --
	(146.80, 94.59) --
	(137.73, 94.59) --
	(137.73, 38.49);

\path[draw=drawColor,line width= 1.2pt,line join=round] (183.10, 51.81) -- (192.18, 51.81);

\path[draw=drawColor,line width= 0.4pt,dash pattern=on 4pt off 4pt ,line join=round,line cap=round] (187.64, 24.24) -- (187.64, 38.49);

\path[draw=drawColor,line width= 0.4pt,dash pattern=on 4pt off 4pt ,line join=round,line cap=round] (187.64,112.32) -- (187.64, 94.59);

\path[draw=drawColor,line width= 0.4pt,line join=round,line cap=round] (185.37, 24.24) -- (189.91, 24.24);

\path[draw=drawColor,line width= 0.4pt,line join=round,line cap=round] (185.37,112.32) -- (189.91,112.32);

\path[draw=drawColor,line width= 0.4pt,line join=round,line cap=round] (183.10, 38.49) --
	(192.18, 38.49) --
	(192.18, 94.59) --
	(183.10, 94.59) --
	(183.10, 38.49);

\path[draw=drawColor,line width= 1.2pt,line join=round] (228.48, 33.04) -- (237.55, 33.04);

\path[draw=drawColor,line width= 0.4pt,dash pattern=on 4pt off 4pt ,line join=round,line cap=round] (233.01, 18.37) -- (233.01, 24.45);

\path[draw=drawColor,line width= 0.4pt,dash pattern=on 4pt off 4pt ,line join=round,line cap=round] (233.01, 63.13) -- (233.01, 55.88);

\path[draw=drawColor,line width= 0.4pt,line join=round,line cap=round] (230.74, 18.37) -- (235.28, 18.37);

\path[draw=drawColor,line width= 0.4pt,line join=round,line cap=round] (230.74, 63.13) -- (235.28, 63.13);

\path[draw=drawColor,line width= 0.4pt,line join=round,line cap=round] (228.48, 24.45) --
	(237.55, 24.45) --
	(237.55, 55.88) --
	(228.48, 55.88) --
	(228.48, 24.45);
\end{scope}
\begin{scope}
\path[clip] (  0.00,  0.00) rectangle (245.72,126.47);
\definecolor{drawColor}{RGB}{0,0,0}

\path[draw=drawColor,line width= 0.4pt,line join=round,line cap=round] ( 25.20, 16.70) -- ( 25.20,117.26);

\path[draw=drawColor,line width= 0.4pt,line join=round,line cap=round] ( 25.20, 16.70) -- ( 21.00, 16.70);

\path[draw=drawColor,line width= 0.4pt,line join=round,line cap=round] ( 25.20, 33.46) -- ( 21.00, 33.46);

\path[draw=drawColor,line width= 0.4pt,line join=round,line cap=round] ( 25.20, 50.22) -- ( 21.00, 50.22);

\path[draw=drawColor,line width= 0.4pt,line join=round,line cap=round] ( 25.20, 66.98) -- ( 21.00, 66.98);

\path[draw=drawColor,line width= 0.4pt,line join=round,line cap=round] ( 25.20, 83.74) -- ( 21.00, 83.74);

\path[draw=drawColor,line width= 0.4pt,line join=round,line cap=round] ( 25.20,100.50) -- ( 21.00,100.50);

\path[draw=drawColor,line width= 0.4pt,line join=round,line cap=round] ( 25.20,117.26) -- ( 21.00,117.26);

\node[text=drawColor,rotate= 90.00,anchor=base,inner sep=0pt, outer sep=0pt, scale=  1.00] at ( 16.80, 16.70) {0};

\node[text=drawColor,rotate= 90.00,anchor=base,inner sep=0pt, outer sep=0pt, scale=  1.00] at ( 16.80, 50.22) {400};

\node[text=drawColor,rotate= 90.00,anchor=base,inner sep=0pt, outer sep=0pt, scale=  1.00] at ( 16.80, 83.74) {800};

\node[text=drawColor,rotate= 90.00,anchor=base,inner sep=0pt, outer sep=0pt, scale=  1.00] at ( 16.80,117.26) {1200};
\end{scope}
\begin{scope}
\path[clip] (  0.00,  0.00) rectangle (245.72,126.47);
\definecolor{drawColor}{RGB}{0,0,0}

\node[text=drawColor,rotate= 90.00,anchor=base,inner sep=0pt, outer sep=0pt, scale=  1.00] at (  6.72, 69.54) {$|E_{\mathrm{out}}|$};
\end{scope}
\begin{scope}
\path[clip] (  0.00,  0.00) rectangle (245.72,126.47);
\definecolor{drawColor}{RGB}{0,0,0}

\path[draw=drawColor,line width= 0.4pt,line join=round,line cap=round] ( 25.20, 14.28) --
	(245.72, 14.28) --
	(245.72,124.79) --
	( 25.20,124.79) --
	( 25.20, 14.28);
\end{scope}
\begin{scope}
\path[clip] ( 25.20, 14.28) rectangle (245.72,124.79);
\definecolor{drawColor}{RGB}{190,190,190}

\path[draw=drawColor,line width= 0.4pt,dash pattern=on 1pt off 3pt ,line join=round,line cap=round] ( 67.40, 14.28) -- ( 67.40,124.79);

\path[draw=drawColor,line width= 0.4pt,dash pattern=on 1pt off 3pt ,line join=round,line cap=round] (112.77, 14.28) -- (112.77,124.79);

\path[draw=drawColor,line width= 0.4pt,dash pattern=on 1pt off 3pt ,line join=round,line cap=round] (158.15, 14.28) -- (158.15,124.79);

\path[draw=drawColor,line width= 0.4pt,dash pattern=on 1pt off 3pt ,line join=round,line cap=round] (203.52, 14.28) -- (203.52,124.79);

\path[draw=drawColor,line width= 0.4pt,dash pattern=on 1pt off 3pt ,line join=round,line cap=round] ( 25.20, 16.70) -- (245.72, 16.70);

\path[draw=drawColor,line width= 0.4pt,dash pattern=on 1pt off 3pt ,line join=round,line cap=round] ( 25.20, 33.46) -- (245.72, 33.46);

\path[draw=drawColor,line width= 0.4pt,dash pattern=on 1pt off 3pt ,line join=round,line cap=round] ( 25.20, 50.22) -- (245.72, 50.22);

\path[draw=drawColor,line width= 0.4pt,dash pattern=on 1pt off 3pt ,line join=round,line cap=round] ( 25.20, 66.98) -- (245.72, 66.98);

\path[draw=drawColor,line width= 0.4pt,dash pattern=on 1pt off 3pt ,line join=round,line cap=round] ( 25.20, 83.74) -- (245.72, 83.74);

\path[draw=drawColor,line width= 0.4pt,dash pattern=on 1pt off 3pt ,line join=round,line cap=round] ( 25.20,100.50) -- (245.72,100.50);

\path[draw=drawColor,line width= 0.4pt,dash pattern=on 1pt off 3pt ,line join=round,line cap=round] ( 25.20,117.26) -- (245.72,117.26);
\definecolor{fillColor}{RGB}{0,0,0}

\path[fill=fillColor] ( 40.56, 59.86) rectangle ( 41.56, 60.86);

\path[fill=fillColor] ( 32.89, 97.82) rectangle ( 33.89, 98.82);

\path[fill=fillColor] ( 36.62,120.20) rectangle ( 37.62,121.20);

\path[fill=fillColor] ( 38.17, 78.30) rectangle ( 39.17, 79.30);

\path[fill=fillColor] ( 33.51, 36.90) rectangle ( 34.51, 37.90);

\path[fill=fillColor] ( 39.63, 44.94) rectangle ( 40.63, 45.94);

\path[fill=fillColor] ( 34.18, 23.91) rectangle ( 35.18, 24.91);

\path[fill=fillColor] ( 36.02, 44.94) rectangle ( 37.02, 45.94);

\path[fill=fillColor] ( 34.69, 41.84) rectangle ( 35.69, 42.84);

\path[fill=fillColor] ( 37.56, 65.31) rectangle ( 38.56, 66.31);

\path[fill=fillColor] ( 82.80, 28.10) rectangle ( 83.80, 29.10);

\path[fill=fillColor] ( 86.39, 40.25) rectangle ( 87.39, 41.25);

\path[fill=fillColor] ( 79.83, 62.63) rectangle ( 80.83, 63.63);

\path[fill=fillColor] ( 83.05, 48.13) rectangle ( 84.05, 49.13);

\path[fill=fillColor] ( 85.83, 32.20) rectangle ( 86.83, 33.20);

\path[fill=fillColor] ( 81.36, 32.87) rectangle ( 82.36, 33.87);

\path[fill=fillColor] ( 78.57, 19.21) rectangle ( 79.57, 20.21);

\path[fill=fillColor] ( 79.02, 21.14) rectangle ( 80.02, 22.14);

\path[fill=fillColor] ( 81.41, 26.76) rectangle ( 82.41, 27.76);

\path[fill=fillColor] ( 81.78, 17.87) rectangle ( 82.78, 18.87);

\path[fill=fillColor] ( 79.23,111.57) rectangle ( 80.23,112.57);

\path[fill=fillColor] (131.05, 58.02) rectangle (132.05, 59.02);

\path[fill=fillColor] (129.79, 72.01) rectangle (130.79, 73.01);

\path[fill=fillColor] (132.21,111.82) rectangle (133.21,112.82);

\path[fill=fillColor] (126.40, 76.37) rectangle (127.40, 77.37);

\path[fill=fillColor] (130.70, 36.90) rectangle (131.70, 37.90);

\path[fill=fillColor] (127.79, 44.61) rectangle (128.79, 45.61);

\path[fill=fillColor] (124.24, 23.74) rectangle (125.24, 24.74);

\path[fill=fillColor] (131.48, 40.75) rectangle (132.48, 41.75);

\path[fill=fillColor] (127.98, 39.08) rectangle (128.98, 40.08);

\path[fill=fillColor] (126.15, 30.78) rectangle (127.15, 31.78);

\path[fill=fillColor] (176.02, 58.02) rectangle (177.02, 59.02);

\path[fill=fillColor] (169.12, 72.01) rectangle (170.12, 73.01);

\path[fill=fillColor] (175.92,111.82) rectangle (176.92,112.82);

\path[fill=fillColor] (177.84, 76.37) rectangle (178.84, 77.37);

\path[fill=fillColor] (174.87, 36.90) rectangle (175.87, 37.90);

\path[fill=fillColor] (170.63, 44.61) rectangle (171.63, 45.61);

\path[fill=fillColor] (173.75, 23.74) rectangle (174.75, 24.74);

\path[fill=fillColor] (173.67, 40.75) rectangle (174.67, 41.75);

\path[fill=fillColor] (174.92, 39.08) rectangle (175.92, 40.08);

\path[fill=fillColor] (169.59, 30.78) rectangle (170.59, 31.78);

\path[fill=fillColor] (215.53, 28.10) rectangle (216.53, 29.10);

\path[fill=fillColor] (217.27, 40.25) rectangle (218.27, 41.25);

\path[fill=fillColor] (214.81, 62.63) rectangle (215.81, 63.63);

\path[fill=fillColor] (214.91, 48.13) rectangle (215.91, 49.13);

\path[fill=fillColor] (217.35, 32.20) rectangle (218.35, 33.20);

\path[fill=fillColor] (215.76, 32.87) rectangle (216.76, 33.87);

\path[fill=fillColor] (215.80, 19.21) rectangle (216.80, 20.21);

\path[fill=fillColor] (223.25, 21.14) rectangle (224.25, 22.14);

\path[fill=fillColor] (221.59, 26.76) rectangle (222.59, 27.76);

\path[fill=fillColor] (223.21, 17.87) rectangle (224.21, 18.87);

\path[fill=fillColor] (216.26,111.57) rectangle (217.26,112.57);
\end{scope}
\begin{scope}
\path[clip] (  0.00,  0.00) rectangle (245.72,126.47);
\definecolor{drawColor}{RGB}{0,0,0}

\path[draw=drawColor,line width= 0.4pt,line join=round,line cap=round] ( 44.71, 14.28) -- (226.21, 14.28);

\path[draw=drawColor,line width= 0.4pt,line join=round,line cap=round] ( 44.71, 14.28) -- ( 44.71, 10.08);

\path[draw=drawColor,line width= 0.4pt,line join=round,line cap=round] ( 90.08, 14.28) -- ( 90.08, 10.08);

\path[draw=drawColor,line width= 0.4pt,line join=round,line cap=round] (135.46, 14.28) -- (135.46, 10.08);

\path[draw=drawColor,line width= 0.4pt,line join=round,line cap=round] (180.83, 14.28) -- (180.83, 10.08);

\path[draw=drawColor,line width= 0.4pt,line join=round,line cap=round] (226.21, 14.28) -- (226.21, 10.08);

\node[text=drawColor,anchor=base,inner sep=0pt, outer sep=0pt, scale=  1.00] at ( 44.71,  0.84) {N};

\node[text=drawColor,anchor=base,inner sep=0pt, outer sep=0pt, scale=  1.00] at ( 90.08,  0.84) {W};

\node[text=drawColor,anchor=base,inner sep=0pt, outer sep=0pt, scale=  1.00] at (135.46,  0.84) {F};

\node[text=drawColor,anchor=base,inner sep=0pt, outer sep=0pt, scale=  1.00] at (180.83,  0.84) {B};

\node[text=drawColor,anchor=base,inner sep=0pt, outer sep=0pt, scale=  1.00] at (226.21,  0.84) {FBW};
\end{scope}
\end{tikzpicture}

%% file: analysis_trans-001.tex
\begin{tikzpicture}[x=1pt,y=1pt]
\definecolor{fillColor}{RGB}{255,255,255}
\path[use as bounding box,fill=fillColor,fill opacity=0.00] (0,0) rectangle (245.72,126.47);
\begin{scope}
\path[clip] ( 25.20, 14.28) rectangle (245.72,124.79);
\definecolor{drawColor}{RGB}{0,0,0}

\path[draw=drawColor,line width= 1.2pt,line join=round] ( 50.87, 60.20) -- ( 62.54, 60.20);

\path[draw=drawColor,line width= 0.4pt,dash pattern=on 4pt off 4pt ,line join=round,line cap=round] ( 56.70, 21.87) -- ( 56.70, 47.39);

\path[draw=drawColor,line width= 0.4pt,dash pattern=on 4pt off 4pt ,line join=round,line cap=round] ( 56.70, 97.50) -- ( 56.70, 67.52);

\path[draw=drawColor,line width= 0.4pt,line join=round,line cap=round] ( 53.79, 21.87) -- ( 59.62, 21.87);

\path[draw=drawColor,line width= 0.4pt,line join=round,line cap=round] ( 53.79, 97.50) -- ( 59.62, 97.50);

\path[draw=drawColor,line width= 0.4pt,line join=round,line cap=round] ( 50.87, 47.39) --
	( 62.54, 47.39) --
	( 62.54, 67.52) --
	( 50.87, 67.52) --
	( 50.87, 47.39);

\path[draw=drawColor,line width= 1.2pt,line join=round] (109.21,113.22) -- (120.87,113.22);

\path[draw=drawColor,line width= 0.4pt,dash pattern=on 4pt off 4pt ,line join=round,line cap=round] (115.04,105.78) -- (115.04,106.51);

\path[draw=drawColor,line width= 0.4pt,dash pattern=on 4pt off 4pt ,line join=round,line cap=round] (115.04,120.70) -- (115.04,118.00);

\path[draw=drawColor,line width= 0.4pt,line join=round,line cap=round] (112.12,105.78) -- (117.96,105.78);

\path[draw=drawColor,line width= 0.4pt,line join=round,line cap=round] (112.12,120.70) -- (117.96,120.70);

\path[draw=drawColor,line width= 0.4pt,line join=round,line cap=round] (109.21,106.51) --
	(120.87,106.51) --
	(120.87,118.00) --
	(109.21,118.00) --
	(109.21,106.51);

\path[draw=drawColor,line width= 1.2pt,line join=round] (167.54,113.22) -- (179.21,113.22);

\path[draw=drawColor,line width= 0.4pt,dash pattern=on 4pt off 4pt ,line join=round,line cap=round] (173.38,105.78) -- (173.38,106.51);

\path[draw=drawColor,line width= 0.4pt,dash pattern=on 4pt off 4pt ,line join=round,line cap=round] (173.38,120.70) -- (173.38,118.00);

\path[draw=drawColor,line width= 0.4pt,line join=round,line cap=round] (170.46,105.78) -- (176.30,105.78);

\path[draw=drawColor,line width= 0.4pt,line join=round,line cap=round] (170.46,120.70) -- (176.30,120.70);

\path[draw=drawColor,line width= 0.4pt,line join=round,line cap=round] (167.54,106.51) --
	(179.21,106.51) --
	(179.21,118.00) --
	(167.54,118.00) --
	(167.54,106.51);

\path[draw=drawColor,line width= 1.2pt,line join=round] (225.88, 60.20) -- (237.55, 60.20);

\path[draw=drawColor,line width= 0.4pt,dash pattern=on 4pt off 4pt ,line join=round,line cap=round] (231.72, 21.87) -- (231.72, 47.39);

\path[draw=drawColor,line width= 0.4pt,dash pattern=on 4pt off 4pt ,line join=round,line cap=round] (231.72, 97.50) -- (231.72, 67.52);

\path[draw=drawColor,line width= 0.4pt,line join=round,line cap=round] (228.80, 21.87) -- (234.63, 21.87);

\path[draw=drawColor,line width= 0.4pt,line join=round,line cap=round] (228.80, 97.50) -- (234.63, 97.50);

\path[draw=drawColor,line width= 0.4pt,line join=round,line cap=round] (225.88, 47.39) --
	(237.55, 47.39) --
	(237.55, 67.52) --
	(225.88, 67.52) --
	(225.88, 47.39);
\end{scope}
\begin{scope}
\path[clip] (  0.00,  0.00) rectangle (245.72,126.47);
\definecolor{drawColor}{RGB}{0,0,0}

\path[draw=drawColor,line width= 0.4pt,line join=round,line cap=round] ( 25.20, 18.37) -- ( 25.20,120.70);

\path[draw=drawColor,line width= 0.4pt,line join=round,line cap=round] ( 25.20, 18.37) -- ( 21.00, 18.37);

\path[draw=drawColor,line width= 0.4pt,line join=round,line cap=round] ( 25.20, 38.84) -- ( 21.00, 38.84);

\path[draw=drawColor,line width= 0.4pt,line join=round,line cap=round] ( 25.20, 59.30) -- ( 21.00, 59.30);

\path[draw=drawColor,line width= 0.4pt,line join=round,line cap=round] ( 25.20, 79.77) -- ( 21.00, 79.77);

\path[draw=drawColor,line width= 0.4pt,line join=round,line cap=round] ( 25.20,100.23) -- ( 21.00,100.23);

\path[draw=drawColor,line width= 0.4pt,line join=round,line cap=round] ( 25.20,120.70) -- ( 21.00,120.70);

\node[text=drawColor,rotate= 90.00,anchor=base,inner sep=0pt, outer sep=0pt, scale=  1.00] at ( 16.80, 18.37) {0.0};

\node[text=drawColor,rotate= 90.00,anchor=base,inner sep=0pt, outer sep=0pt, scale=  1.00] at ( 16.80, 38.84) {0.2};

\node[text=drawColor,rotate= 90.00,anchor=base,inner sep=0pt, outer sep=0pt, scale=  1.00] at ( 16.80, 59.30) {0.4};

\node[text=drawColor,rotate= 90.00,anchor=base,inner sep=0pt, outer sep=0pt, scale=  1.00] at ( 16.80, 79.77) {0.6};

\node[text=drawColor,rotate= 90.00,anchor=base,inner sep=0pt, outer sep=0pt, scale=  1.00] at ( 16.80,100.23) {0.8};

\node[text=drawColor,rotate= 90.00,anchor=base,inner sep=0pt, outer sep=0pt, scale=  1.00] at ( 16.80,120.70) {1.0};
\end{scope}
\begin{scope}
\path[clip] (  0.00,  0.00) rectangle (245.72,126.47);
\definecolor{drawColor}{RGB}{0,0,0}

\node[text=drawColor,rotate= 90.00,anchor=base,inner sep=0pt, outer sep=0pt, scale=  1.00] at (  6.72, 69.54) {$|E_{\mathrm{out}}|/|E_{\mathrm{in}}|$};
\end{scope}
\begin{scope}
\path[clip] (  0.00,  0.00) rectangle (245.72,126.47);
\definecolor{drawColor}{RGB}{0,0,0}

\path[draw=drawColor,line width= 0.4pt,line join=round,line cap=round] ( 25.20, 14.28) --
	(245.72, 14.28) --
	(245.72,124.79) --
	( 25.20,124.79) --
	( 25.20, 14.28);
\end{scope}
\begin{scope}
\path[clip] ( 25.20, 14.28) rectangle (245.72,124.79);
\definecolor{drawColor}{RGB}{190,190,190}

\path[draw=drawColor,line width= 0.4pt,dash pattern=on 1pt off 3pt ,line join=round,line cap=round] ( 77.12, 14.28) -- ( 77.12,124.79);

\path[draw=drawColor,line width= 0.4pt,dash pattern=on 1pt off 3pt ,line join=round,line cap=round] (135.46, 14.28) -- (135.46,124.79);

\path[draw=drawColor,line width= 0.4pt,dash pattern=on 1pt off 3pt ,line join=round,line cap=round] (193.80, 14.28) -- (193.80,124.79);

\path[draw=drawColor,line width= 0.4pt,dash pattern=on 1pt off 3pt ,line join=round,line cap=round] ( 25.20, 18.37) -- (245.72, 18.37);

\path[draw=drawColor,line width= 0.4pt,dash pattern=on 1pt off 3pt ,line join=round,line cap=round] ( 25.20, 38.84) -- (245.72, 38.84);

\path[draw=drawColor,line width= 0.4pt,dash pattern=on 1pt off 3pt ,line join=round,line cap=round] ( 25.20, 59.30) -- (245.72, 59.30);

\path[draw=drawColor,line width= 0.4pt,dash pattern=on 1pt off 3pt ,line join=round,line cap=round] ( 25.20, 79.77) -- (245.72, 79.77);

\path[draw=drawColor,line width= 0.4pt,dash pattern=on 1pt off 3pt ,line join=round,line cap=round] ( 25.20,100.23) -- (245.72,100.23);

\path[draw=drawColor,line width= 0.4pt,dash pattern=on 1pt off 3pt ,line join=round,line cap=round] ( 25.20,120.70) -- (245.72,120.70);
\definecolor{fillColor}{RGB}{0,0,0}

\path[fill=fillColor] ( 36.51, 45.76) rectangle ( 37.51, 46.76);

\path[fill=fillColor] ( 40.69, 48.02) rectangle ( 41.69, 49.02);

\path[fill=fillColor] ( 40.41, 63.55) rectangle ( 41.41, 64.55);

\path[fill=fillColor] ( 35.71, 60.56) rectangle ( 36.71, 61.56);

\path[fill=fillColor] ( 39.64, 70.49) rectangle ( 40.64, 71.49);

\path[fill=fillColor] ( 37.01, 97.00) rectangle ( 38.01, 98.00);

\path[fill=fillColor] ( 40.23, 77.24) rectangle ( 41.23, 78.24);

\path[fill=fillColor] ( 36.35, 57.91) rectangle ( 37.35, 58.91);

\path[fill=fillColor] ( 43.95, 35.47) rectangle ( 44.95, 36.47);

\path[fill=fillColor] ( 39.35, 60.01) rectangle ( 40.35, 61.01);

\path[fill=fillColor] ( 39.51, 21.37) rectangle ( 40.51, 22.37);

\path[fill=fillColor] ( 36.06, 59.40) rectangle ( 37.06, 60.40);

\path[fill=fillColor] ( 95.60,115.88) rectangle ( 96.60,116.88);

\path[fill=fillColor] (100.39, 87.84) rectangle (101.39, 88.84);

\path[fill=fillColor] ( 92.40,111.95) rectangle ( 93.40,112.95);

\path[fill=fillColor] ( 98.81,106.74) rectangle ( 99.81,107.74);

\path[fill=fillColor] ( 98.68,117.02) rectangle ( 99.68,118.02);

\path[fill=fillColor] (102.68,120.20) rectangle (103.68,121.20);

\path[fill=fillColor] (101.95,119.01) rectangle (102.95,120.01);

\path[fill=fillColor] ( 97.31,117.97) rectangle ( 98.31,118.97);

\path[fill=fillColor] ( 94.00,105.28) rectangle ( 95.00,106.28);

\path[fill=fillColor] (101.01,109.16) rectangle (102.01,110.16);

\path[fill=fillColor] ( 97.96, 48.26) rectangle ( 98.96, 49.26);

\path[fill=fillColor] ( 93.93,113.48) rectangle ( 94.93,114.48);

\path[fill=fillColor] (150.11,115.88) rectangle (151.11,116.88);

\path[fill=fillColor] (151.40, 87.84) rectangle (152.40, 88.84);

\path[fill=fillColor] (159.85,111.95) rectangle (160.85,112.95);

\path[fill=fillColor] (156.32,106.74) rectangle (157.32,107.74);

\path[fill=fillColor] (157.17,117.02) rectangle (158.17,118.02);

\path[fill=fillColor] (156.26,120.20) rectangle (157.26,121.20);

\path[fill=fillColor] (152.81,119.01) rectangle (153.81,120.01);

\path[fill=fillColor] (152.87,117.97) rectangle (153.87,118.97);

\path[fill=fillColor] (152.38,105.28) rectangle (153.38,106.28);

\path[fill=fillColor] (154.05,109.16) rectangle (155.05,110.16);

\path[fill=fillColor] (160.21, 48.26) rectangle (161.21, 49.26);

\path[fill=fillColor] (150.46,113.48) rectangle (151.46,114.48);

\path[fill=fillColor] (216.25, 45.76) rectangle (217.25, 46.76);

\path[fill=fillColor] (214.98, 48.02) rectangle (215.98, 49.02);

\path[fill=fillColor] (219.42, 63.55) rectangle (220.42, 64.55);

\path[fill=fillColor] (217.70, 60.56) rectangle (218.70, 61.56);

\path[fill=fillColor] (207.94, 70.49) rectangle (208.94, 71.49);

\path[fill=fillColor] (216.83, 97.00) rectangle (217.83, 98.00);

\path[fill=fillColor] (215.56, 77.24) rectangle (216.56, 78.24);

\path[fill=fillColor] (210.91, 57.91) rectangle (211.91, 58.91);

\path[fill=fillColor] (213.02, 35.47) rectangle (214.02, 36.47);

\path[fill=fillColor] (211.42, 60.01) rectangle (212.42, 61.01);

\path[fill=fillColor] (213.69, 21.37) rectangle (214.69, 22.37);

\path[fill=fillColor] (214.15, 59.40) rectangle (215.15, 60.40);
\end{scope}
\begin{scope}
\path[clip] (  0.00,  0.00) rectangle (245.72,126.47);
\definecolor{drawColor}{RGB}{0,0,0}

\path[draw=drawColor,line width= 0.4pt,line join=round,line cap=round] ( 47.95, 14.28) -- (222.97, 14.28);

\path[draw=drawColor,line width= 0.4pt,line join=round,line cap=round] ( 47.95, 14.28) -- ( 47.95, 10.08);

\path[draw=drawColor,line width= 0.4pt,line join=round,line cap=round] (106.29, 14.28) -- (106.29, 10.08);

\path[draw=drawColor,line width= 0.4pt,line join=round,line cap=round] (164.63, 14.28) -- (164.63, 10.08);

\path[draw=drawColor,line width= 0.4pt,line join=round,line cap=round] (222.97, 14.28) -- (222.97, 10.08);

\node[text=drawColor,anchor=base,inner sep=0pt, outer sep=0pt, scale=  1.00] at ( 47.95,  0.84) {W};

\node[text=drawColor,anchor=base,inner sep=0pt, outer sep=0pt, scale=  1.00] at (106.29,  0.84) {F};

\node[text=drawColor,anchor=base,inner sep=0pt, outer sep=0pt, scale=  1.00] at (164.63,  0.84) {B};

\node[text=drawColor,anchor=base,inner sep=0pt, outer sep=0pt, scale=  1.00] at (222.97,  0.84) {FBW};
\end{scope}
\end{tikzpicture}

%% file: hs-lintimespace.bbl
\begin{thebibliography}{35}
\providecommand{\natexlab}[1]{#1}
\providecommand{\url}[1]{\texttt{#1}}
\providecommand{\urlprefix}{URL }
\expandafter\ifx\csname urlstyle\endcsname\relax
  \providecommand{\doi}[1]{doi:\discretionary{}{}{}#1}\else
  \providecommand{\doi}{doi:\discretionary{}{}{}\begingroup
  \urlstyle{rm}\Url}\fi
\providecommand{\selectlanguage}[1]{\relax}

\bibitem[{Abu-Khzam(2010)}]{Abu10}
F.~N. Abu-Khzam, A kernelization algorithm for {$d$-Hitting Set}, Journal of
  Computer and System Sciences 76 (2010) 524--531,
  \doi{10.1016/j.jcss.2009.09.002}.

\bibitem[{Alber et~al.(2006)Alber, Betzler, and Niedermeier}]{ABN06}
J.~Alber, N.~Betzler, R.~Niedermeier, Experiments on data reduction for optimal
  domination in networks, Annals of Operations Research 146 (2006) 105--117,
  \doi{10.1007/s10479-006-0045-4}.

\bibitem[{Bannach et~al.(2019)Bannach, Heinrich, Reischuk, and Tantau}]{BHRT19}
M.~Bannach, Z.~Heinrich, R.~Reischuk, T.~Tantau, Dynamic kernels for {Hitting
  Set} and {Set Packing}, ECCC preprint TR19-146, 2019,
  \urlprefix\url{https://eccc.weizmann.ac.il/report/2019/146/}.

\bibitem[{Bannach and Tantau(2020)}]{BT20}
M.~Bannach, T.~Tantau, Computing hitting set kernels by {AC$_0$}-circuits,
  Theory of Computing Systems 62 (2020) 374--399,
  \doi{10.1007/s00224-019-09941-z}.

\bibitem[{van Bevern(2014{\natexlab{a}})}]{Bev14c}
R.~van Bevern, Fixed-Parameter Linear-Time Algorithms for NP-hard Graph and
  Hypergraph Problems Arising in Industrial Applications, Universitätsverlag
  der TU Berlin, Berlin, Germany, volume~1 of Foundations of Computing,
  chapter~5, 2014{\natexlab{a}}, pp. 123--150, \doi{10.14279/depositonce-4131}.

\bibitem[{van Bevern(2014{\natexlab{b}})}]{Bev14b}
R.~van Bevern, Towards optimal and expressive kernelization for {$d$-Hitting
  Set}, Algorithmica 70 (2014{\natexlab{b}}) 129--147,
  \doi{10.1007/s00453-013-9774-3}.

\bibitem[{van Bevern et~al.(accepted for publication, 2020)van Bevern,
  Fluschnik, and Tsidulko}]{BFTxx}
R.~van Bevern, T.~Fluschnik, O.~{\relax Yu}. Tsidulko, On approximate data
  reduction for the {Rural Postman Problem}: Theory and experiments, Networks
  (accepted for publication, 2020),
  \urlprefix\url{https://arxiv.org/abs/1812.10131}.

\bibitem[{van Bevern et~al.(2012)van Bevern, Moser, and Niedermeier}]{BMN12}
R.~van Bevern, H.~Moser, R.~Niedermeier, Approximation and tidying---a problem
  kernel for {$s$-Plex Cluster Vertex Deletion}, Algorithmica 62 (2012)
  930--950, \doi{10.1007/s00453-011-9492-7}.

\bibitem[{Bläsius et~al.(2019{\natexlab{a}})Bläsius, Fischbeck, Friedrich,
  and Schirneck}]{BFFS19}
T.~Bläsius, P.~Fischbeck, T.~Friedrich, M.~Schirneck, Understanding the
  effectiveness of data reduction in public transportation networks, in:
  K.~Avrachenkov, P.~Prałat, N.~Ye (Eds.), WAW 2019, Springer, volume 11631 of
  Lecture Notes in Computer Science, 2019{\natexlab{a}}, pp. 87--101,
  \doi{10.1007/978-3-030-25070-6_7}.

\bibitem[{Bläsius et~al.(2019{\natexlab{b}})Bläsius, Friedrich, Lischeid,
  Meeks, and Schirneck}]{BFL+19}
T.~Bläsius, T.~Friedrich, J.~Lischeid, K.~Meeks, M.~Schirneck, Efficiently
  enumerating hitting sets of hypergraphs arising in data profiling, in:
  S.~Kobourov, H.~Meyerhenke (Eds.), 2019 Proceedings of the Meeting on
  Algorithm Engineering and Experiments (ALENEX), SIAM, 2019{\natexlab{b}}, pp.
  130--143, \doi{10.1137/1.9781611975499.11}.

\bibitem[{Brewka et~al.(2019)Brewka, Thimm, and Ulbricht}]{BTU19}
G.~Brewka, M.~Thimm, M.~Ulbricht, Strong inconsistency, Artificial Intelligence
  267 (2019) 78--117, \doi{10.1016/j.artint.2018.11.002}.

\bibitem[{Cormen et~al.(2001)Cormen, Leiserson, Rivest, and Stein}]{CLRS01}
T.~H. Cormen, C.~E. Leiserson, R.~L. Rivest, C.~Stein, Introduction to
  Algorithms, MIT Press, Cambridge, Massachusetts, USA, 2nd edition, 2001.

\bibitem[{Damaschke(2006)}]{Dam06}
P.~Damaschke, Parameterized enumeration, transversals, and imperfect phylogeny
  reconstruction, Theoretical Computer Science 351 (2006) 337--350,
  \doi{10.1016/j.tcs.2005.10.004}.

\bibitem[{Dell and van Melkebeek(2014)}]{DM14}
H.~Dell, D.~van Melkebeek, Satisfiability allows no nontrivial sparsification
  unless the polynomial-time hierarchy collapses, Journal of the ACM 61 (2014)
  23, \doi{10.1145/2629620}.

\bibitem[{Fafianie and Kratsch(2014)}]{FK14}
S.~Fafianie, S.~Kratsch, Streaming kernelization, in: E.~Csuhaj-Varjú,
  M.~Dietzfelbinger, Z.~Ésik (Eds.), MFCS 2014, Springer, volume 8635 of
  Lecture Notes in Computer Science, 2014, pp. 275--286,
  \doi{10.1007/978-3-662-44465-8_24}.

\bibitem[{Fafianie and Kratsch(2015)}]{FK15}
S.~Fafianie, S.~Kratsch, A shortcut to (sun)flowers: Kernels in logarithmic
  space or linear time, in: G.~F. Italiano, G.~Pighizzini, D.~T. Sannella
  (Eds.), MFCS 2015, Springer, volume 9235 of Lecture Notes in Computer
  Science, 2015, pp. 299--310, \doi{10.1007/978-3-662-48054-0_25}.

\bibitem[{Fazekas et~al.(2018)Fazekas, Bacchus, and Biere}]{FBB18}
K.~Fazekas, F.~Bacchus, A.~Biere, Implicit hitting set algorithms for maximum
  satisfiability modulo theories, in: D.~Galmiche, S.~Schulz, R.~Sebastiani
  (Eds.), IJCAR 2018, Springer, volume 10900 of Lecture Notes in Computer
  Science, 2018, pp. 134--151, \doi{10.1007/978-3-319-94205-6_10}.

\bibitem[{Flum and Grohe(2006)}]{FG06}
J.~Flum, M.~Grohe, Parameterized Complexity Theory, Texts in Theoretical
  Computer Science, An EATCS Series, Springer, 2006,
  \doi{10.1007/3-540-29953-X}.

\bibitem[{Fomin et~al.(2019)Fomin, Lokshtanov, Saurabh, and Zehavi}]{FLSZ19}
F.~V. Fomin, D.~Lokshtanov, S.~Saurabh, M.~Zehavi, Kernelization, Cambridge
  University Press, Cambridge, England, 2019, \doi{10.1017/9781107415157}.

\bibitem[{Froese et~al.(2016)Froese, van Bevern, Niedermeier, and
  Sorge}]{FBNS16}
V.~Froese, R.~van Bevern, R.~Niedermeier, M.~Sorge, Exploiting hidden structure
  in selecting dimensions that distinguish vectors, Journal of Computer and
  System Sciences 82 (2016) 521--535, \doi{10.1016/j.jcss.2015.11.011}.

\bibitem[{Hüffner et~al.(2010)Hüffner, Komusiewicz, Moser, and
  Niedermeier}]{HKMN10}
F.~Hüffner, C.~Komusiewicz, H.~Moser, R.~Niedermeier, Fixed-parameter
  algorithms for cluster vertex deletion, Theory of Computing Systems 47 (2010)
  196--217, \doi{10.1007/s00224-008-9150-x}.

\bibitem[{Karp(1972)}]{Kar72}
R.~M. Karp, Reducibility among combinatorial problems, in: R.~E. Miller, J.~W.
  Thatcher, J.~D. Bohlinger (Eds.), Complexity of Computer Computations, The
  {IBM} Research Symposia Series, Springer, 1972, pp. 85--103,
  \doi{10.1007/978-1-4684-2001-2_9}.

\bibitem[{Kratsch(2012)}]{Kra12}
S.~Kratsch, Polynomial kernelizations for {MIN F$^+\Pi_1$} and {MAX~NP},
  Algorithmica 63 (2012) 532--550, \doi{10.1007/s00453-011-9559-5}.

\bibitem[{Mellor et~al.(2010)Mellor, Prieto, Mathieson, and Moscato}]{MPMM10}
D.~Mellor, E.~Prieto, L.~Mathieson, P.~Moscato, A kernelisation approach for
  multiple $d$-hitting set and its application in optimal multi-drug
  therapeutic combinations, PLOS ONE 5 (2010) e0013055,
  \doi{10.1371/journal.pone.0013055}.

\bibitem[{Moreno-Centeno and Karp(2013)}]{MK13}
E.~Moreno-Centeno, R.~M. Karp, The implicit hitting set approach to solve
  combinatorial optimization problems with an application to multigenome
  alignment, Operations Research 61 (2013) 453--468,
  \doi{10.1287/opre.1120.1139}.

\bibitem[{Moser(2010)}]{Mos10}
H.~Moser, Finding Optimal Solutions for Covering and Matching Problems,
  Cuvillier, Göttingen, Germany, 2010,
  \urlprefix\url{https://www.hmoser.info/diss_moser.pdf}.

\bibitem[{Niedermeier and Rossmanith(2000)}]{NR00}
R.~Niedermeier, P.~Rossmanith, A general method to speed up
  fixed-parameter-tractable algorithms, Information Processing Letters 73
  (2000) 125--129, \doi{10.1016/S0020-0190(00)00004-1}.

\bibitem[{Niedermeier and Rossmanith(2003)}]{NR03}
R.~Niedermeier, P.~Rossmanith, An efficient fixed-parameter algorithm for
  {3-Hitting Set}, Journal of Discrete Algorithms 1 (2003) 89--102,
  \doi{10.1016/S1570-8667(03)00009-1}.

\bibitem[{O'Callahan and Choi(2003)}]{OC03}
R.~O'Callahan, J.-D. Choi, Hybrid dynamic data race detection, in:
  R.~Eigenmann, M.~Rinard (Eds.), PPoPP'03, ACM, 2003, pp. 167--178,
  \doi{10.1145/781498.781528}.

\bibitem[{Rahmann et~al.(2007)Rahmann, Wittkop, Baumbach, Martin, Truss, and
  Böcker}]{RWT+07}
S.~Rahmann, T.~Wittkop, J.~Baumbach, M.~Martin, A.~Truss, S.~Böcker, Exact and
  heuristic algorithms for weighted cluster editing, in: P.~Markstein, Y.~Xu
  (Eds.), Computational Systems Bioinformatics, World Scientific Publishing,
  volume~6, 2007, pp. 391--401, \doi{10.1142/9781860948732_0040}.

\bibitem[{Reiter(1987)}]{Rei87}
R.~Reiter, A theory of diagnosis from first principles, Artificial Intelligence
  32 (1987) 57--95, \doi{10.1016/0004-3702(87)90062-2}.

\bibitem[{Smirnov(2018)}]{Smi18}
P.~V. Smirnov, Reduktsiya dannykh dlya zadachi o vershinnom pokrytii gipergrafa
  za lineinoe vremya s lineinoi pamyat'yu, Bachelor's thesis, Novosibirsk State
  University, Novosibirsk, Russian Federation, 2018,
  \urlprefix\url{http://rvb.su/pdf/Smi18.pdf}.

\bibitem[{Sorge et~al.(2014)Sorge, Moser, Niedermeier, and Weller}]{SMNW14}
M.~Sorge, H.~Moser, R.~Niedermeier, M.~Weller, Exploiting a hypergraph model
  for finding {Golomb} rulers, Acta Informatica 51 (2014) 449--471,
  \doi{10.1007/s00236-014-0202-1}.

\bibitem[{Vazquez(2009)}]{Vaz09}
A.~Vazquez, Optimal drug combinations and minimal hitting sets, {BMC} Systems
  Biology 3 (2009) 81, \doi{10.1186/1752-0509-3-81}.

\bibitem[{Weihe(1998)}]{Wei98}
K.~Weihe, Covering trains by stations or the power of data reduction, in:
  R.~Battiti, A.~A. Bertossi (Eds.), Proceedings of Algorithms and Experiments
  (ALEX 1998), 1998, pp. 1--8.

\end{thebibliography}
